\documentclass[11pt]{article}
\usepackage{fullpage}
\usepackage{amsmath}
\usepackage{graphicx}%
\usepackage{amsfonts}%
\usepackage{amssymb}
\usepackage{url}
\usepackage{fancyhdr}
\usepackage{isomath}
\usepackage{amsmath}
\usepackage{amsbsy}
\usepackage{amssymb}
\usepackage{amscd}
\usepackage{amsfonts}
\usepackage{graphicx}
\usepackage{verbatim}
\usepackage{euscript}
\usepackage{alltt}
\usepackage{stmaryrd}
\usepackage{float}
\usepackage{appendix}
\usepackage{caption}
\usepackage[english]{babel}
\usepackage{relsize}

\usepackage{subfigure}
\usepackage{tikz}
\usetikzlibrary{shapes,arrows}

\usepackage{amsmath}
\usepackage{amsbsy}
\usepackage{amssymb}
\usepackage{amscd}
\usepackage{amsfonts}

\newcommand{\bfa}{{\mathbold a}}
\newcommand{\bfb}{{\mathbold b}}
\newcommand{\bfc}{{\mathbold c}}

\newcommand{\bfe}{{\mathbold e}}
\newcommand{\bff}{{\mathbold f}}
\newcommand{\bfg}{{\mathbold g}}

\newcommand{\bfi}{{\mathbold i}}

\newcommand{\bfl}{{\mathbold l}}

\newcommand{\bfn}{{\mathbold n}}

\newcommand{\bfp}{{\mathbold p}}
\newcommand{\bfq}{{\mathbold q}}
\newcommand{\bfr}{{\mathbold r}}
\newcommand{\bfs}{{\mathbold s}}
\newcommand{\bft}{{\mathbold t}}
\newcommand{\bfu}{{\mathbold u}}
\newcommand{\bfv}{{\mathbold v}}
\newcommand{\bfw}{{\mathbold w}}
\newcommand{\bfx}{{\mathbold x}}

\newcommand{\bfA}{{\mathbold A}}
\newcommand{\bfB}{{\mathbold B}}
\newcommand{\bfC}{{\mathbold C}}
\newcommand{\bfD}{{\mathbold D}}
\newcommand{\bfE}{{\mathbold E}}
\newcommand{\bfF}{{\mathbold F}}
\newcommand{\bfG}{{\mathbold G}}
\newcommand{\bfH}{{\mathbold H}}
\newcommand{\bfI}{{\mathbold I}}

\newcommand{\bfR}{{\mathbold R}}
\newcommand{\bfS}{{\mathbold S}}
\newcommand{\bfT}{{\mathbold T}}
\newcommand{\bfU}{{\mathbold U}}

\newcommand{\bfW}{{\mathbold W}}
\newcommand{\bfX}{{\mathbold X}}
\newcommand{\bfY}{{\mathbold Y}}
\newcommand{\bfZ}{{\mathbold Z}}

\newcommand{\chihat}{\hat{\bfchi}}
\newcommand{\chihatl}{\hat{\chi}}

\newcommand{\beq}{\begin{equation}}
\newcommand{\eeq}{\end{equation}}
\newcommand{\beqs}{\begin{eqnarray}}
\newcommand{\eeqs}{\end{eqnarray}}
\newcommand{\beql}{\begin{equation} \label}

\newcommand{\bfchi}{\mathbold{\chi}}
\newcommand{\bfdelta}{\mathbold{\delta}}
\newcommand{\bfsigma}{\mathbold{\sigma}}
\newcommand{\bfepsilon}{\mathbold{\epsilon}}
\newcommand{\bfnu}{\mathbold{\nu}}
\newcommand{\bfalpha}{\mathbold{\alpha}}
\newcommand{\bfomega}{\mathbold{\omega}}

\newcommand{\bfvphi}{\mathbold{\varphi}}

\newcommand{\bftheta}{\mathbold{\theta}}

\newcommand{\bfPi}{\mathbold{\Pi}}

\newcommand{\bfOmega}{\mathbold{\Omega}}

\newcommand{\bfDelta}{\mathbold{\Delta}}

\newcommand{\bfkappa}{\mathbold{\kappa}}
\newcommand{\mbbW}{\mathbb{W}}

\newcommand{\grad}{\mathop{\rm grad}\nolimits}
\newcommand{\divergence}{\mathop{\rm div}\nolimits}
\newcommand{\curl}{\mathop{\rm curl}\nolimits}

\usepackage[margin=1in]{geometry}
\usepackage{titlesec}

\begin{document}

\title{Finite element approximation of the fields of bulk and interfacial line defects}
\author{Chiqun Zhang$^{1}$, Amit Acharya$^{1}$, Saurabh Puri$^{2}$\\
$^{1}$Carnegie Mellon University, Pittsburgh, PA 15213\\ $^{2}$Microstructure Engineering, Portland, OR 97208 \\ }

\date{}
\maketitle


\begin{abstract}

\noindent A generalized disclination (g.disclination) theory \cite{acharya2015continuum} has been recently introduced that goes beyond treating standard translational and rotational Volterra defects in a continuously distributed defects approach; it is capable of treating the kinematics and dynamics of terminating lines of elastic strain and rotation discontinuities.  In this work, a numerical method is developed to solve for the stress and distortion fields of g.disclination systems. Problems of small and finite deformation theory are considered. The fields of a single disclination, a single dislocation treated as a disclination dipole, a tilt grain boundary, a misfitting grain boundary with disconnections, a through twin boundary, a terminating twin boundary, a through grain boundary, a star disclination/penta-twin, a disclination loop (with twist and wedge segments), and a plate, a lenticular, and a needle inclusion are approximated. It is demonstrated that while the far-field topological identity of a dislocation of appropriate strength and a disclination-dipole plus a slip dislocation comprising a disconnection are the same, the latter microstructure is energetically favorable. This underscores the complementary importance of \emph{all} of topology, geometry, and energetics in understanding defect mechanics. It is established that finite element approximations of fields of interfacial and bulk line defects can be achieved in a systematic and routine manner, thus contributing to the study of intricate defect microstructures in the scientific understanding and predictive design of materials. Our work also represents one systematic way of studying the interaction of (g.)disclinations and dislocations as topological defects, a subject of considerable subtlety and conceptual importance \cite{mermin1979topological, kamien_smectics_2017}.

\end{abstract}
     
\section{Introduction} \label{sec:introduction}

In the context of continuum mechanics, the distortion measure is similar to a deformation or a displacement gradient, except such a measure is not the gradient of a vector field in many situations involving material defects. Such a situation arises when the distortion represents, through a non-singular field, the `gradient' of a field that contains a terminating discontinuity on a surface.  If the discontinuity is in the displacement field, the terminating curve is called a dislocation; if the discontinuity is in the rotation field, the terminating curve is called a disclination. In some cases, the discontinuity can arise in the strain field as well, as for instance in the solid-to-solid phase transformation between austenite and martensite. In \cite{acharya2012coupled, acharya2015continuum}, the concept of the disclination is extended to the generalized disclination (g.disclination) to deal with general distortion-discontinuity problems. The g.disclination can be thought of as a discontinuity (along a curve or loop) of a distortion discontinuity (along a surface).

The strain and stress fields of dislocations and disclinations in a linear elastic isotropic body have been studied in \cite{nabarro1985development,Nabarro1967,dewit1973theory}. However, in classical linear elasticity, the stress and strain fields for these defects have singularities at the defect cores, often predicting infinite energies for finite bodies. In \cite{acharya2012coupled, acharya2015continuum}, a continuum model is introduced for the g.disclination static equilibrium as well as dynamic behaviors, where the singularities are well-handled. The Weingarten theorem for g.disclinations established in \cite{acharya2015continuum} is characterized further in \cite{zhang_acharya_2016}, with the derivation of explicit formula for important topological properties of canonical g.disclination configurations. Relationships between the representations of the dislocation, disclination, and the g.disclination from the Weingarten point of view and in g.disclination theory are established therein. Concrete connections are also established between g.disclinations as mathematical objects and the physical ideas of interfacial and bulk line defects like defected grain and phase boundaries, dislocations, and disclinations. The papers \cite{acharya2012coupled,acharya2015continuum, zhang_acharya_2016} explain the theoretical and physical basis for the results obtained in the present work.

This paper focuses on the applications of the g.disclination model through computation. The goal is to show that the g.disclination model is capable of solving various material-defect problems, within both the small and finite deformation settings. Finite element schemes to solve for the stress and energy density fields of g.disclination distributions are proposed, implemented, and verified for the small and finite deformation settings, for a `canonical' class of defect configurations (mentioned in the abstract).

The paper is organized as follows. Section \ref{sec:notation} contains notation and terminology. In Section \ref{sec:overview}, we briefly review elements of g.disclination theory from \cite{acharya2012coupled, acharya2015continuum} that provide the governing equations for this work, rationalize a procedure for defining a g.disclination as data for computation of stress fields, and discuss the stress field of a disclination viewed as an Eshelby cut and weld problem. Section \ref{sec:numerical} proposes numerical schemes based on the Galerkin and Least Squares Finite Element methods to solve for the fields of g.disclinations at small and finite deformations. Section \ref{sec:application} contains results pertaining to twelve illustrative problems (with sub-cases), all modeled by appropriate combinations of g.disclinations, eigenwall fields, and dislocations as data. Section \ref{sec:layer_core} makes contact between the g.disclination model and classical disclination theory of DeWit \cite{dewit1973theory}, under appropriate restriction on specified data. It is also shown here that for identical specified data, g.disclination theory predicts essentially the entire elastic distortion uniquely, while the classical theory uniquely predicts only the elastic strain field, a particularly clear distinction for the special case of both models in which the data specified is only a dislocation density field. Section \ref{sec:conclusion} contains concluding remarks.

\section{Notation and terminology} \label{sec:notation}

The condition that $a$ is defined to be $b$ is indicated by the statement $a := b$. 
The Einstein summation convention is implied unless specified otherwise. $\bfA \bfb$ is denoted as the action of a tensor $\bfA$ on
a vector $\bfb$, producing a vector. A $\cdot$
represents the inner product of two vectors; the symbol $\bfA\bfD$
represents tensor multiplication of the second-order tensors
$\bfA$ and $\bfD$. A third-order tensor is treated as a linear
transformation on vectors to second-order tensors. 

The symbol $div$ represents the divergence, $grad$ represents the
gradient. In this paper, all tensor or vector indices are written with respect to the basis  
$\bfe_i$, $\bfi$=1 to 3, of a rectangular Cartesian coordinate system, unless stated otherwise. In component form, 
\begin{equation*}
 \begin{split}
    \left(\bfA\times\bfv\right)_{im} &= e_{mjk} A_{ij} v_{k}    \\
    \left(\bfB\times\bfv\right)_{irm} &= e_{mjk} B_{irj} v_{k}  \\
    \left(\divergence \bfA\right)_{i} &= A_{ij,j}                \\
    \left(\divergence \bfB\right)_{ij} &= B_{ijk,k}              \\
    \left(\curl\bfA\right)_{im} &= e_{mjk} A_{ik,j}             \\
    \left(\curl\bfB\right)_{irm} &= e_{mjk} B_{irk,j}               \\
 \end{split}
\end{equation*}
where $e_{mjk}$ is a component of the alternating tensor $\bfX$.

The following list describes some of the mathematical symbols we use in this work:

$\bfU^e$: the elastic strain tensor ($2^{nd}$-order).

$\bfF^e$: the elastic distortion tensor. In small deformation, $\bfF^e = \bfI + \bfU^e$ ($2^{nd}$-order).

$\bfW$: the inverse-elastic (i-elastic) 1-distortion tensor. $\bfW = \left(\bfF^e\right)^{-1}$ ($2^{nd}$-order).

$\hat{\bfF}^e$: the closest-well elastic distortion tensor ($2^{nd}$-order).  

$\hat{\bfW}$: the closest-well-inverse-elastic (cwi-elastic) 1-distortion tensor. $\hat{\bfW•} = \left(\hat{\bfF}^e\right)^{-1}$ ($2^{nd}$-order).

$\bfS$: the eigenwall tensor ($3^{rd}$-order).

$\bfY$: the i-elastic 2-distortion tensor ($3^{rd}$-order). 

$\bfalpha$: the dislocation density tensor ($2^{nd}$-order).

$\bfPi$: the g.disclination density tensor ($3^{rd}$-order).

The normalized difference between two stress fields $\bfsigma_A$ and $\bfsigma_B$ is denoted as $\delta\sigma_{A,B}$, defined as 
\begin{equation} \label{eqn:difference}
\delta\sigma_{A,B} = \frac{|\bfsigma_A -\bfsigma_B|}{|\bfsigma_A|},
\end{equation}
where $|\cdot|$ represents the $l^2$-norm of a matrix. The mean of $\delta\sigma_{A,B}$ is defined as the volume average of the field $\delta\sigma_{A,B}$ over the entire body. \emph{Note that, by definition, whenever such comparisons are presented, they represent differences between the tensors involved and not that of any specific components.} 


\section{Elements of g.disclination theory}\label{sec:overview}

We recapitulate the basic theory for g.disclination statics from \cite{acharya2012coupled, acharya2015continuum} for the sake of completeness and provide the arguments for defining individual g.disclination cores for work in subsequent sections.

\begin{figure}
\centering
\includegraphics[width=0.4\textwidth]{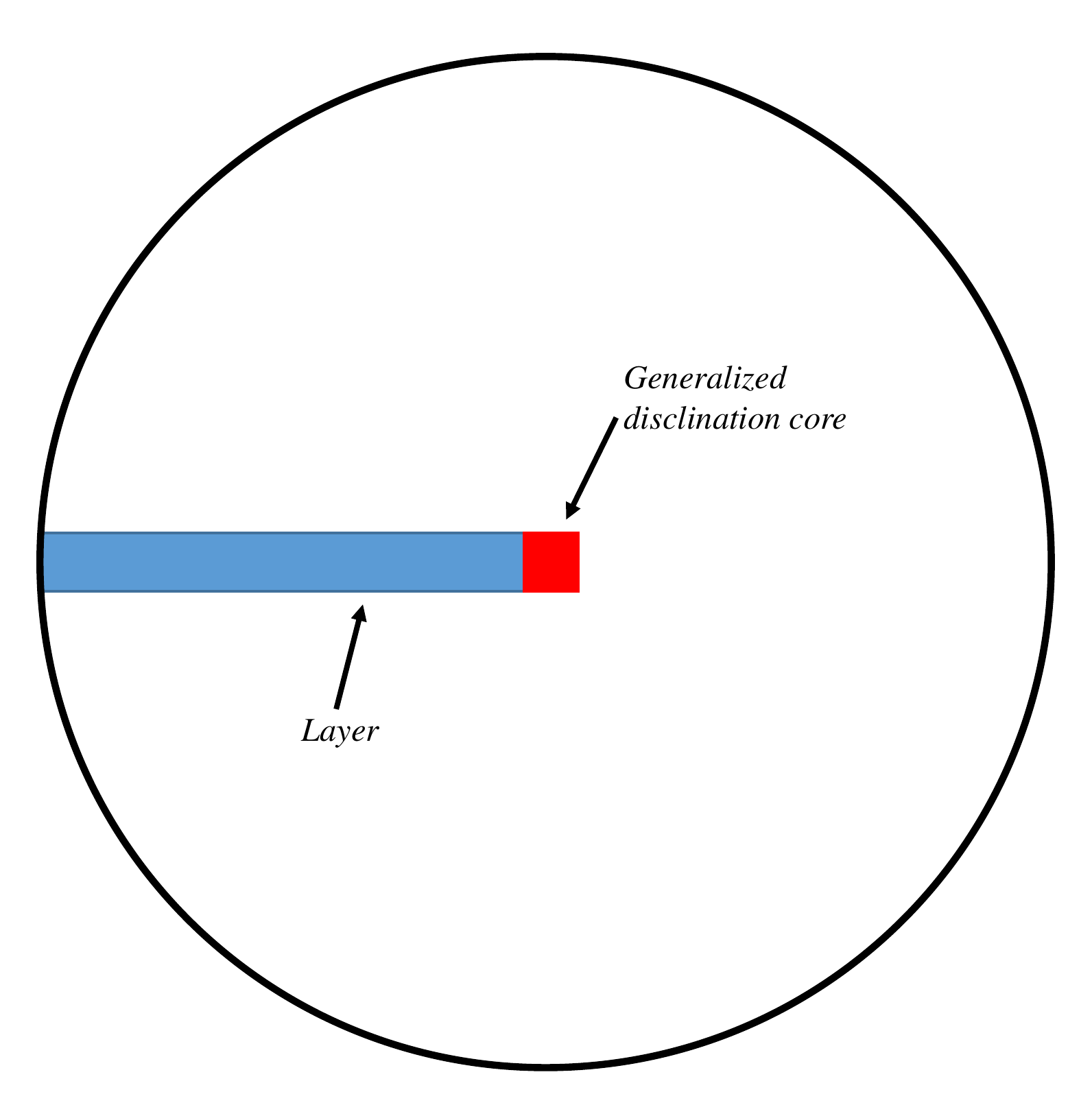}
\caption{Physical regularization of a classical terminating discontinuity of a vector/tensor field. Treat its distortion discontinuity as a field localized inside the layer.}
\label{fig:g_theory}
\end{figure}

Developed as a generalization of eigenstrain theory of Kr\"{o}ner, Mura, and deWit, an individual g.disclination is a curve that terminates a discontinuity of elastic distortion on a surface. The distortion discontinuity is modeled by a field with support within a layer \cite{acharya2015continuum}, as shown in Figure \ref{fig:g_theory}. The termination is considered as continuous over the core within the layer. The core is the support of the g.disclination density field. The strength of an individual g.disclination is simply the difference of the distortions forming the distortion discontinuity terminated by it. One way of setting up the 3-order g.disclination density tensor is to assign the tensor product of the strength tensor and the tangent direction vector of the g.disclination curve as a uniformly distributed field within the g.disclination core, and zero outside it - further details are provided below in (\ref{eqn:density_initial})-(\ref{eqn:Dij}).

The fundamental kinematic decomposition of g.disclination theory is
\begin{equation}\label{eqn:Y-decomp}
\bfY = \grad\, \bfW + \bfS,
\end{equation}
where $\bfW$ is the i-elastic 1-distortion and $\bfS$ is the eigenwall field.

With this decomposition of $\bfY$, a natural measure of the g.disclination density is
\begin{equation}\label{Pi-measure}
\curl\left( \bfY - \grad \bfW \right) = \curl \bfS =: \bfPi,
\end{equation}
since it characterizes the closure failure of integrating $\bfY$ on closed contours in the body:
\begin{equation*}
\int_A \bfPi \bfn da = \int_C \bfY d\bfx,
\end{equation*}
where $A$ is any area patch with closed boundary contour $C$. Physically, it is to be interpreted as a density of lines (threading areas) in the current configuration, carrying a tensorial attribute that reflects a jump in the values of $\bfW$ across the layer representing a phase/grain boundary.

The dislocation density is defined as
\begin{equation}
\bfalpha := \bfY : \bfX = \left( \bfS + \grad\bfW \right) : \bfX.
\label{eqn:dislocation}
\end{equation}
When there is no discontinuity of elastic distortion across a layer, namely $\bfS = \bf0$, (\ref{eqn:dislocation}) becomes $\bfalpha = - \curl\bfW$, since $\curl \bfA = -\grad \bfA :\bfX$ for any smooth tensor field $\bfA$. We utilize a Stokes-Helmholtz-like orthogonal decomposition of the field $\bfS$ into compatible and incompatible parts, 
\begin{equation}\label{eqn:s_decompose}
\bfS = \bfS^{\perp} + \grad\bfZ^{s}.
\end{equation}

For the problems of g.disclination statics considered in this paper, $\bfalpha$ and either $\bfPi$ or $\bfS$ need to be prescribed as data. In the case where $\bfalpha$ and $\bfPi$ are prescribed, we take $\bfZ^s= -\bfI$ and $\bfS = \bfS^{\perp}$ with $\bfS^{\perp}$ determined by the system
\begin{eqnarray}\label{eqn:S_decomposition}
\begin{aligned}
\curl \bfS^{\perp} &= \bfPi \\
\divergence\bfS^{\perp} &= \textbf{0} \\  
\mbox{with} \ \ \bfS^{\perp}\bfn &= \textbf{0} \ \ \mbox{on boundary of the body},
\end{aligned}
\end{eqnarray}
which guarantees that the field $\bfS^\perp$ is vanishing if and only if $\bfPi = \bf0$.

Defining a new field $\bfH^s$ as the deviation of $-\bfZ^s$ from the identity so that 
\begin{equation}\label{eqn:Hs}
\bfH^s := -(\bfZ^s + \bfI) \ \ \mbox{and} \ \ \bfS = \bfS^{\perp} - \grad \bfH^s,
\end{equation}
when $\bfalpha$ and $\bfS$ are prescribed, $\bfZ^s$ is determined from
\begin{equation}\label{eqn:zs}
\divergence(\bfS) = \divergence(\grad\bfZ^s) = -\divergence(\grad\bfH^s)
\end{equation}
with the value of $\bfH^s = \bf0$ at a single point of the body. 

Then, given $\bfalpha$, $\bfS$, and $\bfH^s$, the i-elastic distortion field $\bfW$ is determined from the system
\begin{eqnarray}\label{eqn:summary}
\begin{aligned}
\bfalpha &= \left( \bfS + \grad\bfW \right) : \bfX \\ 
\hat{\bfW} &= \bfW - \bfH^s \\
\divergence [\bfT(\hat{\bfW})] &= \textbf{0}\\
\bfT \bfn & = \bft \ \ \mbox{on the boundary},
\end{aligned}
\end{eqnarray}
where $\bfT$ (symmetric) is the stress field depending on $\hat{\bfW}$ (and the unstressed elastic reference from which $\hat{\bfW}$ is measured). $\bft$ is a prescribed, statically consistent traction field on the boundary of the body. \emph{For all computations in this paper we will assume $\bft = \bf0$, unless otherwise specified}, but this implies no loss of essential generality in the formulation or in the computational work.

We view the i-elastic distortion $\bfW(\bfx)$ as a mapping between a local configuration, around the generic point $\bfx$ in the generally stressed configuration, and a fixed (over all $\bfx$) local stress-free configuration; how the local configuration around each point $\bfx$ of the current configuration is to be understood, at least in principle, is described in Appendix \ref{sec:append_3}. In our model there is some freedom in making the choice of the fixed local stress-free configuration; for instance, it may be associated with the stress-free state of a particular phase of the material, e.g. the high-temperature/symmetry austenite phase. In this paper, we associate it with the stress-free local configuration of a particular point in the body (that would represent one of the phases of the material, say a martensite variant); the point is the one where $\bfH^s$ is specified (see the discussion surrounding \eqref{eqn:zs}). The cwi-elastic distortion, $\hat{\bfW}(\bfx)$, on the other hand represents the mapping between a local configuration around the generic point $\bfx$ in the stressed configuration and the unstressed configuration it would attain when (conceptually) released from all loads on it. The motivation and detailed discussions for the dependency of $\bfT$ on $\hat{\bfW}$ are presented in Sections \ref{sec:twin_grain} and \ref{sec:layer_core}. An example for developing intuition for some qualitative differences between these fields in the context of a through and terminating twin boundary is also provided in Appendix \ref{sec:append_3}.

We obtain the governing equations for the small deformation case by defining the tensors $\bfU^e$ and $\hat{\bfU}^e$ through the approximations $\bfW = \bfI - \bfU^e$ and $\hat{\bfU}^e := \bfI - \hat{\bfW} = \bfU^e+\bfH^s$ with $\bfT = \bfC: \hat{\bfU^e}$. Substituting in \eqref{eqn:summary} and using \eqref{eqn:Hs} we have
\begin{eqnarray}\label{eqn:small_deformation_summary_hat}
\begin{aligned}
\curl \hat{\bfU^e} &=\bfalpha - \bfS^{\perp} : \bfX \\ 
\divergence [\bfC:\hat{\bfU}^e] &= \textbf{0}\\
[\bfC:\hat{\bfU}^e] \bfn & = \bft \ \ \mbox{on the boundary}.
\end{aligned}
\end{eqnarray}
with $\bfS^\perp$ satisfying \eqref{eqn:S_decomposition}. We refer to the symmetric part of $\hat{\bfU}^e$, $\hat{\bfU}^e_{sym} =: \hat{\bfepsilon}^e$, as the closest-well elastic strain and the skew-symmetric part, $\hat{\bfU}^e_{skw} =: \hat{\bfOmega}^e$, as the closest-well elastic rotation tensor. We similarly define the elastic strain $\bfepsilon^e$ and the elastic rotation $\bfOmega^e$ tensor fields from $\bfU^e$.

It is important to note that if the defect fields $\bfPi$, $\bfalpha$, and $\bfZ^s$ transform to $\bfR \bfPi$, $\bfR \bfalpha$, and $\bfR \, \bfZ^s$ for a spatially constant (on the current configuration), rotation field $\bfR$ representing a change in the point-wise unstressed elastic reference, then the solution $\bfW$ to (\ref{eqn:summary})$_1$ transforms as $\bfR{\bfW}$ and hence $\hat{\bfW}$ transforms as $\bfR \hat{\bfW}$. The corresponding closest-well elastic distortion field is $\hat{\bfF}^{e} \bfR^T$ measured from the point-wise rotated, closest-well, unstressed reference. Since elastic constitutive equations for stress from two different  reference configurations, say $1$ and $2$, necessarily have the property that $\bfT^{(2)}\left(\bfF^{(2)}\right) = \bfT^{(1)} \left(\bfF^{(1)}\right)$, where $\bfF^{(2)} \bfG = \bfF^{(1)}$ and $\bfG$ is the invertible tensor mapping reference $1$ to $2$ (pointwise), we have $\bfT^{(2)}\left(\bfF^{(2)}\right) = \bfT^{(1)}\left(\bfF^{(2)} \bfG \right)$  $\forall$ invertible $\bfF^{(2)}$, and this implies that, for  $\bfG = \bfR$ and $2$ representing the rotated unstressed reference, $\bfT^{(2)}\left(\hat{\bfW}^{-1} \bfR^T \right) = \bfT^{(1)} \left(\hat{\bfW}^{-1} \right) =: \bfT(\hat{\bfW})$ and therefore the stress prediction on the current configuration from (\ref{eqn:summary}) is invariant to the choice of unstressed elastic reference.

We will assume $\bfZ^s = -\bfI$ for many problems considered in this paper where $\bfalpha$ and $\bfPi$ are prescribed as data. Sections \ref{sec:twin_grain}-\ref{sec:layer_core} are exceptions where $\bfalpha$ and $\bfS$ are specified. Our model ensures that, at least with respect to the $L^2$-norm on the space of third-order tensor fields, the stresses generated are only in response to the prescribed g.disclination (and dislocation) density fields, with no other sources involved. It also allows the realistic representation of terminating grain/phase boundaries with an eigenwall field $\bfS$ specified in a layer as in Fig. \ref{fig:g_theory}, with the concomitant recovery of classical results of defect theory related to dislocation and disclination stress fields. The use of the field $\bfZ^s$ ($\bfH^s$) is essential for this purpose, as it is impossible to represent a through or terminating grain/phase boundary interface by setting $\bfS = \bfS^\perp$, with $\bfS^\perp$ determined from the g.disclination density field (possibly vanishing). Details of these situations are discussed in Sections \ref{sec:twin_grain} and \ref{sec:layer_core}.

\subsection{Modeling a $\bfPi$ field representing an individual g.disclination core}\label{sec:Pi_layer_def}
\begin{figure}
\centering
\includegraphics[width=0.8\textwidth]{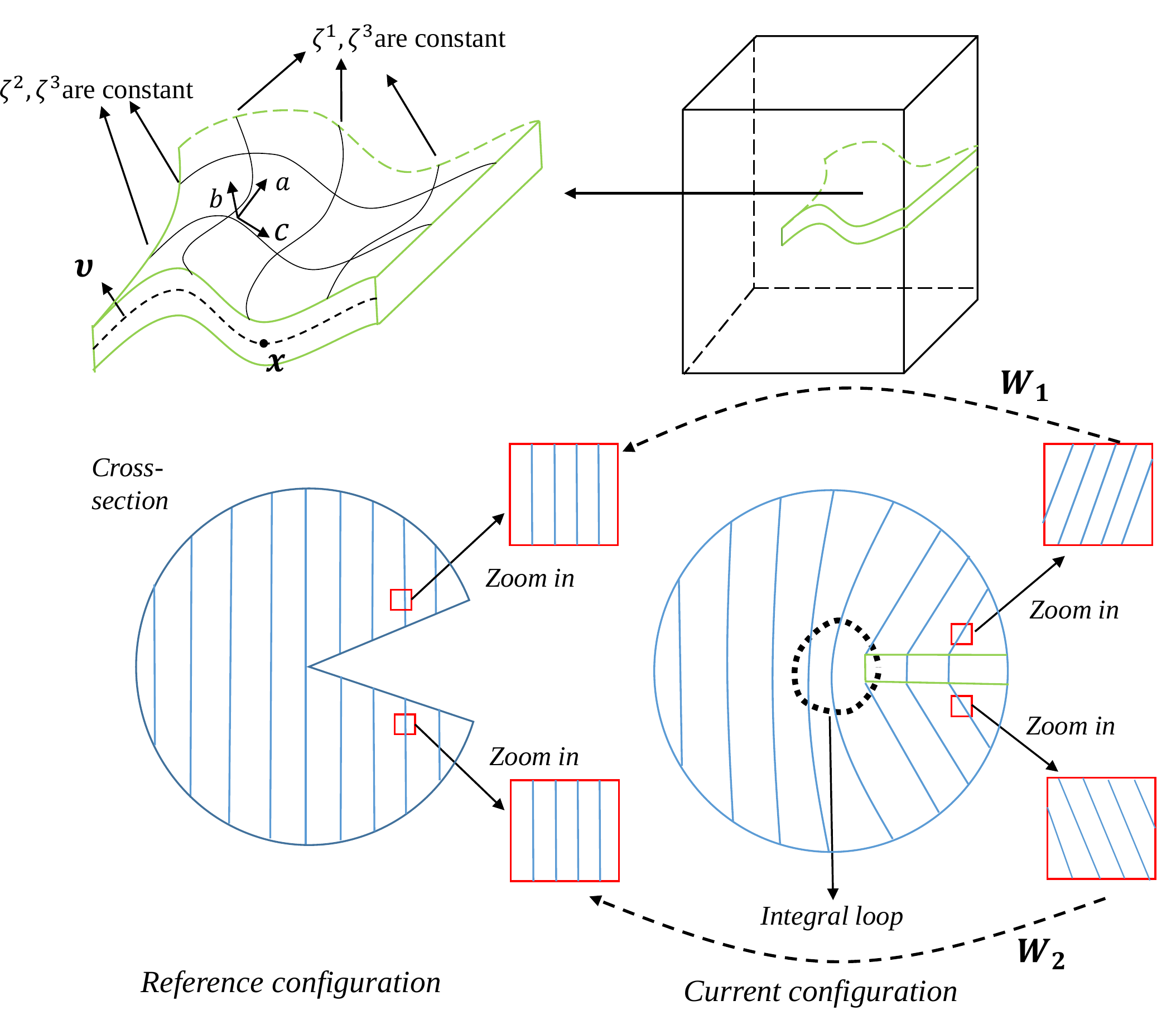}
\caption{The geometric definition of the layer. $\bfa$, $\bfb$ and $\bfc$ are natural basis vectors for a parametrization of the layer by coordinates $\xi^2$, $\xi^3$, $\xi^1$. The two lower sketches conceptualize the formation of a wedge disclination by the closing and welding of the gap in the unstretched reference configuration to form the stressed current configuration.}
\label{fig:disclination_ini}
\end{figure}

The tensor $\bfPi$ for a discrete g.disclination can be defined for prescription as given data as follows. Figure \ref{fig:disclination_ini} shows an eigenwall field $\bfS$ supported in a layer, whose termination represents the g.disclination core. The layer is, in general, `non-planar' and its termination not a straight `line'. We assume the layer to be amenable to the description
\begin{equation}\label{eqn:append_s}
\bfs(\xi^1, \xi^2, \xi^3) = \bfx(\xi^1, \xi^2) + \xi^3\bfnu(\xi^1, \xi^2),
\end{equation}
where $\bfx$ is the `mid-surface' of the layer, parametrized by curvilinear coordinates $(\xi^1, \xi^2)$, and $\bfnu$, the unit normal field to the mid-surface, is defined as 
\begin{equation}\label{eqn:normal}
\bfnu(\xi^1, \xi^2) = \frac{\frac{\partial \bfx}{\partial \xi^1} \times \frac{\partial \bfx}{\partial \xi^2}}{| \frac{\partial \bfx}{\partial \xi^1} \times \frac{\partial \bfx}{\partial \xi^2}|}.
\end{equation}
$\xi^3$ serves as the remaining coordinate parametrizing the 3-dimensional layer. The parametrization (in the vicinity of the g.disclination core) is such that the surface $\xi^1 = 0, -\frac{t}{2} \leq \xi^3 \leq \frac{t}{2}$ coincides with the layer termination within the body, and the surfaces $\xi^3 = \pm \frac{t}{2}$ are the top and bottom surfaces of the layer, respectively. The layer mid-surface (and therefore the $\bfnu$ field), is assumed known (e.g. from observations) for the definition of the $\bfS$ and $\bfPi$ fields in this static setting. Denote the i-elastic distortion field (the inverse rotation field in the disclination case) of the upper part as $\bfW_1$; the i-elastic distortion of the lower part is denoted as $\bfW_2$. The thickness of the layer is $t$ in the normal direction to the layer. The eigenwall field $\bfS$ in the layer is defined as
\begin{equation}\label{eqn:density_initial}
\bfS = a(\xi^1)\dfrac{\left(\bfW_1-\bfW_2\right)}{t}\otimes\bfnu,
\end{equation}
where $a(\xi^1)$ is a scalar function indicating the longitudinal extent of the core of the g.disclination; a candidate we utilize is
\begin{equation}\label{eqn:a_candidate}
a(\xi^1) =
\begin{cases}
0 \qquad \xi^1<0 \\
\frac{1}{c}\xi^1 \qquad 0\le \xi^1<c \\
1 \qquad \xi^1\ge c,
\end{cases}
\end{equation}
with $c$ being the core width. The field $\bfS$ is assumed to vanish outside the layer. In general, $\bfW_1$ and $\bfW_2$ could be spatially varying along the longitudinal directions of the layer, while being always uniform in the transverse direction. Here we assume that $\bfW_1$ and $\bfW_2$, viewed as fields in the layer, are constant (In Section \ref{sec:martensite_micro} we encounter a curved twin boundary of a lenticular inclusion where this is not the case; we comment on this after (\ref{eqn:Dij})).  Then $\bfPi = \curl \bfS$ and is nonzero only in the core, given from Appendix \ref{sec:append_pi} as
\begin{equation}\label{eqn:Pi}
\bfPi = \dfrac{\left(\bfW_1-\bfW_2\right)}{t}\otimes(\grad\,a \times \bfnu) \qquad \text{in the core}.
\end{equation}
As discussed in Appendix \ref{sec:append_pi}, $\bfPi$ has support only in the layer and for a single g.disclination, only in the core. 

We note here that defining $\bfPi$ is essential for many problems where the notion of a g.disclination with a prescribed strength makes sense without the notion of a corresponding physical interface, e.g. a pentagon-heptagon pair in a graphene monolayer, where the strength can be inferred without recourse to a distortion discontinuity. In situations where the axis of a g.disclination core cylinder is a general space curve, the procedure we have outlined above involving a layer field is still useful for defining the corresponding $\bfPi$ field.

The strength of a single disclination defined by $\bfPi$ given in (\ref{eqn:Pi}) is obtained by integrating $\bfPi$ over any area patch $A$ enclosing the core, such as any whose bounding curve is given by the black dashed line in Figure \ref{fig:disclination_ini}:
\begin{equation}\label{eqn:D}
\bfD := \int_{A} \bfPi d\bfa = \int_{core} \bfPi d\bfa,
\end{equation}
and as shown in Appendix \ref{sec:append_pi} this is given by $\bfW_1-\bfW_2$, which also corresponds to the line integral of $\bfY$ on any circuit encircling the core cylinder, since $\curl \bfY = \bfPi$. For a planar layer with $\bfnu = \bfe_2$ and $\xi^1 = x_1$,
\begin{equation}\label{eqn:straight_S}
\bfPi = \frac{\bfW_1 - \bfW_2}{ct} \otimes \bfe_3,
\end{equation}
and choosing the area patch to be one with normal in the $\bfe_3$ direction, we have
\begin{equation}\label{eqn:Dij}
D_{ij} = c t \frac{(W_1-W_2)_{ij}}{c t} = (W_1-W_2)_{ij},
\end{equation}
on the orthonormal basis $(\bfe_i), i = 1,2,3$.

If $(\bfW_1 - \bfW_2)$ is not a constant along the interface, then there is an additional contribution to $\bfPi$, as can be seen from the derivation of (\ref{eqn:append_pi}).

\subsection{Disclinations in small and finite deformation theory}\label{sec:relationship_classical}
\begin{figure}[H]
\centering
\includegraphics[width = 0.5\textwidth]{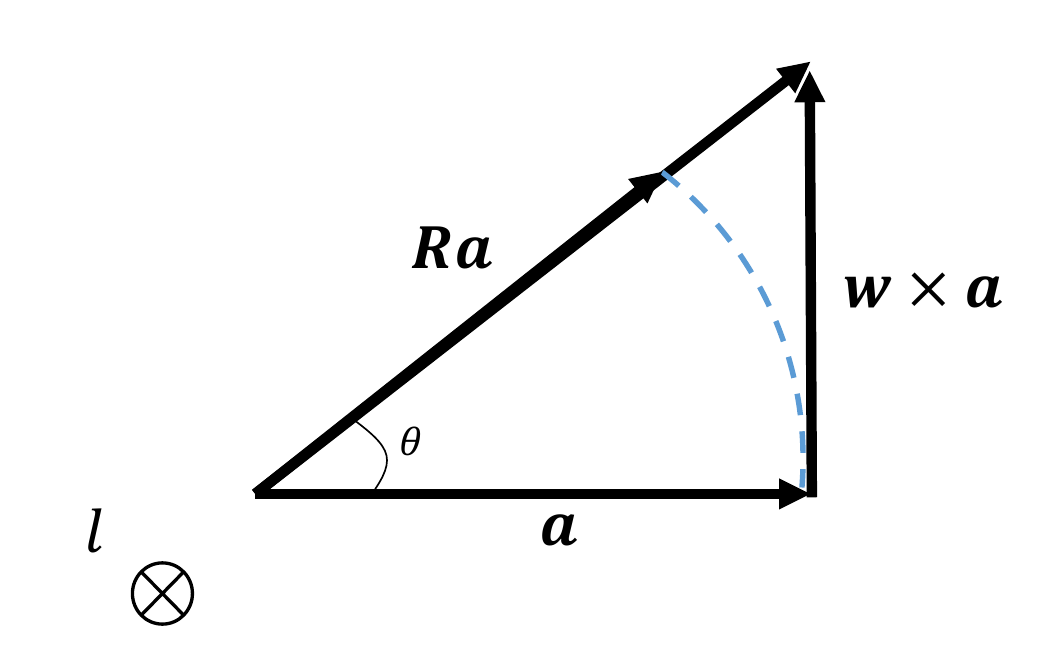}
\caption{The difference in the action of a finite rotation, $\bfR$, and its spin, $\bfw$, with axis $\bfl$ on a vector $\bfa$.}
\label{fig:skew_vector}
\end{figure}

Consider an interface across which $\bfW_1 = \bfR_1$ and $\bfW_2 = \bfR_2$ are rotation tensors. For a given rotation tensor $\bfR$ corresponding to a rotation by an angle $\theta$ about an axis $\bfl$, one associates a skew tensor $\mathbb{W}$, which we shall refer to as the spin of the rotation in this paper, and its axial vector $\bfw$ such that
\[
\bfR \bfa \approx \bfa + \mathbb{W} \bfa = \bfa + \bfw \times \bfa
\]
for all vectors $\bfa$ in the plane normal to $\bfl$ when $\theta$ is small, as shown in Fig. \ref{fig:skew_vector}.
The axial vector $\bfw$ is given by
\[
\bfw = \tan \theta \, \bfl, 
\] 
and it follows that in an orthonormal basis
\[
\mathbb{W}_{ij} = e_{imj}l_m \tan \theta.
\]
For $\bfl = \bfe_3$, the only non-zero components of $\mathbb{W}$ are $\mathbb{W}_{21} = \tan \theta = - \mathbb{W}_{12}$. 

Thus the small deformation approximation of the difference of two rotation tensors $\bfR_1$ and $\bfR_2$ corresponding to angles and axes of rotation $(\theta_1, \bfl_1)$ and $(\theta_2, \bfl_2)$ is given, in the first instance, by  $\mathbb{W}_1 - \mathbb{W}_2$ with components
\[
(\mbbW_1)_{ij} - (\mbbW_2)_{ij} = e_{imj} \left[ (l_1)_m \tan \theta_1 - (l_2)_m \tan \theta_2 \right].
\]

In linear disclination theory \cite{dewit1973theory}, the plastic bend-twist tensor  arises when the skew-symmetric part of the plastic distortion tensor $\mbbW^p$, which we shall refer to here as the plastic spin, exhibits discontinuities such that its gradient field is not well-defined in the whole body as integrable functions. DeWit \cite{dewit1973theory} replaces the gradient of the axial vector of the plastic spin in such circumstances by the \emph{plastic bend-twist} tensor, $\bfkappa^P$, which is not irrotational (i.e. $\curl$-free) in the whole domain to reflect the possibility of the singularities of the plastic spin field, even when $\bfkappa^P$ is smooth. DeWit further defines the \emph{Frank vector} of a closed curve $\partial A$ to be 
\begin{eqnarray*}
\Omega_q = -\int_{\partial A} \kappa_{kq}^P dx_k = - \int_A \epsilon_{pmk}\kappa_{kq,m}^P n_p da,
\end{eqnarray*}
where $A$ is any area patch whose boundary is $\partial A$, and $\bfn$ is the unit normal field on $A$.

For a single disclination, $\bfOmega \neq \bf0$ in the core. Following the arguments in \cite{zhang_acharya_2016},  one can create a non-simply connected domain by excluding the core cylinder/curve from the overall simply-connected body. By making an appropriate cut one can then render the body without the core simply-connected again (but not continuously deformable to the original body with the core). On this cut-induced simply connected domain one can construct a spin field $\mbbW$, the gradient of whose axial vector field matches the given plastic bend-twist field, even though every cut-surface corresponds, in general, to a different spin field. However, for $\bfOmega \neq \bf0$, each such spin field displays a \emph{constant} jump (discontinuity) across its corresponding cut-surface and, moreover, this jump is constant regardless of the spin field (and corresponding cut-surface) involved. Let us denote this constant jump for a single disclination as $\llbracket \mbbW \rrbracket$ and it can be shown, following the arguments in \cite{zhang_acharya_2016}, that 
\begin{equation}\label{eqn:omega}
\Omega_q  = -\frac{1}{2}\epsilon_{lqr}\llbracket \mbbW \rrbracket_{rl}.
\end{equation} 
 
\begin{figure}
\centering
\includegraphics[width = 0.5\textwidth]{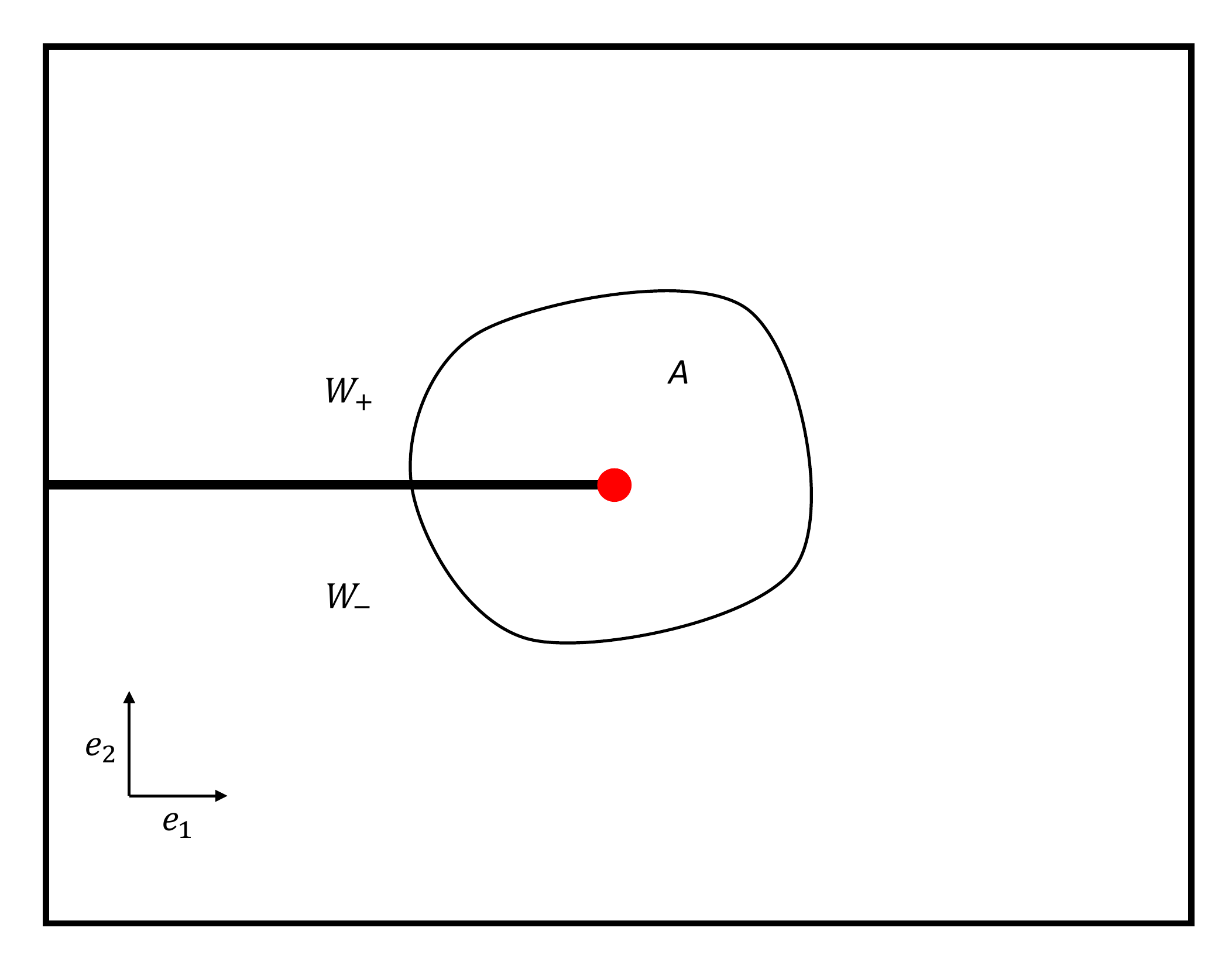}
\caption{A single disclination characterized by the difference in the spin tensors, $\bfW_+$ and $\bfW_-$. `$A$' is an area patch enclosing the core.}
\label{fig:strength_construct}
\end{figure}

As illustration of these concepts, consider a single, straight, disclination through the plane of the paper as shown in Figure \ref{fig:strength_construct}. 
The red point is the disclination core. For the cut-surface shown, $\mbbW_+$ and $\mbbW_-$ represent the limiting values, from the top and bottom respectively, of the constructed spin field $\mbbW$ on the surface and they have the same rotation axis ($\bfe_3$). Assuming the Frank vector is specified as $|\bfOmega| \bfe_3$, (\ref{eqn:omega}) implies
\begin{equation}\label{eqn:strength}
|\bfOmega| = tan\theta_1 - \tan \theta_2 \approx \tan(\theta_1 - \theta_2) \approx \theta_1 - \theta_2, \ \  \mbox{when} \ \ |\theta_1| \ll 1, |\theta_2| \ll 1.
\end{equation}
Thus,  when the angles $\theta_1$ and $\theta_2$ are small, then the magnitude of DeWit's Frank vector may be interpreted as the \emph{misorientation} across any interface terminated by the disclination.

Finally, an observation on stress fields of single disclinations (involving large rotations, in general) is in order. Due to the lack of full rotational invariance of the linear elastic stress constitutive assumption, it is natural to expect large differences between results of small and finite deformation theory for single disclinations with large misorientations. This can be appreciated by noting that if $\bfT$ is the nonlinear elastic stress response function out of some reference configuration
\[
\bfT(\bfF) = \bfT(\bfI) + D\bfT(\bfI) [\bfF-\bfI] + H.O.T,
\]
where $H.O.T$ stands for higher order terms and $D\bfT$ is the  derivative of the stress function, and we assume that $\bfF$ is measured from a stress-free reference. Let $D\bfT(I) = \bfC$, the $4^{th}$-order tensor of elastic moduli (with minor symmetries). Frame-indifference implies that $\bfT(\bfR) = \bf0$ for all rotations $\bfR$. Then it is valid to write
\begin{equation}\label{eqn:lin_stress}
\bf0 = \bfT(\bfR) = \bfC[\bfR - \bfI - \mbbW] + H. O.T.,
\end{equation}
where $\mbbW$ is the spin of $\bfR$. In problems where the elastic distortion field attempts to attain locally large rotations (e.g. the field of a single disclination), it is clear that the linear elastic stress-approximation to such deformations, given by the first term on the rhs of (\ref{eqn:lin_stress}),  degrades as the angle of rotation increases. This is so since the argument involves (spurious) stretching of vectors (see Fig. \ref{fig:skew_vector}) and therefore, strain, and this is sensed by the linear elastic moduli.

\section{Numerical scheme} \label{sec:numerical}

The standard Galerkin method is not adequate for solving the div-curl system (\ref{eqn:S_decomposition}) \cite{jiang1998least}. Instead, we utilize the Least Squares Finite Element Method \cite{jiang1998least} adapting the ideas in \cite{roy2005finite} for calculating fields of line defects in solids. The scheme for solving the entire system (\ref{eqn:summary}) is divided into three steps. 

If $\bfPi$ is prescribed as data, the first step is to solve for the incompatible part $\bfS^{\perp}$ given the g.disclination density field $\bfPi$. If $\bfS$ is prescribed as data, the first step is to solve for the compatible part $\bfZ^s$ given the eigenwall field $\bfS$. The second step is to solve for the i-elastic 1-distortion tensor $\bfW$ from (\ref{eqn:summary}), with $\bfH^s = \bf0$ and $\bfS := \bfS^\perp$ from the first step substituted in (\ref{eqn:summary})$_1$ if $\bfPi$ is data. In the second step, different numerical schemes are applied to solve for force equilibrium (\ref{eqn:summary})$_{3,4}$ depending on whether a `small' or `finite' deformation result is desired. In the following, the symbol $\bfdelta\left(\cdot\right)$  represents a variation associated with the field $\left(\cdot\right)$ in a class of functions.

When $\bfPi$ is prescribed, the equations to be solved for calculating $\bfS^{\perp}$ are 
\begin{eqnarray*}\label{first_part}
\curl \bfS^{\perp} &=& \bfPi \\
\divergence\bfS^{\perp} &=& \textbf{0} \\  
\mbox{with} \ \ \bfS^{\perp}\bfn &=& \textbf{0} \ \ \mbox{on the boundary},  
\end{eqnarray*}
where $\bfPi$ is a given 3rd-order tensor field. In an orthonormal basis, the weak form for the above equations is given by
\begin{equation} \label{eqn:part1_weakform}
\int_V e_{ijk}{\delta}S^{\perp}_{rsk,j}\left(e_{imn}S^{\perp}_{rsn,m} - \pi_{rsi}\right)dv + \int_V{\delta}S^{\perp}_{isj,j}S^{\perp}_{ism,m}dv = 0.
\end{equation}
The essential boundary condition $\bfS^{\perp}\bfn = \bf0$ needs to be imposed. Also, (\ref{eqn:part1_weakform}) should hold for all possible variations $\bfdelta\bfS^{\perp}$ satisfying the essential boundary condition. The variational statement is obtained by looking for critical points of the least squares functional
\begin{equation*}
\int_V \left(\frac{1}{2} \left\Vert \curl \bfS^{\perp} - \bfPi \right\Vert^2 + \frac{1}{2} \left\Vert \divergence \bfS^{\perp} \right \Vert^2\right) dv.
\end{equation*}

When $\bfS$ is prescribed, the equation for calculating $\bfH^s$ is 
\[
- \divergence(\grad \bfH
^s) = \divergence(\bfS),
\]
where $\bfS$ is the prescribed eigenwall field and with $\bfH^s = \bf0$ prescribed at one point of the body. The weak form for the above equation is given by
\begin{equation}\label{eqn:Zs_solve}
\int_V {\delta}H^s_{ij,k}\left(H^s_{ij,k} + S_{ijk}\right)dv =  \textbf{0}.
\end{equation}

Noting that regardless of the prescribed data we now have $\bfS$ and $\bfH^s$ defined by the above rules, the following equations need to be solved in the second step:
\begin{eqnarray*}\label{second_part_all}
\bfA &:=& \bfS : \bfX - \bfalpha \\
\curl \bfW &=& \bfA \\ 
\divergence \left[ \bfT\left(\hat{\bfW}\right)\right] &=& \textbf{0}. 
\end{eqnarray*}
where $\bfT(\hat{\bfW})$ represents the stress response with $\hat{\bfW} = \bfW - \bfH^s$. To solve this system, the small and finite deformation cases are separately dealt with.

\subsection{Small deformation}\label{eqn:small_def}
On writing $\bfW \approx \bfI - \bfU^e$ and expressing $\bfU^e = \bfchi + \grad\, \bff$ and $\bfT = \bfC (\bfU^e +\bfH^s)$, $\bfchi$ is solved from the following equations:
\begin{eqnarray*} 
\curl\bfchi &=& -\bfA \\ 
\divergence\bfchi &=& \textbf{0}\\
\bfchi\bfn &=& \textbf{0} \quad\text{on the boundary},
\end{eqnarray*}
where $\bfn$ is the unit normal vector on the boundary. The weak form of these equations is 
\begin{equation} \label{eqn:part2_small_chi}
\int_V e_{ijk}{\delta}\chi_{rk,j}\left(e_{imn}\chi_{rn,m} +A_{ri}\right)dv + \int_V \delta\chi_{ij,j}\chi_{im,m}dv = 0, 
\end{equation}
with boundary condition $\chi_{ij} n_j = 0$. In the small deformation case, the governing equation for $\bff$ is given by
\begin{equation} 
\divergence\left[\bfC : \left(\grad\bff + \bfchi + \bfH^s \right)\right] = \bf0, \\
\end{equation}
where $\bfC$ is the possibly anisotropic, 4-order tensor of linear elastic moduli. Its corresponding weak form is 

\begin{equation} \label{eqn:part2_small_f}
\int_V {\delta}f_{i,j}\left(C_{ijkl}f_{k.l} + C_{ijkl}\chi_{kl}  + C_{ijkl}H^s_{kl} \right)dv - \int_{\partial V_t} \delta f_{i} t_i da= 0
\end{equation}
where $\partial V_t$ represents the set of point on the boundary where the tractions $t_i$ are specified. Also, the standard essential boundary condition on $\bff$ are implemented to remove the rigid deformation mode. Given the generalized disclination density $\bfPi$ and the dislocation density $\bfalpha$, the discretized weak forms (\ref{eqn:part1_weakform}), (\ref{eqn:part2_small_chi}), and (\ref{eqn:part2_small_f}) yield the static solutions of a g.disclination problem for the small deformation case. When $\bfS$ and $\bfalpha$ are prescribed, (\ref{eqn:Zs_solve}), (\ref{eqn:part2_small_chi}), and (\ref{eqn:part2_small_f}) form the corresponding governing equations.

\subsection{Finite deformation}\label{sec:numerical_finite}
In the finite deformation case one needs to solve $\hat{\bfchi}$ from
\begin{eqnarray} \nonumber
\curl\chihat &=& \bfA \\ \nonumber
\divergence\chihat &=& \textbf{0}\\
\chihat\bfn &=& \textbf{0}  \qquad \text{on the boundary,} \nonumber
\end{eqnarray}
see \cite{acharya2006size}. The corresponding weak form is \cite{puri2009modeling} 
\begin{equation}\label{eqn:chihat_solve}
\int_Be_{ijk}\delta\chihatl_{rk,j}\left(e_{imn}\chihatl_{rn,m} - A_{ri}\right)dv + \int_B\delta\chihatl_{ij,j}\chihatl_{im,m}dv = 0,
\end{equation}
and the boundary condition $\hat{\chi}_{ij}n_{j} = 0$ for all $i=1,2,3$ on the boundary, $\bfn$ being the normal vector on the boundary. In addition, we need to solve the following equations:
\begin{eqnarray} \label{eqn:disc_num_finite}
\begin{aligned} 
\bfW &= \chihat + \grad\hat{\bff} \\ 
\hat{\bfW} &= \bfW - \bfH^s \\
\bfE^e &=  \frac{1}{2}\left(\hat{\bfW}^{-T}\hat{\bfW}^{-1} - \bfI\right) \\ 
\bfT &= \hat{\bfW}^{-1}\left[\bfC : \bfE^e\right]\hat{\bfW}^{-T} \\ 
\divergence\bfT &= \textbf{0}, 
\end{aligned}
\end{eqnarray}
where (\ref{eqn:disc_num_finite})$_3$ represents a St. Venant-Kirchhoff constitutive assumption for the stress, with $\bfC$ being the linear elastic moduli for the material(our basic methodology is, of course, not restricted to this choice). Also, essential boundary conditions on $\hat{\bff}$ are required to eliminate the rigid deformation mode.

Since the governing equation $\divergence\,\bfT = \textbf{0}$ is nonlinear in $\hat{\bff}$, we apply the Newton Raphson method to solve the problem utilizing the scheme in \cite{puri2009modeling}. We find that the initial guess for $\hat{\bff}$ is crucial for success in solving problems of g.disclination theory. One contribution of this work is the development of a systematic strategy for generating this initial guess, as described in the following.

The initial guess for $\hat{\bff}$ is denoted as $\hat{\bff}_0$. A good candidate for $\hat{\bff}_0$ is based on the solution $\bff$ from the small deformation theory. Namely, to obtain $\hat{\bff}_0$, we solve $\bff$ from the small deformation theory equations exactly as given in Section \ref{eqn:small_def}. Then we set
\[
\hat{\bff}_0 = \bfX - \bff \ \ \ \mbox{as the initial guess for}\ \  \hat{\bff}  \ \ \mbox{in the finite deformation theory,}
\]
following the justification in \cite[Sec. 5, p.1707]{acharya2006size}.

With this initial guess for $\hat{\bff_0}$ and the solution for $\chihat$ obtained from solving (\ref{eqn:chihat_solve}), we solve the weak form of (\ref{eqn:disc_num_finite})$_4$ for $\hat{\bff}$. The discrete residual is formed from the variational statement for (\ref{eqn:disc_num_finite})$_4$,
\begin{equation}\label{eqn:finite_res}
\int_B \delta \hat{f}_{i,j} T_{ij} dv = 0,
\end{equation}
and is given by
\begin{equation*}
R^A_i = \int_BT_{ij}\dfrac{\partial {N^A}}{\partial {x_j}}dv,
\end{equation*}
where $N^A$ is the shape function corresponding to the finite element mesh node $A$, and $R^A_i$ is the discrete residual for the $(A,i)$ degree of freedom.

The tangent stiffness for the problem is obtained by taking a variation of the residual (\ref{eqn:finite_res}) in a direction $d\hat{\bff}$; the discrete form of the Jacobian matrix corresponding to the degree-of-freedom pair $\left\{(A,a),(B,b)\right\}$ is 
\begin{equation*}
J^{AB}_{ab} = \int_B \frac{\partial N^A}{\partial x_j} \frac{\partial T_{aj}}{\partial F^e_{mn}}\frac{\partial F^e_{mn}}{\partial \hat{W}_{ru}}\frac{\partial  \hat{W}_{ru} }{\partial (\grad \hat{f})_{bc}}\frac{\partial N^B}{\partial x_c} dv.
\end{equation*}

To summarize, the algorithm for the finite deformation scheme is 

\begin{itemize}
\item Make a guess for $\hat{\bff}_0$. $\hat{\bff}_0$ is based on the solution $\bff$ from small deformation theory, given as $\hat{\bff}_0 = \bfX-\bff$.
\item Solve for $\hat{\bfchi}$.
\item Solve for $\hat{\bff}$ using the equilibrium equation, $\divergence \,\bfT = \textbf{0}$. This equation is nonlinear, and solved using the Newton-Raphson method.
\item Obtain $\hat{\bfW}= \hat{\bfchi}+\grad \hat{\bff} - \bfH^s$; $\bfE^e=\frac{1}{2}(\hat{\bfW}^{-T}\hat{\bfW}^{-1}-\bfI)$; $\bfT = \hat{\bfW}^{-1}[\bfC:\bfE^e]\hat{\bfW}^{-T}$.
\end{itemize}

\section{Applications} \label{sec:application}
In this section, an extensive list of model problems are solved to demonstrate the capability and features of our theoretical-computational model. Most problems are solved within both the small and finite deformation settings. In all 2D problems, the body is meshed with quadriltateral, bilinear elements. In this work, all stress fields are non-dimensionalized by the shear modulus $G$. All length variables are non-dimensionalized by the core/layer height $t$. Unless otherwise specified, the elasticity tensor $\bfC$ is assumed to be isotropic with $E = 2.6 G$, $\nu = 0.3$, where $E$ is the Young's Modulus, $G$ is shear modulus and $\nu$ is the Poisson's ratio. For all but two of the problems dealt with in this work, $\bfalpha$ is set zero; the use of dislocations is explicitly mentioned, when it arises. The calculations in this section are conducted within the PETSc package on a 16-core computer. 

In all figures in this work the horizontal axis represents the $\bfe_1$ direction and the vertical axis represents the $\bfe_2$ direction, unless otherwise specified. For all disclination problems treated here, given the misorientation angle $\theta$, the eigenwall field $\bfS$ and the g.disclination density field $\bfPi$ are defined from \eqref{eqn:density_initial} and \eqref{eqn:Pi} in Section \ref{sec:Pi_layer_def}, with $\bfW_2$ assumed as $\bfI$ and $\bfW_1$ to be
\begin{itemize}
\item 
\[
\begin{bmatrix}
    1 & tan \theta \\
    -tan \theta & 1
\end{bmatrix} 
\] 
for the small deformation case and
\item 
\[
\begin{bmatrix}
    \cos \theta & \sin \theta \\
    -\sin \theta & \cos \theta
\end{bmatrix} 
\]
for finite deformation.
\end{itemize}
We discuss a further point related to the definitions of $\bfS$ and $\bfPi$ in Sec. \ref{sec:delta_w} after the discussion of the Eshelby cut-weld problem.

The stress comparisons between the small deformation and the finite deformation settings in this section are for all stress components followed the identical definition of the stress difference given in (\ref{eqn:difference}). Denoting $\bfsigma_{s}$ as the stress field from the small deformation setting and $\bfsigma_f$ as the stress field from the finite deformation setting, the difference of the stress fields between the small deformation setting and the finite deformation setting is denoted as $\delta\sigma_{s,f}$. 

\subsection{A single disclination viewed as an Eshelby cut-and-weld problem}\label{sec:eshelby}
The stress field of a single disclination can be interpreted as a non-standard problem of nonlinear elasticity by adapting Eshelby's cut-and-weld procedures \cite{eshelby1957determination, eshelby1956continuum}. As will be evident, this is certainly not the most efficient methodology for dealing with disclinations, in particular, when they appear in collections of more than one; nevertheless, the example helps to develop intuition and we describe below the basis of our computation of the analogy.

\begin{figure}
  \centering
  \subfigure[Configuration C1 with a gap wedge.]{
    \label{fig:eshelby_illu_1} 
    \includegraphics[width=0.4\linewidth]{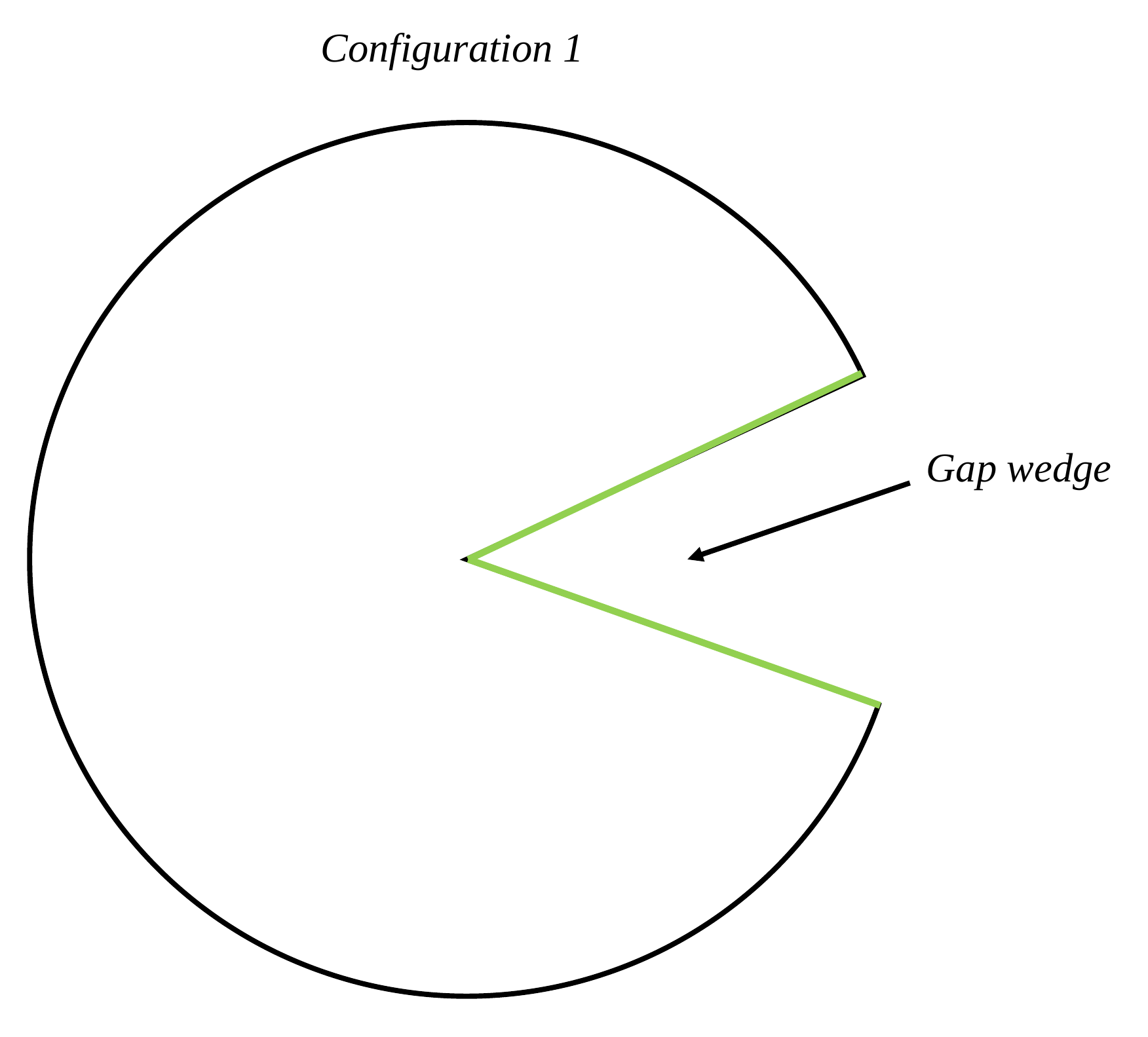}}\qquad
  \subfigure[Configuration C2 with a closed wedge. The configuration after welding the two edges is denoted as $\widetilde{C2}$.]{
    \label{fig:eshelby_illu_2} 
    \includegraphics[width=0.4\linewidth]{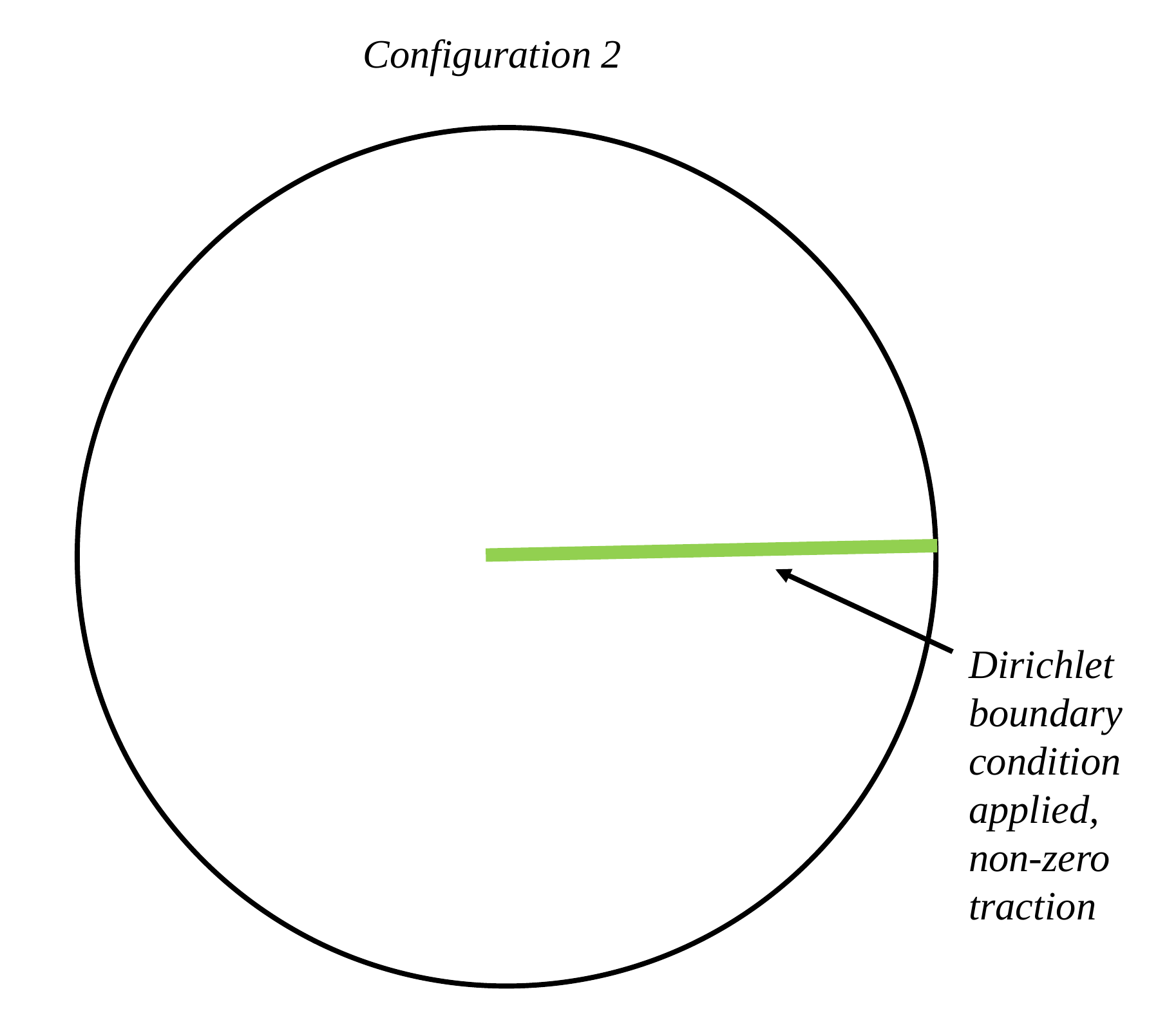}}
 \subfigure[Schematic of possible configuration C3 after welding the wedge and relaxing the body.]{
    \label{fig:eshelby_illu_3} 
    \includegraphics[width=0.4\linewidth]{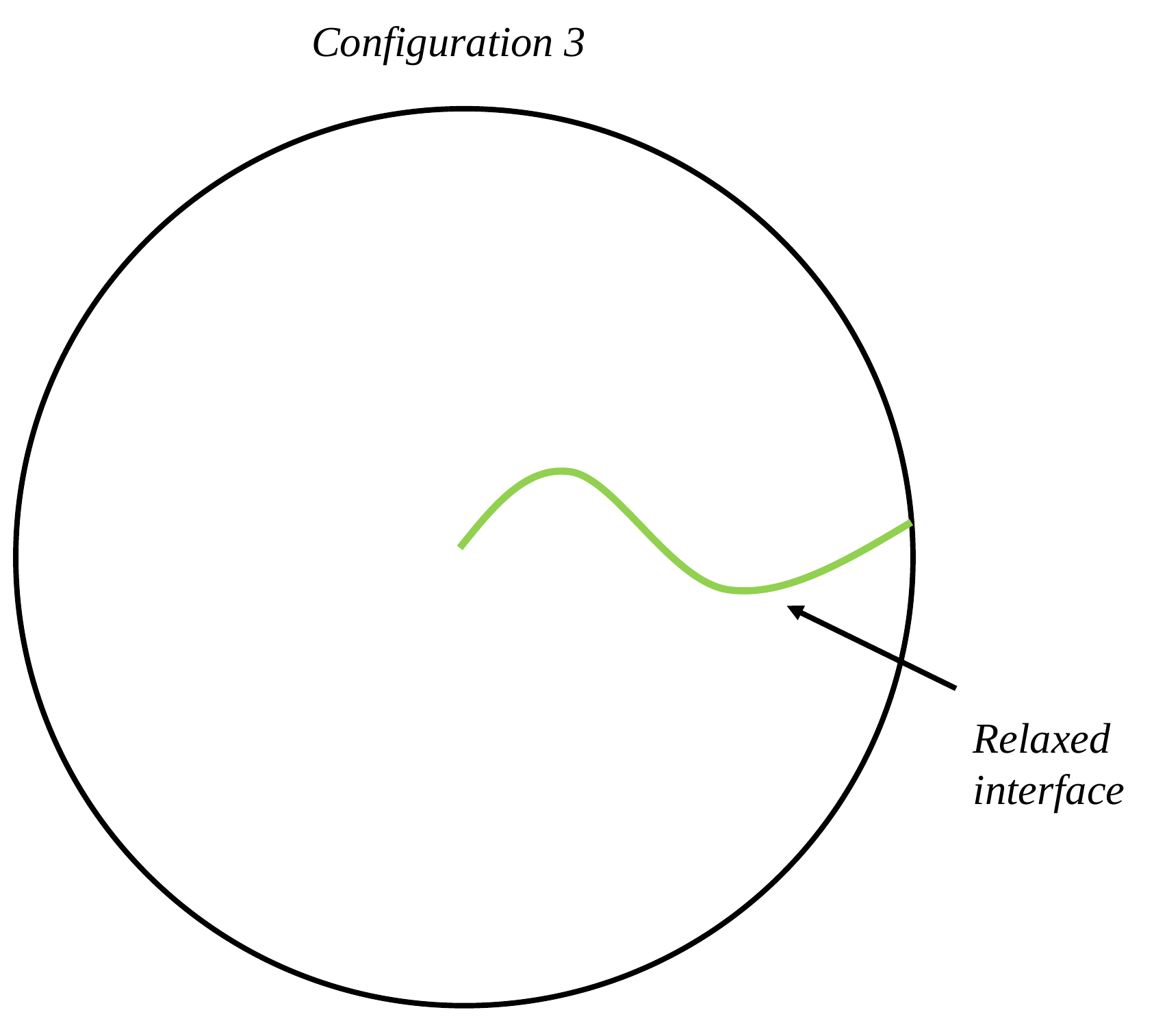}}
  \caption{An Eshelby cut-weld process to form a single positive disclination. After applying Dirichlet boundary conditions, the gap wedge in $C1$ is closed and the resultant traction along the interface in $C2$ is non-zero. The two edges are welded to generate the configuration $\widetilde{C2}$. $C3$ is attained by applying the negative of the obtained resultant traction along the interface on $\widetilde{C2}$ and solving for equilibrium of forces. }
  \label{fig:eshelby_illu} 
\end{figure}

With reference to Fig. \ref{fig:eshelby_illu} we first consider the following thought experiment. In Step 1 the edges of a gap wedge (the green lines in Figure \ref{fig:eshelby_illu_1}) in $C1$, a stress-free configuration, are brought together  to close the gap, resulting in the configuration $C2$ (Figure \ref{fig:eshelby_illu_2}). This is achieved by applying appropriate displacement boundary conditions to the edges of the gap wedge. Clearly, non-zero (reaction) tractions exist along both adjoining edges on $C2$. In Step 2,  imagine welding the edges to generate the configuration $\widetilde{C2}$ and removing from them the reaction tractions generated in Step 1, letting the welded body relax to the configuration $C3$. Concretely, the act of welding generating $\widetilde{C2}$ amounts to thinking that all further deformations of $C2$ are continuous on the surface in it along which the adjoining edges overlap. $C2$ has `two additional' boundary surfaces than $\widetilde{C2}$. The act of relaxation implies that the stressed configuration $\widetilde{C2}$, now  connected along the surface formed by the overlapping edges, is subjected to no internal, singular body force fields. Due to the removal of the reaction tractions on the edges, the stress field arising from the deformation in Step 1 no longer satisfies equilibrium on $\widetilde{C2}$, but the body now can only deform through a compatible deformation of $\widetilde{C2}$ to achieve the configuration $C3$ where it is in (force) equilibrium with no applied tractions or body forces. 

\begin{figure}
  \centering
  \subfigure[Stress field of the `Eshelby disclination' in the finite deformation setting.]{
    \label{fig:eshelby_stress} 
    \includegraphics[width=0.4\linewidth]{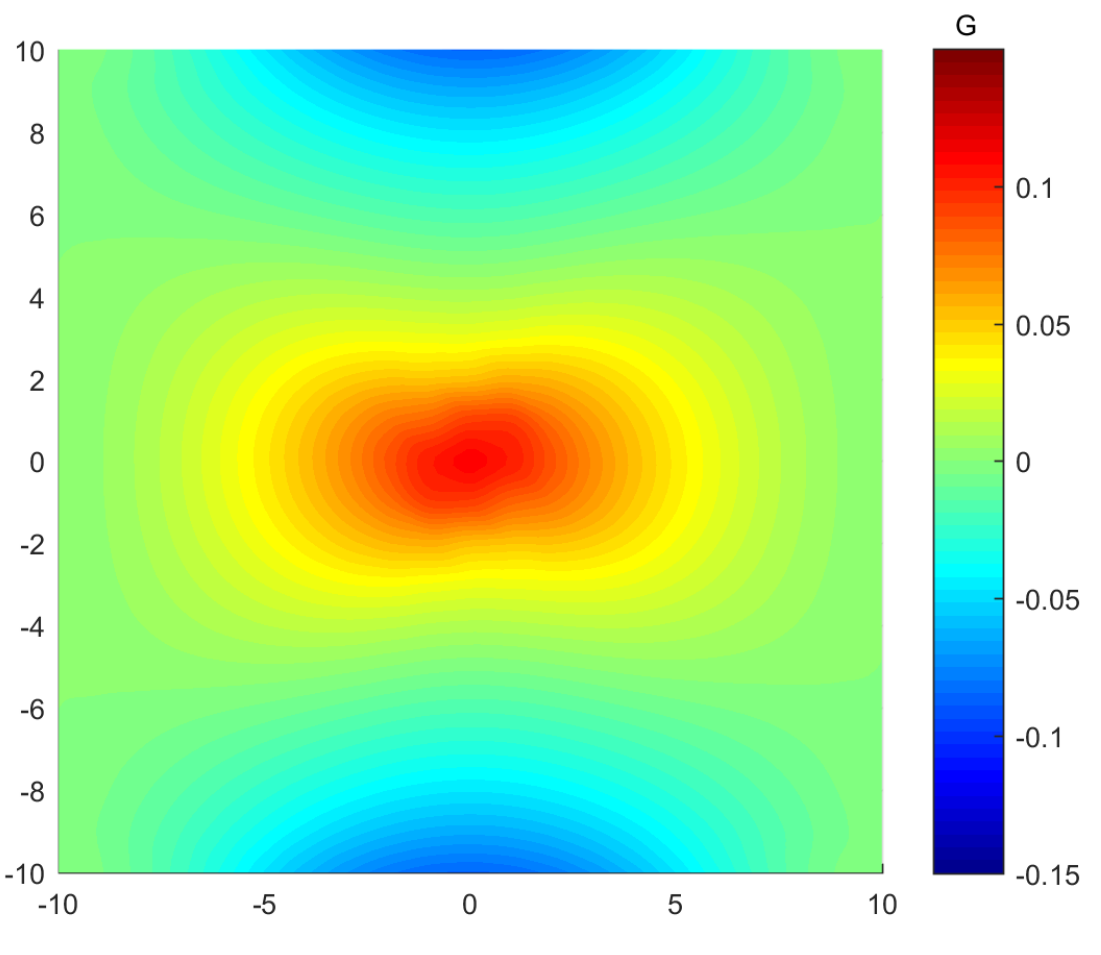}}\qquad
  \subfigure[$\delta \sigma_{e,g}$ between Eshelby process and the g.disclination model.]{
    \label{fig:eshelby_difference} 
    \includegraphics[width=0.4\linewidth]{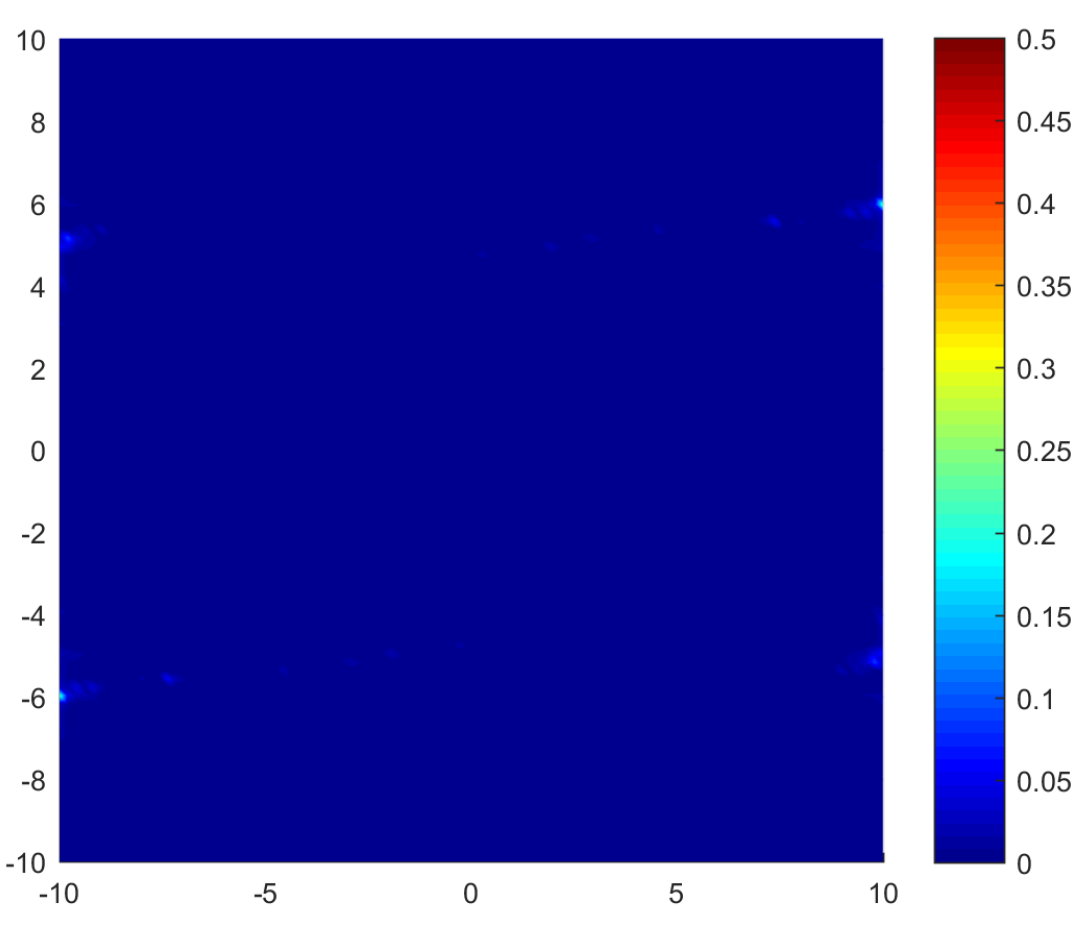}}
  \caption{Stress field $\sigma_{11}$ for a single disclination viewed as an Eshelby process in the finite deformation setting. The maximum of $\delta \sigma_{e,g}$ is less than $5\%$.}
  \label{fig:eshelby_stress_error} 
\end{figure}

We approximate the solution of the above problem with the algorithms described in Section \ref{sec:numerical} as follows. We assume the configuration $\widetilde{C2}$ as known (the domain in Fig. \ref{fig:eshelby_stress}) and first determine the stress-free configuration $C1$. This is done by viewing the intersection of the positive $x$-axis and the body as two surfaces on which are applied appropriate Dirichlet boundary conditions to represent the (inverse) deformation  of these surfaces to their positions on the otherwise unknown unstressed reference configuration $C1$. On the rest of the boundary, traction-free boundary conditions are imposed. The solution is obtained by solving (\ref{eqn:disc_num_finite}) for $\hat{\bff}$ with $\hat{\bfchi} = \bf0$ and $\bfH^s = \bf0$. Let the deformation gradient of $C1$ with respect to $C2$, the latter with the slit, be denoted as $\bfW^{(1)}$. Let the continuous deformation from $\widetilde{C2}$ to the unknown configuration $C3$ be denoted as $\bfg$ and the inverse of its deformation gradient as $\bfW^{(2)}$. Then, defining $\grad\, \hat{\bff}$ as $\grad_3 \, \hat{\bff} = \bfW^{(1)} (\grad_2\, \bfg)^{-1} = \bfW^{(1)}\bfW^{(2)}$, we solve (\ref{eqn:disc_num_finite}) for $\bfg$ with $\hat{\bfchi} = \bf0$ and $\bfH^s = \bf0$; the subscripts $3$ and $2$ are included to indicate the fact that the spatial derivatives are w.r.t the configurations $C3$ and $\widetilde{C2}$, respectively, and the $\divergence$ in (\ref{eqn:disc_num_finite}) is to be understood as $\divergence_3$ as well. As this is simply a motivational example, in the solutions shown in Fig. \ref{fig:eshelby_stress}, we assume for simplicity that $\divergence_3 \approx  div_2$ which may be justified for $|\grad_2 \, \bfg - \bfI| \ll 1$ (in the context of nonlinear finite element computations, this approximation is not essential in any way).

Figure \ref{fig:eshelby_stress} shows the $\sigma_{11}$ field of a $45^\circ$ positive disclination computed from the Eshelby process described above; the Dirichlet b.c. in Step 1 corresponds to the geometry of setting up a $45^{\circ}$ gap-wedge between C1 and C2. The stress field $\sigma_{11}$ of a disclination of the same strength on the configuration C2 is computed by setting up the g.disclination density field according to (\ref{eqn:Dij}).

The system (\ref{eqn:disc_num_finite}) is solved with $\bfZ^s = \bf0$ and $\divergence = \divergence_2$ and the result is shown in Figure \ref{fig:disclination_finite}. Figure \ref{fig:eshelby_difference} shows the difference $\delta \sigma_{e,g}$ following the definition in (\ref{eqn:difference}), where the subscript $e$ denotes the stress field from the Eshelby process and the subscript $g$ denotes the stress field from g.disclination model. The maximum of $\delta \sigma_{e,g}$ is less than $5\%$.

We note here that both the Eshelby cut-weld problem and the g.disclination problem are solved on a FE mesh with the same refinement and cannot represent singularities. It is most likely that the exact solution for the Eshelby cut-weld problem actually has a stress singularity at the origin which would be evident with mesh refinement. On the other hand, the g.disclination problem of the same strength does not have a singularity due to the definition of a well-defined core defined by the parameter $c$ (that is expected to emerge in more comprehensive modeling from energetics). The far-field correspondence of the results however is expected to remain as shown in Fig. \ref{fig:eshelby_difference}.

\subsection{Approximation in $\bfS$ prescription} \label{sec:delta_w}
The considerations above related to the Eshelby cut-weld problem also make clear an important issue in the definition of the strength of a disclination; namely, that the definition of the difference $(\bfW_1 - \bfW_2)$ in the strength of a g.disclination in (\ref{eqn:Pi}), (\ref{eqn:Dij}), strictly speaking, cannot simply be achieved from the knowledge of the geometry of the gap/overlap wedge to be eliminated. Instead, it also requires knowledge of the additional tensor field $\bfW^{(2)}$ along the `weld' surface. In principle, this is not a problem when physical observations are at hand defining the details of the interface and the question is to compute the elastic fields on the whole body, or when a full problem of evolution is solved, in which case the g.disclination density $\bfPi$, the eigenwall field $\bfS$, and their elastic fields are predicted quantities. Denote the (geometrically, or otherwise) inferred i-elastic distortion fields across the interface as $\bfW^i_1$ and $\bfW^i_2$. Then we have
\[
\bfW_1-\bfW_2 = (\bfW^i_1-\bfW^i_2)\bfW^{(2)}.
\] 
On defining $\Delta \bfW :=\bfW_1 - \bfW_2$ and $\Delta \bfW^i := \bfW^i_1 - \bfW^i_2$, we have 
\[
\Delta \bfW - \Delta \bfW^i = \Delta \bfW^i (\bfW^{(2)} - \bfI).
\]
In most problems solved in this paper, we assume the $\bfW^{(2)}$ field to be approximately the identity tensor for the purpose of \emph{defining} the g.disclination strength, the eigenwall fields and the dislocation density along interfaces (that serves as specified data), and approximate $\Delta \bfW$ as $\Delta \bfW^i$.

\subsection{Field of a single disclination: comparison with the classical theory}\label{sec:single_disclination_classical}

In the linear elastic, small deformation theory \cite{dewit1973theory4}, the 2-d stress field at $\bfx$ for a straight disclination along the $x_3$ direction passing through the coordinate origin is given as
\begin{equation*}
\begin{aligned}
\sigma_{11} =& \frac{G\Omega_3}{2\pi(1-\nu)}\left[ \ln \rho + \frac{x_2^2}{\rho^2} + \frac{\nu}{1-2\nu}\right]\\
\sigma_{22} =& \frac{G\Omega_3}{2\pi(1-\nu)}\left[ \ln \rho + \frac{x_1^2}{\rho^2} + \frac{\nu}{1-2\nu}\right]\\
\sigma_{12} =& - \frac{G\Omega_3x_1x_2}{2\pi(1-\nu)\rho^2},
\end{aligned}
\end{equation*}
where $\rho = \sqrt{x_1^2 + x_2^2}$. With reference to Figure \ref{fig:strength_construct}, and a misorientation angle $\theta$ of $5^\circ$ , we have 
\begin{equation}\label{eqn:Frank}
\bfOmega = 0.0875 \bfe_3,
\end{equation}
from (\ref{eqn:strength}).

The g.disclination density is defined from (\ref{eqn:straight_S}) as
\begin{equation}\label{eqn:Pi2}
\bfPi = 
\begin{cases} \frac{\Delta W_{ij}}{ct} \bfe_i \otimes \bfe_j \otimes \bfe_3 & \text{where $|x_1|\le\frac{c}{2}$ and $|x_2|\le \frac{t}{2}$} \\
\bf0 & \text{otherwise},
\end{cases}
\end{equation}
where $i,j=1,2$, $c$ is the core width, $t$ is the layer thickness and the matrix $[\Delta W]$ is given as 
\[
\begin{bmatrix}
    0 & 0.0875  \\
    -0.0875 & 0
  \end{bmatrix}.
\]

\begin{figure}[H]
\centering
\subfigure[$\sigma_{11}$ for a single disclination from classical linear elasticity.]{
\includegraphics[width = 0.4\textwidth]{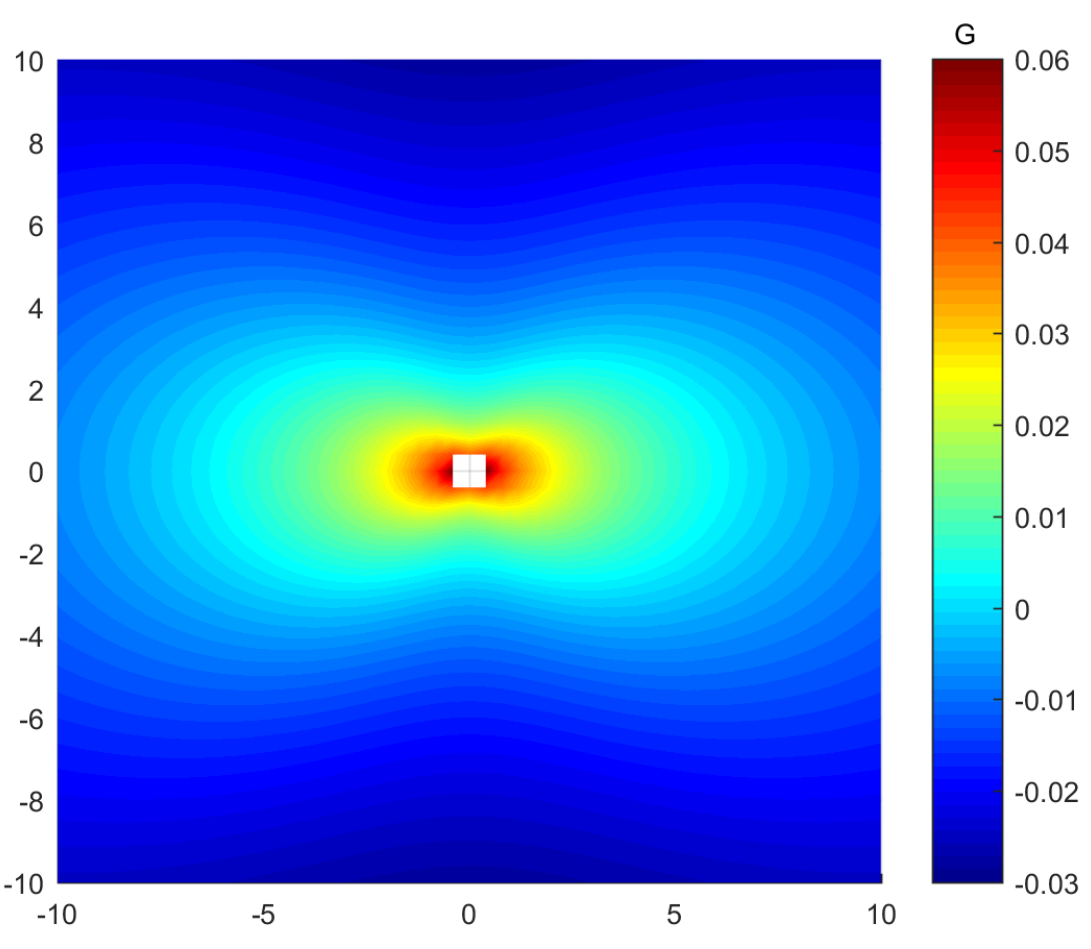}
\label{fig:analytical_single}
}\qquad 
\subfigure[$\sigma_{11}$ for a single disclination from the g.disclination model.]{
\includegraphics[width = 0.4\textwidth]{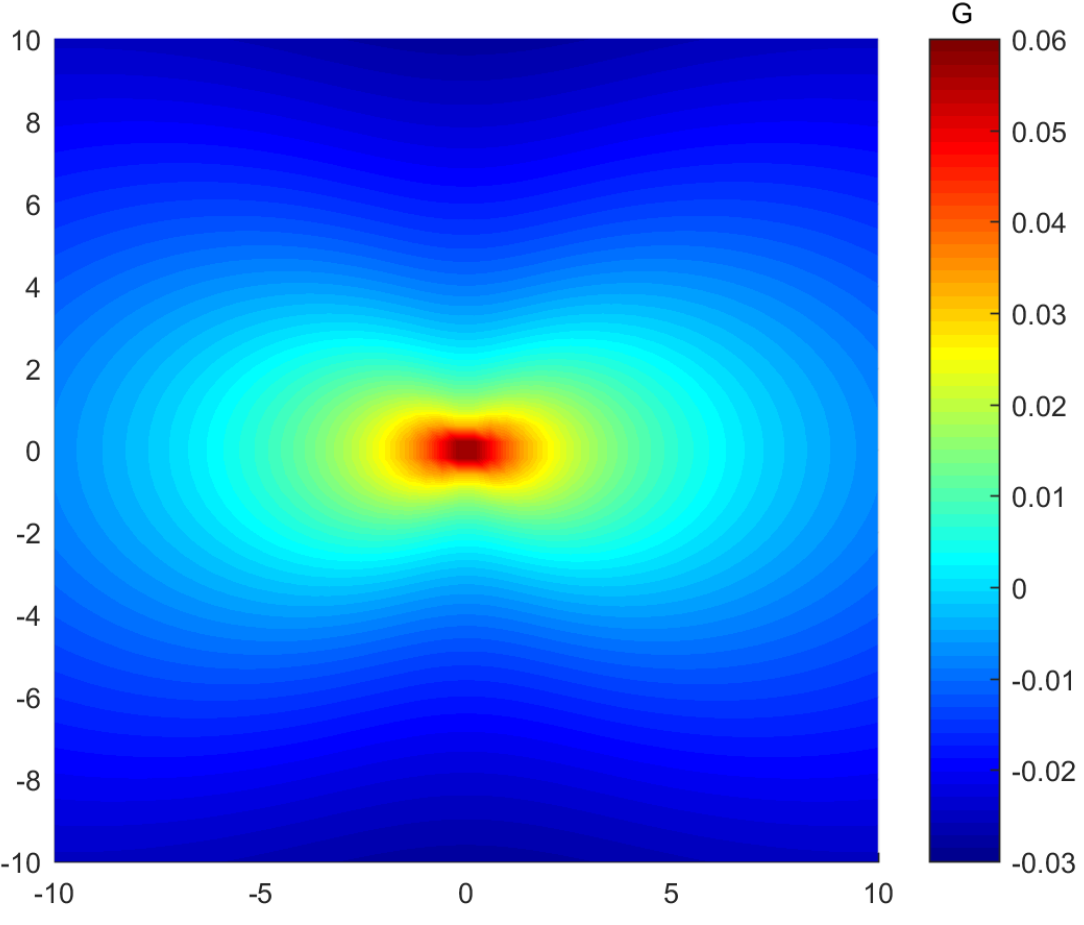}
\label{fig:g.disclination_single}
}
\subfigure[$\delta\sigma_{a,g}$ between classical linear elasticity and the g.disclination model.]{
\includegraphics[width = 0.5\textwidth]{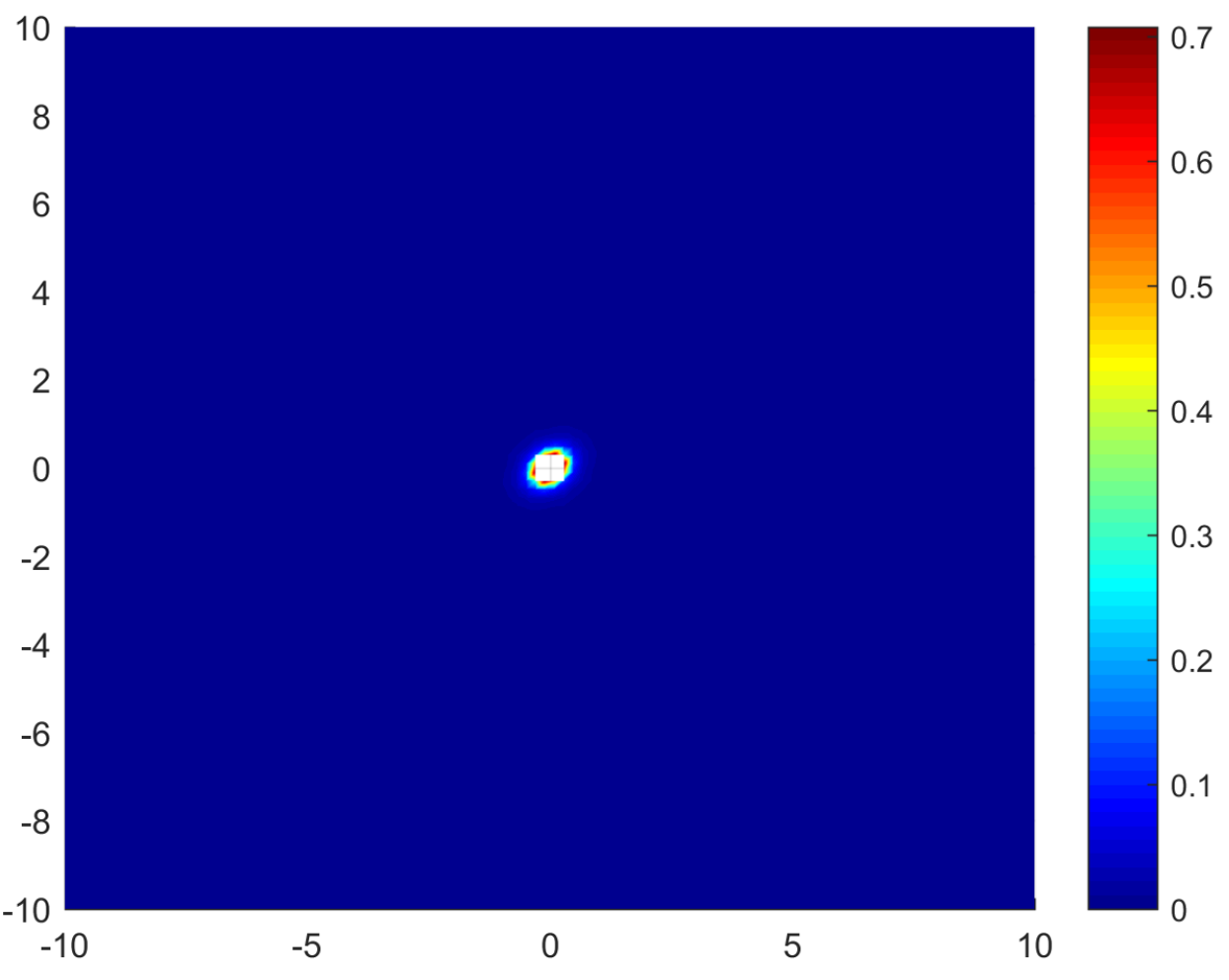}
\label{fig:compare_single}
}
\caption{The stress field $\sigma_{11}$ and the comparison $\delta\sigma_{a,g}$ for a single disclination. The result from the g.disclination model matches well with the linear elasticity solution, with the $\delta\sigma_{a,g}$ maximum outside the core being less than $1\%$.}
\end{figure}

The size of the body is $10\times10$ and the size of the disclination core is $0.5 \times 0.5$ (in units of $t$, the core height). To compare our numerical solution with DeWit's infinite-medium solutions, the following Neumann boundary conditions are utilized. Considering the body in our model as a patch in an infinite domain, the traction field on the boundary of the corresponding patch from the infinite-medium solution is applied. Figure \ref{fig:analytical_single} is the stress field $\sigma_{11}$ from the DeWit solution with Frank vector (\ref{eqn:Frank}) and Figure \ref{fig:g.disclination_single} is the stress field $\sigma_{11}$ from small deformation g.disclination theory with the g.disclination density (\ref{eqn:Pi2}). Here, we denote $\bfsigma_a$ as the stress field from the analytical solution and $\bfsigma_g$ as the stress field from the g.disclination model. The difference between the analytical solution and the g.disclination solution is denoted as $\delta\sigma_{a,g}$ following the definition (\ref{eqn:difference}).

Figure \ref{fig:compare_single} shows the defined difference of the stress field; the computed stress field from g.disclination theory matches with the DeWit solution very well. Outside the core, the defined difference is less than $1\%$. 

\subsection{A single disclination with large misorientation} \label{sec:single_disclination}
\begin{figure}
  \centering
  \subfigure[Stress field $\sigma_{11}$ from small deformation setting.]{
    \label{fig:disclination_linear} 
    \includegraphics[width=0.4\linewidth]{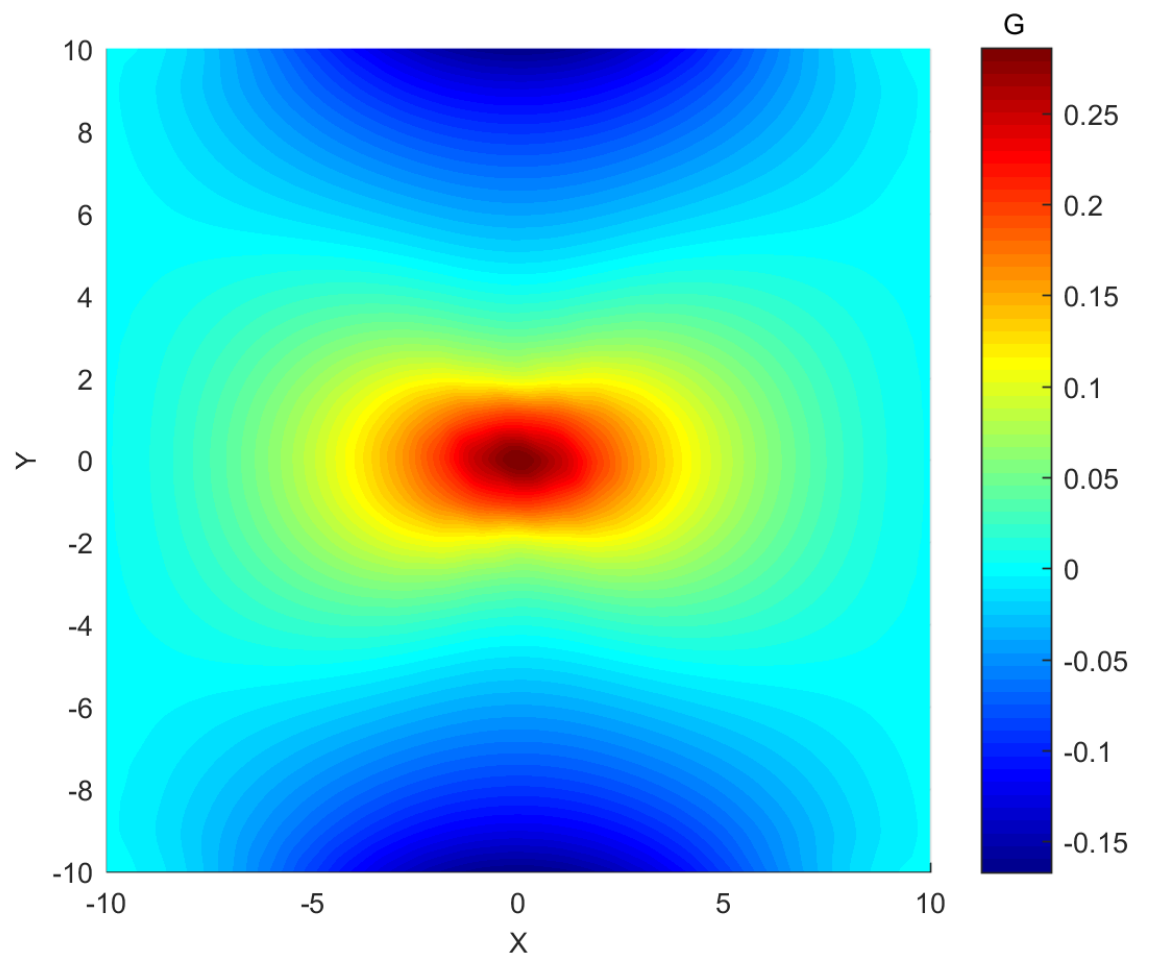}}\qquad
  \subfigure[Stress field $\sigma_{11}$ from finite deformation setting.]{
    \label{fig:disclination_finite} 
    \includegraphics[width=0.4\linewidth]{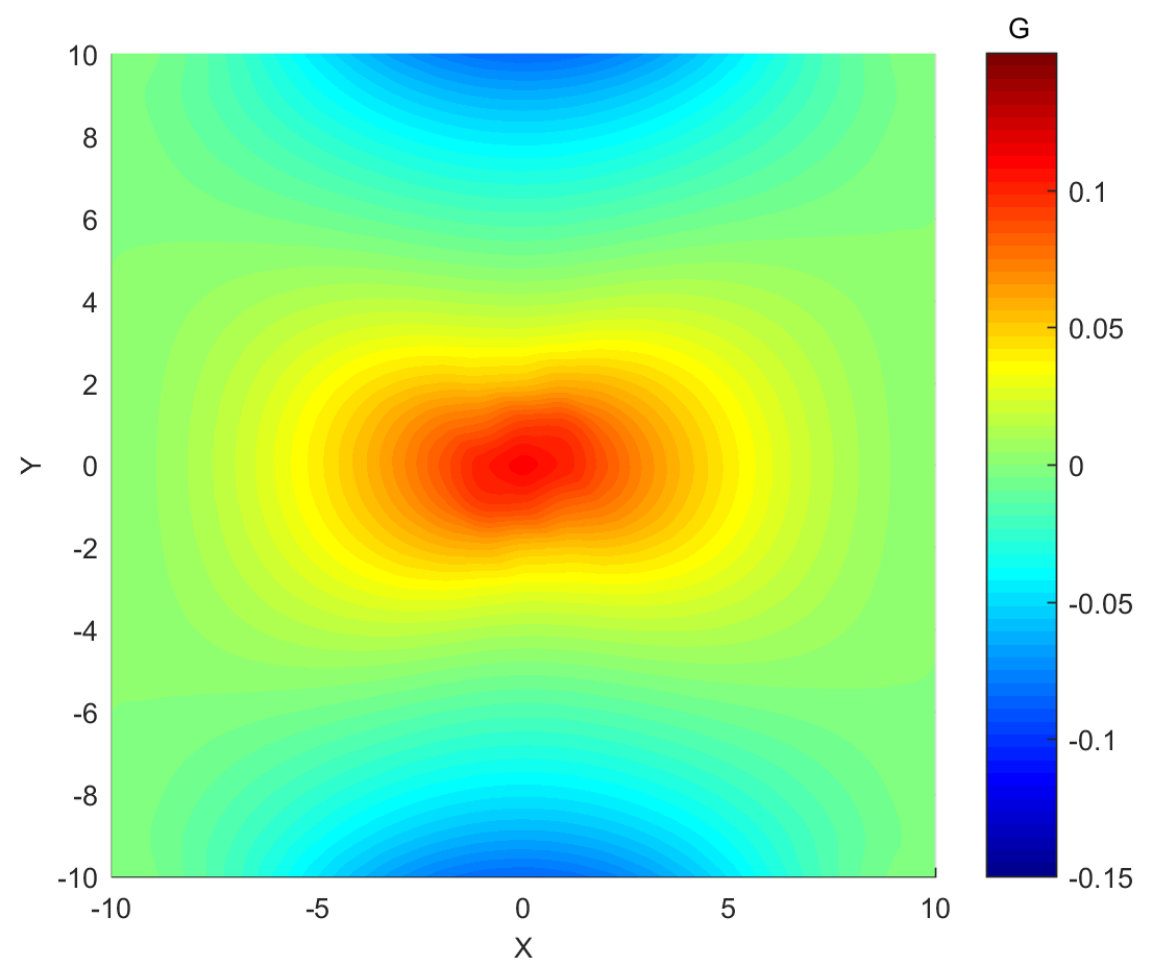}}
 \subfigure[$\delta\sigma_{s,f}$ between the small and finite deformation settings.]{
    \label{fig:disclination_error} 
    \includegraphics[width=0.4\linewidth]{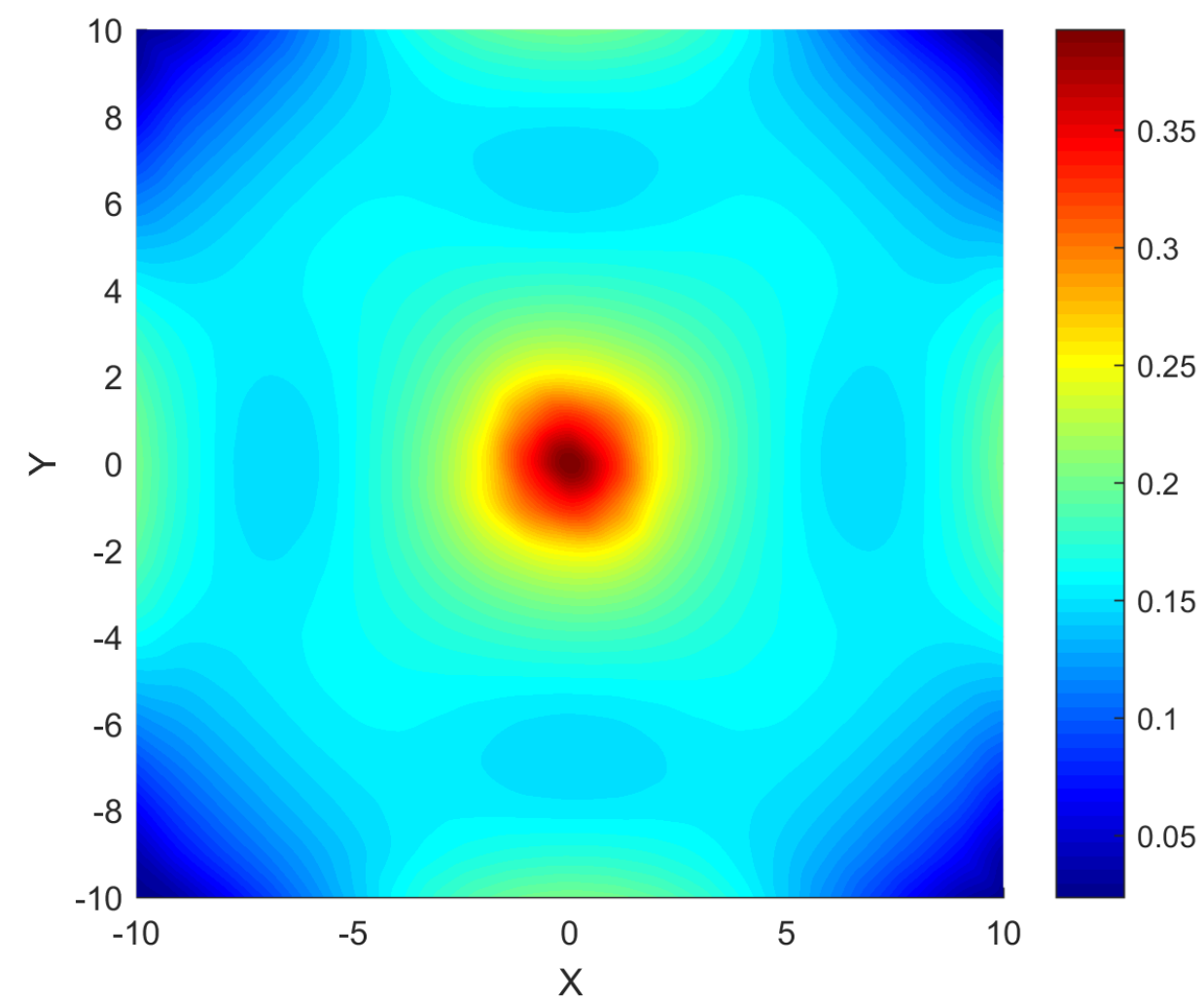}}
  \caption{Stress field $\sigma_{11}$ for a single disclination from both the small and finite deformation settings. The maximum of $\delta\sigma_{s,f}$ is about $40\%$ and the mean of $\delta\sigma_{s,f}$ is $1.39\%$.}
  \label{fig:disclination_lf} 
\end{figure}

We examine the difference between the stress fields from the small and finite deformation settings arising from a single disclination representing a high  misorientation. For the small deformation problem, we assume the misorientation magnitude to be represented by $\tan \theta$, where $\theta$ is the misorientation, following (\ref{eqn:strength}). We set up a single disclination with a $45^{\circ}$ misorientation and apply traction-free boundary conditions. Figure \ref{fig:disclination_linear} is the stress field from the small deformation setting and Figure \ref{fig:disclination_finite} is that from the finite deformation setting. Figure \ref{fig:disclination_error} is the plot of the difference $\delta\sigma_{s,f}$, whose maximum is about $40\%$ and the mean of $\delta\sigma_{s,f}$ is $1.39\%$. It is clear that for large misorientations like the one shown (which is more than the commonly believed threshold of $ > 11^{\circ}$), there are significant differences between the small and finite deformation results. 

\subsection{Single dislocation}
Here we solve an edge dislocation problem, interpreted as a g.disclination dipole, as discussed in \cite[Sec. 4.3]{zhang_acharya_2016}. In this context, two opposite-sign g.disclinations are prescribed with the distortion differences as pure rotation differences (g.disclinations become pure disclinations), with Frank vector $\bfOmega$ and $-\bfOmega$ respectively. Based on the results in \cite{zhang_acharya_2016}, the Burgers vector $\bfb$ for this disclination dipole in small deformation theory is given as $\bfb = \bfOmega \times \delta \bfr$, where $\delta \bfr$ is the dipole vector (the vector that separates the two disclinations in the dipole).
\begin{figure}
  \centering
  \subfigure[Stress field $\sigma_{11}$ from the g.disclination dipole model.]{
    \label{fig:dislocation_comp_1} 
    \includegraphics[width=0.45\linewidth]{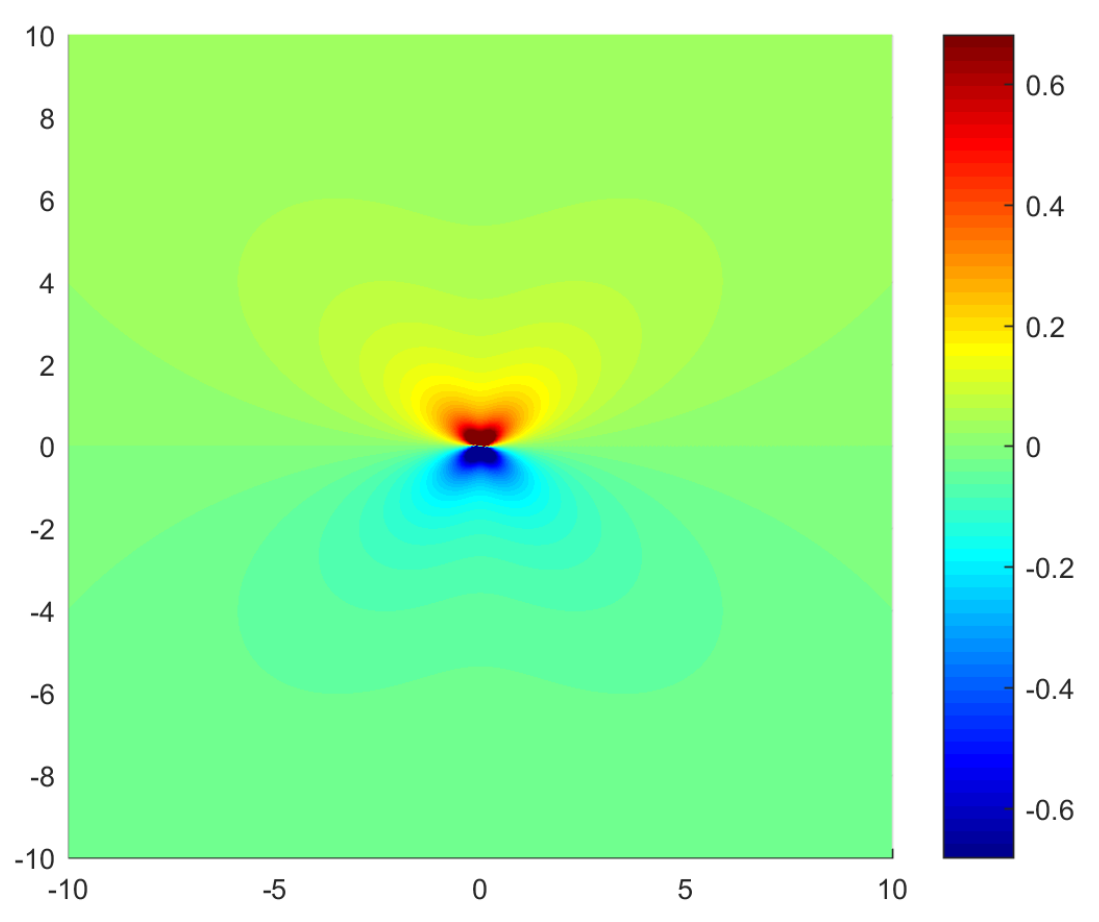}}
	\subfigure[Stress field $\sigma_{11}$ from linear elasticity.]{
    \label{fig:dislocation_comp_4} 
    \includegraphics[width=0.45\linewidth]{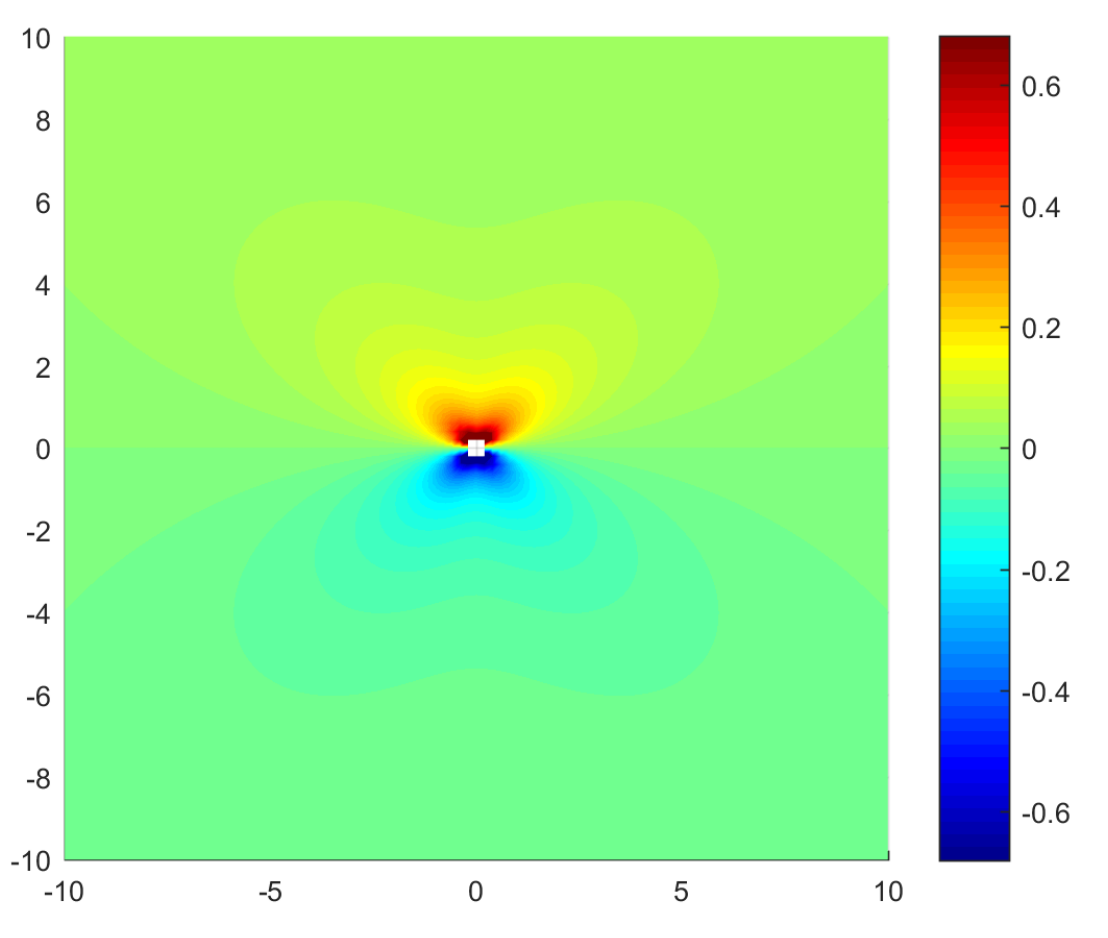}}
   \subfigure[$\delta\sigma_{a,d}$ between the g.disclination dipole model and linear elasticity. ]{
    \label{fig:dislocation_comp_3} 
    \includegraphics[width=0.45\linewidth]{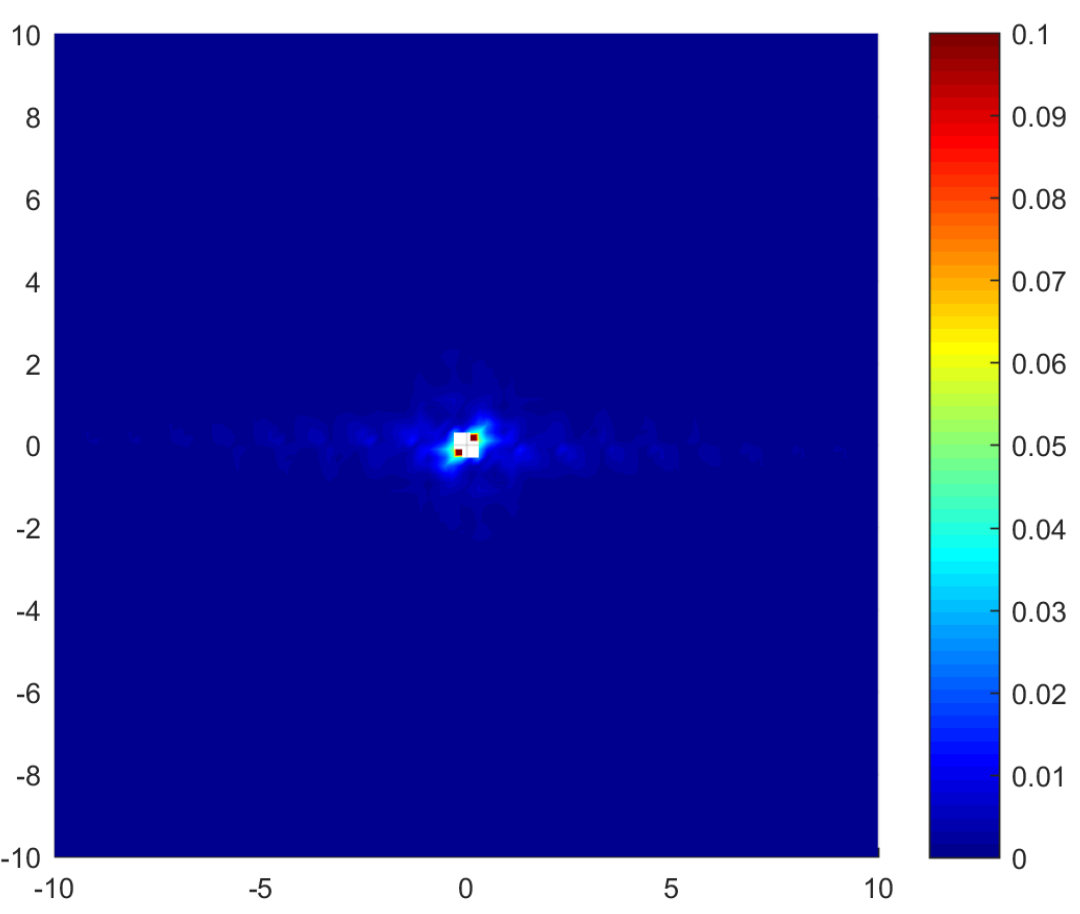}}
  \caption{Stress fields $\sigma_{11}$ of a single dislocation from the g.disclination dipole model and linear elasticity, respectively. Outside the core, the difference $\delta\sigma_{a,d}$ is less than $3\%$. Inside core, the stress field from linear elasticity blows up.}
  \label{fig:dislocation_comp} 
\end{figure}

Figure \ref{fig:dislocation_comp_1} is the stress field $\sigma_{11}$ from the g.disclination dipole model and Figure \ref{fig:dislocation_comp_4} is the stress field $\sigma_{11}$ for the classical linear elastic dislocation with the corresponding Burgers vector $\bfb = \bfOmega \times \delta \bfr$. The traction boundary condition in the g.disclination dipole model is set to be that arising from the stress field of the corresponding classical linear elastic dislocation, following identical logic as in Section \ref{sec:single_disclination_classical}. $\bfsigma_a$ denotes the stress field of the classical linear elastic edge dislocation and $\bfsigma_{d}$ is the stress field from the g.disclination dipole model. The difference between the classical linear elasticity and the g.disclination dipole model is denoted as $\delta\sigma_{a,d}$ following definition (\ref{eqn:difference}). Figure \ref{fig:dislocation_comp_3} shows $\delta\sigma_{a,d}$. Outside the core, the stress fields from the g.disclination model match the one from the classical linear elastic dislocation very well. 

\subsection{High-angle grain boundaries}

As discussed in \cite{zhang_acharya_2016}, a grain boundary can be interpreted as a series of disclination dipoles. The elastic field of such a high-angle grain boundary is computed in this section. Also computed are the fields of a tilt grain boundary with disclination dipoles as well as with additional dislocations.

\subsubsection{High-angle grain boundary modeled by g.disclination dipoles}
\begin {figure}
\centering
\includegraphics[width=15cm]{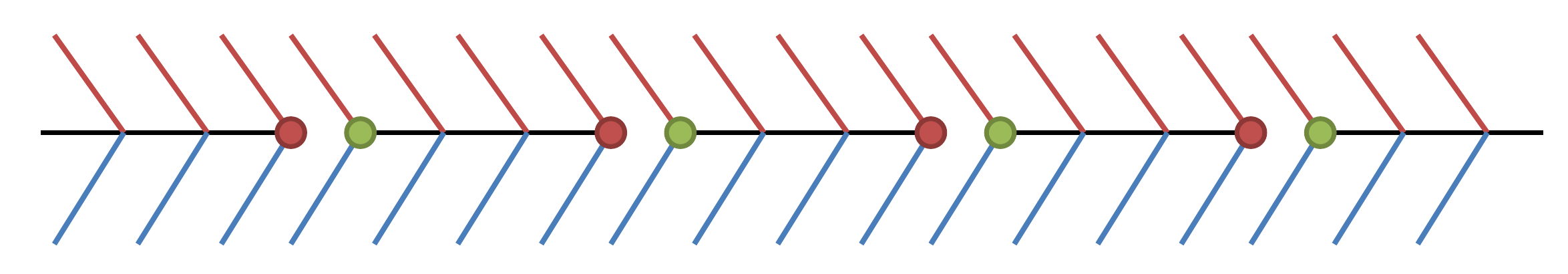}
\caption{A grain boundary interpreted as disclination dipoles equally spaced along the boundary interface. The red lines represent one grain while the blue lines represent another grain. Red points are positive disclinations and green points are negative disclinations.}
\label{fig:grain_boundary_wall}
\end {figure}

Consider a grain boundary interpreted as four disclination dipoles equally spaced along the boundary interface, as illustrated in Figure \ref{fig:grain_boundary_wall}. The individual misorientation magnitude of the disclinations involved in each dipole is $45^{\circ}$. The resulting grain boundary has the same misorientation magnitude.

\begin {figure}
\centering
\subfigure[Stress field $\sigma_{22}$ for a grain boundary wall from the small deformation setting.]{
\includegraphics[width=0.45\linewidth]{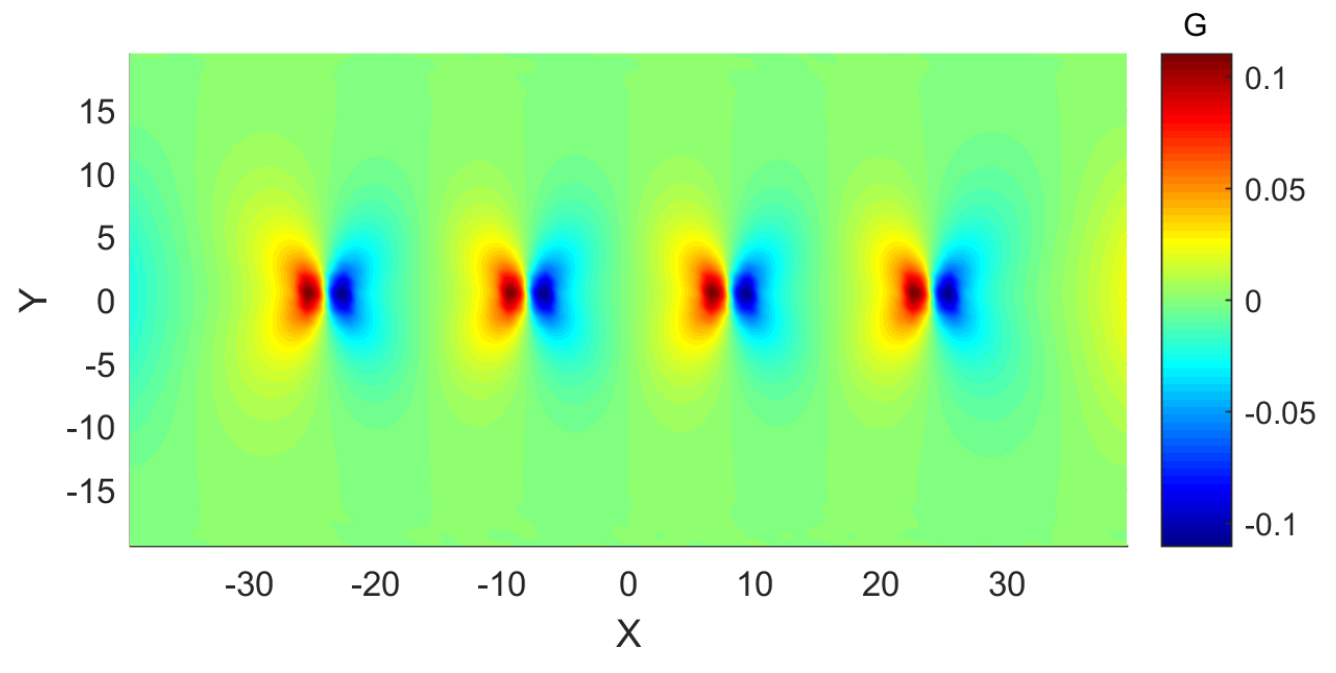}
\label{fig:grain_stress_linear}} \qquad
\subfigure[Stress field $\sigma_{22}$ for the grain boundary wall from the finite deformation setting.]{
\includegraphics[width=0.45\linewidth]{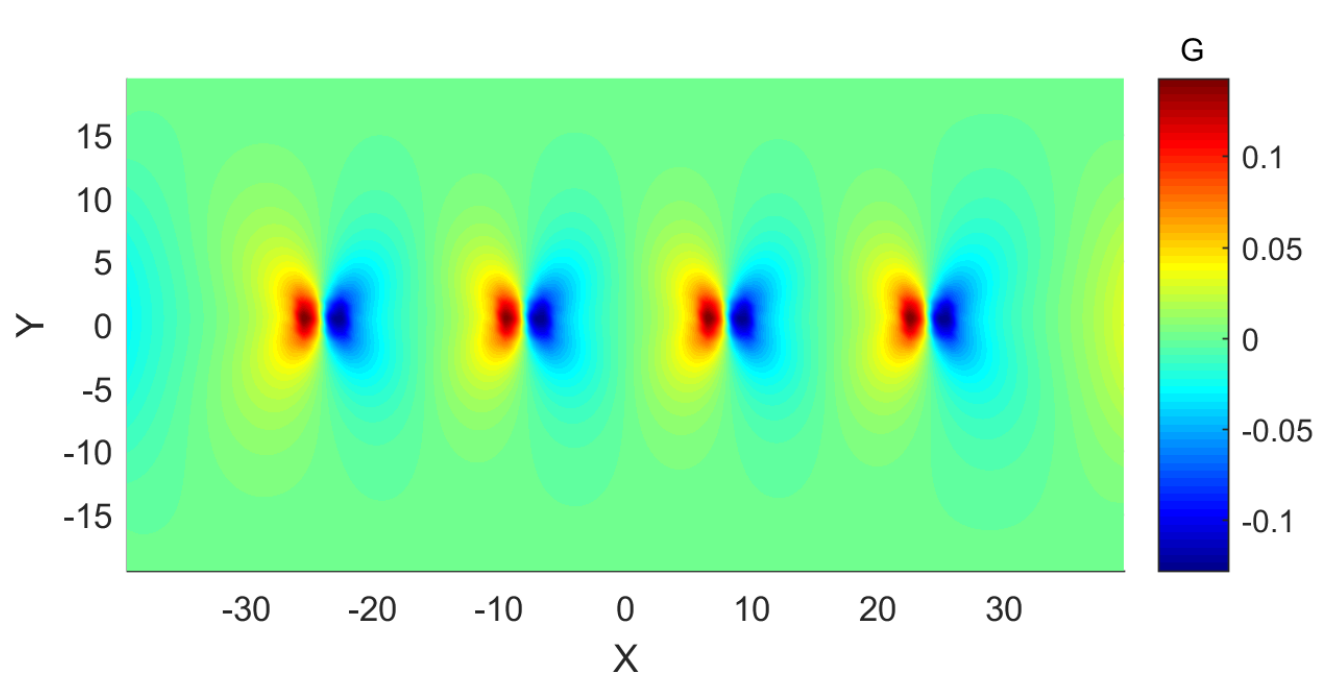}
\label{fig:grain_stress_finite}}
\subfigure[$\delta\sigma_{s,f}$ between the small and finite deformation settings.]{
\includegraphics[width=0.45\linewidth]{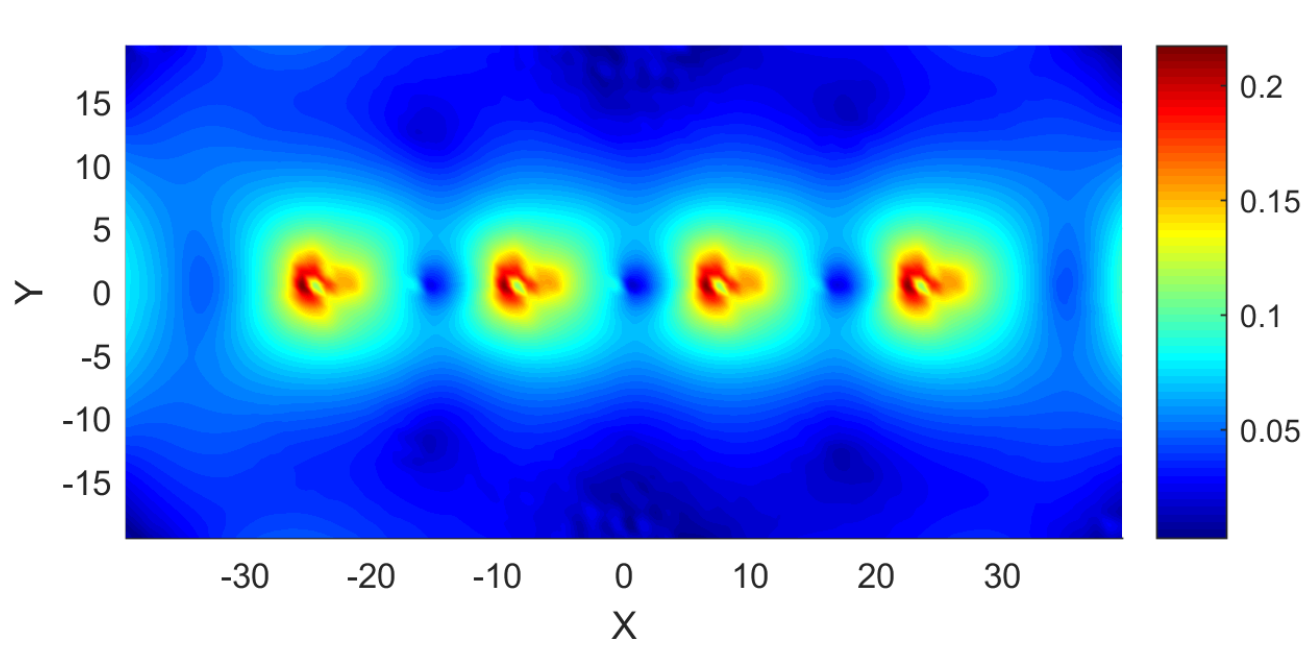}
\label{fig:grain_stress_error}}
\caption{Stress field $\sigma_{22}$ for a grain boundary represented by a series of disclination dipoles. The maximum of $\delta\sigma_{s,f}$ is about $20\%$ and the mean of $\delta\sigma_{s,f}$ is $0.57\%$.}
\end {figure}

Figure \ref{fig:grain_stress_linear} and Figure \ref{fig:grain_stress_finite} show the $\sigma_{22}$ stress fields for the grain boundary in Figure \ref{fig:grain_boundary_wall} from the small and finite deformation settings, respectively. Figure \ref{fig:grain_stress_error} is the plot of the defined difference between the two deformation settings. The maximum of $\delta\sigma_{s,f}$ is about $20\%$ and the mean of $\delta\sigma_{s,f}$ is $0.57\%$.

\subsubsection{Tilt grain boundary comprising disclination dipoles and dislocations}

\begin{figure}
\centering
\includegraphics[width=0.6\linewidth]{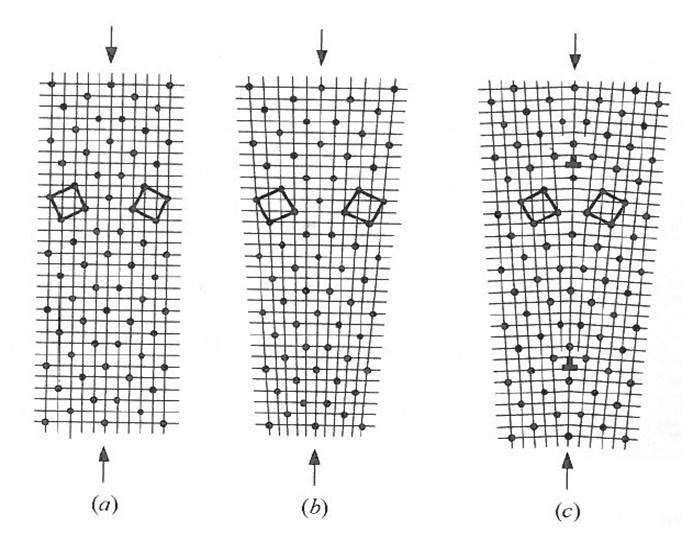}
\caption{(a) A common tilt grain boundary with a $53.1$ degree tilt angle. (b) The configuration after applying a little additional tilt angle on the grain boundary in (a), without any rearrangement, which will have far field stress. (c) The configuration with some dislocations introduced along the interface to eliminate far field stress. (Figures reproduced from  \cite{balluffi2005kinetics} with permission from John Wiley and Sons.)}
\label{fig:tilt_phy}
\end{figure}

In some circumstances, dislocations and disclination dipoles both exist along a boundary interface, as shown in Figure \ref{fig:tilt_phy} from \cite{balluffi2005kinetics}. Figure \ref{fig:tilt_phy}(a) shows a large-angle, symmetric tilt grain boundary with a $53.1^{\circ}$ misorientation. A slightly increased tilt angle is established by a bending load while maintaining the grain boundary structure intact, as shown in Figure \ref{fig:tilt_phy}(b). In Figure \ref{fig:tilt_phy}(c), dislocations are introduced to eliminate the long-range stresses generated in Figure \ref{fig:tilt_phy}(b), i.e. the configuration with the additional tilt can be supported with no bending loads in the presence of the added dislocations; such a configuration is actually observed in reality \cite{balluffi2005kinetics}. 

\begin{figure}
\centering
\subfigure[Defect prescription for tilt grain boundary without dislocations.]{
\includegraphics[width=0.25\linewidth]{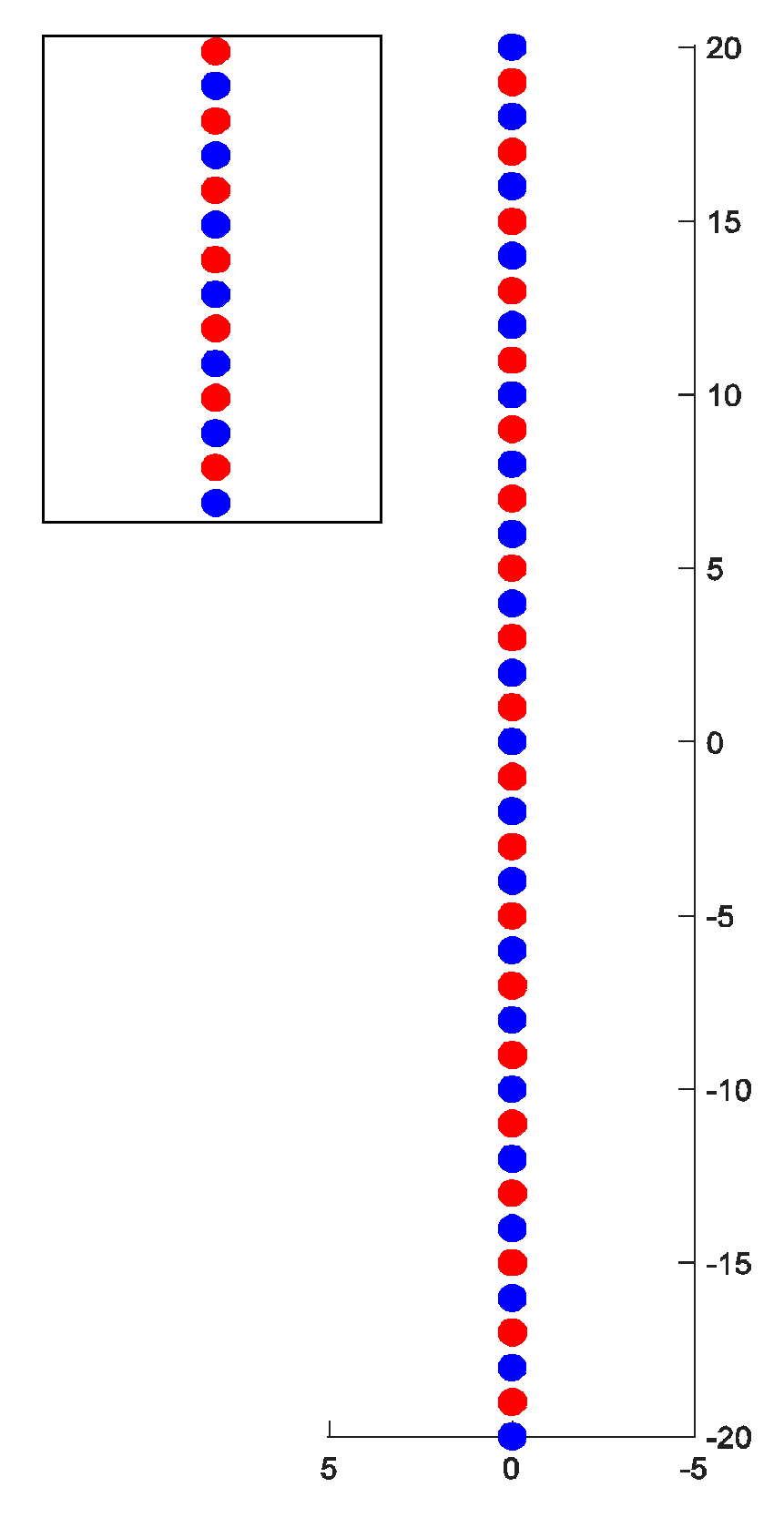}
\label{fig:defect_tilt_2}}\qquad
\subfigure[Defect prescription for tilt grain boundary with dislocations.]{
\includegraphics[width=0.25\linewidth]{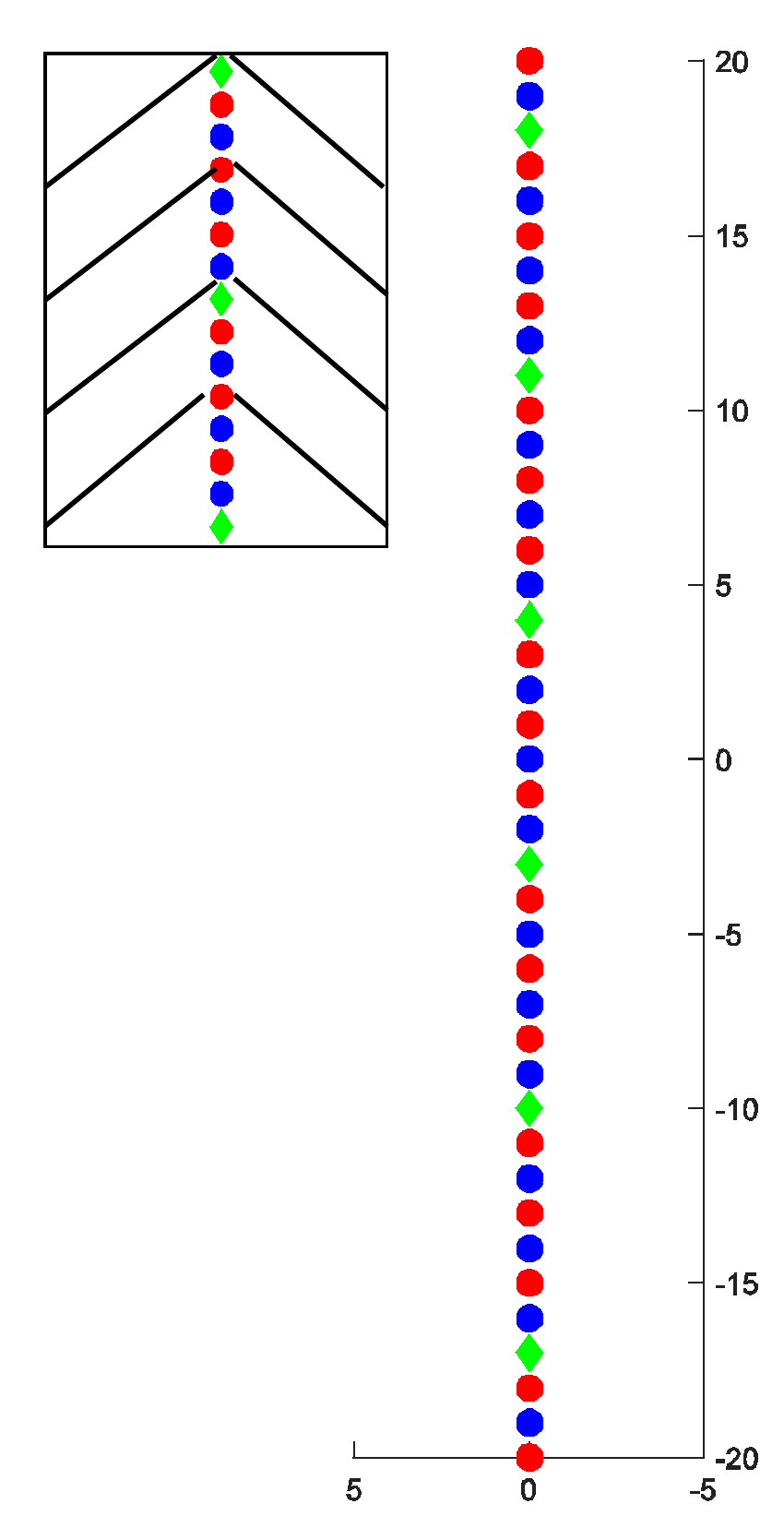}
\label{fig:defect_tilt_1}}
\caption{Defect illustrations for tilt grain boundary. In both with-dislocation case and without-dislocation case, the red dots represent positive disclinations; the blue dots represent negative disclinations and the green diamonds represent dislocations.}
\label{fig:defect_tilt}
\end{figure}

\begin{figure}
\centering
\subfigure[Stress $\sigma_{11}$ for a tilt grain boundary from with-dislocation model in the small deformation setting.]{
\label{fig:tilt_stress_linear}
\includegraphics[width=0.3\linewidth]{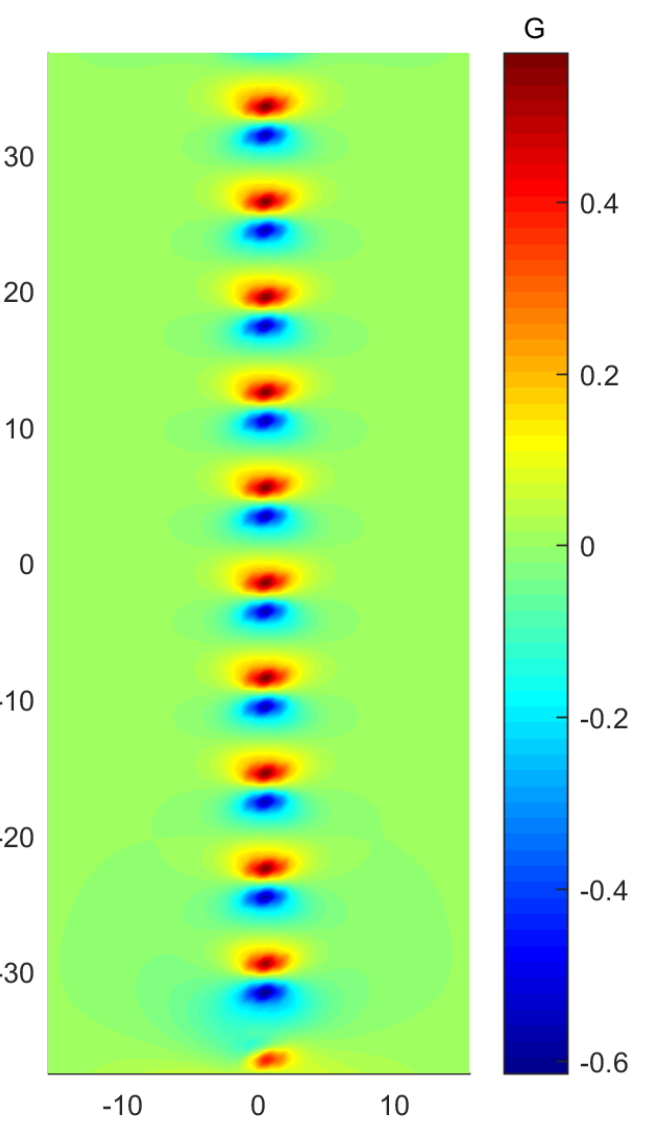}}\qquad
\subfigure[Stress $\sigma_{11}$ for a tilt grain boundary from with-dislocation model in the finite deformation setting.]{
\label{fig:tilt_stress_fe}
\includegraphics[width=0.3\linewidth]{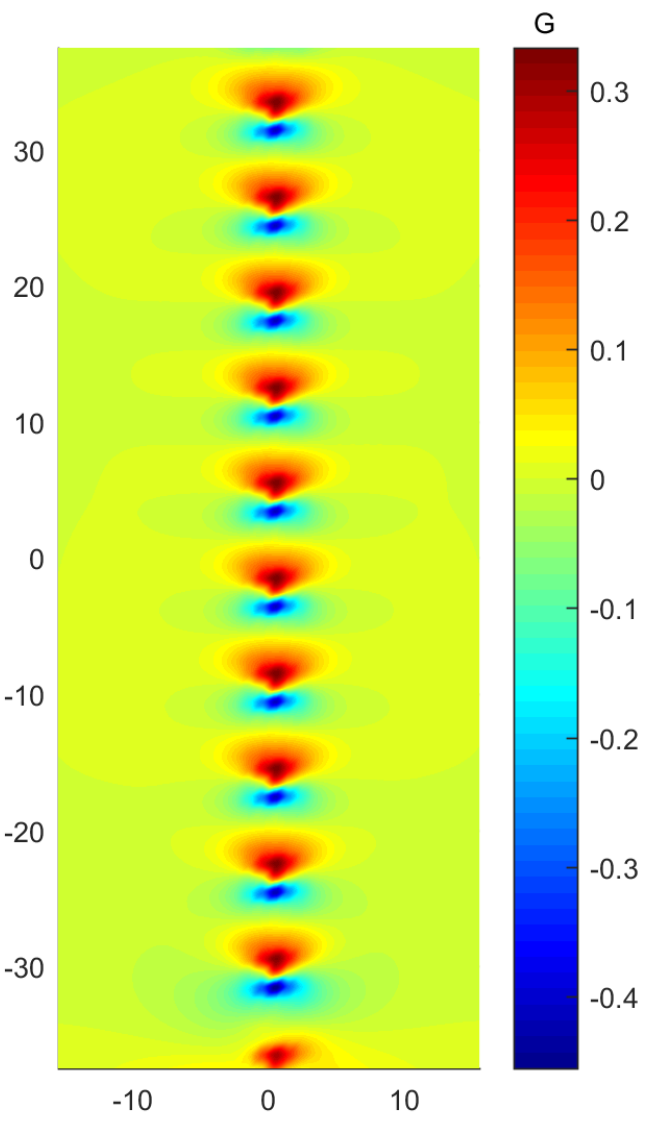}} \\
\subfigure[$\delta\sigma_{s,f}$ between small and finite deformation settings. The colormap is plotted in logarithmic scale.]{
\label{fig:tilt_stress_error} 
\includegraphics[width=0.3\linewidth]{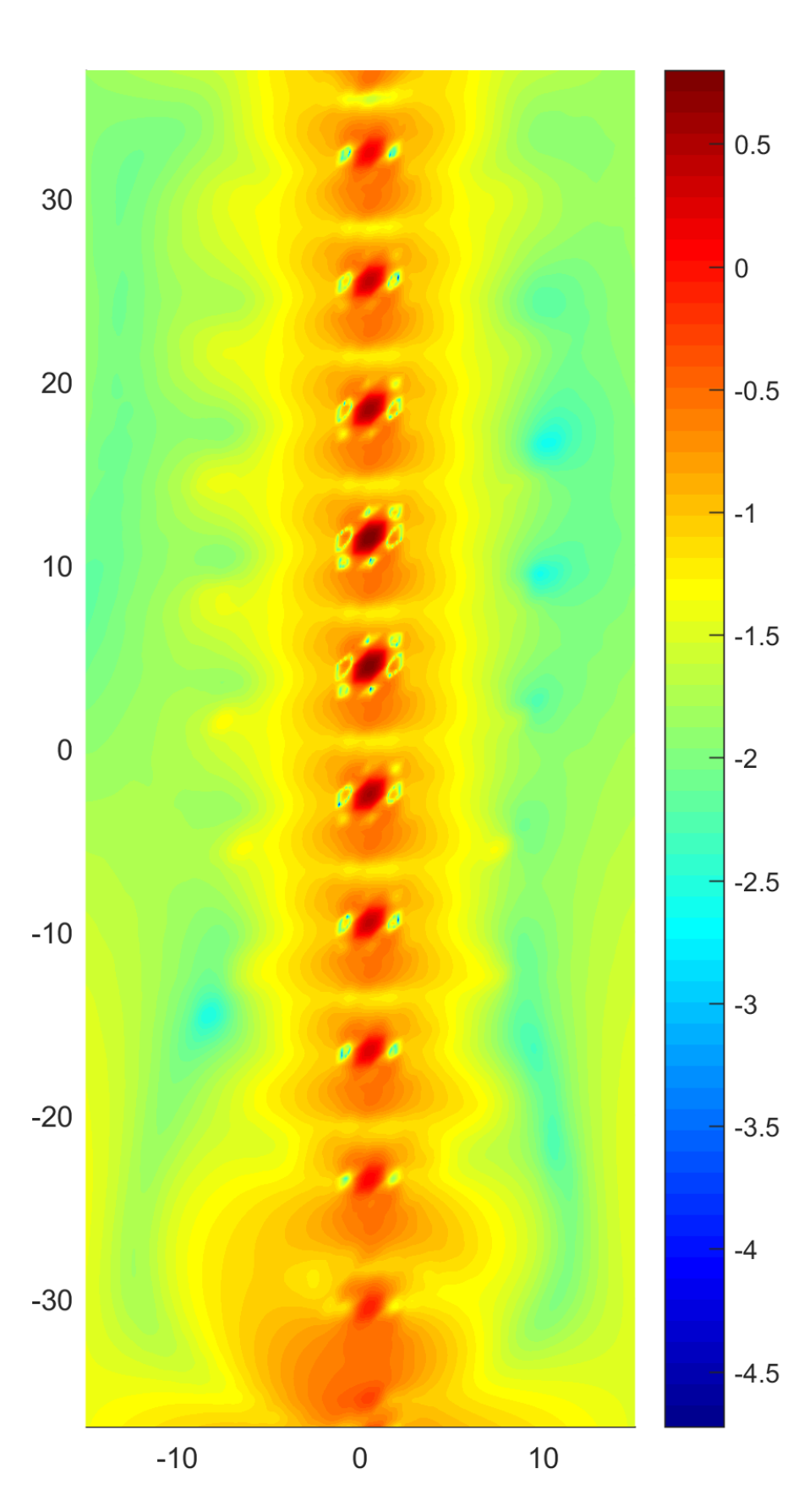}}\qquad
\subfigure[Stress field $\sigma_{11}$ for dislocation-free case.]{
\includegraphics[width=0.3\linewidth]{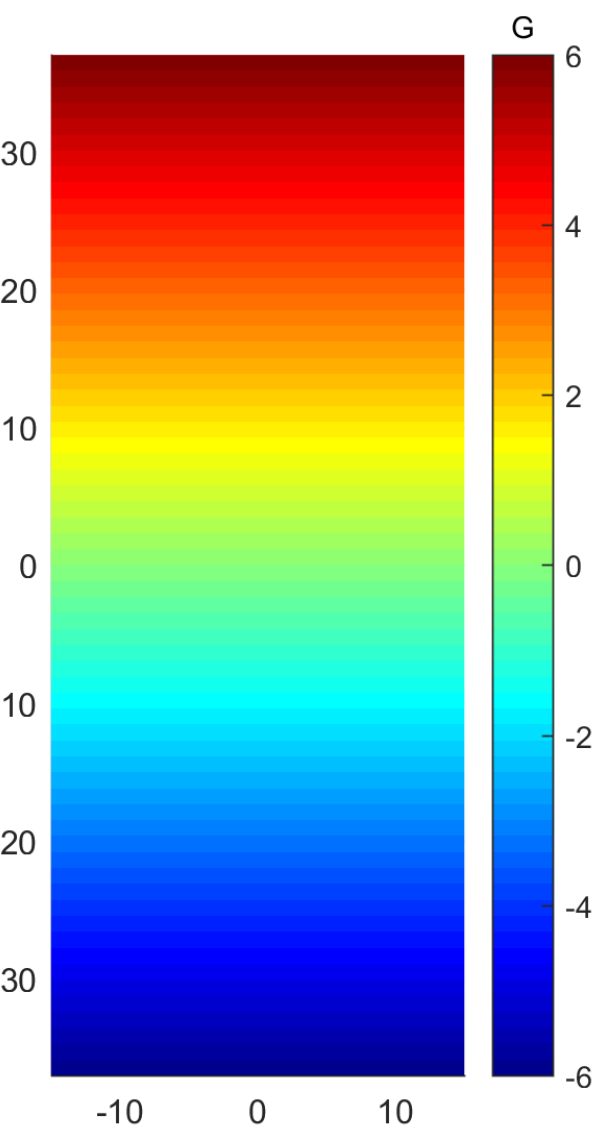}
\label{fig:tilt_comp_linear}}
\caption{Stress fields $\sigma_{11}$ of the tilt grain boundary with and without introduced dislocations, from both small and finite deformation settings. The maximum of $\delta\sigma_{s,f}$ is about $53\%$ and the mean of $\delta\sigma_{s,f}$ is $1.62\%$.}
\end{figure}

We now calculate the fields of a tilt grain boundary without dislocations as in Figure \ref{fig:tilt_phy}(b) and the tilt grain boundary with dislocations as in Figure \ref{fig:tilt_phy}(c), aiming to prove that the tilt grain boundary with dislocations in this case is a preferred state with lower energy. The crystal rotation field with respect to the interface of both sides far away from the interface in Figure \ref{fig:tilt_phy}(b) is the same as the one in Figure \ref{fig:tilt_phy}(c). To model the configuration in Figure \ref{fig:tilt_phy}(b), the grain boundary is modeled as a series of disclination dipoles as shown in Figure \ref{fig:defect_tilt_2}, where the red points represent positive disclinations and the blue points represent the negative disclinations. A Dirichlet boundary condition is applied, equivalent to a bending deformation due to an increased angle of $5^{\circ}$. Namely, the dislocation-free case in Figure \ref{fig:tilt_phy}(b) can be treated as a superposition of a grain boundary problem and an elastic bending problem. The grain boundary interface in Figure \ref{fig:tilt_phy}(c) is modeled as an array of disclination dipoles with dislocations being inserted between every three dipoles, as shown in Figure \ref{fig:defect_tilt_1}. The magnitude of the Burgers vector of the inserted dislocations is obtained from the Frank-Bilby formula $|\bfb| = \theta / d$ where $\theta$ is the additional tilt angle ($5^{\circ}$ in this problem) and $d$ is the dislocation spacing. Thus, the additional title angle is generated by the extra half planes introduced by the inserted dislocations, instead of additional elastic bending. In Figure \ref{fig:defect_tilt_1}, the red points represent positive disclinations, the blue points represent negative disclinations, and the green diamonds represent dislocations. The stress fields $\sigma_{11}$ of the with-dislocation configuration in Figure \ref{fig:tilt_phy}(c) from the small and finite deformation settings are shown in Figure \ref{fig:tilt_stress_linear} and \ref{fig:tilt_stress_fe} respectively. Figure \ref{fig:tilt_stress_error} shows $\delta\sigma_{s,f}$ between the two deformation settings. The maximum of $\delta\sigma_{s,f}$ is $53\%$ and the mean of $\delta\sigma_{s,f}$ is $1.62\%$. The stress field $\sigma_{11}$ of the dislocation-free case in Figure \ref{fig:tilt_phy}(b) is shown in Figure \ref{fig:tilt_comp_linear} and the total energy of the dislocation-free problem is $10^3$ times larger than the one in the with-dislocation case. Thus, this calculation indicates that with-dislocation case is the preferred state because of its lower total energy.

\subsection{Disconnection on a grain boundary}\label{sec:disconnection_single}

\begin{figure}
\centering
\includegraphics[width=0.8\textwidth]{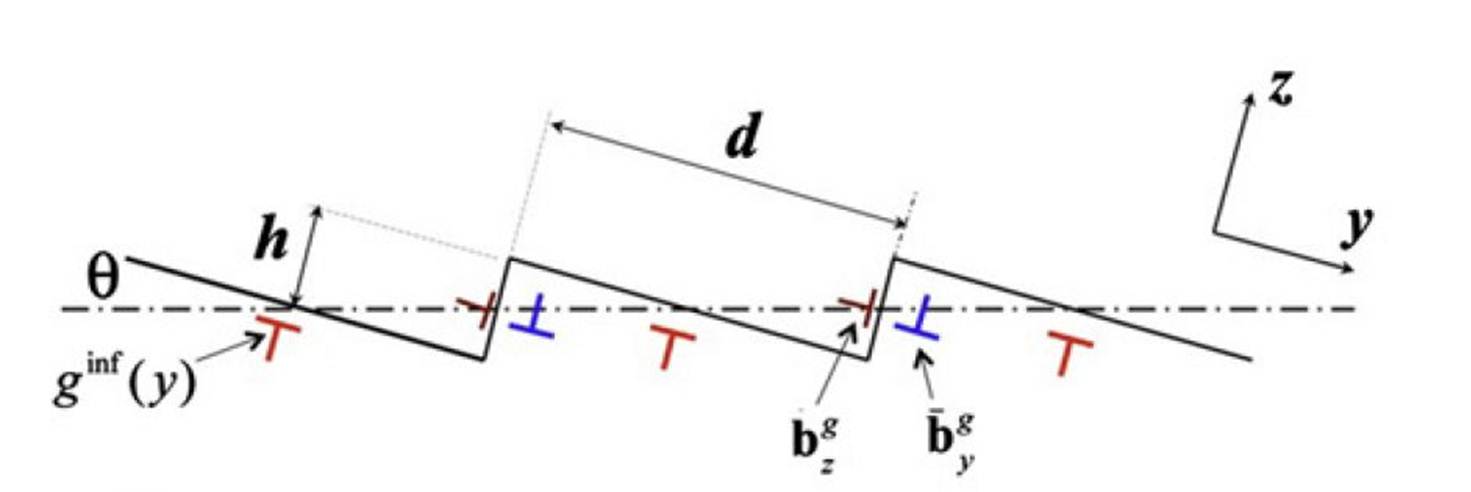}
\caption{The disconnection is modeled as a dislocation whose Burgers vector can be decomposed into the $y$ and $z$ directions. The brown dislocation represents the component in the $z$ direction, while the blue one represents the component in the $y$ direction. The red dislocations along the interface are the interface dislocations. (Figures reproduced from \cite{hirth2013interface} with permission from Elsevier.)}
\label{fig:pond_phy}
\end{figure}

\begin{figure}
\centering
\subfigure[The schematic illustration of two crystals before bonding.]{
\includegraphics[width=0.45\textwidth]{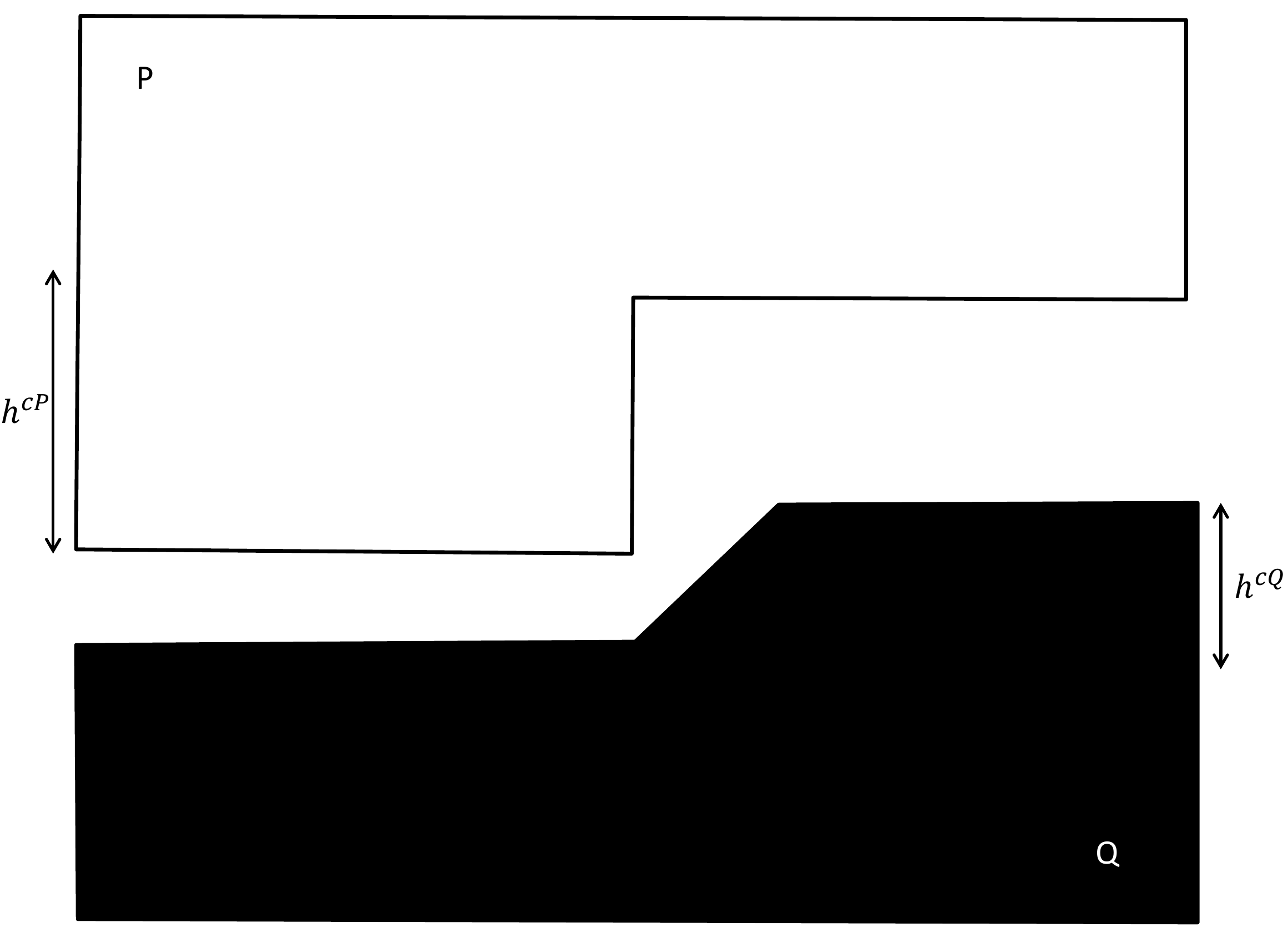}
\label{fig:pond_terrace_1}}\qquad 
\subfigure[The configuration after bonding, with a disconnection formed.]{
\includegraphics[width=0.45\textwidth]{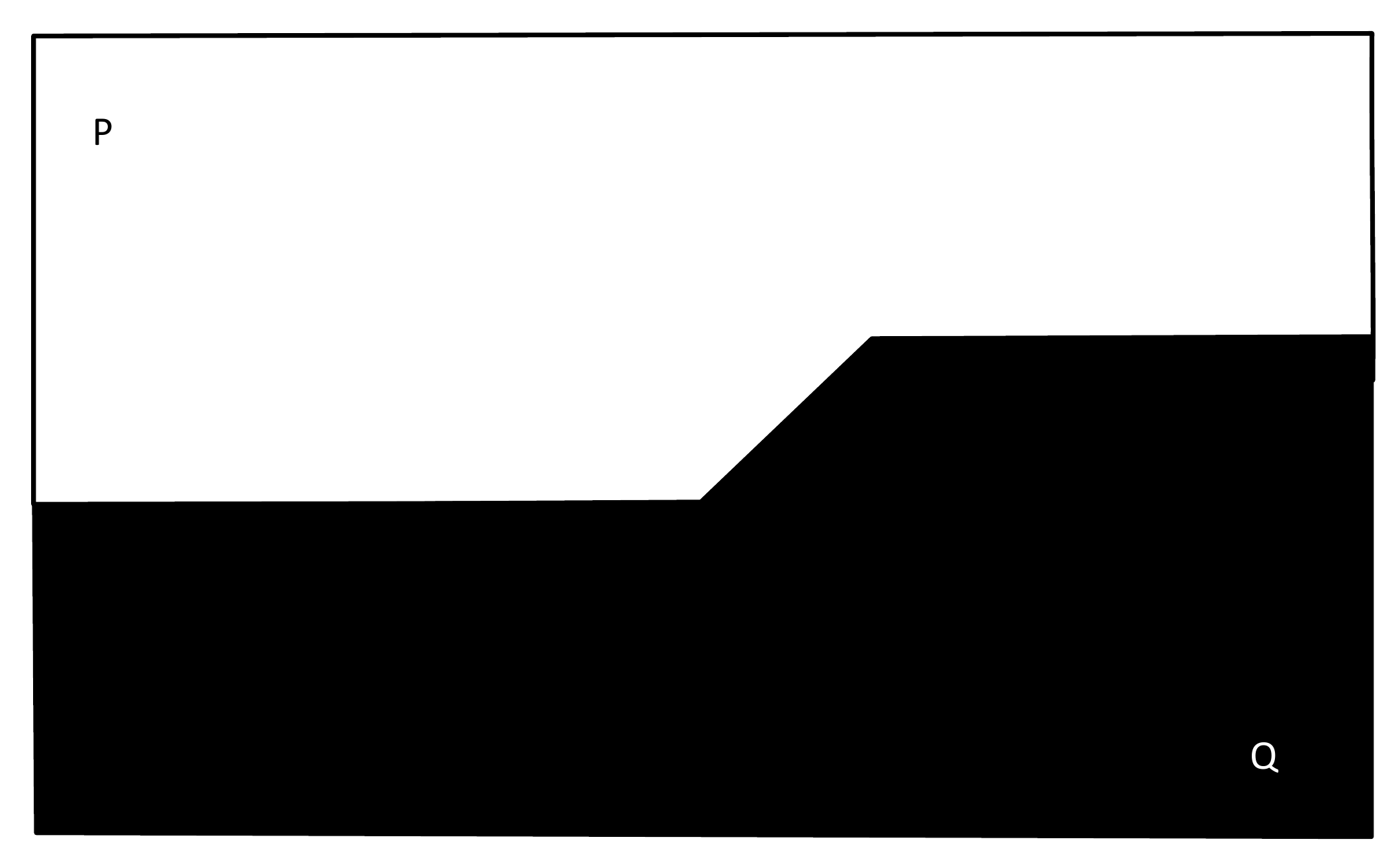}
\label{fig:pond_terrace_2}}
\caption{Schematic of a grain boundary with terraces and a disconnection. Since the lattice vectors of the two crystals do not match, a step is formed after bonding. (Figures reprinted from \cite{hirth2013interface} with permission from Elsevier.)}
\label{fig:pond_terrace}
\end{figure}

A disconnection is the region that connects two parallel grain boundary segments, referred to as terraces, that do not belong to a common plane. Extensive work on grain boundary disconnections have been done by Hirth, Pond and co-workers \cite{hirth2013interface, hirth2006disconnections}. They described the entire grain boundary as a series of terraces joined by disconnections. Figure \ref{fig:pond_phy} from \cite{hirth2013interface} shows the terrace model and Figure \ref{fig:pond_terrace} shows a schematic for understanding the reason for the occurrence of a disconnection. The terraces are assumed to contain misfit dislocations, and the disconnections are interpreted as additional dislocations located at the steps joining the terraces.

\begin{figure}
\centering
\subfigure[The current configuration of a disconnection on the grain boundary, represented by a disclination dipole and a dislocation.]{
\label{fig:disconnection_1}
\includegraphics[width=0.45\linewidth]{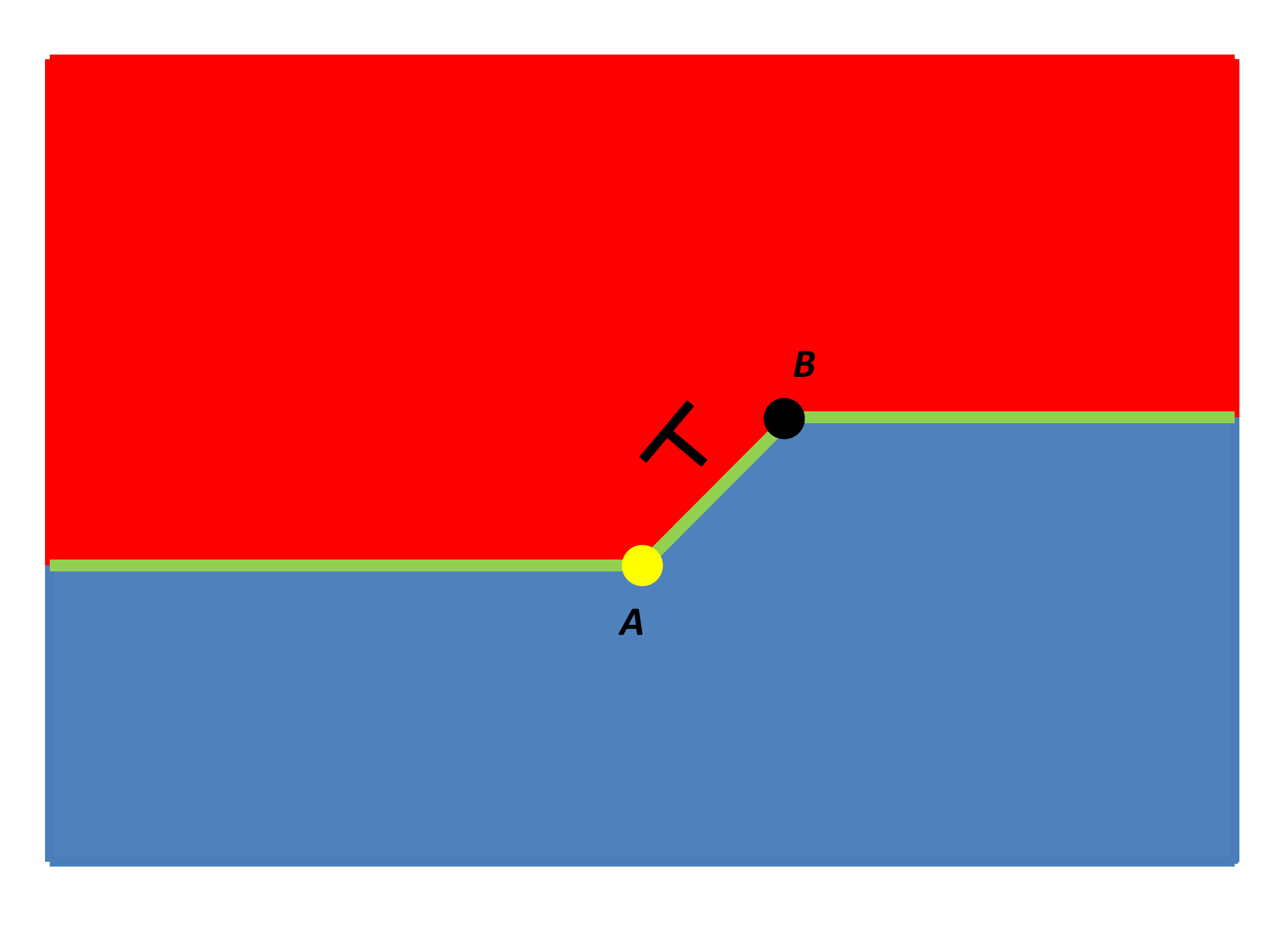}}\qquad
\subfigure[Cut the interface from the right end to $B$ and relax the negative disclination. An overlap wedge appears with angle $\Omega$.]{
\label{fig:disconnection_2}
\includegraphics[width=0.45\linewidth]{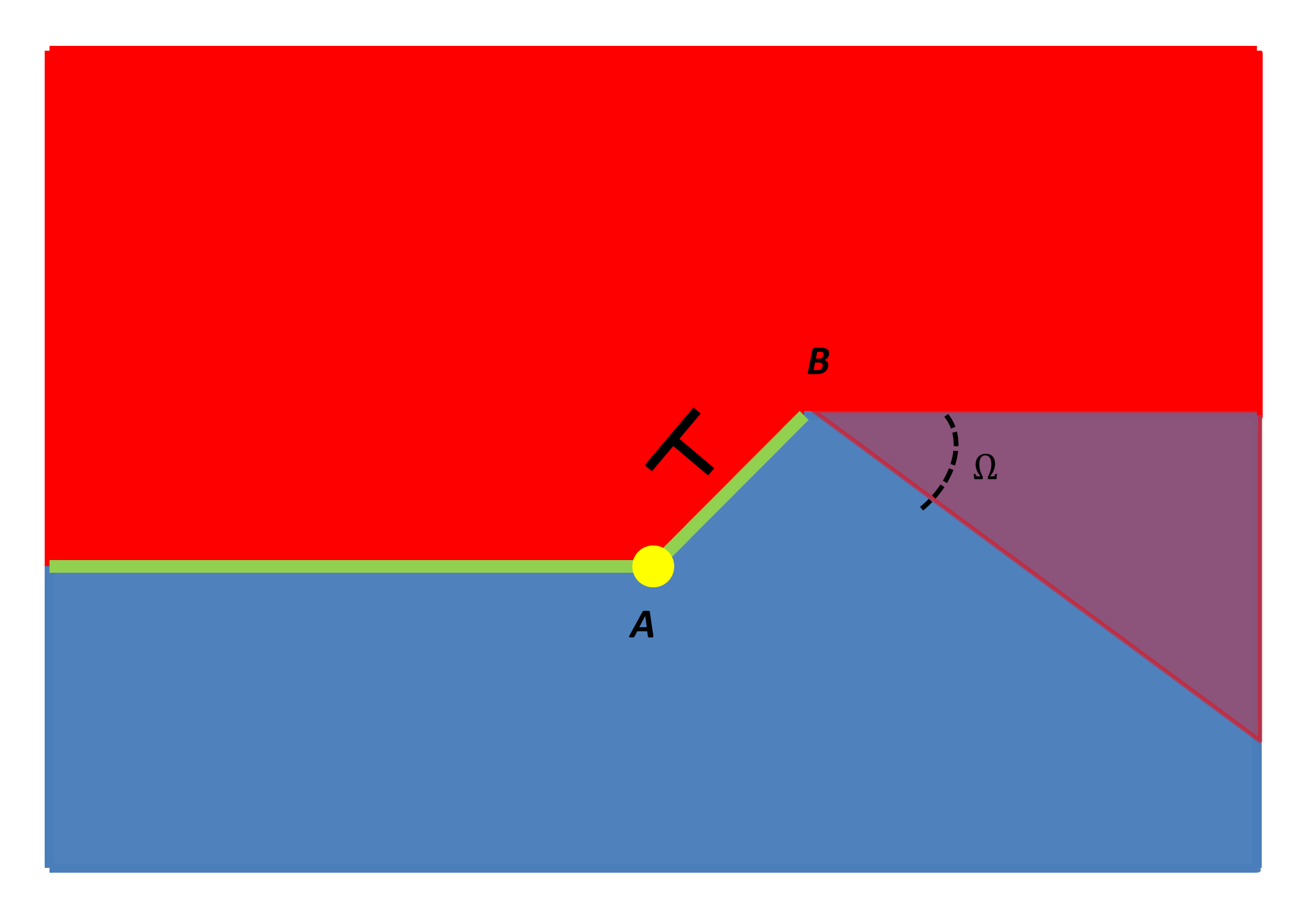}}
\subfigure[Cut the interface from $B$ to $A$ and relax the positive disclination.The overlap wedge is eliminated while a gap wedge is formed.]{
\label{fig:disconnection_3}
\includegraphics[width=0.45\linewidth]{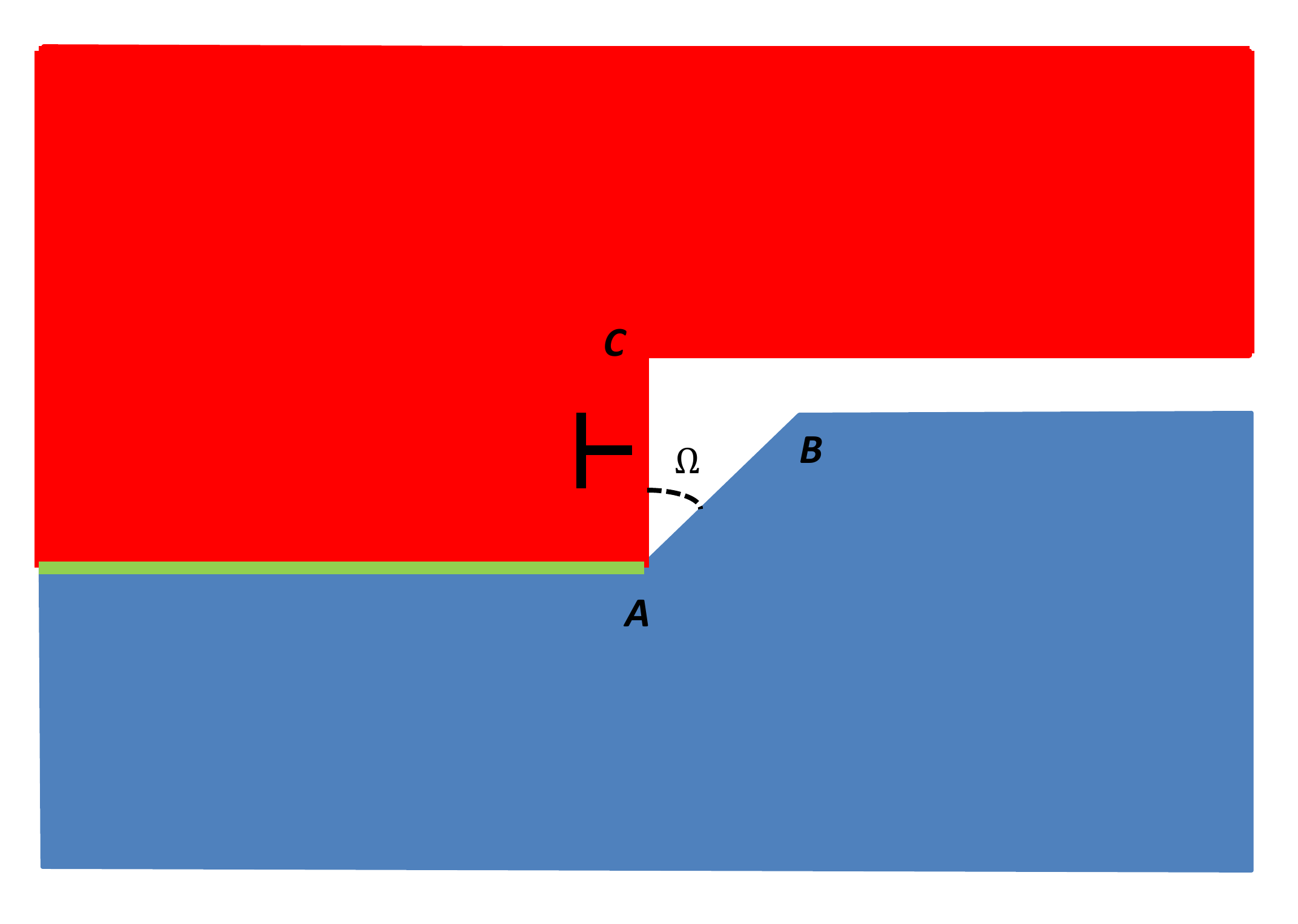}}\qquad
\subfigure[Relax the dislocation at $C$. The red part moves upwards.]{
\label{fig:disconnection_4}
\includegraphics[width=0.45\linewidth]{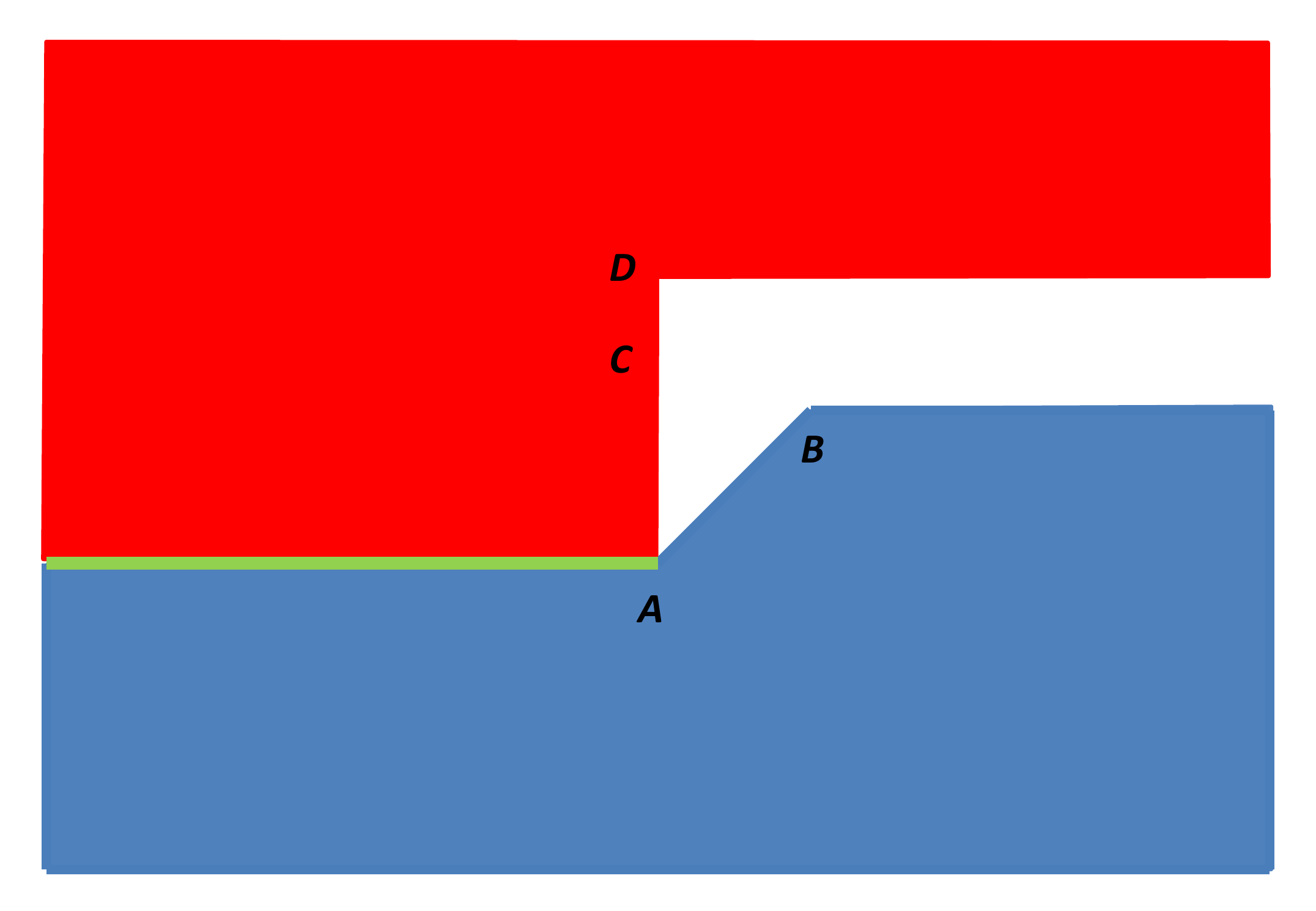}}
\caption{The mechanism to represent a disconnection by a disclination dipole and a dislocation, starting from the current configuration to the reference configuration.}
\label{fig:disconnection_dipole}
\end{figure}

In this work, a disclination dipole model is introduced  and computed to describe the grain boundary disconnection discussed in \cite{hirth2013interface} shown in Figure \ref{fig:pond_terrace}. Figure \ref{fig:disconnection_dipole} shows the thought experiment for representing the disconnection by a disclination dipole and a dislocation. According to g.disclination theory, we start from the current configuration of a disconnection that is represented by a disclination dipole and a dislocation, as shown in Figure \ref{fig:disconnection_1}. The red part is one grain and the blue part is another grain. The black dot at $A$ represents a negative disclination and the yellow dot at $B$ represents a positive disclination. Both disclinations have the same Frank vector magnitude $\Omega$ with opposite signs. The disclination density for each disclination is assumed to be derived from the difference of two (inverse) rotation matrices. The dislocation is located at $B$. The green lines represents the interface of the grain boundary. To get the reference configuration (the stress-free configuration shown in Figure \ref{fig:pond_terrace}(a), we need to relax the body by the following steps: 

\begin{itemize}
\item Cut the interface from the right end to $B$ and relax the negative disclination at $B$. Thus, the red part rotates clockwise by $\Omega$, generating an overlap wedge. The configuration after this step is shown in Figure \ref{fig:disconnection_2}.
\item Cut the interface from the $B$ to $A$ and relax the positive disclination at $A$. The red crystal rotates anticlockwise by $\Omega$. Therefore, the point $B$ on the red crystal moves to $C$ and there is now a gap wedge $CAB$. Furthermore, the overlap wedge generated by relaxing the negative disclination is counteracted by the opposite rotation in this step, as shown in Figure \ref{fig:disconnection_3}.
\item We now assume that the (true, F-S) Burgers vector of the dislocation at $B$ in Fig. \ref{fig:disconnection_1} measured on the relaxed configuration  is given by the vector joining $C$ to $D$ in Fig. \ref{fig:disconnection_4}. We now relax this dislocation.
\end{itemize}

\begin{figure}
\centering
\subfigure[The composite Burgers vector diagram of a disconnection from a disclination dipole and a dislocation.]{
\label{fig:disconnection_burgers_1}
\includegraphics[width=0.45\linewidth]{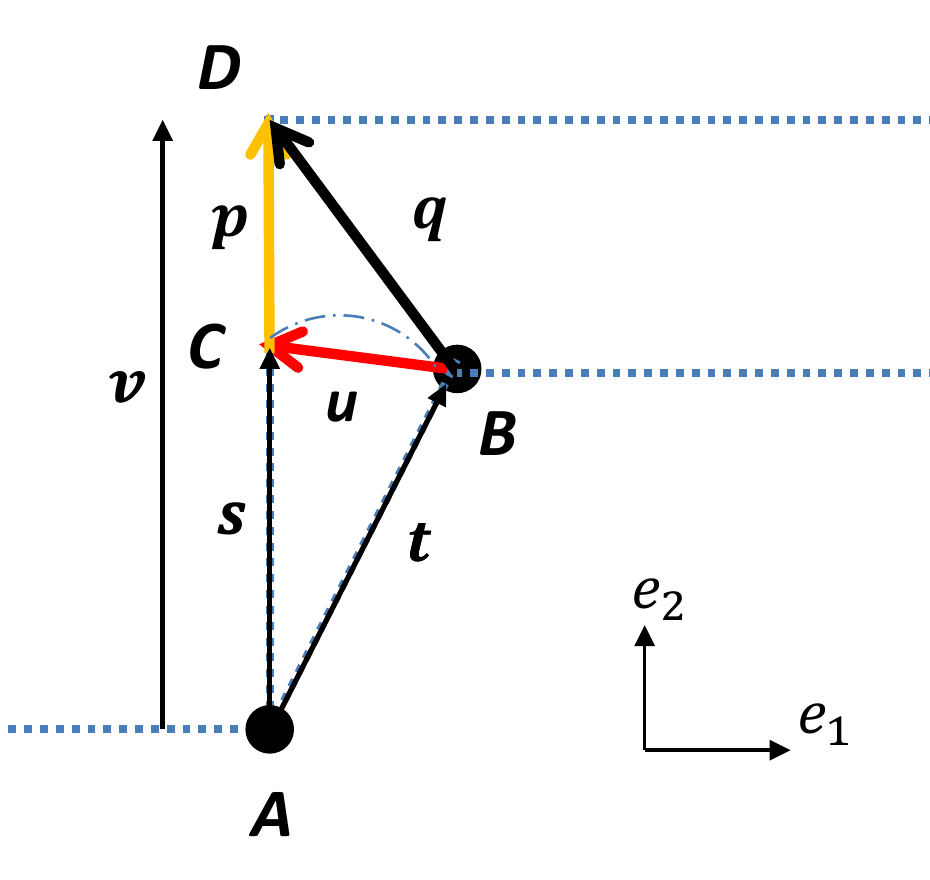}}
\qquad
\subfigure[The composite Burgers vector diagram for the small deformation case.]{
\label{fig:disconnection_burgers_2}
\includegraphics[width=0.45\linewidth]{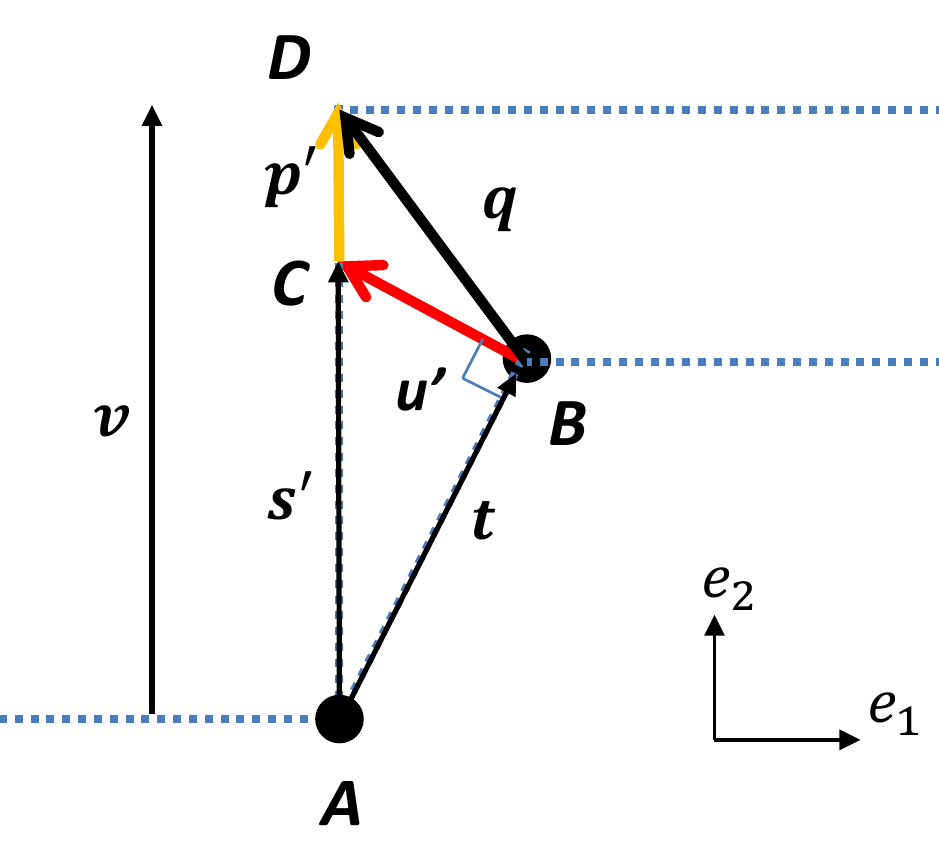}}
\caption{The composite Burgers vector diagram of a disconnection. The composite Burgers vector from the disclination dipole and the dislocation give the same vector as in \cite{hirth2013interface}.}
\label{fig:disconnection_dipole_burgers}
\end{figure}

Figure \ref{fig:disconnection_dipole_burgers} shows the composite Burgers vector obtained from the above relaxation as a superposition of the Burgers vectors of the disclination dipole and the dislocation. It turns out that the Burgers vector from our model matches with the Burgers vector from \cite{hirth2013interface}. Figure \ref{fig:disconnection_burgers_1} is the Burgers vector diagram based on g.disclination theory - disclination densities are based on the inverse rotation matrices. Considering finite deformations, $\bft$ is rotated to $\bfs$ by relaxing the disclinations and $\bfb_{dipole}^f$ is denoted as the Burgers vector of the disclination dipole. Denote the i-elastic 1-distortion difference of the positive disclination as $\bfDelta$. The dipole separation vector in Figure \ref{fig:disconnection_burgers_1} is $\bft$. Based on a result in \cite[Eqn(33)]{zhang_acharya_2016}, the Burgers vector of the g.disclination dipole $\bfb_{dipole}^f$ is given as 
\[
\bfb_{dipole}^f = \bfDelta \bft.
\]

Assuming a completely in-plane problem, denote the rotation tensor acting on $\bft$ to produce $\bfs$ in Figure \ref{fig:disconnection_burgers_1} as $\bfR$. The vector $\bft$ on the blue crystal is assumed to  remain unchanged under the whole relaxation. Thus the i-elastic, 1-distortion difference is given by
\[
\bfDelta = \bfR-\bfI,
\]
with the matrix of $\bfR$ (in any orthonormal basis) given by
\[
\begin{bmatrix}
\cos \Omega && -\sin \Omega \\
\sin \Omega && \cos \Omega
\end{bmatrix}.
\]

With reference to Fig. \ref{fig:disconnection_burgers_1}, $\bfb_{dipole}^f$ can be written as 
\[
\bfb_{dipole}^f = (\bfR-\bfI)\bft = \bfR\bft - \bft = \bfs - \bft = \bfu
\]

From the description shown in Figure \ref{fig:disconnection_dipole}, the Burgers vector of the dislocation, $\bfb_{dislocation}^f$, can be written as $\bfb_{dislocation}^f = \bfv - \bfs = \bfp$. Therefore, the total Burgers vector of the disconnection is given by
\[
\bfb_{total}^f = \bfb_{dipole}^f+\bfb_{dislocation}^f = \bfu + \bfp = \bfq,
\]
matching the result from \cite{hirth2013interface}. Figure \ref{fig:disconnection_burgers_2} is the Burgers vector diagram for the small deformation case. In this approximation, the dipole Burgers vector, $\bfb_{dipole}^s$, is given by \cite[Eqn(7)]{zhang_acharya_2016}
\[
\bfb_{dipole}^s = \bfOmega\times\bft, 
\]
where $\bfOmega$ is the Frank vector of the positive disclination and is given by $tan \Omega \bfe_3$ by (\ref{eqn:strength}). Then, $\bfb_{dipole}^s$ can be written as 
\[
\bfb_{dipole}^s  = \bfu',
\]
where $\bfu'$ is a vector perpendicular to $\bft$ with length $tan \Omega |t|$. The `rotated' image of $\bft$ is $\bfs' = \bft+\bfu'$. The Burgers vector of the dislocation $\bfb_{dislocation}^s$ is ,
\[
\bfb_{dislocation}^s = \bfv - \bfs'  = \bfp'
\]
Thus, the total Burgers vector $\bfb_{total}^s$ is given as 
\[
\bfb_{total}^s = \bfb_{dipole}^s+\bfb_{dislocation}^s = \bfu' + \bfp'.
\]

Note that if we use the $\bfb_{dislocation}^f$ as the Burgers vector of the dislocation in the small deformation case, since $\bfp \ne \bfp'$, $\bfb_{total}^s = \bfu'+\bfp \neq \bfq$. Writing $\bft=t_1\bfe_1 + t_2 \bfe_2$ w.r.t any orthonormal basis, we have
\[
\bfu = \bfR \bft - \bft = (\cos\Omega t_1 - \sin \Omega t_2 - t_1)\bfe_1 + (\sin \Omega t_1 + \cos \Omega t_2 - t_2)\bfe_2.
\]
For $|\Omega| \ll 1$, $\cos \Omega \approx 1$ and $\sin \Omega \approx \tan \Omega$, we have 
\[
\bfu \approx -tan \Omega t_2 \bfe_1 + tan \Omega t_1 \bfe_2 = \bfu'.
\]
In addition, we can have the following approximations
\begin{eqnarray*}
\bfs' = \bfu' + \bft \approx \bfu+\bft = \bfs \\
\bfp' = \bfv - \bfs' \approx \bfv - \bfs = \bfp \\
\bfu' + \bfp' \approx \bfu' + \bfp \approx \bfu + \bfp = \bfq.
\end{eqnarray*}
and the total Burgers vector of the disconnection in the small deformation setting closely approximates the finite deformation result for small disclination strengths $\Omega$.

\subsubsection{`Topological equaivalence $\neq$ energetic equivalence'}
\begin{figure}
\centering
\includegraphics[width=0.95\linewidth]{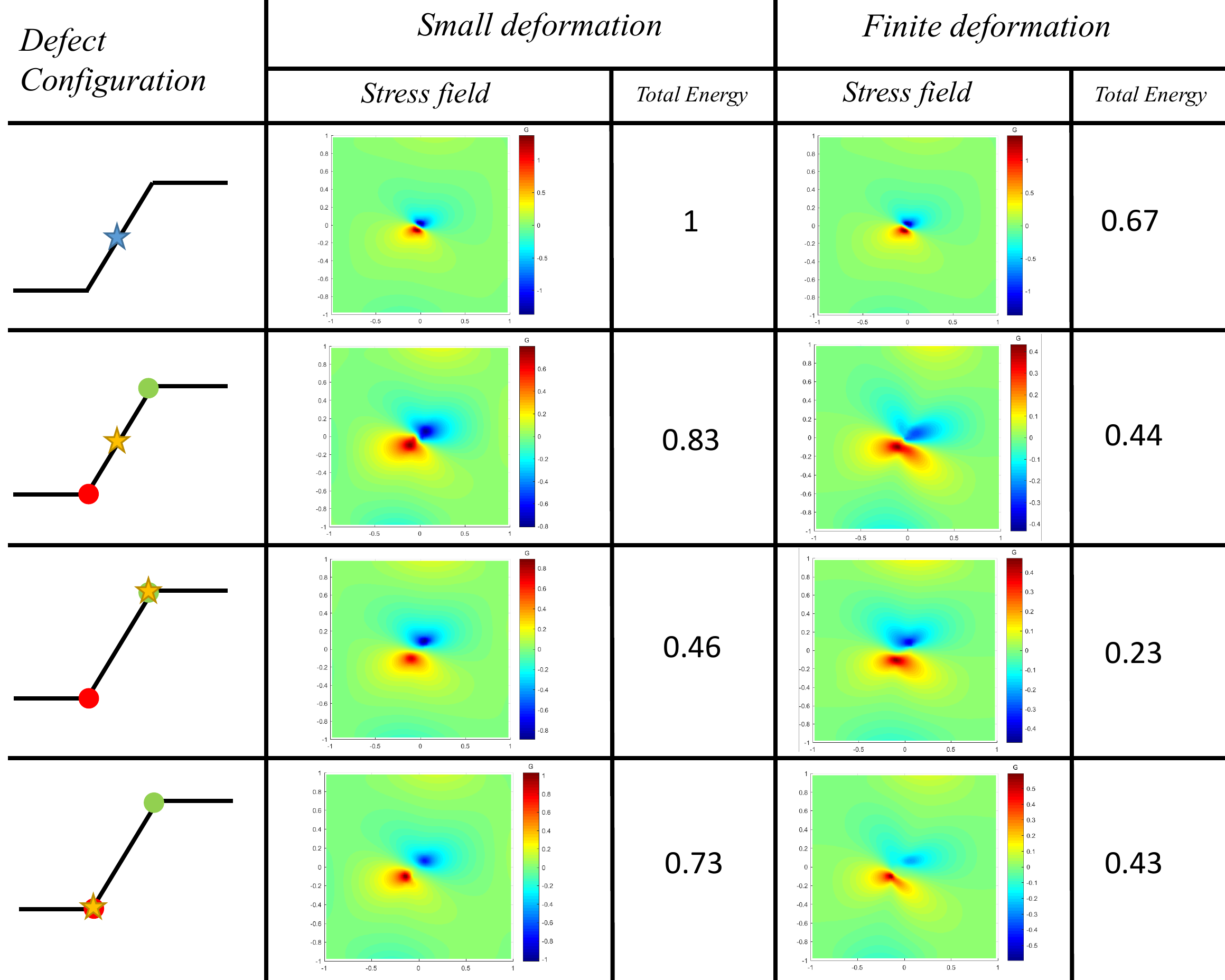}
\caption{A comparison of the stress fields $\sigma_{11}$ and the total energies from both the small and finite deformation settings for different defect configurations. The red dots are positive disclinations, the green dots are negative disclinations and the yellow star are dislocations. The blue star is the dislocation with the same overall Burgers vector as other cases. Although the overall Burgers vectors are same in all cases, the stress fields and the total energies are quite different.}
\label{fig:core_table}
\end{figure}

Figure \ref{fig:core_table} shows the stress field and total energy comparisons between the disconnection represented by an effective dislocation with Burgers vector $\bfq$ (Fig. \ref{fig:disconnection_dipole_burgers}), and three different disclination dipole-dislocation representations of the disconnection where the dislocation is prescribed at different locations along the disconnection step. The total Burgers vector is identical for all cases involved.

In Figure \ref{fig:core_table}, the green points are negative disclinations, the red points are positive disclinations, the yellow stars are the disconnection dislocations and the blue star is the dislocation of strength equal to the overall disconnection Burgers vector. In the cases with the disclination dipole (the second, third, and fourth rows in Figure \ref{fig:core_table}), the misorientations of all disclinations are $45^\circ$ and the magnitude of Burgers vector of the dislocation is $2$ lattice constants. In the case without the disclination dipole, namely the first row of Figure \ref{fig:core_table}, the magnitude of the Burgers vector is $5$ lattice constants, based on the explanation in Figure \ref{fig:disconnection_dipole_burgers}. The first column of Figure \ref{fig:core_table} shows different defect configurations; the second column of Figure \ref{fig:core_table} is the stress field $\sigma_{11}$ and the total energy from the small deformation setting; and the last column is the stress field $\sigma_{11}$ and the total energy from the finite deformation setting. These results show that although the Burgers vectors, for every circuit encircling the disconnection step, for all four cases are the same, the stress fields and the total energy are quite different. Furthermore, the total energy of the configuration with the dislocation being coincident with the negative disclination is the lowest.

In general, we find that the (outside core) topologically equivalent, dislocation-only configuration is the highest energy configuration, this being similar to the finding of \cite{kamien_smectics_2017} in the context of smectics.

\subsection{A disconnected grain boundary with misfit dislocations on terraces}\label{sec:disconnected_grain_misfit}

\begin{figure}
\centering
\subfigure[The misfit configuration of two grains with different lattice parameters.]{
\includegraphics[width=0.45\textwidth]{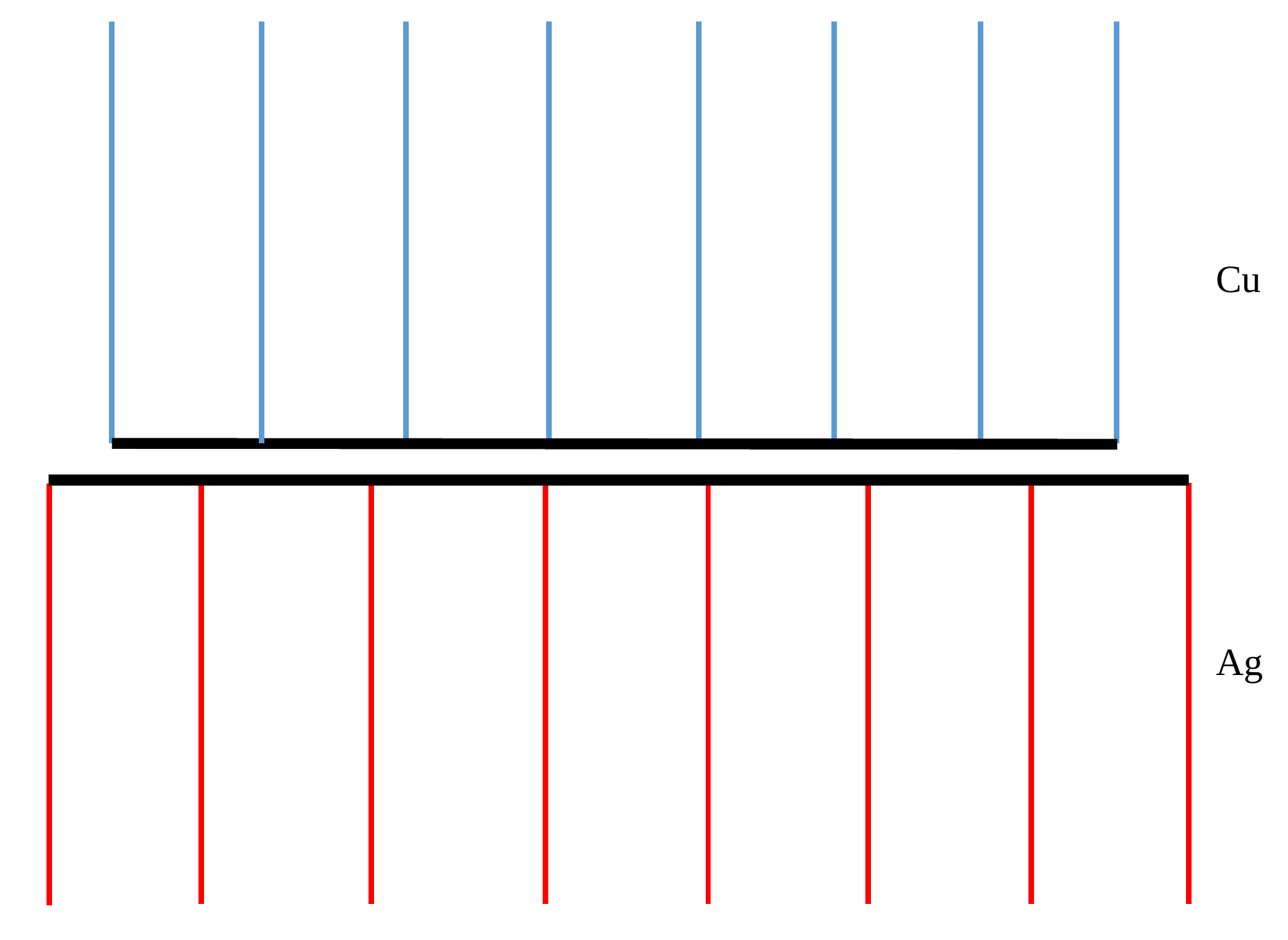}
\label{fig:misfit_1}
} \qquad
\subfigure[The configuration after introducing an extra half plane (a dislocation).]{
\includegraphics[width=0.45\textwidth]{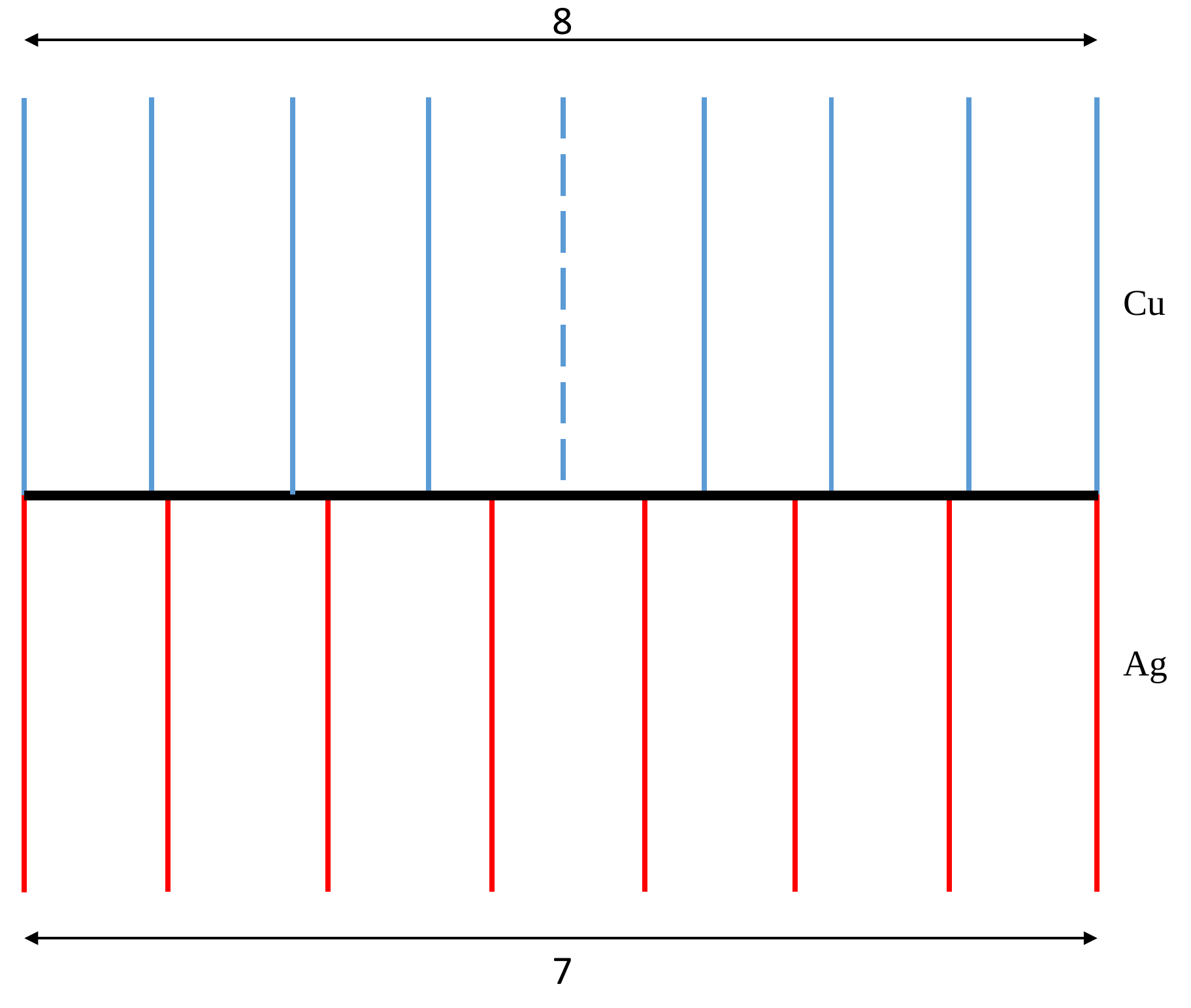}
\label{fig:misfit_2}
}
\caption{Defining a misfit dislocation. By introducing a dislocation, the far field incoherency strain of two misfit grains is eliminated.}
\label{fig:misfit}
\end{figure}

\begin{figure}
\centering
\includegraphics[width=0.5\linewidth]{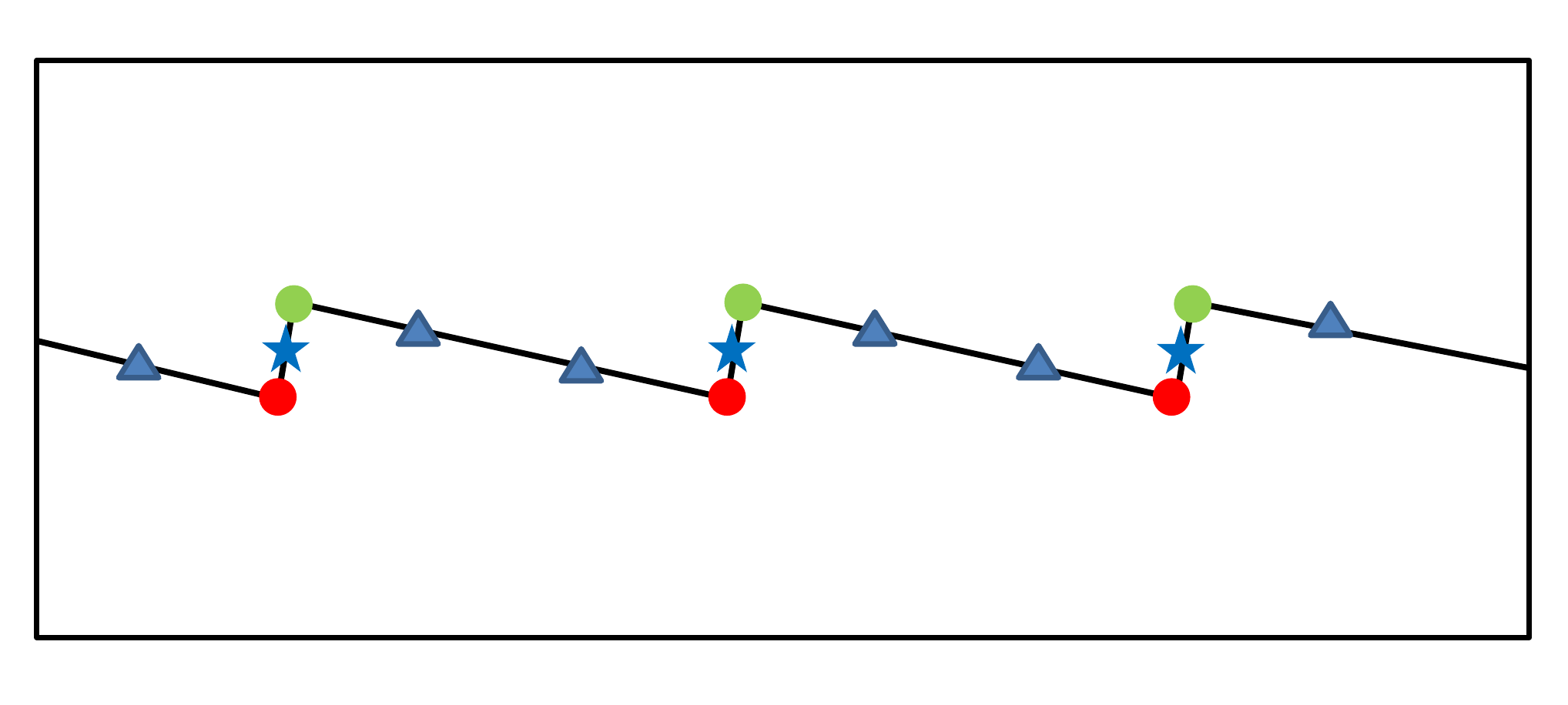}
\caption{Defect configuration of the incoherent grain boundary disconnection. The black lines within the body are grain boundary interfaces; the red dots are positive disclinations; the green dots are negative disclinations; the blue stars are disconnection dislocations; and the blue triangles are misfit dislocations.}
\label{fig:pond_defect_ incoherent}
\end{figure}

\begin{figure}
\centering
\subfigure[Stress $\sigma_{11}$ for the incoherent grain boundary disconnection in the small deformation setting.]{
\label{fig:pond_incoherent_stress_small}
\includegraphics[width=0.45\linewidth]{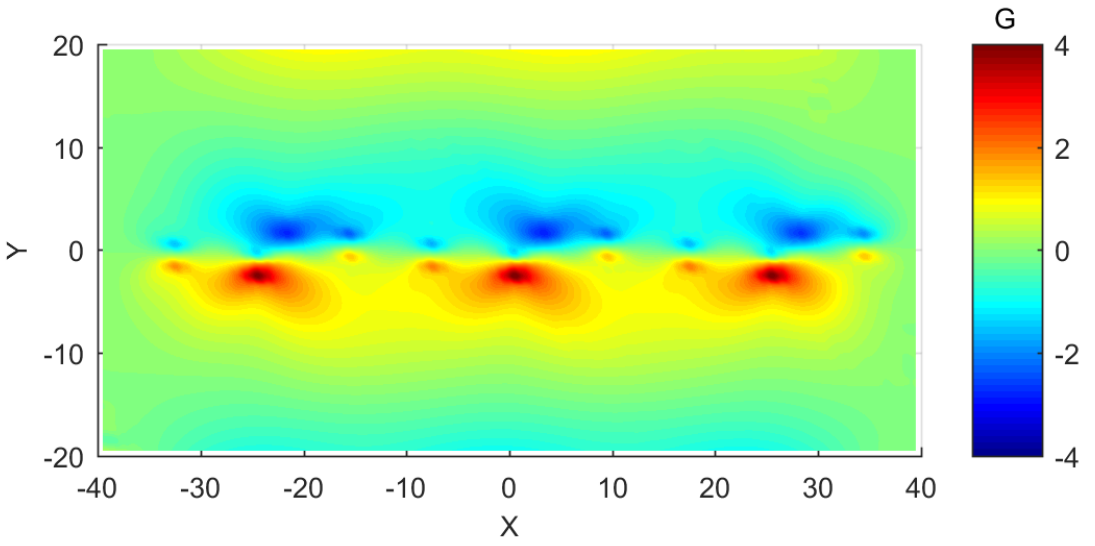}}\qquad
\subfigure[Stress $\sigma_{11}$ for the incoherent grain boundary disconnection in the finite deformation setting.]{
\label{fig:pond_incoherent_stress_finite}
\includegraphics[width=0.45\linewidth]{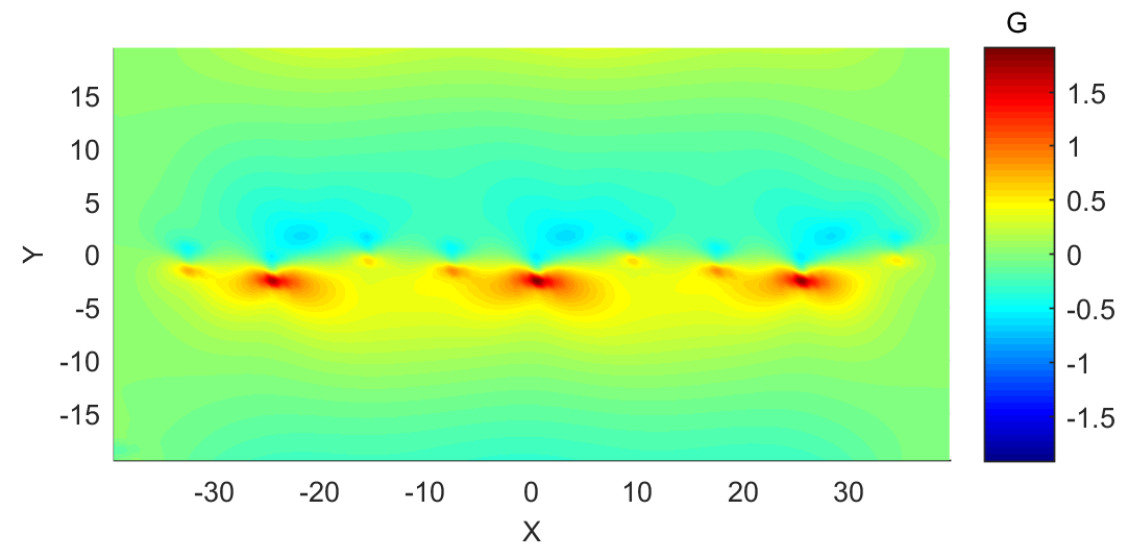}}
\subfigure[$\delta\sigma_{s,f}$ between the small and finite deformation settings.]{
\label{fig:pond_incoherent_stress_error}
\includegraphics[width=0.45\linewidth]{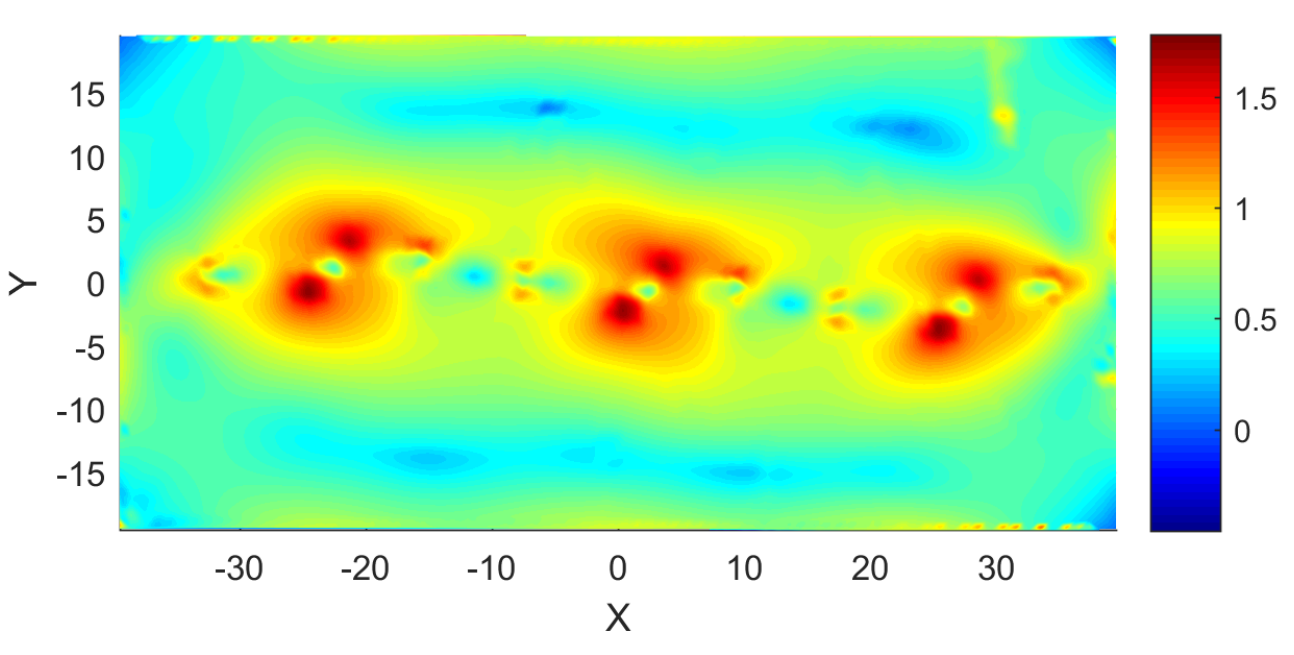}}
\caption{Stress field $\sigma_{11}$ for the incoherent grain boundary disconnection and the stress field difference $\delta\sigma_{s,f}$. The maximum of $\delta\sigma_{s,f}$ is about $170\%$ and the mean of $\delta\sigma_{s,f}$ is $0.67\%$.}
\label{fig:pond_incoherent_stress}
\end{figure}

We utilize the arguments of Section \ref{sec:disconnection_single} to model a grain boundary with a series of disconnections. The disconnections are represented as a series of disclination dipoles with the dislocations. The misorientation angle for every disclination is set to be $45^\circ$. The Burgers vector of the dislocation in each disconnection is assumed to be $\bfb=-0.5\bfe_1 -0.5\bfe_2$ (that in reality is to be determined by the crystal structure of the constituent crystals forming the interface). In addition, we consider the terraces as containing misfit dislocations. Figure \ref{fig:pond_defect_ incoherent} is the defect configuration of the incoherent grain boundary with the disconnections, where the incoherency is represented by the misfit dislocations whose Burgers vectors are determined by the crystal structure of the interface. In this calculation, we assume the two grain materials are $Cu$ and $Ag$, with the ratio of the lattice parameter being $a_{Cu}/a_{Ag} = 36/41$ based on \cite{wang2012nucleation}. Figure \ref{fig:misfit_1} shows two grains before bonding together, where the top is $Cu$ and the bottom is $Ag$. Based on the lattice parameter ratio, it can be shown that the far field incoherency strain can be eliminated by introducing an extra half $Cu$ plane every seven lattice constants, as shown in Figure \ref{fig:misfit_2}. Therefore, the Burgers vector of the misfit dislocation is one lattice constant and the interval distance between the misfit dislocations is seven lattice constants. In Figure \ref{fig:pond_defect_ incoherent}, the black lines represent the terraces; the red points are positive disclinations; the green points are negative disclinations; the blue stars are the disconnection dislocations; and the blue triangles are the misfit dislocations.  Figure \ref{fig:pond_incoherent_stress} displays the stress field $\sigma_{11}$ for this configuration in both the small and finite deformation settings. The maximum of $\delta\sigma_{s,f}$ is about $170\%$ and the mean of $\delta\sigma_{s,f}$ is $0.67\%$.

Since defect dynamics depends upon the local stress field, such difference may be expected to have significant impacts for kinetics.

\subsubsection{Disconnected grain boundary with misfit dislocations on terrace separating anisotropic bulk crystals}\label{sec:anisotropic_grain}
Anisotropic elasticity is the physically natural elastic response of single crystals across a grain boundary. Here, we study a grain boundary with anisotropic elastic bulk response and compare the results with the isotropic case. Consider a grain boundary where the misdistortion across the $Cu-Ag$ interface is the same as the one in Sec. \ref{sec:disconnected_grain_misfit}. The specification of the anisotropic stiffness tensors for the top and bottom crystals is described in Appendix \ref{sec:append_2}. Figure \ref{fig:pond_incoherent_stress_anisotropic} shows the stress field $\sigma_{11}$ with anisotropy from the finite deformation settings and Figure \ref{fig:pond_incoherent_stress_anisotropic_diff} shows the difference between the isotropic finite deformation stress and the anisotropic finite deformation stress, $\delta\sigma_{iso,aniso}$, following the definition (\ref{eqn:difference}). The maximum of $\delta\sigma_{iso,aniso}$ is $320\%$, and the mean of $\delta\sigma_{iso,aniso}$ is $12.3\%$.

\begin{figure}
\centering
\subfigure[Stress $\sigma_{11}$ for a grain boundary with dislocations and disconnections separating crystals with anisotropic bulk elastic properties in the  finite deformation setting.]{
\label{fig:pond_incoherent_stress_anisotropic}
\includegraphics[width=0.45\linewidth]{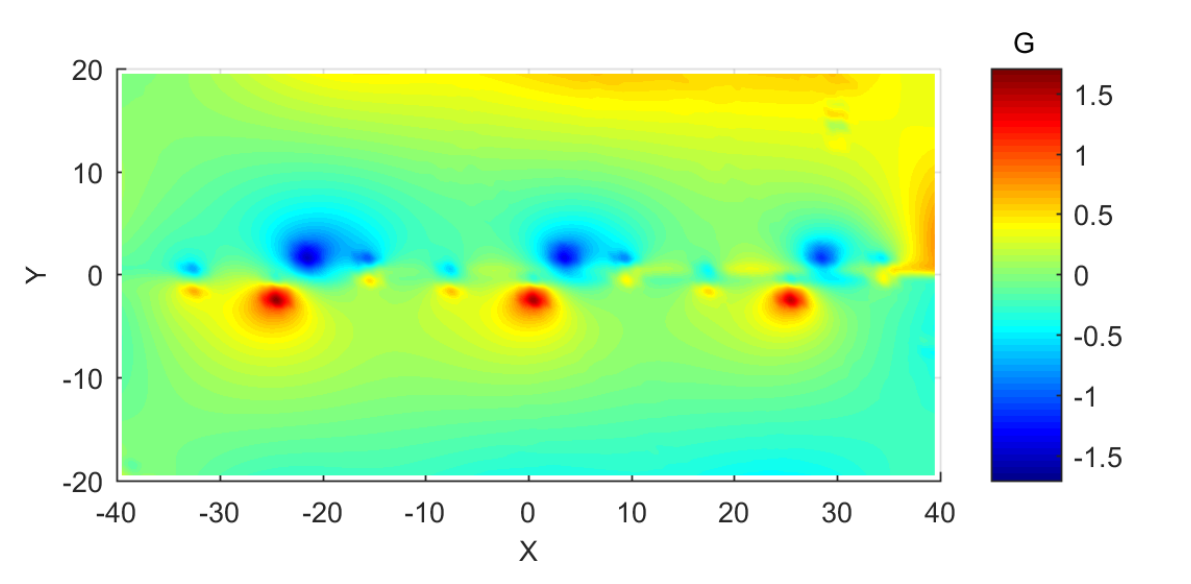}}\qquad 
\subfigure[The difference between the isotropic case and the anisotropic case, $\delta\sigma_{iso,aniso}$.]{
\label{fig:pond_incoherent_stress_anisotropic_diff}
\includegraphics[width=0.45\linewidth]{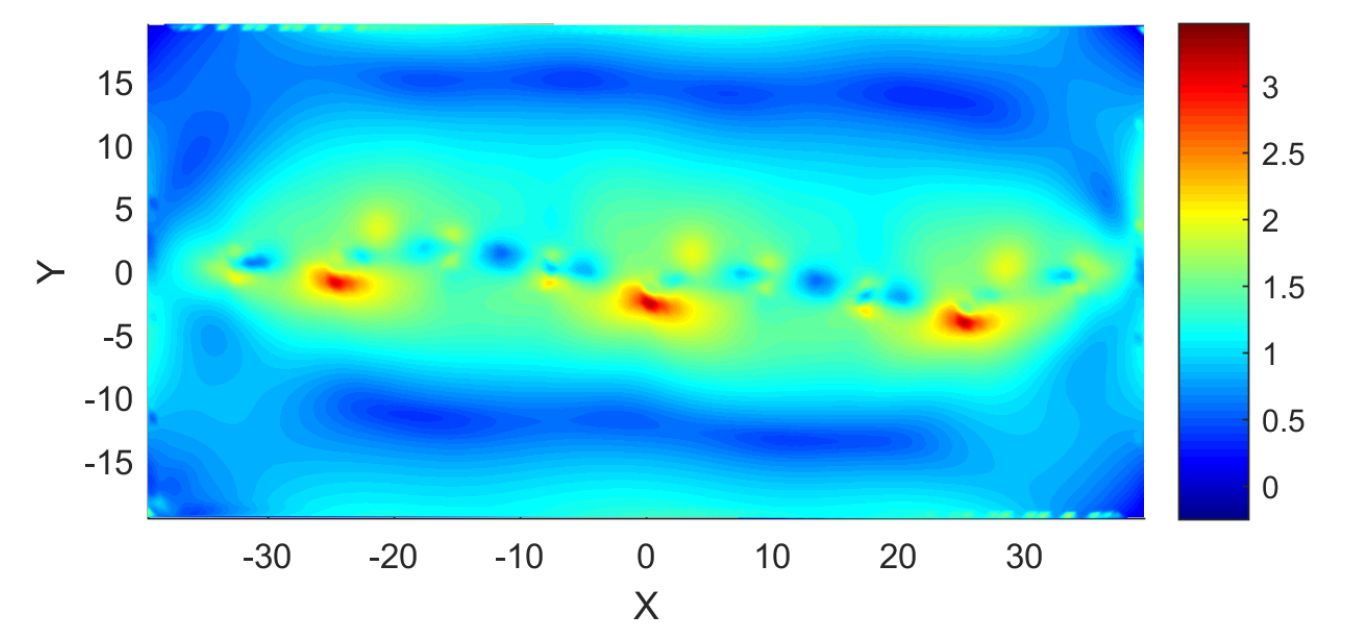}}
\caption{Stress field $\sigma_{11}$ for an incoherent grain boundary with dislocations and disconnections in the finite deformation setting, and a comparison between the isotropic case and anisotropic case. The maximum of $\delta\sigma_{iso,aniso}$ is $320\%$, and the mean of $\delta\sigma_{iso,aniso}$ is $320\%$ is $12.3\%$.}
\label{fig:pond_incoherent_anisotropic}
\end{figure}

\subsection{Flat, through, and terminating twin and grain boundaries}\label{sec:twin_grain}
In this section we explain the implications of our model with regard to the modeling of elastic fields of flat interfaces. We consider both the case of a twin and a grain boundary.

Before considering grain and phase boundaries separately, we note a feature of our model pertaining to both of them. With regard to stress, the governing equations common to both situations are given by
\begin{equation}\label{eqn:fb_stress}
\begin{split}
& \curl \, \hat{\bfW} = \bfS^{\perp}:\bfX - \bfalpha\\
& \bfT = \bfT(\hat{\bfW})\\
& \divergence\, \bfT(\hat{\bfW})  = \bf0 \\
& \bfT \bfn = \bft \qquad \text{on the boundary},
\end{split}
\end{equation}
as implied by (\ref{eqn:Hs}) and (\ref{eqn:summary}), where $\bfT$ is the stress. With statically admissible traction boundary conditions (and assuming for the sake of argument the traction b.c.s to vanish), this implies that the stress field on the body is solely determined by the fields $\bfalpha$ and $\bfS^{\perp}: \bfX$ - in the linear case, such uniqueness is proven in Appendix \ref{sec:append_uniqueness}. An important implication of this fact is that two different $\bfS$ fields lead to the same stress field as long as their g. disclination fields $\bfPi = \curl \bfS$ are identical, since $\bfS^{\perp}$ is uniquely determined from $\bfPi$. We return to this issue in Section \ref{sec:layer_core}.

\subsubsection{The through twin} \label{sec:twin_through}
In order to model a twin boundary it is imperative to predict an elastic distortion field that is a gradient of a vector field (i.e. compatible) representing a shear of one crystal with respect to the other, which nevertheless results in a stress-free state. In our model, a flat twin boundary can be represented by an eigenwall field $\bfS$ with support in a layer along the interface and of the form $\bfa \otimes \bfn \otimes \bfn$ where $\bfa$ is a vector parallel to the interface plane with magnitude determined by the amount of shearing involved, and $\bfn$ is the unit normal vector of the interface. This is motivated from the fact that the inverse deformation for a twin is continuous at the interface.

\begin{figure}
\centering
\subfigure[The eigenwall field is prescribed in a layer that does not terminate in the body.]{
\includegraphics[width = 0.4\linewidth]{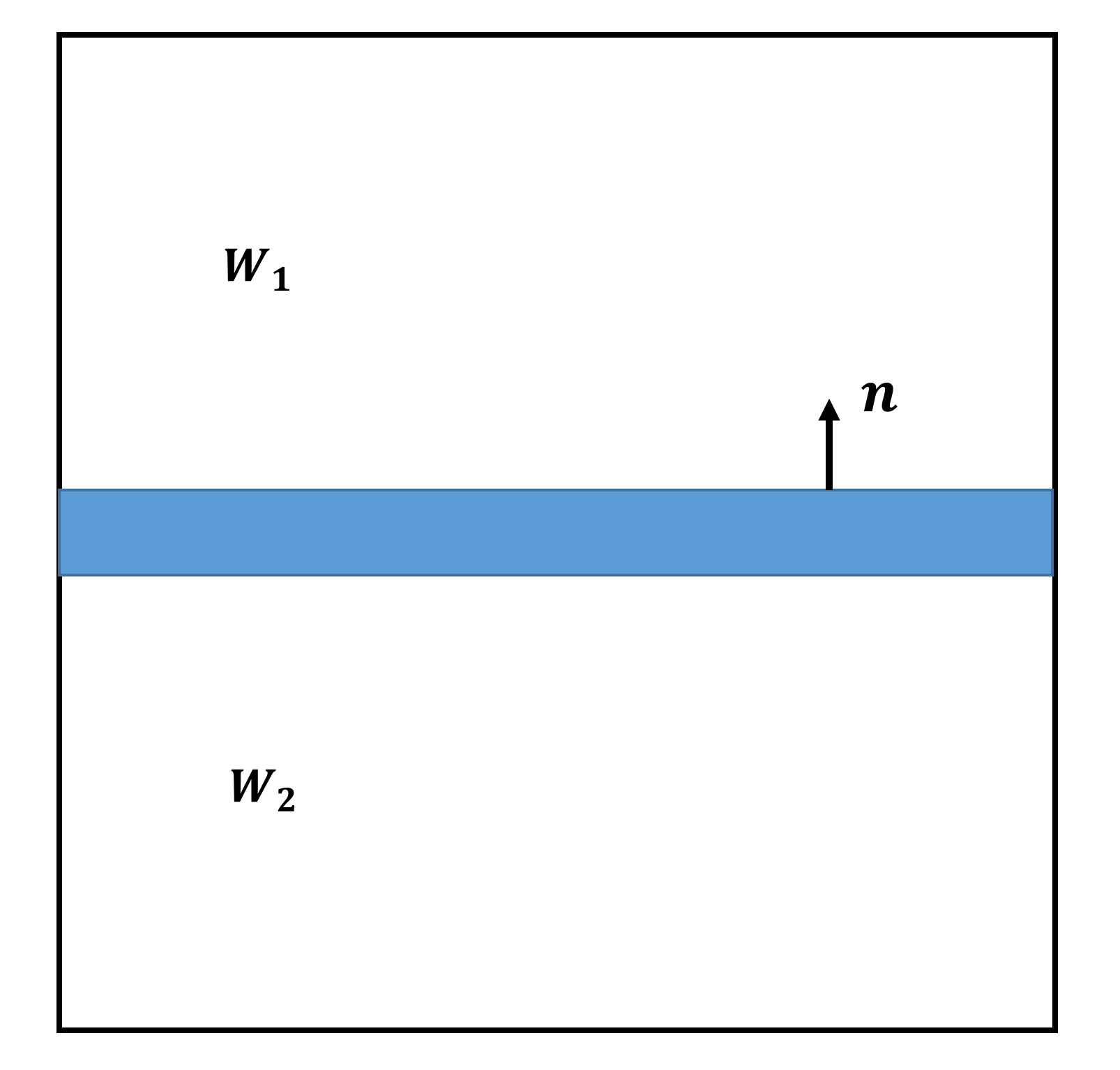}
\label{fig:layer_skew_through_defect}
}\qquad
\subfigure[The magnitude of the stress field $\bfsigma$. The stress field is zero for the eigenwall field in (a).]{
\includegraphics[width = 0.4\linewidth]{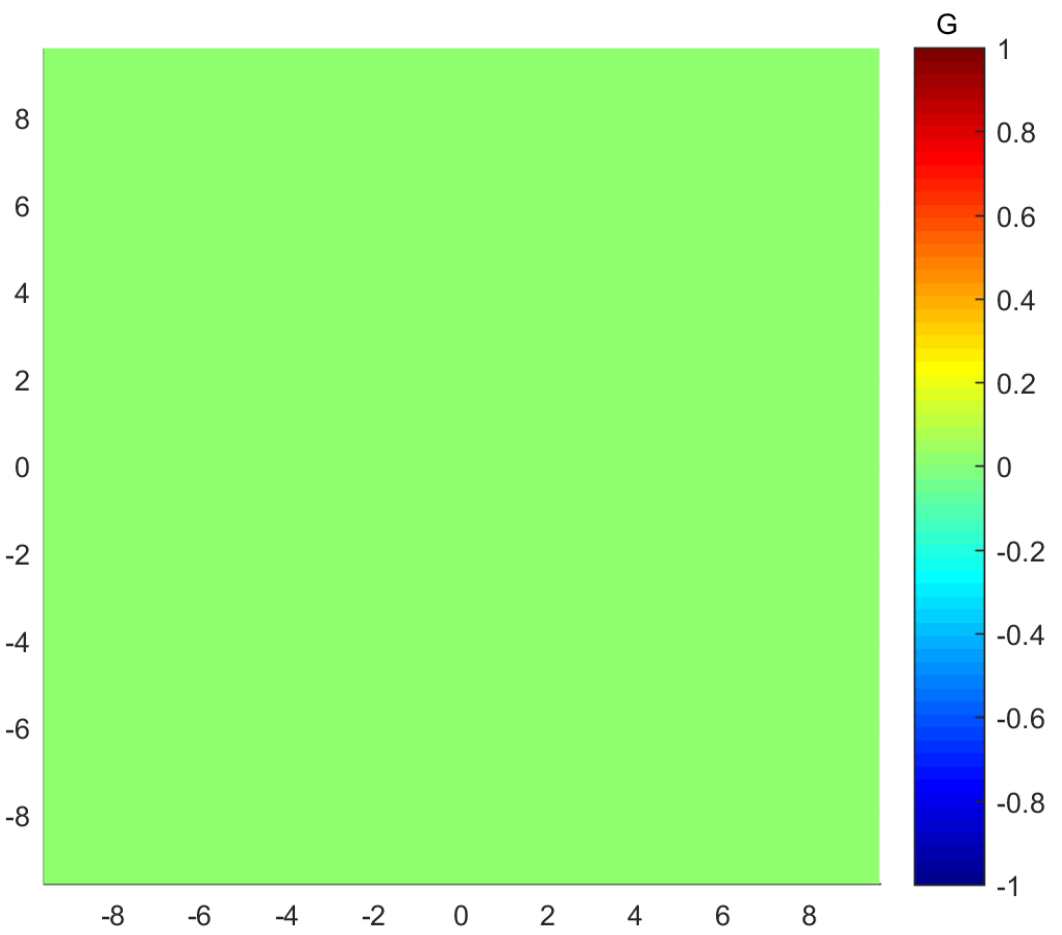}
\label{fig:layer_skew_through_stress}
}
\subfigure[The reference configuration for the through twin boundary.]{
\includegraphics[width = 0.4\linewidth]{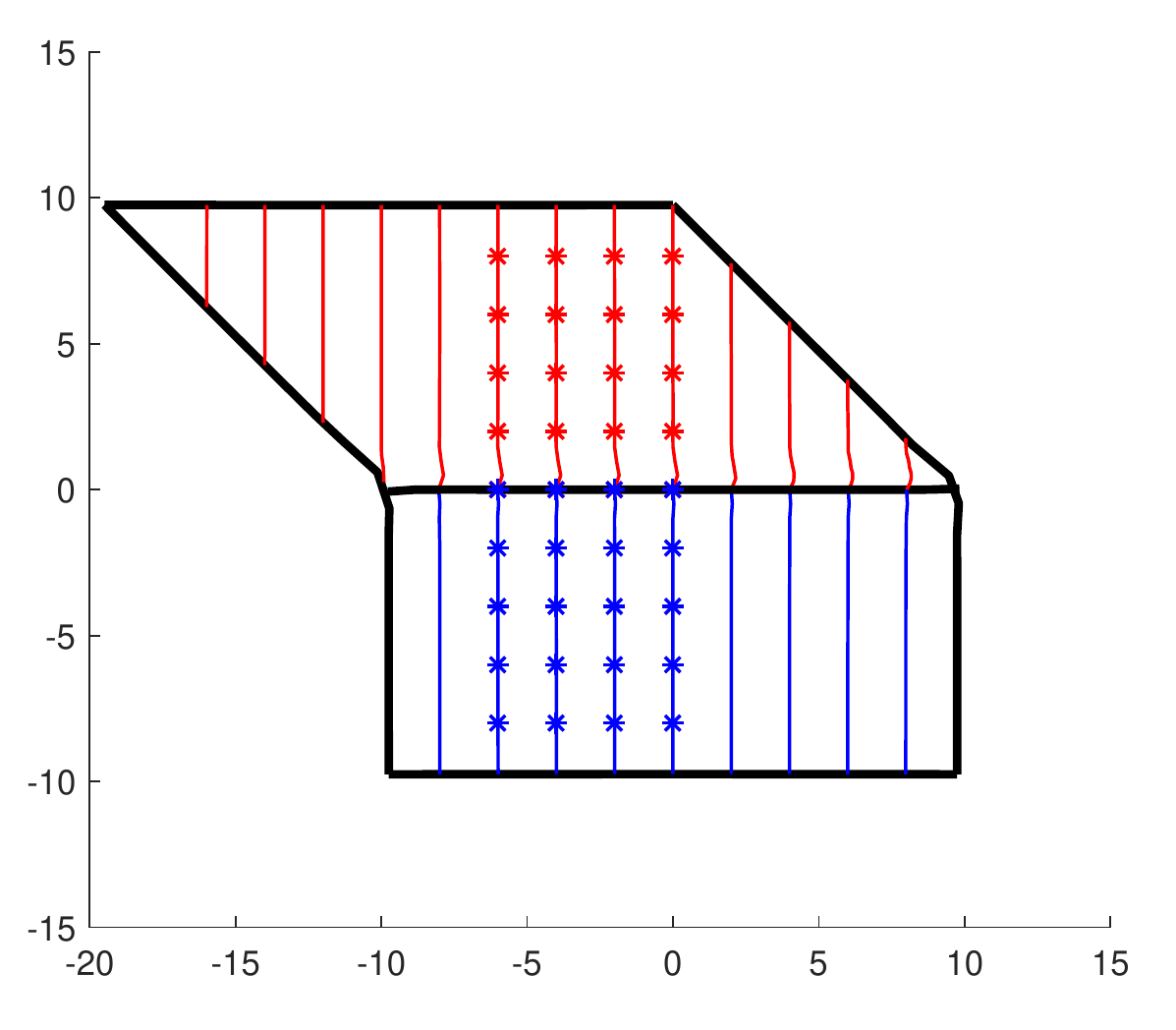}
\label{fig:layer_skew_through_deformation_3}
}\qquad
\subfigure[The current configuration for the through twin boundary.]{
\includegraphics[width = 0.4\linewidth]{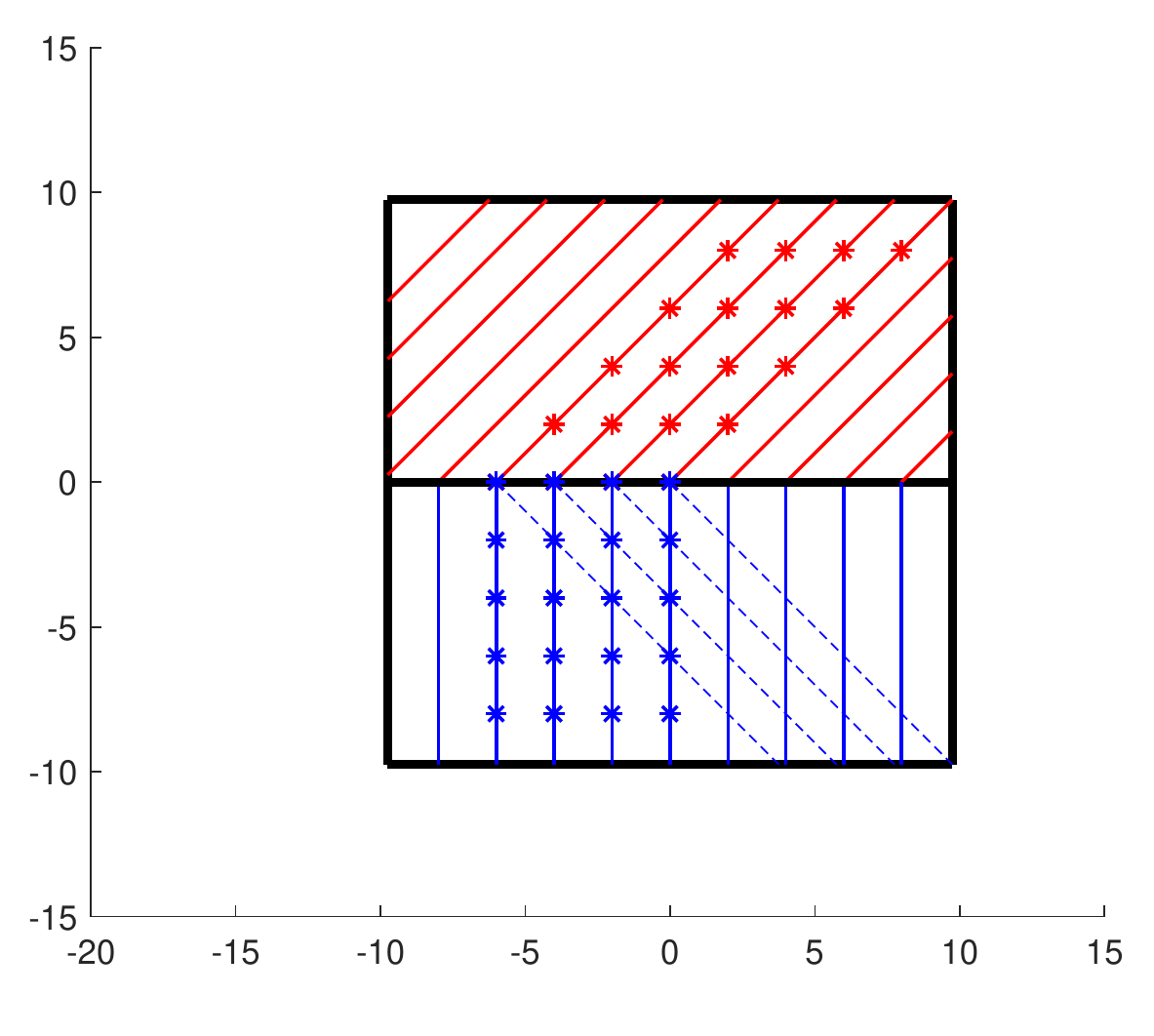}
\label{fig:layer_skew_through_deformation_2}
}
\caption{The eigenwall field prescription of a through twin boundary and its corresponding stress as well as the inverse deformation fields. Red lines and blue lines represent different lattice orientations, and the deformation fields indicate a shear difference cross the boundary interface. The stars on blue dashed lines and the red lines in (d) are the mirrored images of lattice sites across this twin boundary.}
\label{fig:layer_skew_through}
\end{figure}

Recall from (\ref{eqn:dislocation}) that when $\bfalpha = \bf0$, the i-elastic 1-distortion $\bfW$ satisfies
\[
\curl \bfW = \bfS:\bfX.
\]
Given the configuration shown in Figure \ref{fig:layer_skew_through_defect}, if $\bfS$ is prescribed in the form $\bfa \otimes \bfn \otimes \bfn$ in the layer and vanishing outside it, then $\bfS:\bfX = \bf0$ due to the symmetry of $\bfS$ in its last two indices. Thus $\curl \bfW = \bf0$.

Since the through boundary has a constant distribution of $\bfS$ along it,
\begin{equation*}
\begin{split}
&\curl \bfS = \curl \bfS^{\perp} = \bfPi = \bf0 \\
&\divergence \bfS^{\perp} = \bf0 \\
&\bfS^{\perp} \bfn =\bf0 ,
\end{split}
\end{equation*}
indicating $\bfS^{\perp}=\bf0$. With this observation, and the discussion surrounding (\ref{eqn:fb_stress}), we have $\hat{\bfW}= \bfI$ (with appropriate boundary conditions imposed on $\bff$ to eliminate rigid deformation from the current configuration) and the stress vanishes. Also, since $\curl \bfW = \bf0$ in this case, (\ref{eqn:summary})$_2$ implies $\curl \bfH^s = \bf0$ and (\ref{eqn:zs}) implies that the i-elastic distortion $\bfW = \bfI + \bfH^s = -\bfZ^s$ is indeed a non-trivial gradient. 

Our computations recover this exact result; Figure \ref{fig:layer_skew_through_stress} shows the $L^2$-norm of the stress field $|\bfsigma|$ and it turns out the full stress tensor field vanishes for this prescribed $\bfS$ field. The compatible deformation due to the i-elastic distortion $\bfW$ is shown in Figure \ref{fig:layer_skew_through_deformation_2} and \ref{fig:layer_skew_through_deformation_3}. Figure \ref{fig:layer_skew_through_deformation_2} is the current configuration with a twin boundary. Figure \ref{fig:layer_skew_through_deformation_3} is the reference configuration containing the image of the twin mapped by $\bfW^{-1}$ (the mirror planes for this twin boundary are marked as blue dash lines and the red lines in Fig. \ref{fig:layer_skew_through_deformation_2}). In Figure \ref{fig:layer_skew_through_deformation_3}, the inverse deformation at the left bottom corner and the vertical inverse deformation at the right bottom corner are fixed. With this particular Dirichlet boundary condition, $\hat{\bfW}=\bfI$.

It should be noted that if the stress response function was simply a function of $\bfW$ instead of $\hat{\bfW}$, it would not have been possible to predict the non-trivial twinning deformation corresponding to the stress-free state.

\subsubsection{The terminating twin}\label{sec:twin_terminate}

\begin{figure}
\centering
\subfigure[The eigenwall field is prescribed within the layer and terminates inside the body.]{
\includegraphics[width = 0.4\linewidth]{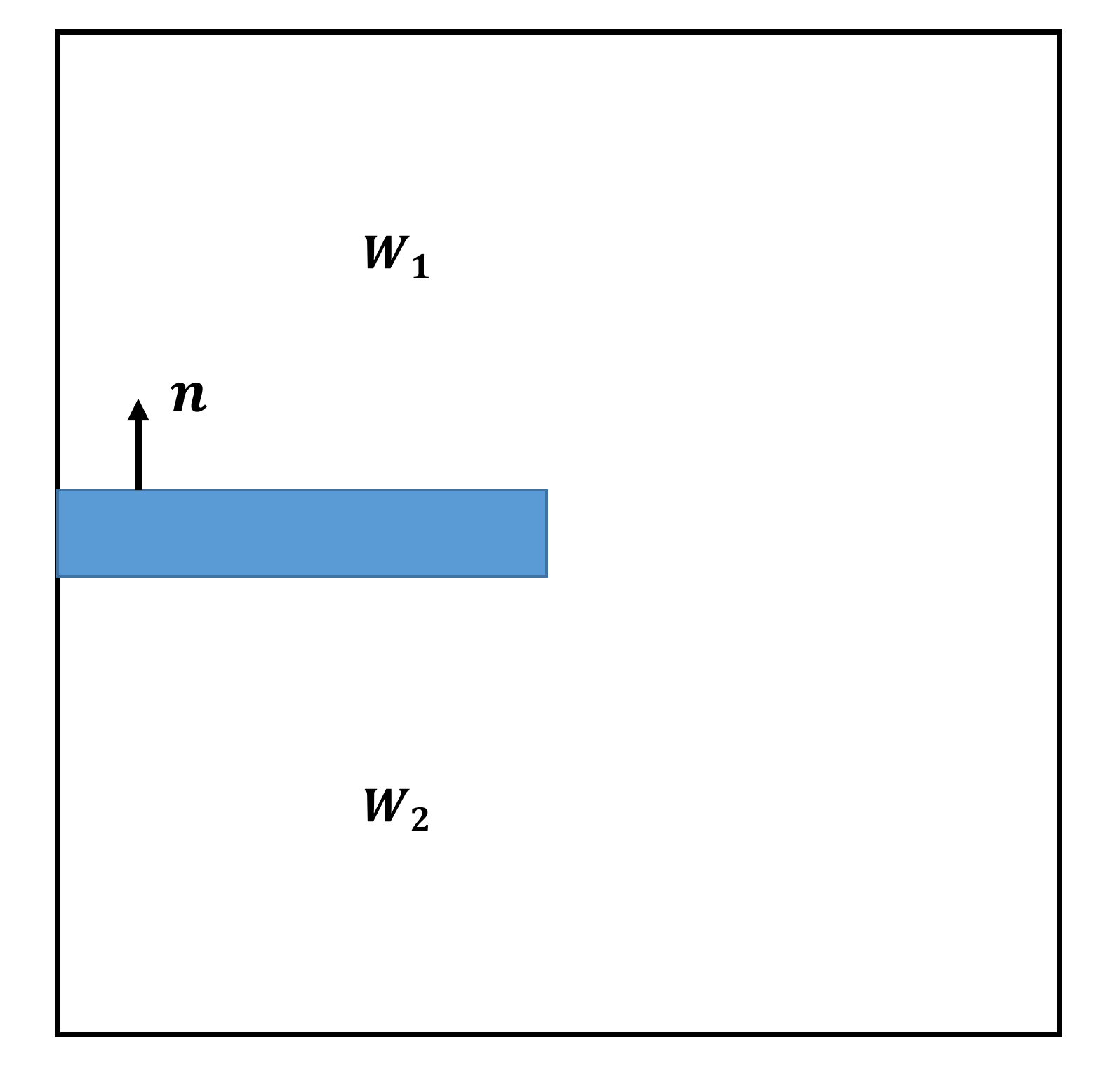}
\label{fig:layer_skew_end_defect}
}\qquad
\subfigure[The stress field $\sigma_{11}$ is non-zero for the corresponding eigenwall field in (a).]{
\includegraphics[width = 0.4\linewidth]{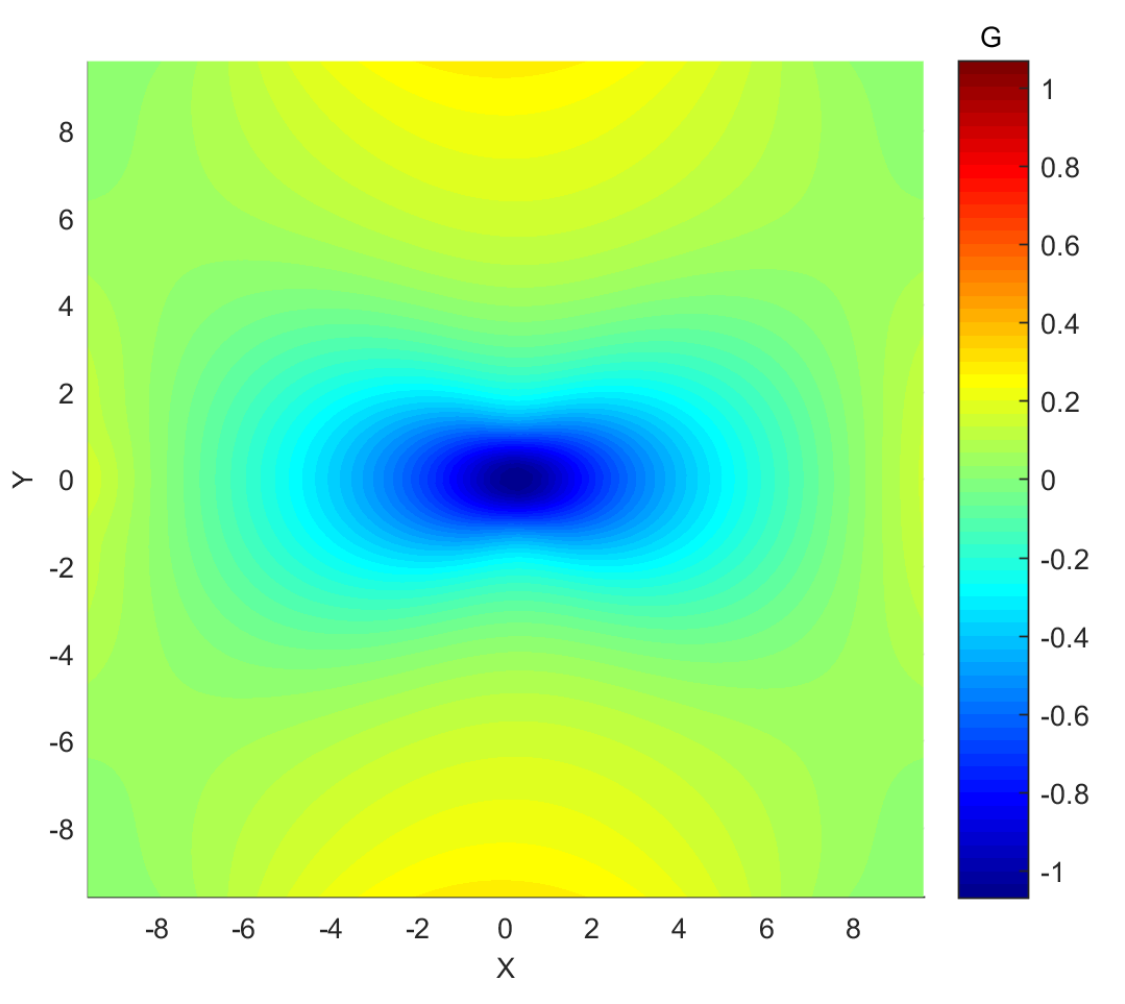}
\label{fig:layer_skew_end_stress}
}
\caption{The eigenwall field prescription for the terminating twin boundary and its corresponding stress field $\sigma_{11}$.}
\label{fig:layer_skew_end}
\end{figure}

Consider now a terminating twin boundary. The specification of the $\bfS$ field is the same as before in the layer, but the layer does not go through the body, as shown in Figure \ref{fig:layer_skew_end_defect}. The terminating twin calculated in this part is equivalent to a negative g.disclination problem. The field $\bfS:\bfX = \bf0$ on the body as before; however, $\curl \bfS = \bfPi \neq \bf0$ and (\ref{eqn:fb_stress}) implies there is a non-vanishing stress field now. 

As for the i-elastic distortion, we note first that, for $\bfalpha = \bf0$, (\ref{eqn:dislocation}) implies that $\curl \bfW = \bf0$ so that the i-elastic distortion is compatible. This can alternatively be understood from the fact that $\bfS = \bfS^{\perp} - \grad\, \bfH^s$ so that $\bfS:\bfX = \bf0$ and (\ref{eqn:fb_stress})$_1$ imply that $\curl \bfH^s = -  \bfS^{\perp}:\bfX = - \curl \hat{\bfW}$ and (\ref{eqn:summary})$_2$ then implies that $\bfW$ is compatible.

Figure \ref{fig:layer_skew_end_stress} shows the stress field $\sigma_{11}$ of the defect configuration in Figure \ref{fig:layer_skew_end_defect} with the stress function given as $\bfT(\hat{\bfW})$. 

We note that had the stress response been taken as simply a function of $\bfW$, then we would have $\curl \bfW = \bf0$ from (\ref{eqn:dislocation}) and this associated with $\divergence\bfT(\bfW)=\bf0$ would yield the erroneous result that the stress field vanishes.

\subsubsection{The through grain boundary}
For the through grain boundary, the $\bfS$ distribution is specified much like in the case of the through twin, except now $\bfS:\bfX \neq \bf0$, but for the same reasons as for the through twin, $\bfS^{\perp} = \bf0$. Since the misdistortion at a grain boundary involves a difference in rotations, it cannot be represented in the form of a rank-one tensor. Thus, the interface is incompatible and, in the absence of g.disclinations, a dislocation density field must be located along the interface. In general, the dislocation density should be measured and prescribed from experiments. Here we we approximate the interfacial dislocation density as 
\begin{equation}\label{eqn:alpha_prescribe}
\bfalpha = \left(\frac{\bfW_1-\bfW_2}{t}\otimes \bfn\right) :\bfX,
\end{equation}
with $t$ being the layer width and $\bfn$ the interface unit normal. 

\begin{figure}
\centering
\includegraphics[width = 0.4\linewidth]{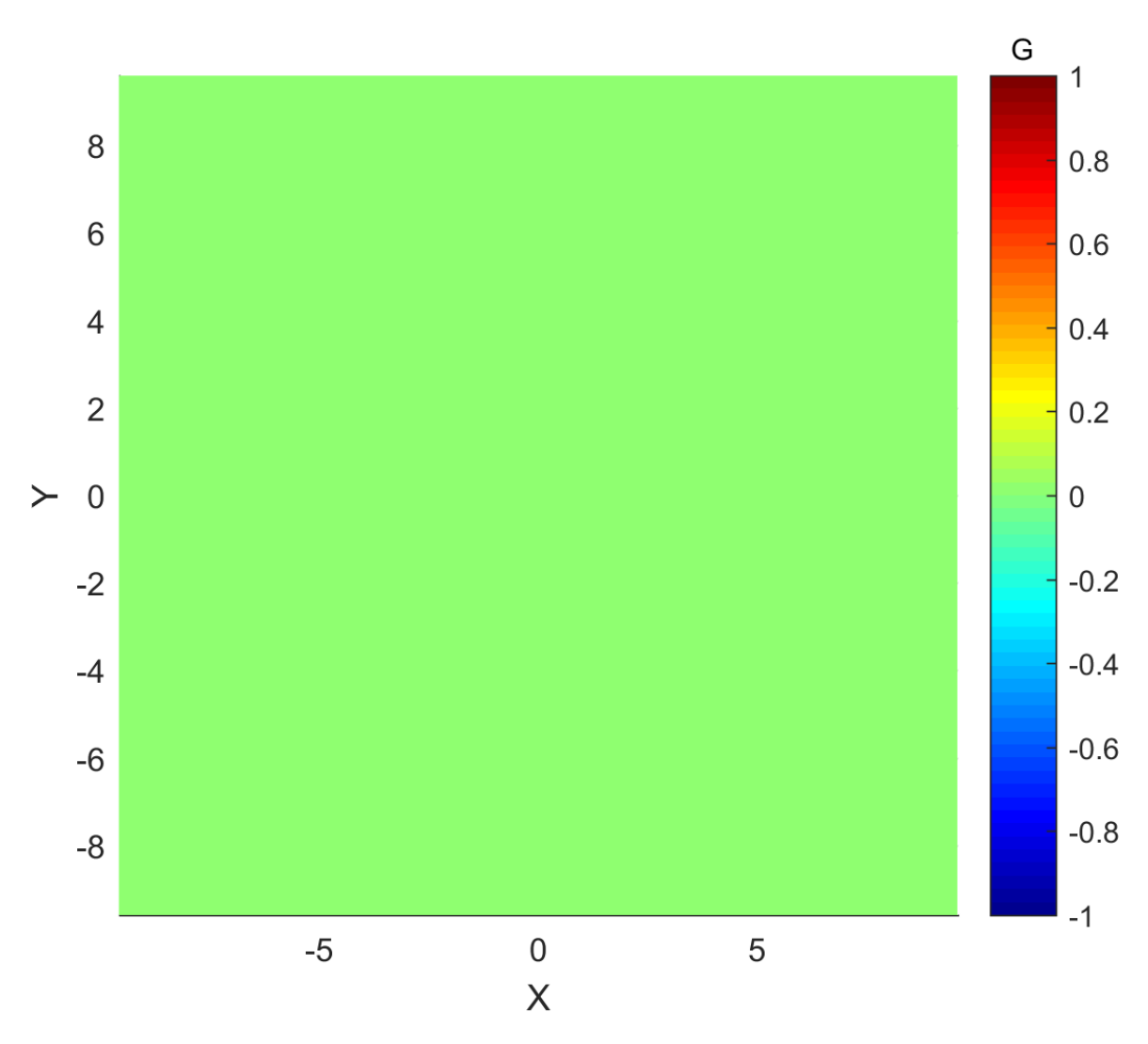}
\caption{The magnitude of the stress field $\bfsigma$ for a through grain boundary in the small deformation setting. The stress field vanishes.}
\label{fig:grain_through_stress}
\end{figure}

As shown in Figure \ref{fig:layer_skew_through_defect}, an eigenwall field $\bfS$ is prescribed along the interface through the body and the dislocation density field \eqref{eqn:alpha_prescribe} is also prescribed in the layer; $\bfW_1 - \bfW_2$ in the expression represents a misorientation of $10^{\circ}$. {Figure \ref{fig:grain_through_stress} shows the $L^2$-norm of the stress field $\bfsigma$ of the prescribed grain boundary in the small deformation setting. Since $\bfalpha$ is calculated from a skew matrix, the stress field is zero in the small deformation case.}

In the finite deformation setting, the interfacial dislocation density specification \eqref{eqn:alpha_prescribe} results in a non-vanishing stress field. However, an alternate prescription of the $\bfalpha$ field in the layer can be generated from an interpolation of the two (constant) finite rotations $\bfW_1$ and $\bfW_2$ by a pure rotation field across the layer and subsequently taking a $curl$ of this field. In such a case, the $\hat{\bfW}$ solution to \eqref{eqn:fb_stress} would be an inhomogeneous rotation field, resulting in vanishing stress everywhere.

\subsubsection{When is stress induced by flat grain/phase boundaries?}\label{sec:stress_free_flat}
By the considerations presented in this Section \ref{sec:twin_grain}, we have obtained the interesting result that for flat twins and grain boundaries that do not induce a g.disclination density along them due to gradients of misorientation/misdistortion, there is no stress in the body. In addition, the elastic distortion for such twins are compatible whereas for grain boundaries they are not, agreeing with classical notions that twin boundaries result in compatible deformations \cite{bhattacharya2003microstructure} and that a strain-free elastic distortion field is necessarily consistent only with a spatially uniform rotation field \cite{shield1973rotation}, a property not satisfied by a body containing a through grain boundary. Moreover, we see the vanishing-stress result of through boundaries as a justification for many works on grain boundary microstructure evolution \cite{de1972partial, bieler2012role, hefferan2012observation} that do not involve the notion of stress at all in the first instance.

Another interesting feature of our model with respect to the modeling of twin boundaries is the fact that regardless of the distribution of flat twin boundaries (possibly terminating) in a body, all individually modeled by an in-layer distribution of the type $\bfa \otimes \bfn \otimes \bfn$, (\ref{eqn:dislocation}) implies that the i-elastic distortion $\bfW$ is curl-free. However, as the case of the terminating twin and the considerations of the next section (\ref{sec:star_dis_free}) show, there can still be induced stresses due to terminating twin boundaries, picked up by a different condition related to the incompatibility of the $\hat{\bfW}$ field, sourced by $\bfS^{\perp}$ that is in turn sourced by the g.disclination density field $\bfPi$. This is reminiscent of the additional condition for compatibility beyond the twinning equation that needs to be satisfied for the occurrence of stress-free crossing twins \cite[p. 83-84, Sec. 5.10]{bhattacharya2003microstructure} - in our case an additional condition is the vanishing of $\bfPi$, beyond the $\bfS$ field being, pointwise, representable as $\sum_i \bfa_i \otimes \bfn_i \otimes \bfn_i$ (with range of $i$ possibly varying from point to point). 

\begin {figure}
\centering
\includegraphics[width=0.65\linewidth]{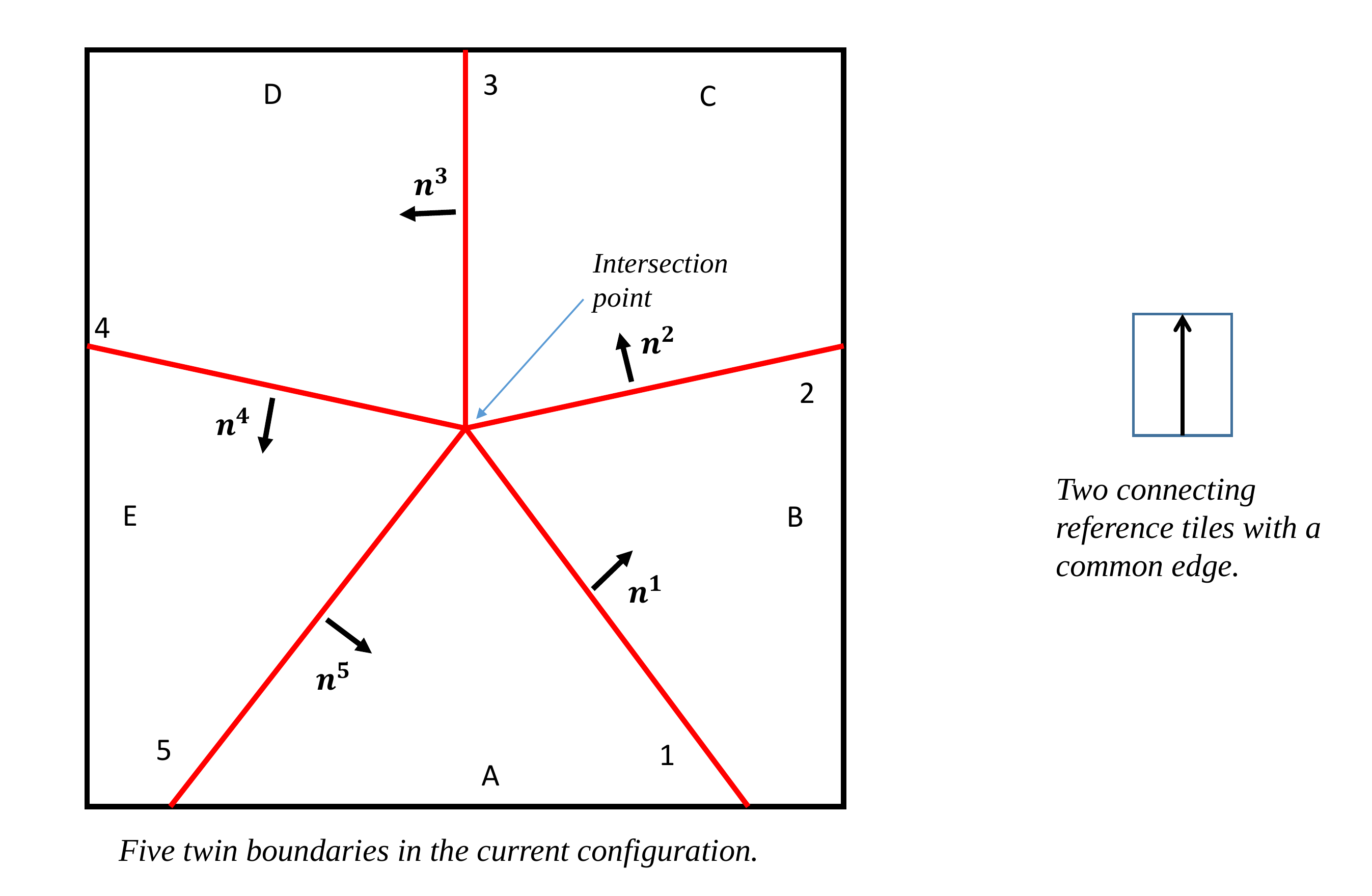}
\caption{An illustration of the configuration of five intersecting a-twin boundaries. The red lines show five a-twin boundaries with index $1$ to $5$. The right part shows two reference tiles sharing a common edge (the black vector). Each reference tile is a rectangle.}
\label{fig:sf_penta_ill}
\end {figure}

\begin {figure}
\centering
\subfigure[The magnitude of the stress field for the penta-a-twin configuration.]{
\includegraphics[width=0.45\linewidth]{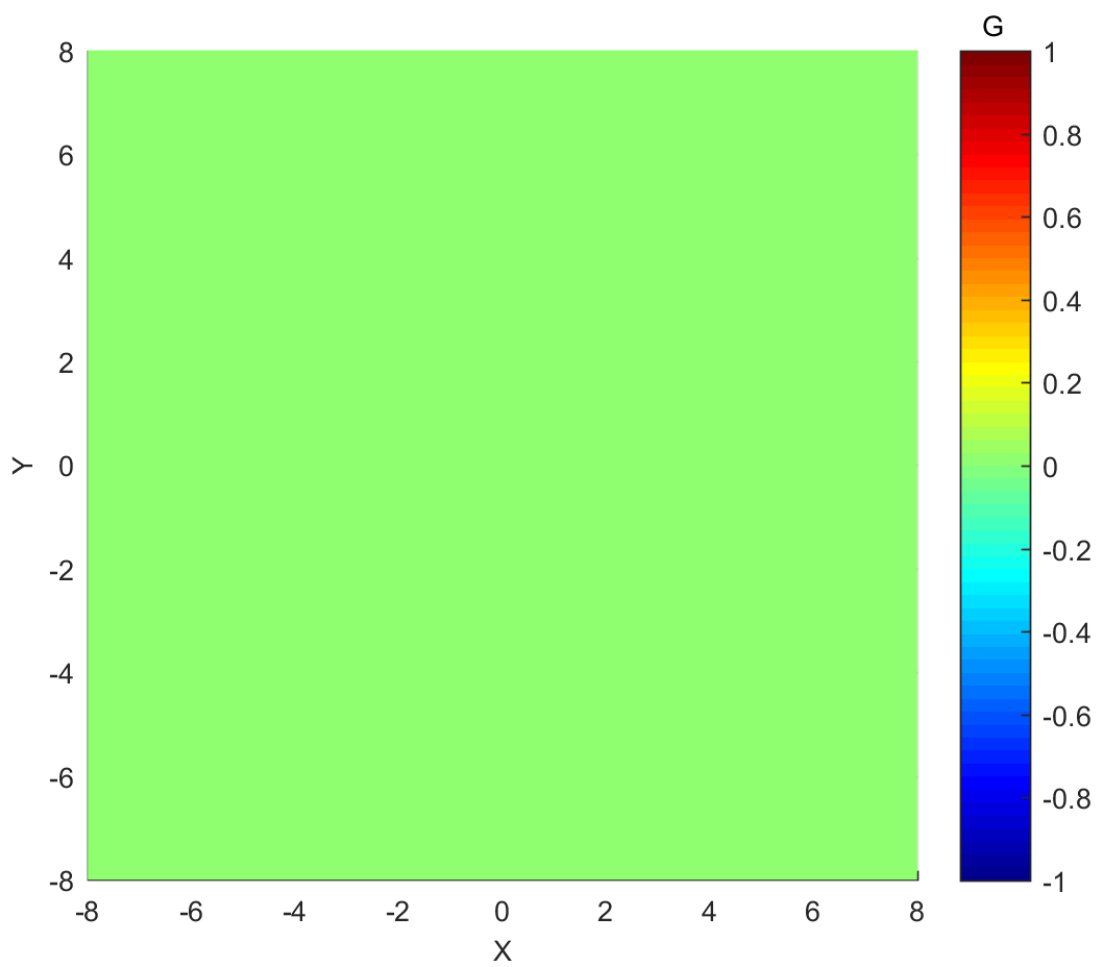}
\label{fig:sf_penta_stress}}\\
\subfigure[The reference tiles across a-twin boundaries with contiguous edges in the reference configuration.]{
\includegraphics[width=0.4\linewidth]{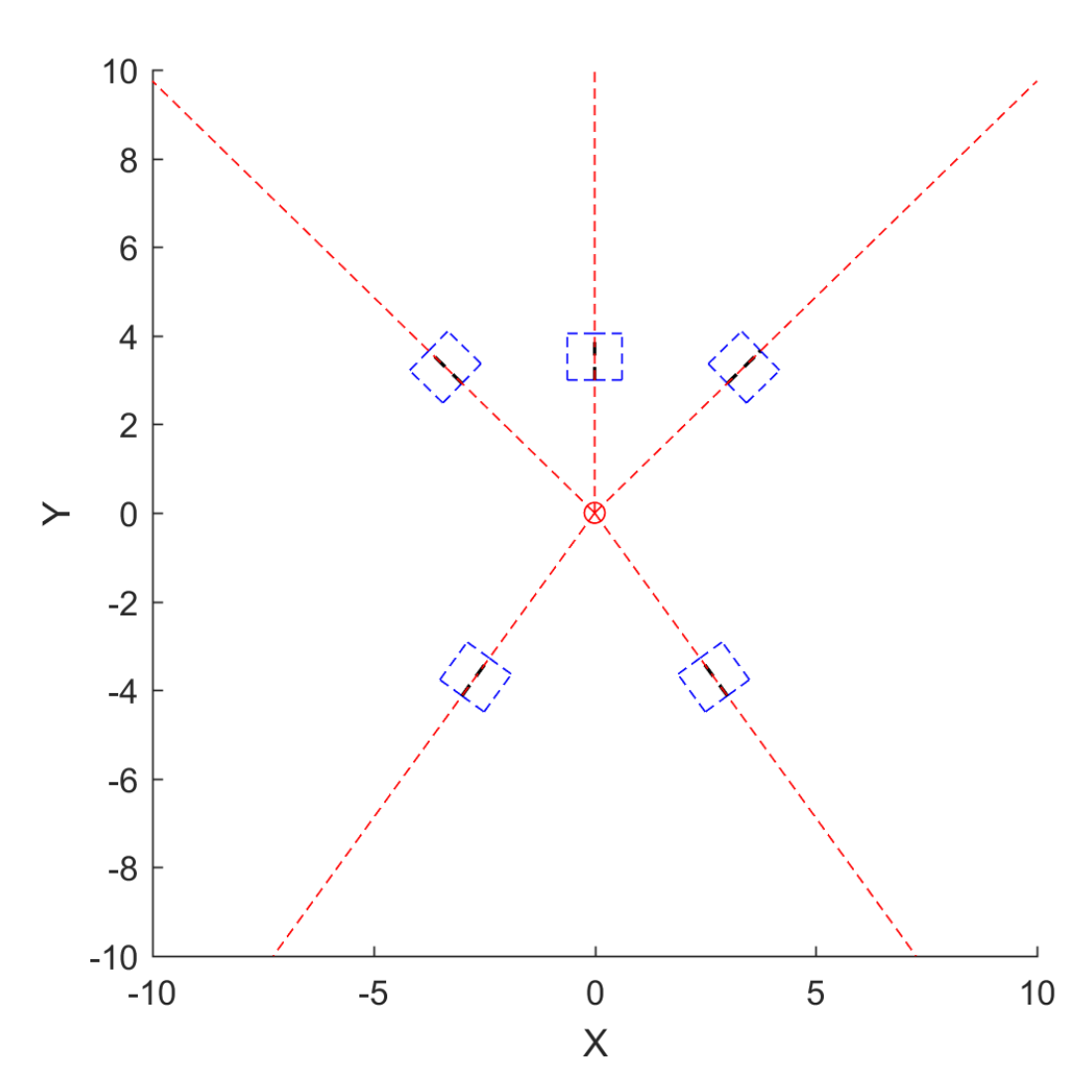}
\label{fig:sf_penta_lat}} \qquad
\subfigure[The rendition of the reference tiles in the current configuration mapped by $\bfW^{-1}$.]{
\includegraphics[width=0.4\linewidth]{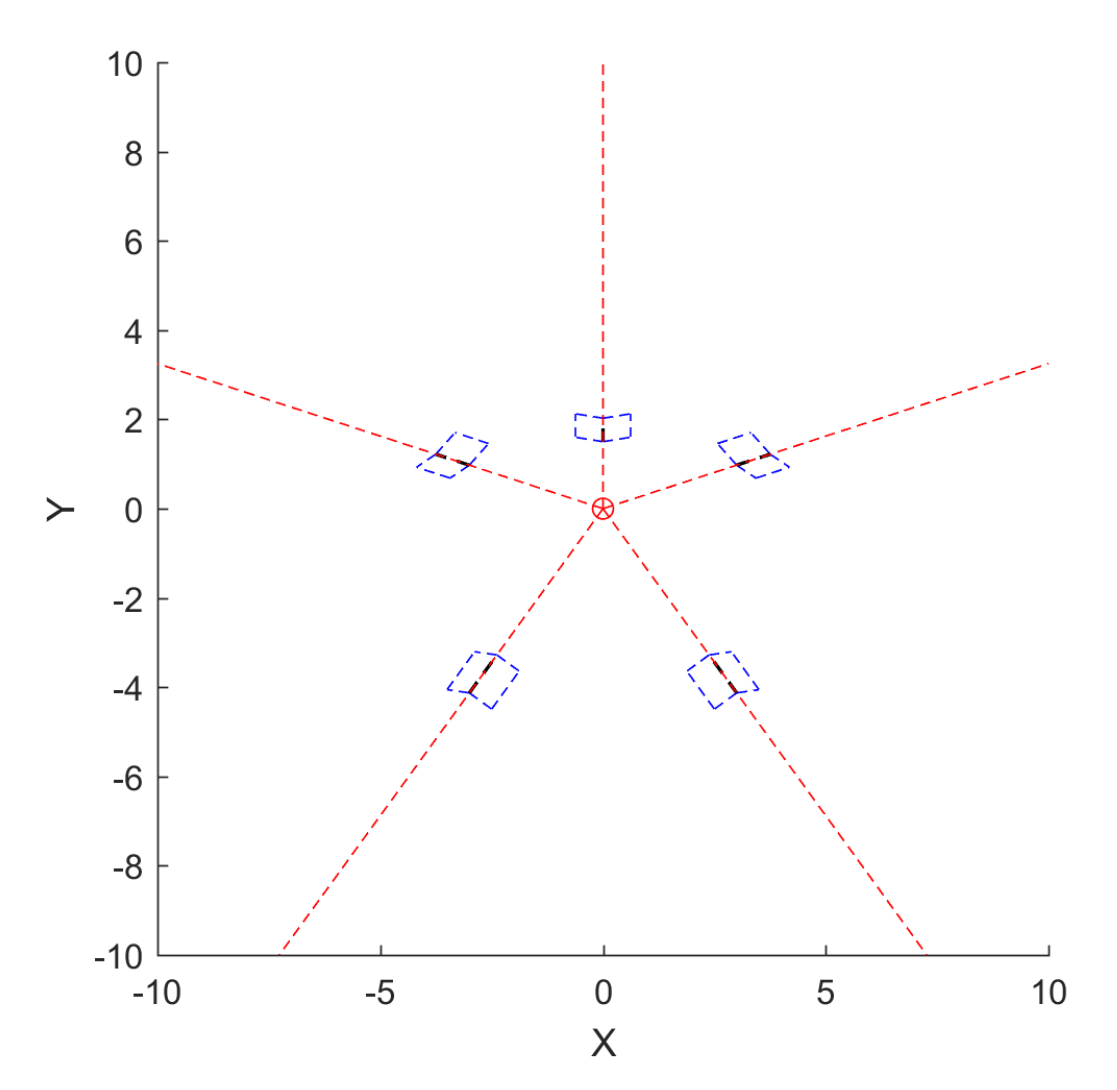}
\label{fig:sf_penta_cur}}
\caption{The zero stress field and the rendition of reference tiles for the penta-a-twin configuration. The misorientation angle for each a-twin boundary is $72^\circ$. The connectivity for each pair of reference tiles across the a-twin boundary remains intact.}
\label{fig:sf_penta}
\end {figure}

As an example, we demonstrate the stress field of a hypothetical configuration of five compatible phase boundaries converging at a point, modeled after a penta-twin configuration \cite{de1972partial}. We refer to this idealized configuration as a `penta-a-twin' configuration and the boundaries as a-twin boundaries, the `a' standing for almost. Each a-twin interface involves a $72^\circ$ misorientation, resulting in prefect compatibility at their junction. Fig. \ref{fig:sf_penta_ill} shows the configuration of five intersecting a-twin boundaries and two reference tiles (see Appendix \ref{sec:append_3}) sharing a common edge (the black vector). The i-elastic distortion difference between two parts $X$ and $Y$, denoted as $\delta \bfW^{X,Y}$, is defined as $\delta \bfW^{X,Y} := \bfW^X-\bfW^Y$. For a compatible phase boundary, $\delta \bfW^{X,Y}$ can be written in the form 
\begin{equation}\label{eqn:comp_boundary}
\delta \bfW^{X,Y} = s \bfa \otimes \bfn^i,
\end{equation}
where $i$ is the a-twin boundary index, $s$ represents the shear strain of one part relative to its adjoining part across the boundary in question, $\bfn^i$ is the unit normal vector field for each a-twin boundary in the current configuration, and $\bfa$ is a unit vector parallel to the interface in the reference configuration. Following the interpretation in Appendix \ref{sec:append_3}, we assume the reference tile at any point $\bfx$ to be the same rectangle up to a rigid rotation, as shown in Fig. \ref{fig:sf_penta_ill}. With the assumed reference tile, any unit vector parallel to the a-twin boundary interface in the current configuration is mapped to the black vector ($\bfe_2$) in the reference configuration, indicating $\bfa$ to be the black vector in the reference tile. Therefore, the contribution to the eigenwall field $\bfS$ from each a-twin boundary $i$ (in its region of support is) is 
\begin{equation}\label{eqn:penta_s}
\bfS^i = \frac{s}{t} \bfe_2 \otimes \bfn^i \otimes \bfn^i \ \ \ \mbox{(no sum)}.
\end{equation}
where $t$ is the layer thickness for each a-twin boundary. The unit normal vector for each a-twin boundary is specified as $\bfn^i = \cos(\alpha^i) \bfe_1 + \sin(\alpha^i) \bfe_2$, with $\alpha^i$ given as follows
\begin{center}
  \begin{tabular}{c | c | c | c | c | c }
    \hline
     i & 1 & 2 & 3 & 4 & 5 \\ \hline
    $\alpha^i$ & $0.2\pi$ & $0.6\pi$ & $\pi$ & $1.4\pi$ & $1.8\pi$ \\
    \hline
  \end{tabular}.
\end{center}
The numbers $1,2,..,5$ correspond to the indices in Figure \ref{fig:sf_penta_ill}. The total eigenwall field $\bfS$ is the superimposition of contributions from all five a-twin boundaries, $\bfS = \sum_i \chi^i  \bfS^i$, where $\chi^i$ represents the characteristic function of the $i^{th}$ a-twin boundary. In the region of overlap of the boundaries, based on \eqref{eqn:Pi}, $\bfPi$ can be written as
\begin{equation} \label{eqn:pi_penta}
\bfPi = \mathlarger{\mathlarger{\sum}}_i \frac{\bfS^i \bfn^i}{c} \otimes \bfe_3 = \frac{s}{ct} \bfe_2 \otimes (\sum_i  \bfn^i),
\end{equation}
where $c$ is the width of the overlap region. Since $\sum_i  \bfn^i = \bf0$ for the prescribed five a-twin boundaries, $\bfW$ corresponding to this $\bfS$ field is is compatible and $\hat{\bfW}$ should be as well since $\bfPi = \bf0$. Indeed, in our modeling we find that both fields $\bfW$ and $\hat{\bfW}$ are curl-free for this problem and we demonstrate the stress-free body in Fig. \ref{fig:sf_penta_stress}. Fig. \ref{fig:sf_penta_lat} shows the reference tiles across each a-twin boundary in the compatible reference configuration. Given the rectangular reference tile shown in Fig. \ref{fig:sf_penta_ill}, we rotate the reference tile such that the contiguous edge of two reference tiles matches the a-twin boundary interface direction, as shown in Fig. \ref{fig:sf_penta_lat}. Fig. \ref{fig:sf_penta_cur} is a rendition of the deformed image of these reference tiles in the current configuration under the elastic distortion $\bfW^{-1}$. The red dashed lines represent the a-twin boundary interfaces. The blue dashed lines show the connecting shapes in the reference and current configurations; the black dashed lines are the contiguous edges for each pair of shapes across an a-twin boundary. Since $\bfW$ is compatible, the connectivity of each pair remains intact.  

\begin {figure}
\centering
\subfigure[The body in the current configuration.]{
\includegraphics[width=0.45\linewidth]{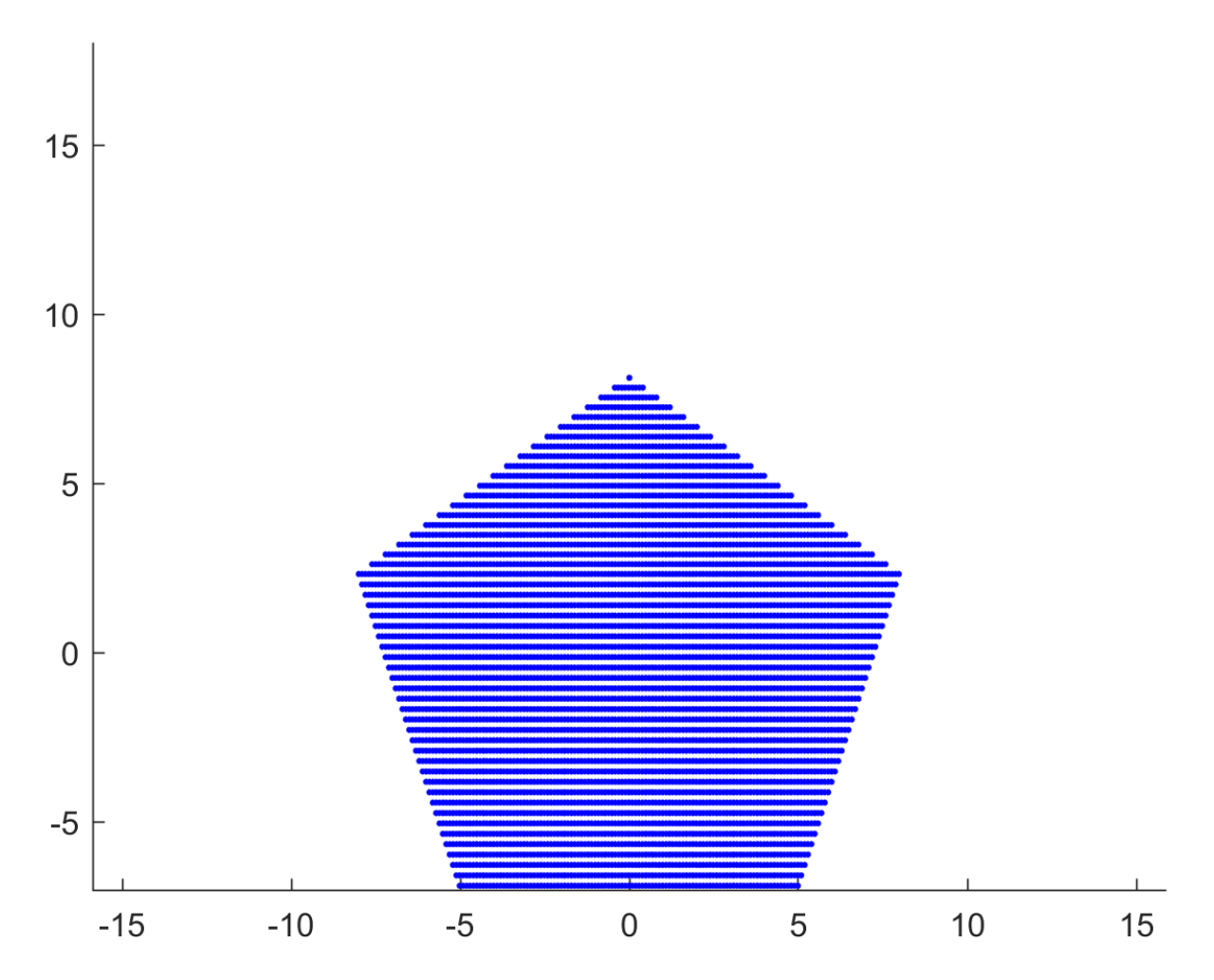}
\label{fig:boundary_current}}
\subfigure[The body in the reference configuration with $\bfa = \bfe_2$.]{
\includegraphics[width=0.45\linewidth]{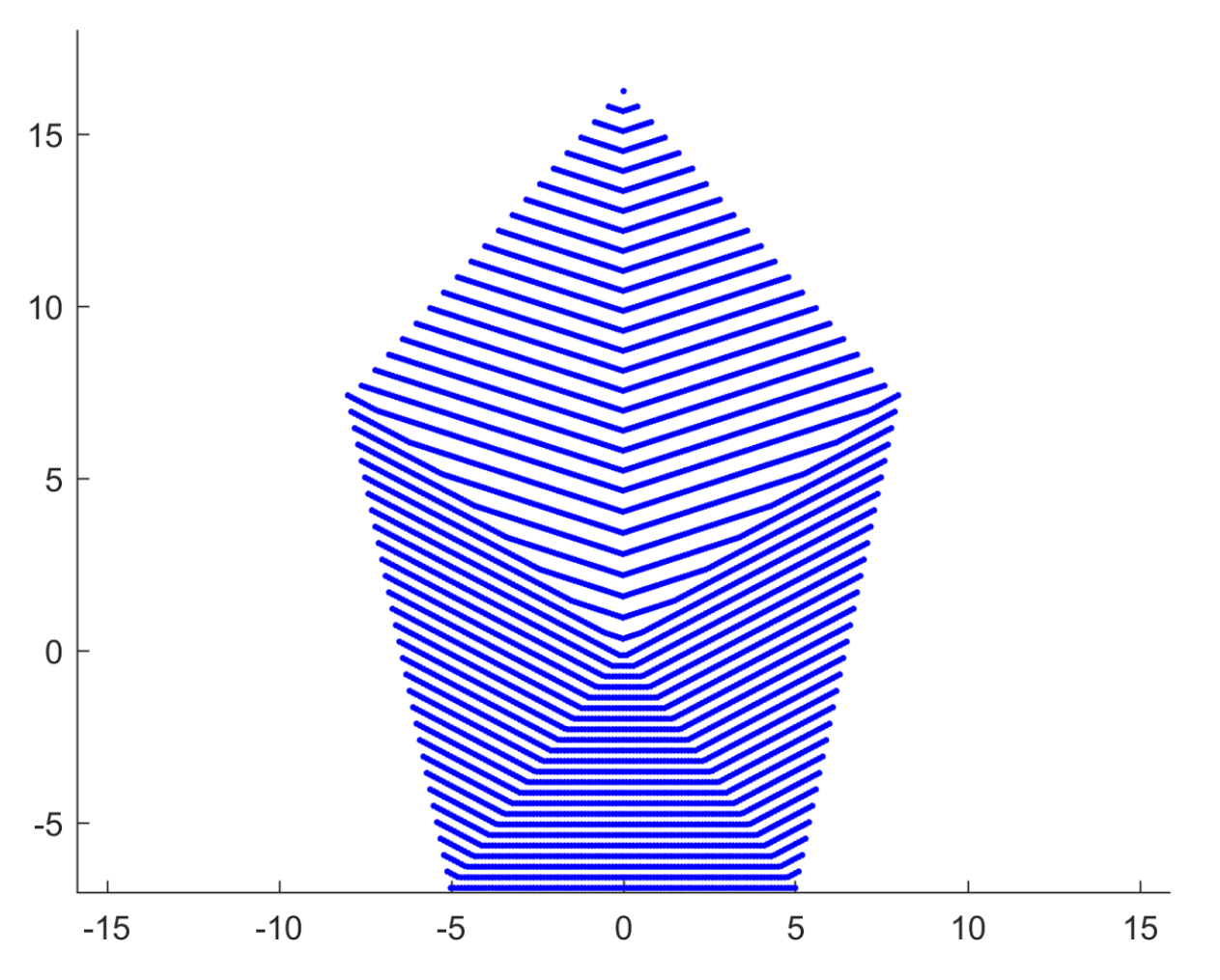}
\label{fig:boundary_vertical}}
\subfigure[The body in the reference configuration with $\bfa = \cos(0.7\pi)\bfe_1+\sin(0.7\pi)\bfe_2$.]{
\includegraphics[width=0.45\linewidth]{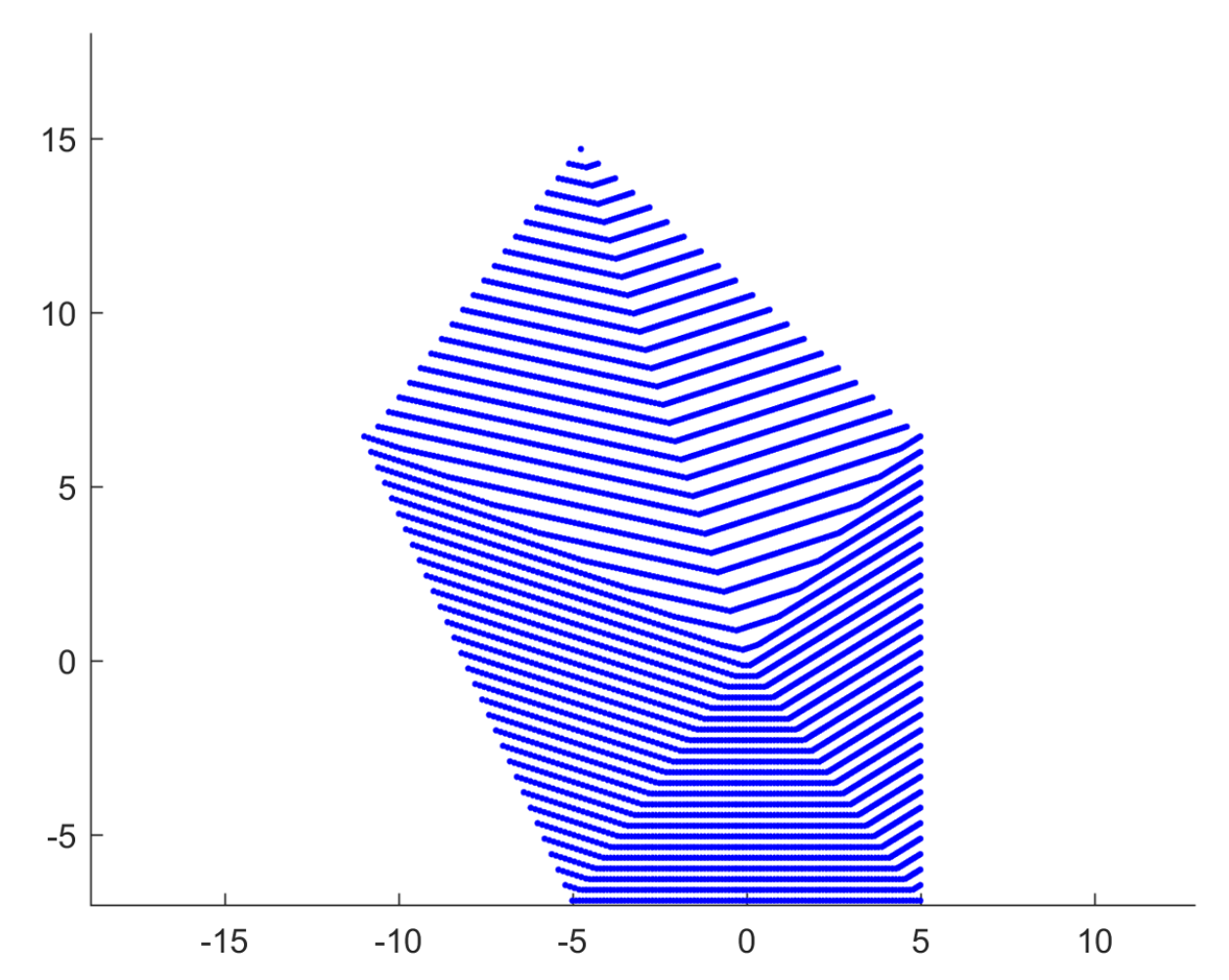}
\label{fig:boundary_incline}}
\caption{The renditions of the body in the reference configuration mapped by the compatible ielastic distortion field $\bfW$ with different prescribed $\bfa$ \eqref{eqn:comp_boundary}.}
\label{fig:penta_boundary}
\end {figure}

Given different reference tiles (to be decided by crystallography), the corresponding prescribed eigenwall fields $\bfS$ are different, leading to different i-elastic distortions $\bfW$. Fig. \ref{fig:boundary_current} is the body in the current configuration. Fig. \ref{fig:boundary_vertical} shows the rendition of the body in the reference configuration mapped by $\bfW$ with $\bfa$ being $\bfe_2$; Fig. \ref{fig:boundary_incline} is the body in the reference configuration mapped by $\bfW$ with $\bfa$ being $\cos(0.7\pi)\bfe_1+\sin(0.7\pi)\bfe_2$. 

\subsection{A stress-inducing almost penta-twin}\label{sec:star_dis_free}
\begin {figure}
\centering
\includegraphics[width= 0.45\textwidth]{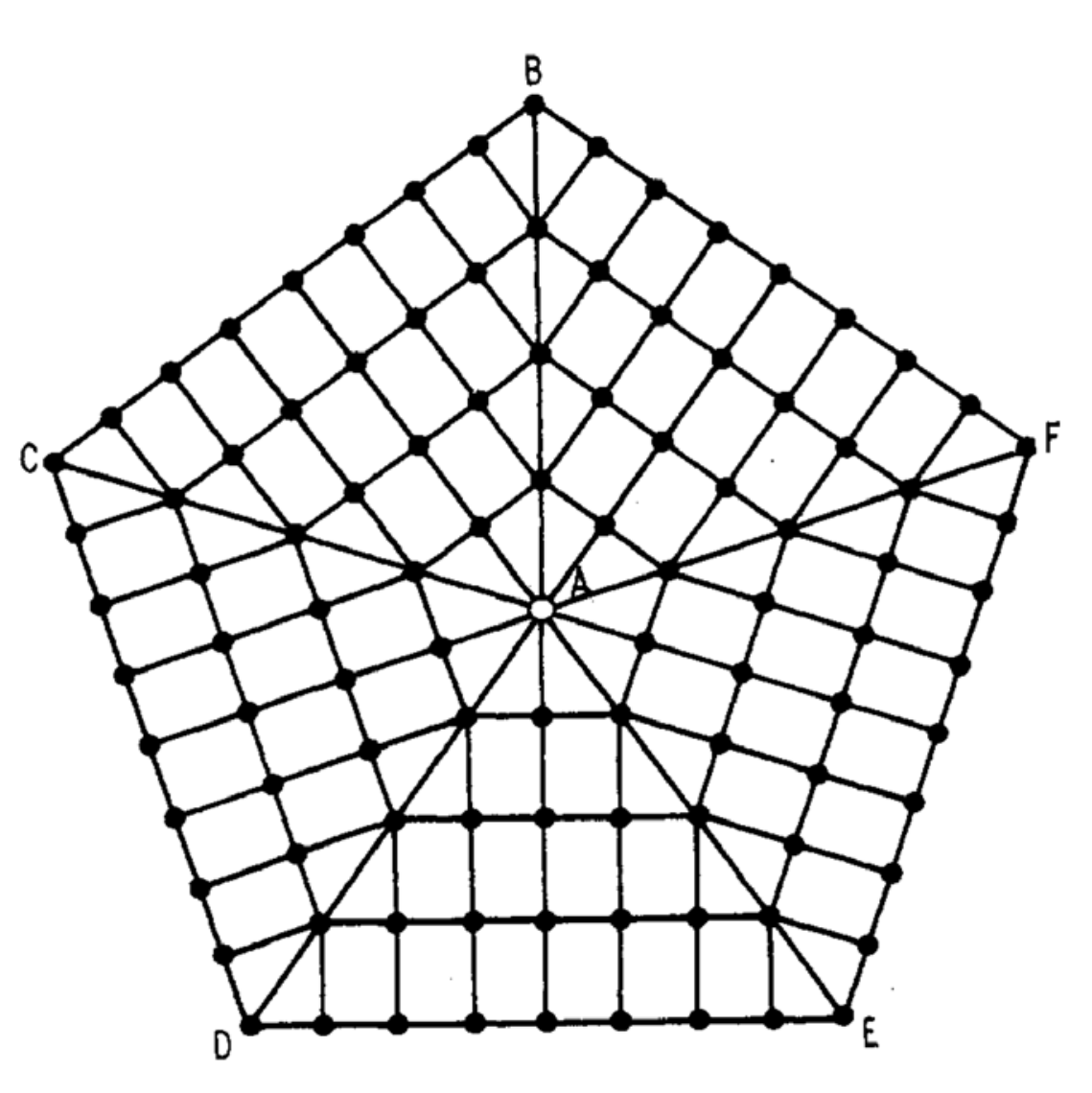}
\caption{The star disclination configuration with five twin boundaries intersecting at point A. Each twin boundary has misorientation angle $70^{\circ}32'$. (Figure reproduced from \cite{de1972partial} with permission from IOP Publishing.)}
\label{fig:star_disclination1}
\end {figure}

\begin{figure}
\centering
\includegraphics[width= 0.5\textwidth]{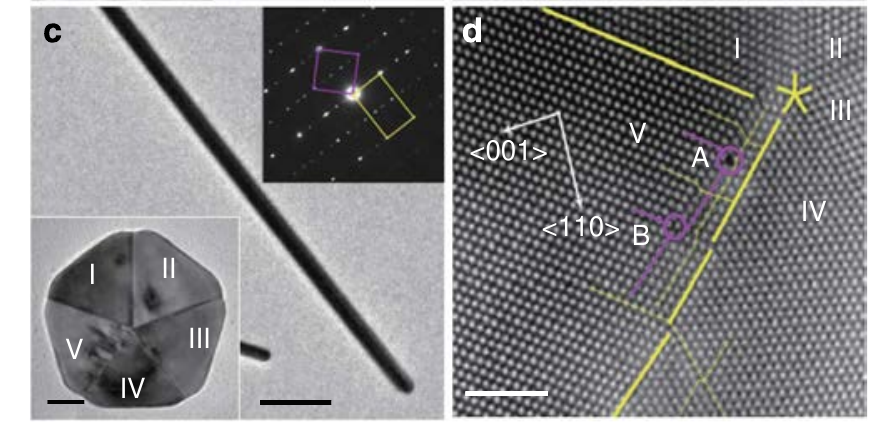}
\caption{Experimental observation of the star disclination, indicating the five twin interfaces are straight. (Figure reproduced from \cite{qin2015recoverable} with permission from Nature Publishing Group of article under an open-access Creative Commons license.)}
\label{fig:star_disclination_interface}
\end{figure}

A special stress-inducing almost penta-twin (with short form \emph{i-a penta twin standing for incompatible almost penta-twin}) is studied in the context of the g.disclination model, serving as an analog of star disclination \cite{de1972partial, gao2005silver}. A star disclination is an observed configuration consisting of five flat twin boundary interfaces converging at the same point, as shown in Figure \ref{fig:star_disclination1} \cite{de1972partial, gao2005silver}. The five twin boundary interfaces appear as straight lines in observations \cite{qin2015recoverable}, as shown in Figure \ref{fig:star_disclination_interface}. The misorientation angle for each twin boundary is $70^{\circ}32'$. The resulting 'stress-free multicrystal' therefore has a gap wedge of $7^{\circ}20'$. 

\begin {figure}
\centering
\includegraphics[width=0.45\linewidth]{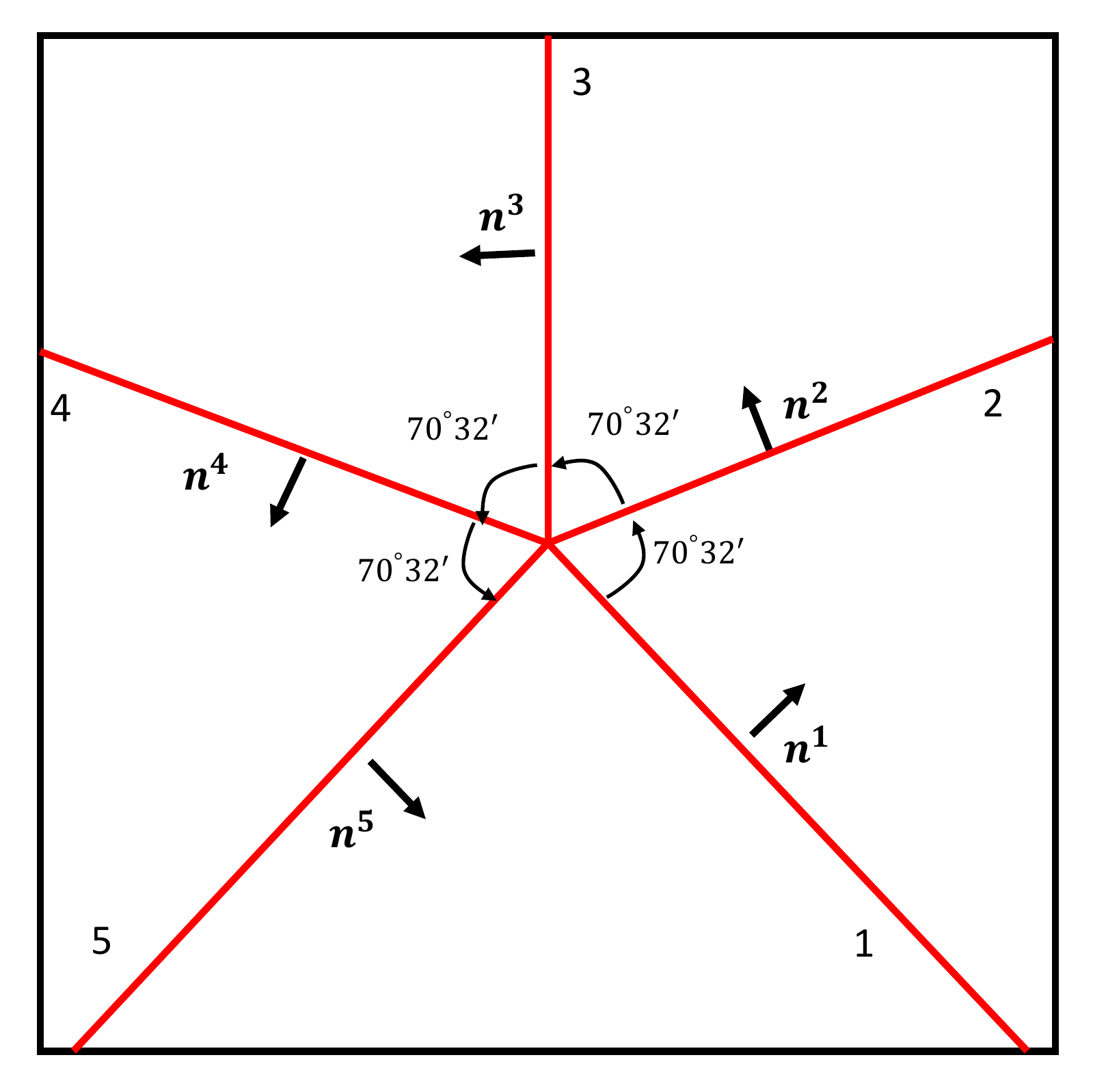}
\caption{A schematic of the configuration of i-a penta-twin with misorientation angle $70^{\circ}32'$. The red lines show five a-twin boundary interfaces where $\bfS$ has support.}
\label{fig:cross_ill}
\end {figure}

\begin {figure}
\centering
\subfigure[Stress $\sigma_{11}$ from small deformation setting.]{
\includegraphics[width=0.45\linewidth]{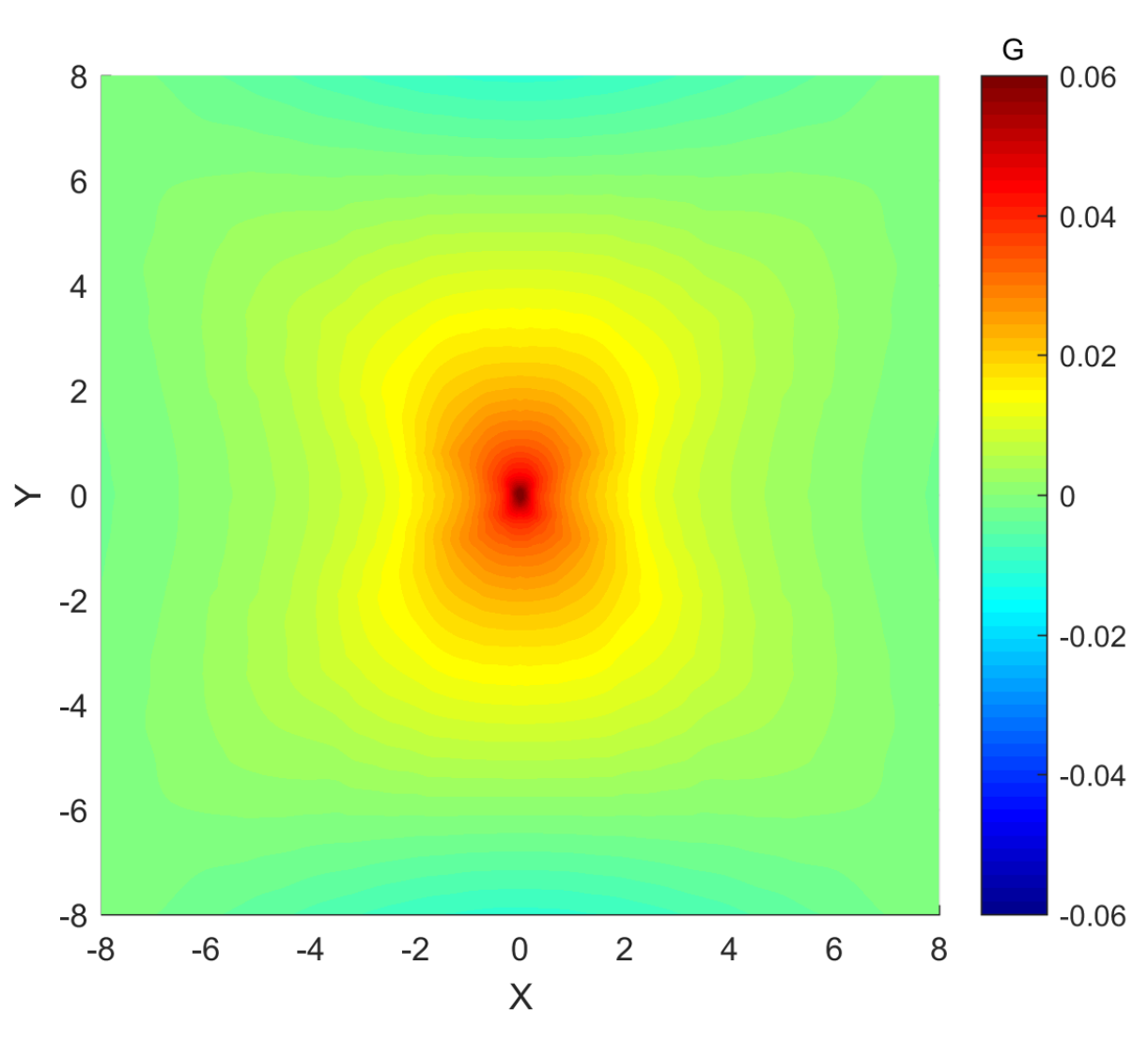}
\label{fig:star_twin_stress_small}}
\subfigure[Stress $\sigma_{11}$ from finite deformation setting.]{
\includegraphics[width=0.45\linewidth]{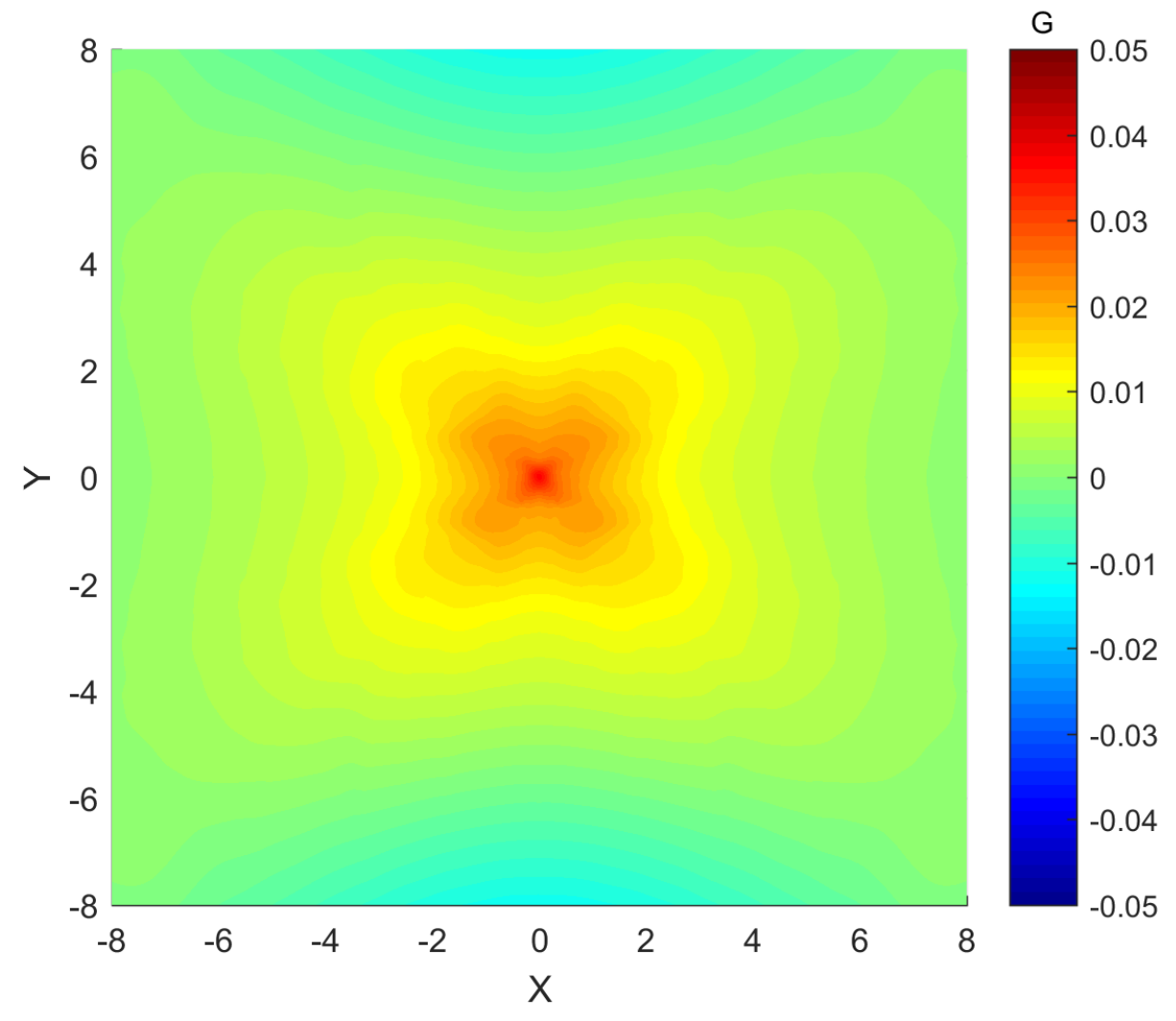}
\label{fig:star_twin_stress_finite}}
\caption{Stress field $\sigma_{11}$ of an i-a penta-twin from both small deformation and finite deformation settings.}
\label{fig:star_twin_stress}
\end {figure}

Motivated by the star disclination, we set up an analogous problem by putting five a-twin boundaries as follows.
\begin{itemize}
\item Put an a-twin boundary indexed as $1$ in Fig. \ref{fig:cross_ill}.
\item Rotate $70^{\circ}32'$ anti-clockwise from a-twin boundary $1$ and put another a-twin boundary $2$.
\item Rotate $70^{\circ}32'$ anti-clockwise from a-twin boundary $2$ and put another a-twin boundary $3$.
\item Rotate $70^{\circ}32'$ anti-clockwise from a-twin boundary $3$ and put another a-twin boundary $4$.
\item Rotate $70^{\circ}32'$ anti-clockwise from a-twin boundary $4$ and put another a-twin boundary $5$.
\end{itemize}
The misorientation angle for all prescribed a-twin boundaries is $70^{\circ}32'$. The eigenwall field $\bfS^i$ for each a-twin boundary has support within the interface layers as shown in Fig. \ref{fig:cross_ill}, and is specified through \eqref{eqn:penta_s}, with vectors $\bfa = \bfe_2$ and $\alpha^i$ as follows:
\begin{center}
  \begin{tabular}{c | c | c | c | c | c }
    \hline
     i & 1 & 2 & 3 & 4 & 5 \\ \hline
    $\alpha^i$ & $0.212\pi$ & $0.606\pi$ & $\pi$ & $1.394\pi$ & $1.788\pi$ \\
    \hline
  \end{tabular}.
\end{center}
The numbers $1,2,..,5$ correspond to the indices in Figure \ref{fig:cross_ill}. Recall \eqref{eqn:pi_penta} 
\[
\bfPi = \frac{s}{ct} \bfe_2 \otimes (\sum_i  \bfn^i),
\]
it can be verified that $\sum_i  \bfn^i$ is no longer $\bf0$ for the prescribed i-a penta-twin. Thus, $\bfPi$ is non-zero. Although $\bfW$ is still compatible for the same reason discussed in the penta-a-twin case, $\hat{\bfW}$ will not be zero and this produces stress. Figure \ref{fig:star_twin_stress} shows the stress fields $\sigma_{11}$ from the small and finite deformation settings, respectively. 


\subsection{Incompatible almost penta-twin with dislocations: stress shielding}

Here we allow for dislocations to be present to maximally shield the stress field produced by the i-a penta-twin configuration in  Sec. \ref{sec:star_dis_free} and explore the resulting stress field and lattice orientation. The main idea is to introduce a dislocation density field to exactly be the excess content in $\bfS^{\perp}:\bfX$ beyond its projection on $\curl$s of rotation fields, where $\bfS^{\perp}$ is the incompatible part of the eigenwall field $\bfS$ obtained by solving the dislocation-free problem of Sec. \ref{sec:star_dis_free}. The obtained rotation field is denoted as $\tilde{\bfW}$. Given the incompatible $\bfS^{\perp}$, $\tilde{\bfW}$ is obtained by 
\begin{gather}\label{eqn:star_rotation}
\tilde{\bfvphi} := arg\min\limits_{\bfvphi}\int_B\frac{1}{2}\left(\curl ((\bfe^{\bfvphi})^{-1}) - \bfS^{\perp} : \bfX\right)^2 dv \\ \nonumber
\tilde{\bfW}^{-1} = \bfe^{\tilde{\bfvphi}},
\end{gather}
where $\bfvphi$ is the rotation vector and $\bfe^{\bfvphi}$ is the exponential map of the same, producing the corresponding orthogonal tensor (an alternative is to require $\tilde{\bfW}$ as an exponential map). By requiring $\tilde{\bfW}^{-1}=\bfe^{\tilde{\bfvphi}}$, the i-elastic distortion field is required to be a rotation matrix. It can be shown that $\tilde{\bfW}$ obtained from \eqref{eqn:star_rotation} is one solution to the g.disclination theory as follows.

The introduced dislocation density is defined as 
\begin{equation} \label{eqn:app_star_1}
\bfalpha := \bfS^{\perp}:\bfX-\curl\tilde{\bfW}.
\end{equation}
Recalling (\ref{eqn:dislocation}),
\begin{equation}\label{eqn:app_star_2}
\begin{aligned} 
\bfalpha = \bfS:\bfX- \curl\bfW = \bfS^{\perp}:\bfX - \curl \hat{\bfW}.
\end{aligned}
\end{equation}
We now substitute (\ref{eqn:app_star_1}) into (\ref{eqn:app_star_2}),
\begin{equation}  \label{eqn:app_star_3}
\bfS^{\perp}:\bfX=(\bfS^{\perp}:\bfX- \curl\tilde{\bfW}) + \curl \hat{\bfW}
\end{equation}
to obtain
\begin{equation} \nonumber
\curl\hat{\bfW}= \curl\tilde{\bfW},
\end{equation}
which implies $\tilde{\bfW}$ is a solution for $\hat{\bfW}$ in generalized disclination theory for this problem.

\begin{figure}
\centering
\includegraphics[width=0.4\linewidth]{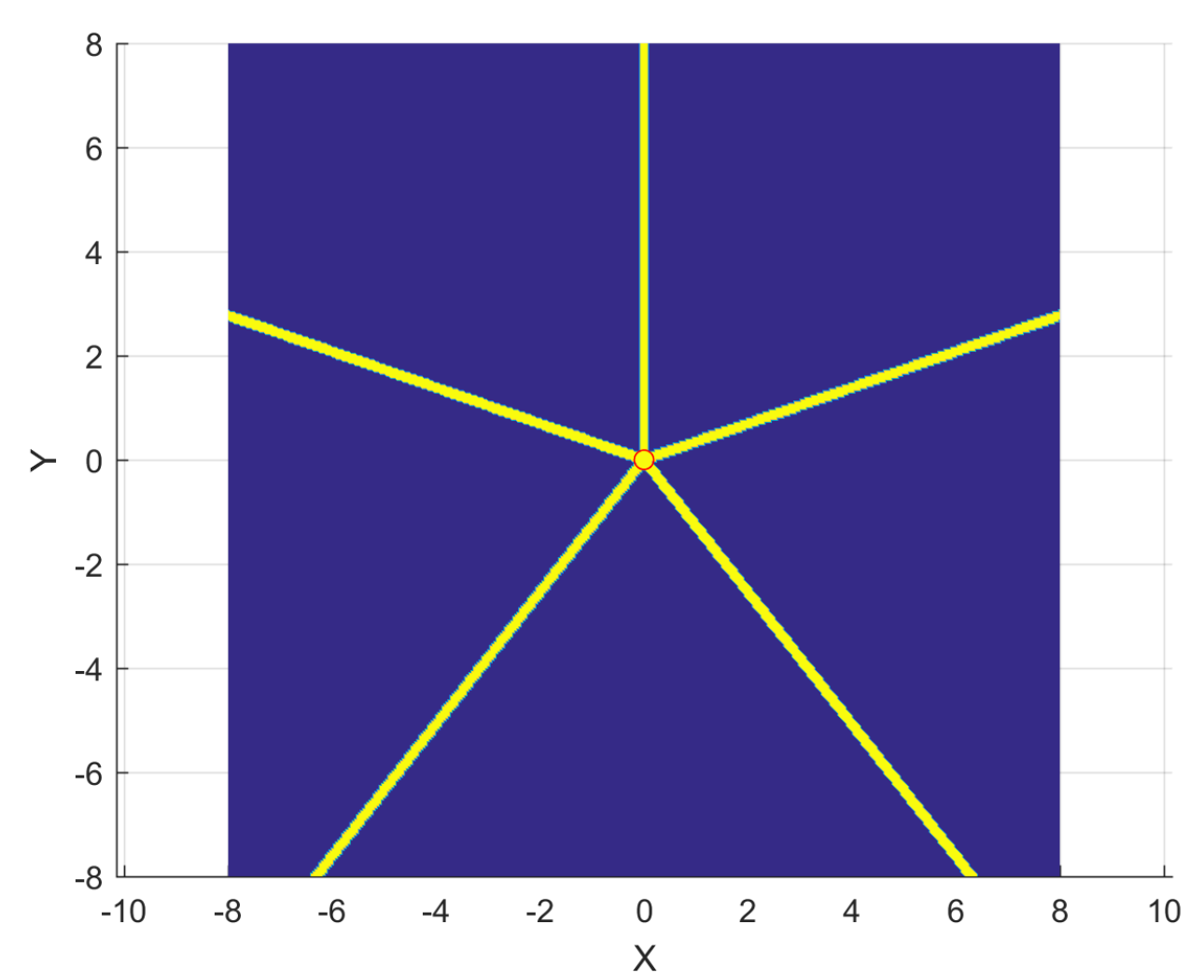}
\caption{Dislocation distribution for the stress-shielded i-a penta-twin. The dislocation densities are localized along the five a-twin boundary interfaces with the identical magnitude.}
\label{fig:star_defect}
\end{figure}

In the with-dislocation case, we find that the dislocation density field defined by (\ref{eqn:app_star_1}) is localized along the five a-twin boundary interfaces (Figure \ref{fig:star_defect}). Furthermore, the norm of the dislocation density along all five a-twin boundary interfaces is the same. 

\begin{figure}
\centering
\subfigure[i-elastic distortion field in dislocation-free case mapped by $\hat{\bfW}$.]{
\label{fig:star_rotation_2}
\includegraphics[width=0.45\linewidth]{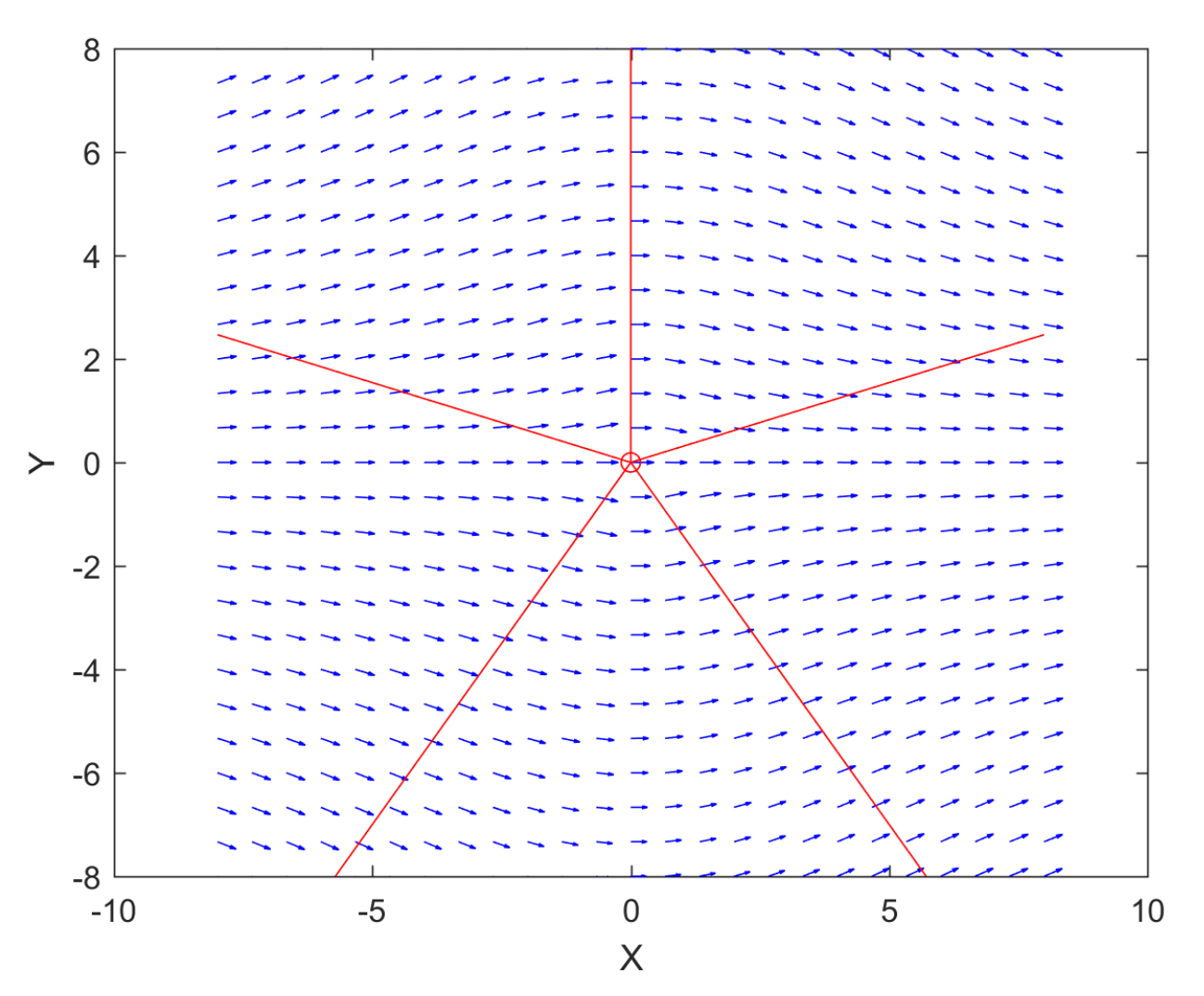}}\qquad
\subfigure[The rendition of unit cell shapes in the with-dislocation case mapped by $\tilde{\bfW}$.]{
\label{fig:star_rotation_4}
\includegraphics[width=0.4\linewidth]{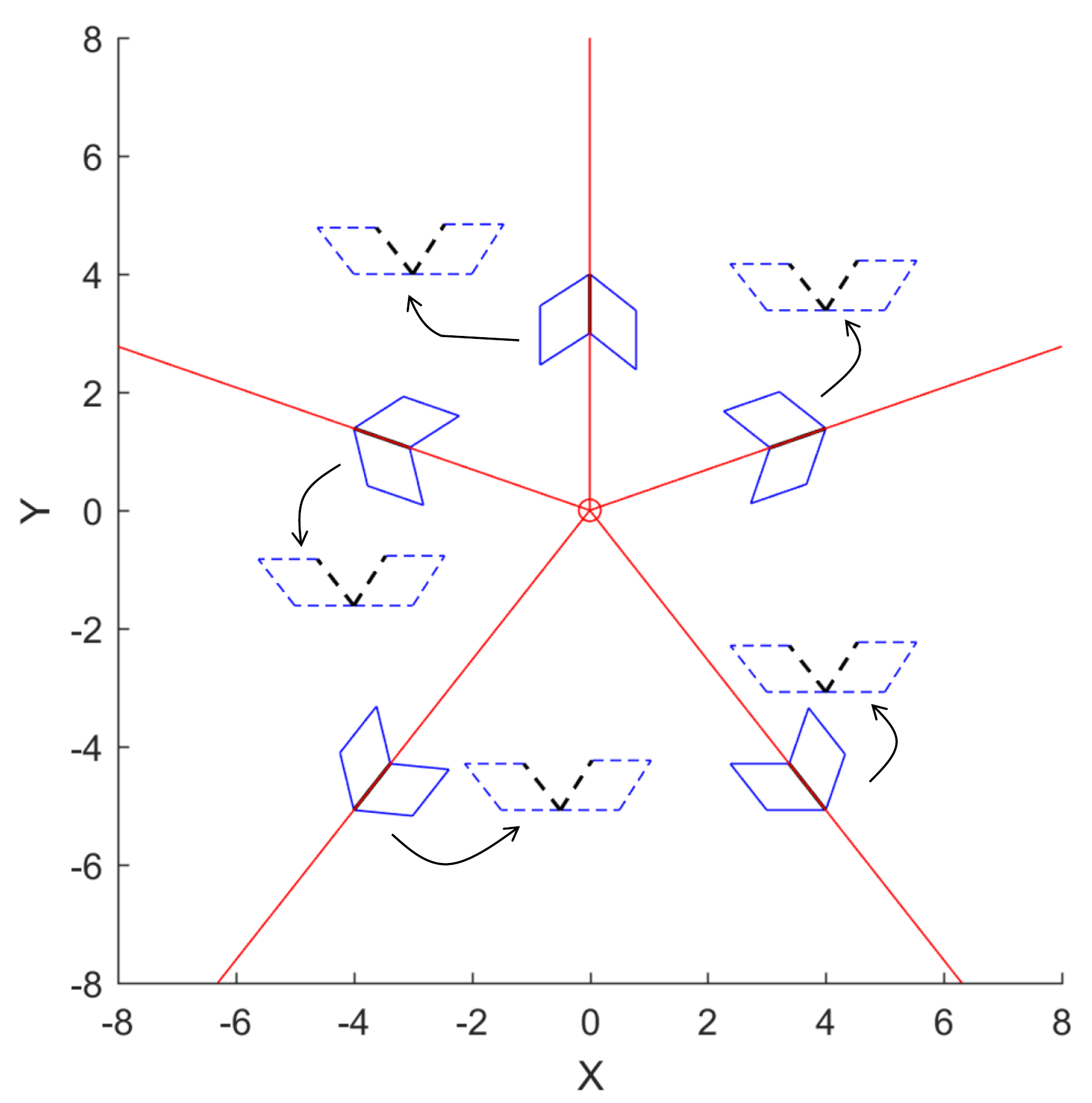}}
\caption{The i-distortion fields of the dislocation-free case represented by a vector field, and the rendition of unit cell shapes for the with-dislocation i-a penta-twin case. In the dislocation-free case, the distortion field involves elastic strain. In the with-dislocation case, the distortion field is a stress-free rotation field, which is incompatible. The connectivity at the black contiguous edges do not persist.}
\label{fig:star_rotation}
\end{figure}

Fig. \ref{fig:star_rotation_2} shows the lattice vectors of the dislocation-free case mapped by $\hat{\bfW}^{-1}$ from a uniformly oriented reference. Fig. \ref{fig:star_rotation_4} shows the lattice shapes of the with-dislocation case mapped by $\tilde{\bfW}$. The dislocation densities eliminate the stress as well as far-field distortion caused by the disclination at the center of the domain. In Figure \ref{fig:star_rotation_4}, the lattice shapes in the current configuration are chosen to be the same ones in Fig. \ref{fig:sf_penta_lat}. Since $\tilde{\bfW}$ is a spatially inhomogeneous rotation field, it cannot be compatible and the connectivity for each pair across the a-twin boundaries does not persist, as shown in Fig. \ref{fig:star_rotation_4}. In Fig. \ref{fig:star_rotation_4}, the black contiguous edges in the current configuration do not remain connected in the reference. 

This example emphasizes the need for dynamics as it is physically reasonable to expect that the production of the maximal supply of dislocations to shield the stress field of the i-a penta-twin should be subject to kinetic constraints.

\subsection{3-D fields: disclination loop and lenticular, plate, and lath microstructures}\label{sec:loop_lenticular_plate}

Problems that have to be posed in three-dimensional domains are now solved. We apply g.disclination theory to study a disclination loop, and lenticular, plate, and lath microstructures. All the results presented in this section are solved within with the finite deformation setting. The body is assumed to be a brick with dimensions of $10\times10\times10$ and eight-node, hexahedral, bilinear finite elements are used with size $0.1\times 0.1 \times 0.1$ (recall that lengths are in terms of the layer width for the eigenwall distributions involved).

\subsubsection{Disclination loop}
\begin{figure}
\centering
\includegraphics[width=0.5\textwidth]{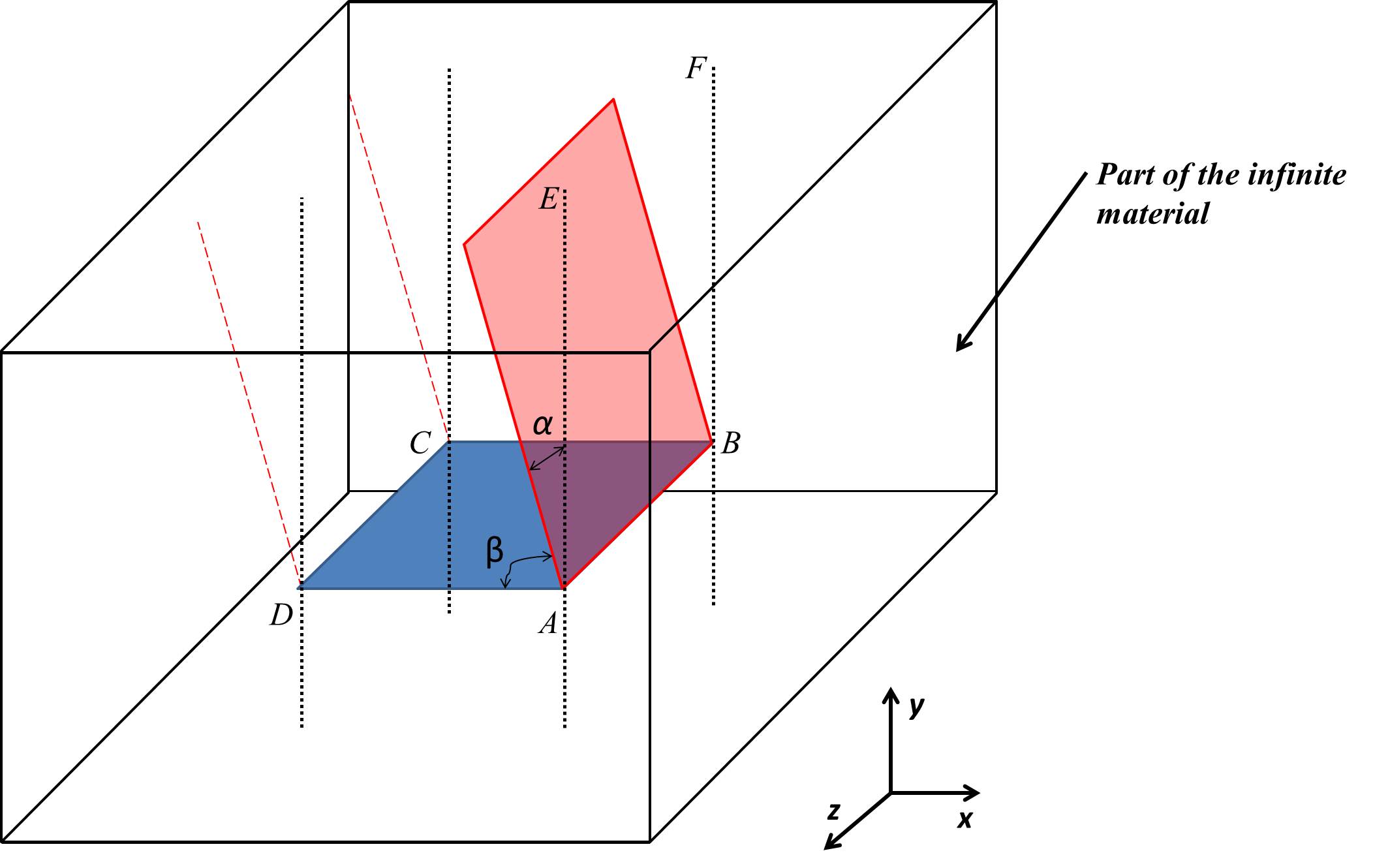}
\caption{Disclination loop configuration in 3d case. The misorientation angle is $\alpha$. $AB$ and $CD$ are wedge disclinations while $AD$ and $BC$ are twist disclinations.}
\label{fig:loop_config}
\end{figure}

\begin{figure}
\centering
\includegraphics[width=0.45\textwidth]{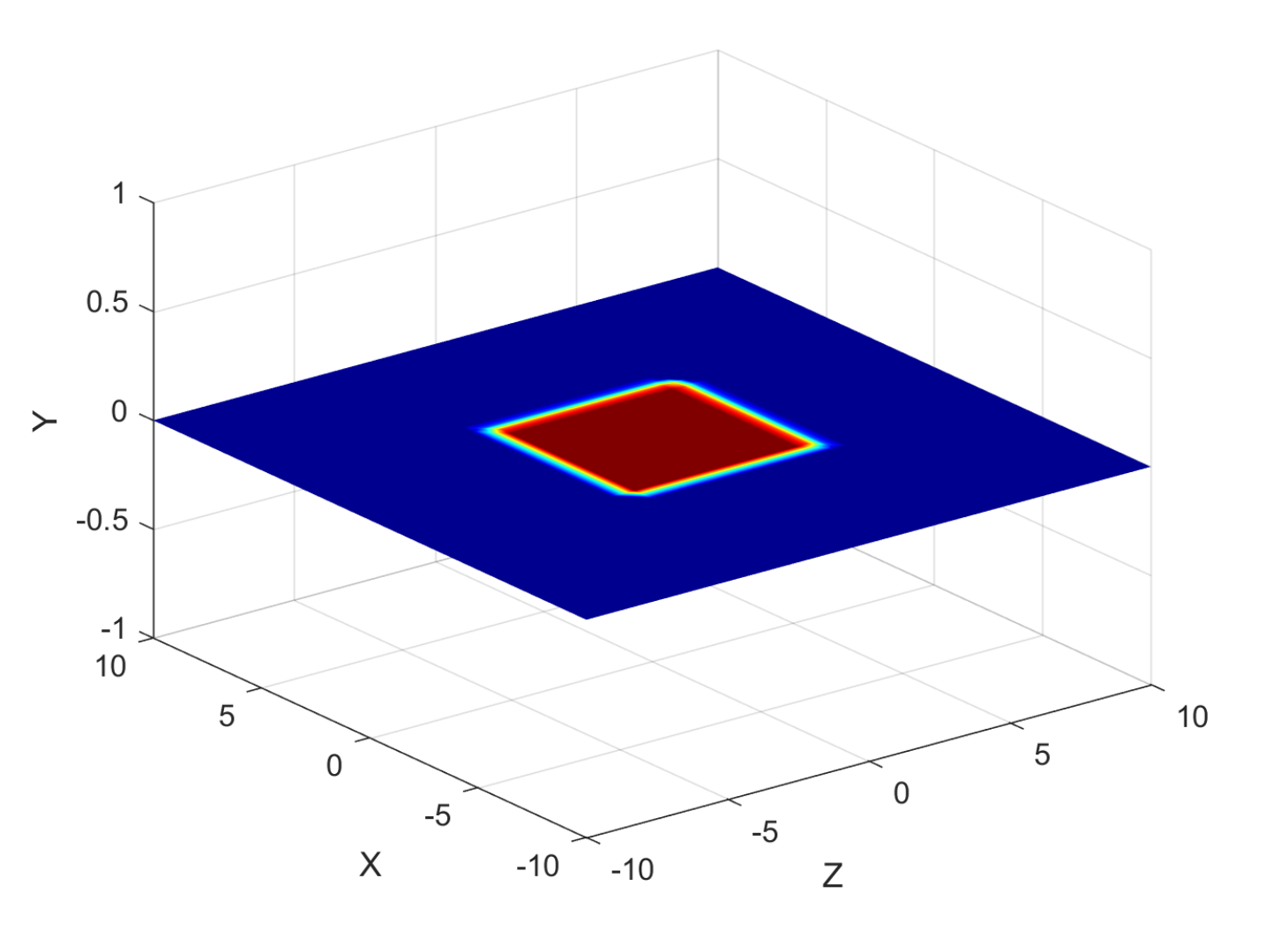}
\caption{The eigenwall field $\bfS$ is constant with support in the layer.}
\label{fig:loop_defect}
\end{figure}

Consider a disclination loop in a 3d domain that is discussed in \cite{zhang_acharya_2016}. The configuration of the disclination loop is shown in Figure \ref{fig:loop_config}, where $AB$ and $CD$ are wedge disclinations while $AD$ and $BC$ are twist disclinations. In this problem, we assume that the $\bfS$ comprises a rotation discontinuity with a $45^{\circ}$ misorientation angle along the $z$ axis, constant in the layer,  as shown in Figure \ref{fig:loop_defect}. 

After assuming the matrix as the reference configuration, the prescribed eigenwall field is defined as
\[
\bfS = 
\begin{cases} \Delta G_{ij}\bfe_i \otimes \bfe_j \otimes \bfe_3 & \text{$|y|\le 1, |x| \le 3$ and $|z| \le 3$} \\
\bf0 & \text{otherwise},
\end{cases}
\]
where $i,j=1,2,3$ and $\Delta G$ is given as 
\[
\begin{bmatrix}
    \cos 45^\circ -1 & \sin 45^\circ & 0 \\
    -\sin 45^\circ & \cos 45^\circ - 1 & 0 \\
    0 & 0 & 0
\end{bmatrix}.
\]

\begin{figure}
\centering
\subfigure[Stress $\sigma_{11}$ for disclination loop viewed on $z=0$ plane in the finite deformation setting.]{
\label{fig:loop_stress_finite_11}
\includegraphics[width=0.45\linewidth]{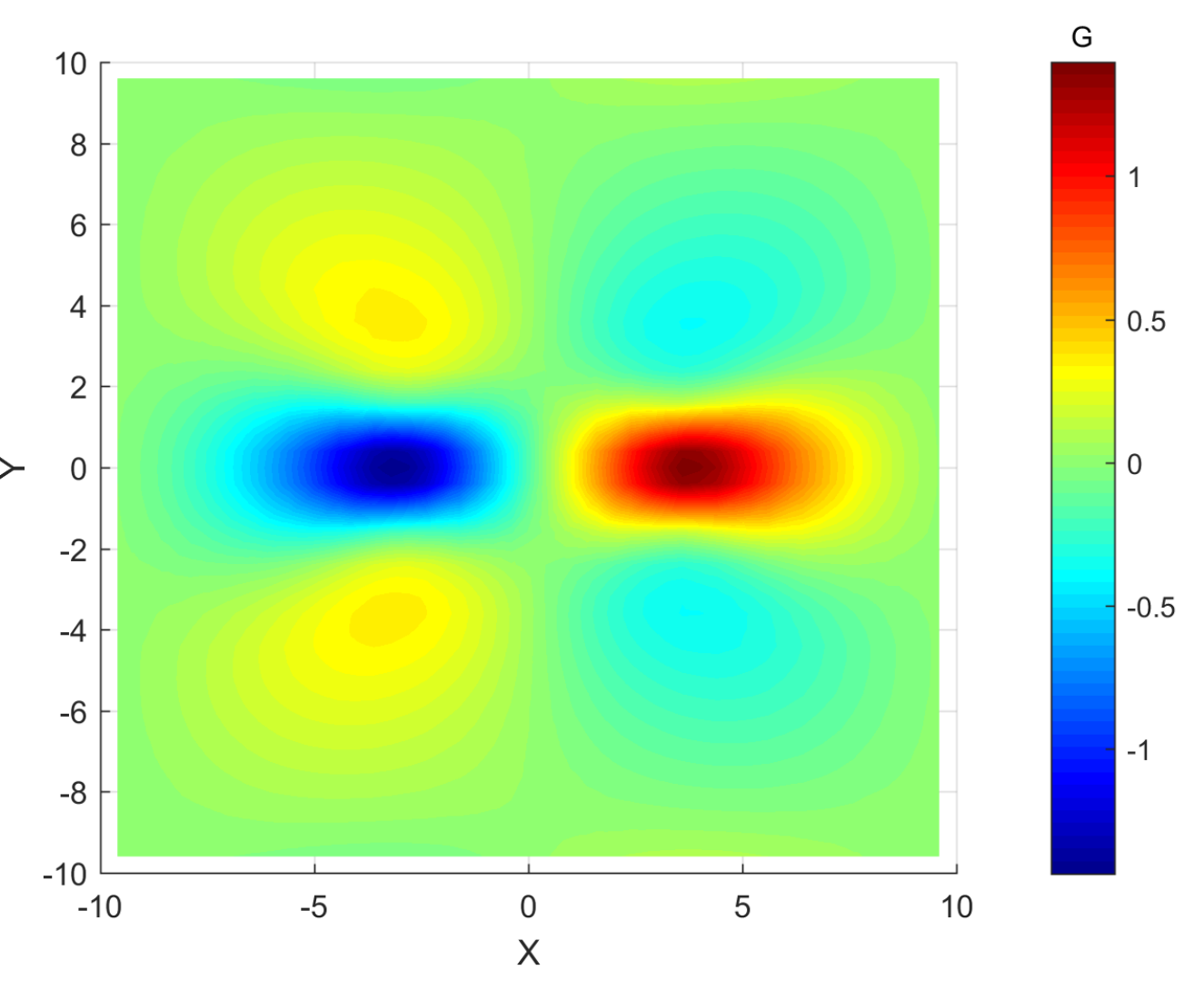}}\qquad
\subfigure[Stress $\sigma_{13}$ for disclination loop viewed on $x=0$ plane in the finite deformation setting.]{
\label{fig:loop_stress_finite_13}
\includegraphics[width=0.45\linewidth]{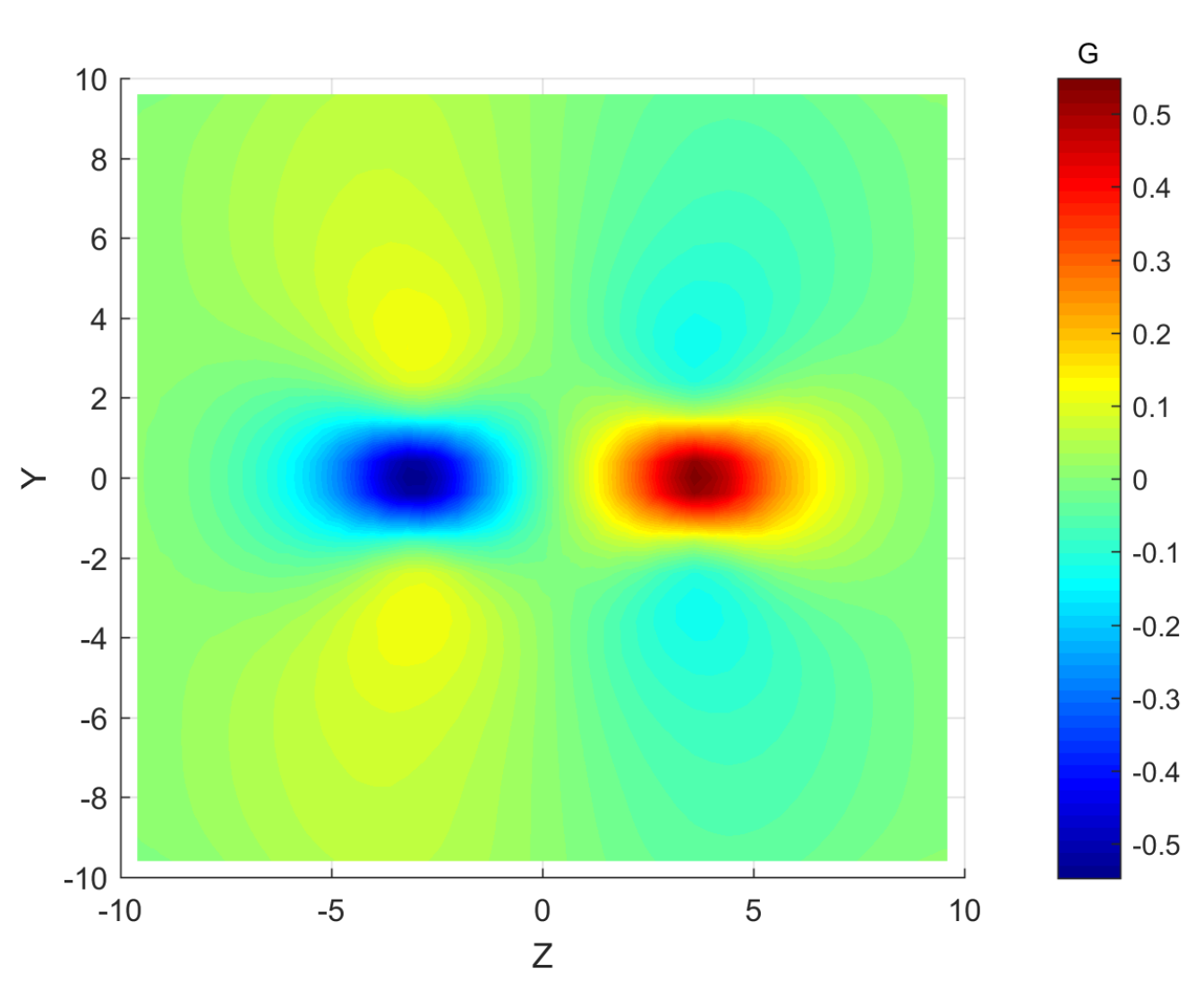}}
\caption{Stress fields $\sigma_{11}$ on $z=0$ plane and $\sigma_{13}$ on $x=0$ plane in the finite deformation setting.}
\end{figure}

Figure \ref{fig:loop_stress_finite_11} and Figure \ref{fig:loop_stress_finite_13} are the stress fields $\sigma_{11}$ on the $z=0$ plane and $\sigma_{13}$ on $x=0$ plane. The stress fields physically match with the description of the disclination loop in \cite{zhang_acharya_2016} that the disclination lines $AB$ and $CD$ parallel to $z$ axis are wedge disclinations ($\sigma_{11}$ is concentrated along $AB$ and $CD$) and the disclination lines $AD$ and $BC$ parallel to $x$ axis are twist disclinations ($\sigma_{13}$ is concentrated along $AD$ and $BC$).

\subsubsection{Stress-inducing inclusion microstructures}\label{sec:martensite_micro}
In this Section we consider four different scenarios by which phase inclusions may induce stresses.

\begin{figure}
\centering
\subfigure[Configuration of a lenticular inclusion in a matrix.]{
\includegraphics[width = 0.45\textwidth]{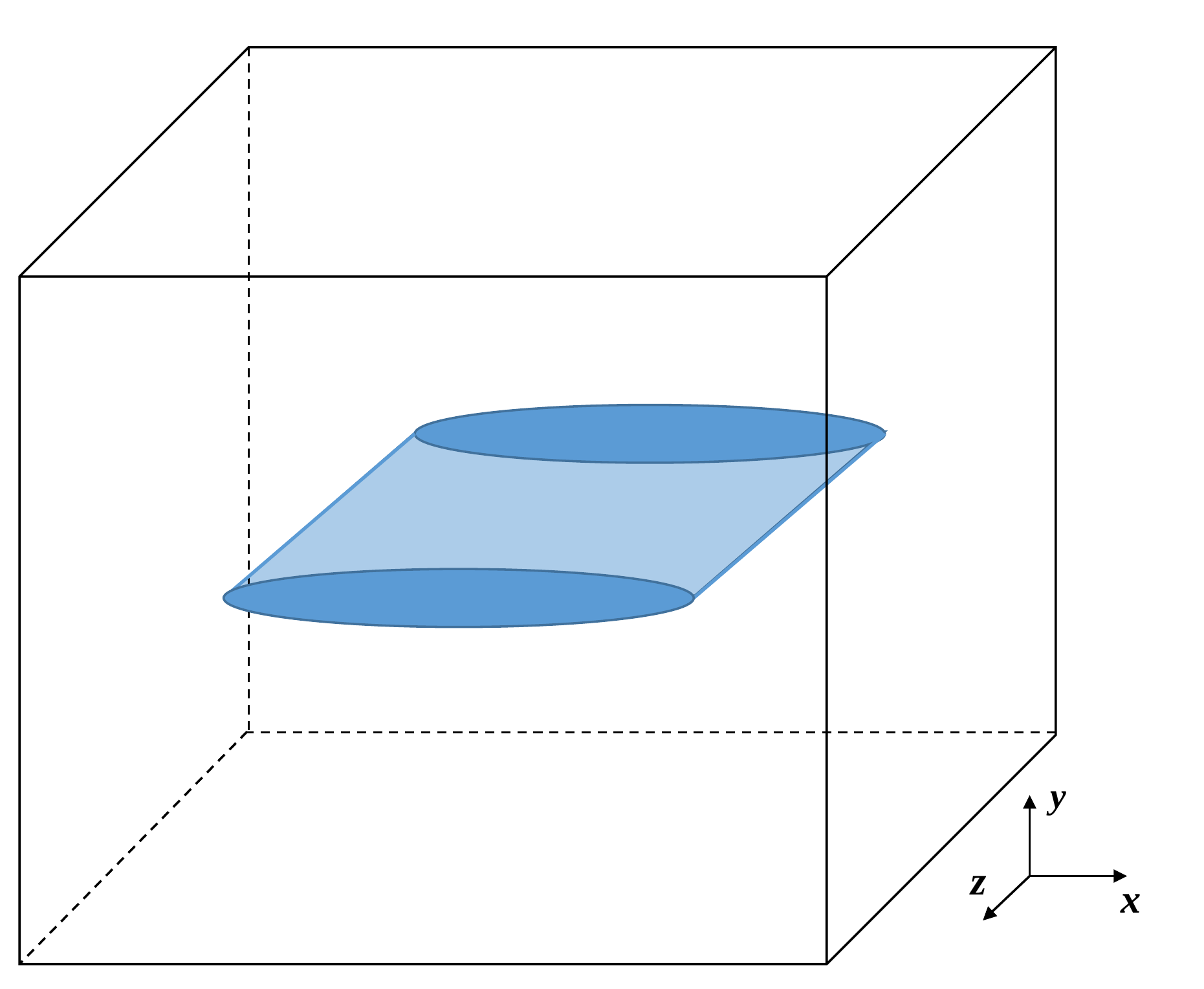}
\label{fig:lenticular_illu}
}
\subfigure[Configuration of a plate inclusion in a matrix.]{
\includegraphics[width = 0.45\textwidth]{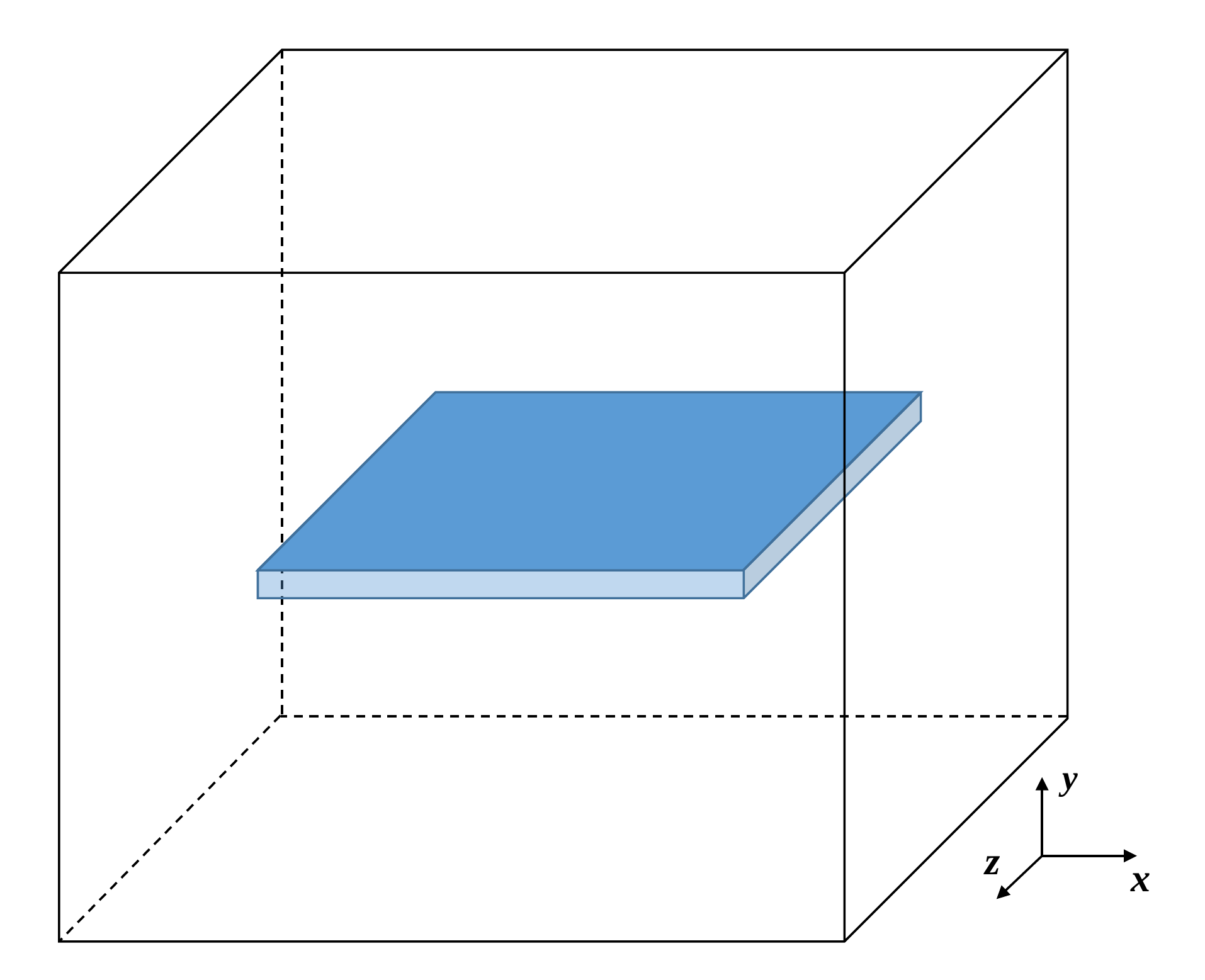}
\label{fig:plate_illu}
}
\caption{Illustrations of the lenticular and plate inclusions in a 3d matrix. In both cases, the inclusions are surrounded, on all sides, by the matrix.}
\label{fig:lenticular_plate_illu}
\end{figure}

Figures \ref{fig:lenticular_illu} and \ref{fig:plate_illu} show the configurations of a lenticular inclusion and a plate inclusion. In all cases, the eigenwall fields $\bfS$ are prescribed along the top and bottom planes of the inclusions; a dislocation density field $\bfalpha$ is also prescribed when the interface is incompatible.

\begin{figure}
\centering
\subfigure[Stress $\sigma_{11}$ for the plate inclusion viewed on $z=0$ plane in the finite deformation setting.]{
\label{fig:plate_stress_finite_11}
\includegraphics[width=0.45\linewidth]{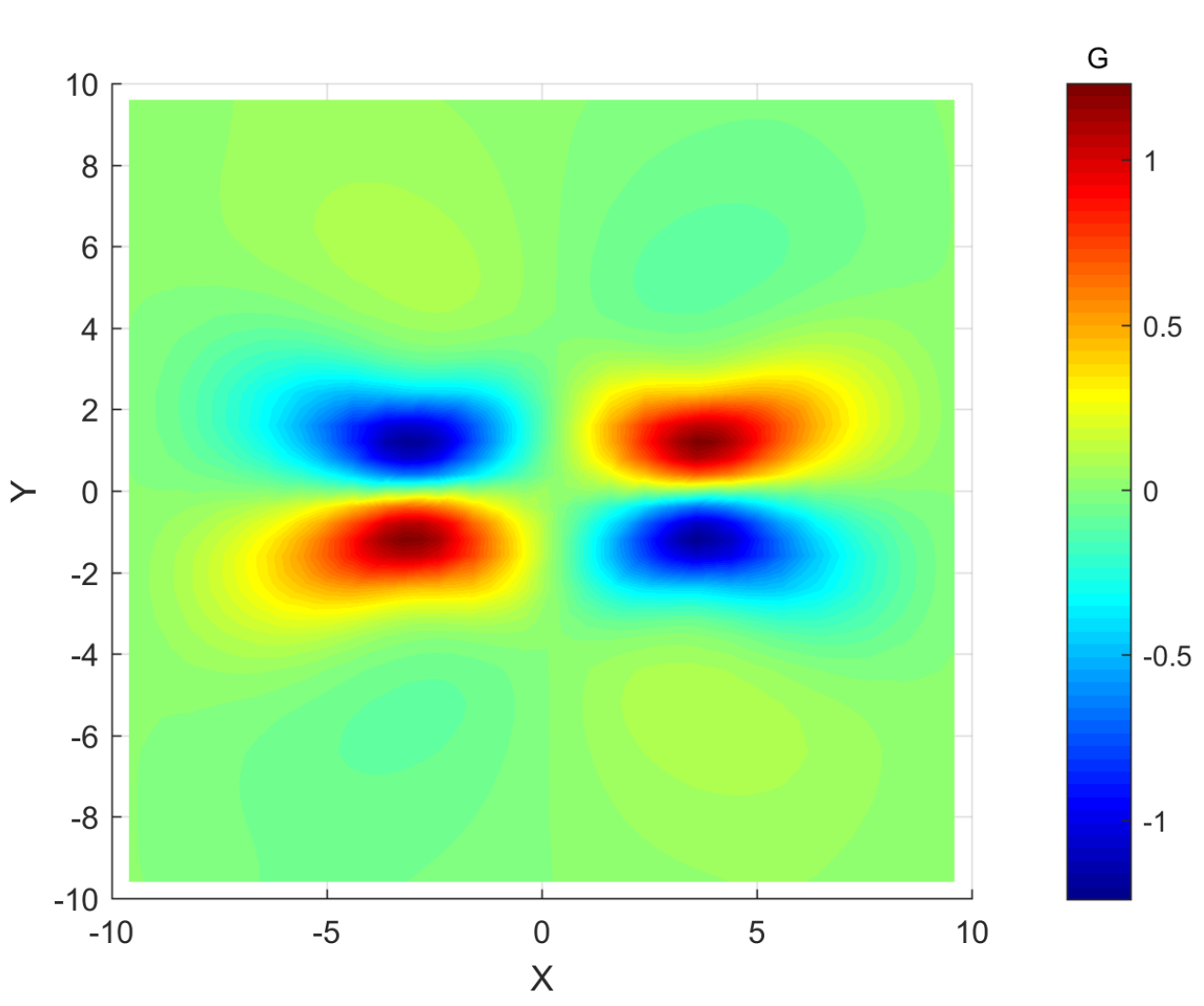}}\qquad
\subfigure[Stress $\sigma_{13}$ for the plate inclusion viewed on $x=0$ plane in the finite deformation setting.]{
\label{fig:plate_stress_finite_13}
\includegraphics[width=0.45\linewidth]{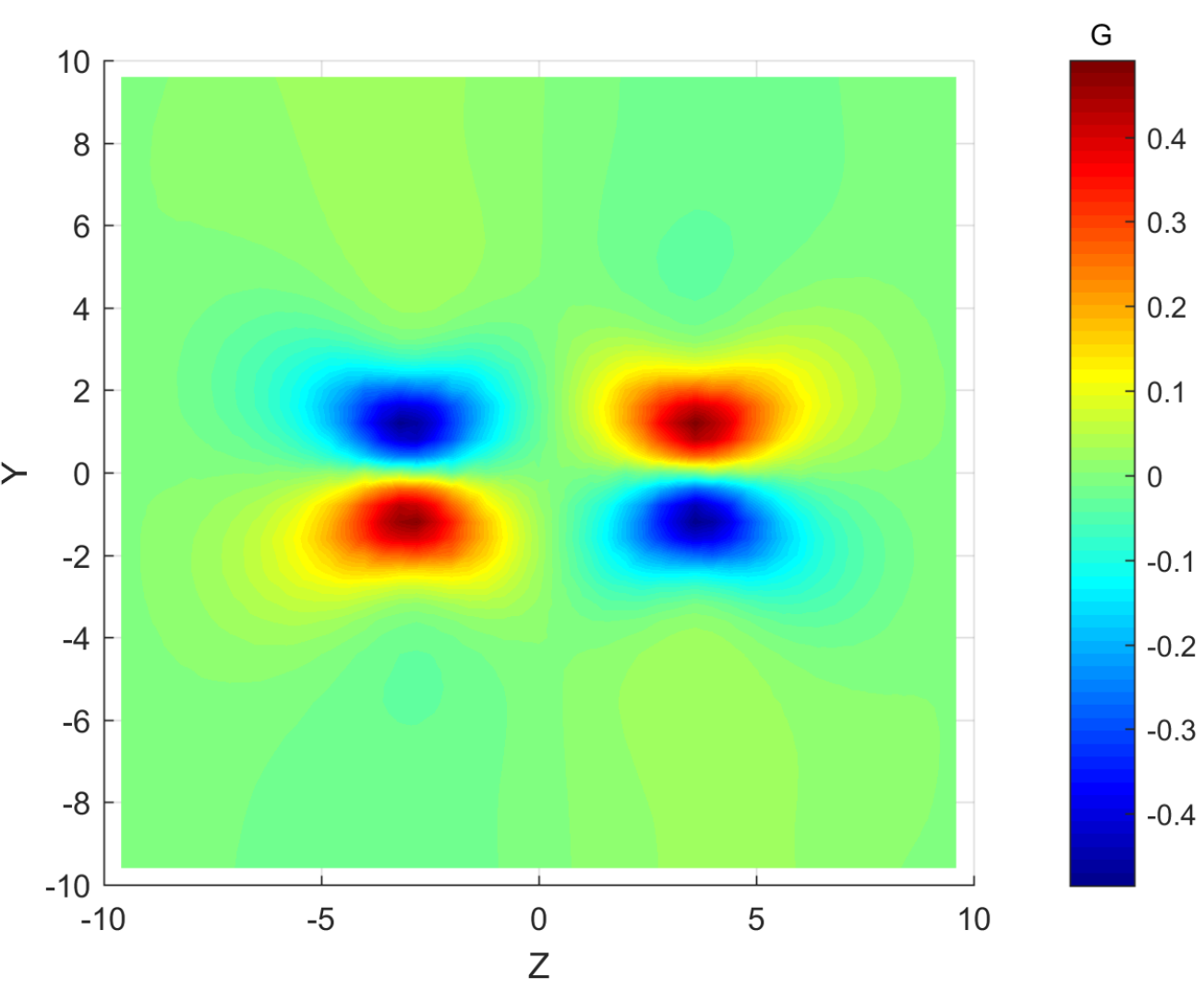}}
\caption{Stress fields $\sigma_{11}$ on $z=0$ plane and $\sigma_{13}$ on $x=0$ plane for the plate inclusion problem.}
\label{fig:plate_stress}
\end{figure}

In the calculations for the plate inclusion, we consider martensite variant transformation problems where the distortions comprising the discontinuity represented by $\bfS$ are based on \cite{levitas1998continuum}. The i-elastic distortion $\bfW_1$ of the martensite inclusion and $\bfW_2$ of the matrix are given as 
\begin{equation} \label{eqn:material_para_martensite}
\bfW_1 = 
\begin{bmatrix}
1 && -0.195  && 0\\
0 && 0.975 && 0 \\
0 && 0 && 1
\end{bmatrix}
\qquad 
\bfW_2 = 
\begin{bmatrix}
1 && 0  && 0\\
0 && 1 && 0 \\
0 && 0 && 1
\end{bmatrix}.
\end{equation}
The thickness of the top and bottom layers comprising the boundaries of the inclusion is $1$. Figure \ref{fig:plate_stress} shows the stress components $\sigma_{11}$ on the $z=0$ plane and $\sigma_{13}$ on the $x=0$ plane for the plate inclusion. For the plate inclusion, the top and bottom interfaces are flat so the g.disclination density field $\bfPi$ as well as the stress field is localized at the terminating cores. Another commonly observed microstructural unit is a lath that can be easily modeled within our setting as a very thin and tall plate inclusion.

A second case we consider is a lenticular martensite inclusion with the transformation distortion of NiTi martensite-austenite adopted from \cite[Sec. 4.1]{bhattacharya2003microstructure} as follows
\[
\bfe^m_i = \bfF \bfe^a_i,
\]
where $\bfe^m_i$ is the image, in the martensite, of $\bfe^a_i$ a lattice vector in the austenite, and $\bfF^e$ is the austenite-martensite transformation distortion. $\bfF^e$ is given as 
\[
\bfF^e =
 \begin{bmatrix}
    0.985 & -0.825 & -0.825 \\
    0 & 9.284 & 0.5 \\
    0 & 0.5 & 9.284
\end{bmatrix}.
\]
In this situation, there does not exist a normal direction to a single interface such that $\bfI - \bfF^{e-1}$ can be represented in rank-one form. Consequently, there have to be dislocations along the interface. Assuming the austenite matrix as the reference configuration and following \eqref{eqn:density_initial}, we have
\[
\bfS = \frac{\bfI - \bfF^{e-1}}{t} \otimes \bfn,
\]
where $t=1$ is the layer thickness, and $\bfn$ is the layer normal pointing outwards from the inclusion. Since the misdistortion (and the eigenwall field) is constant along the interface, the g.disclination density $\bfPi$ is only non-zero at the terminating cores as discussed in Sec. \ref{sec:Pi_layer_def}. In addition, the interface for this martensite-austenite transformation is incompatible and a dislocation density field needs to be prescribed along the interface. In this calculation, we approximate the dislocation density $\bfalpha$ following \eqref{eqn:alpha_prescribe},
\[
\bfalpha = \left(\frac{\bfI - \bfF^{e-1}}{t}\otimes \bfn\right):\bfX.  
\]

\begin{figure}
\centering
\subfigure[Stress $\sigma_{11}$ for the martensite lenticular inclusion in the austenite matrix viewed on $z=0$ plane in the finite deformation setting.]{
\label{fig:lenticular_stress__ma_finite_11}
\includegraphics[width=0.45\linewidth]{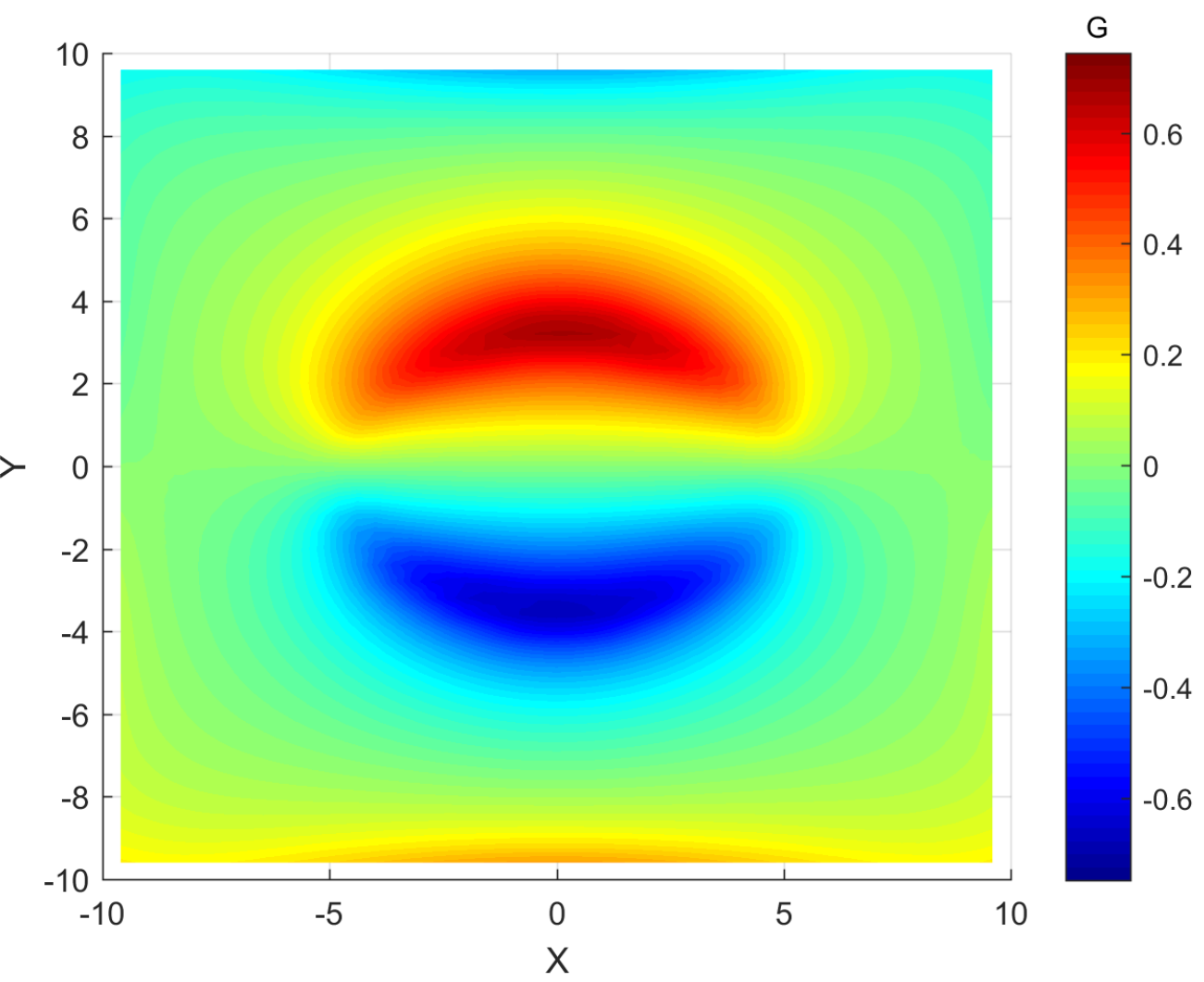}}\qquad
\subfigure[Stress $\sigma_{13}$ for the martensite lenticular inclusion in the austenite matrix viewed on $x=0$ plane in the finite deformation setting.]{
\label{fig:lenticular_stress_ma_finite_13}
\includegraphics[width=0.42\linewidth]{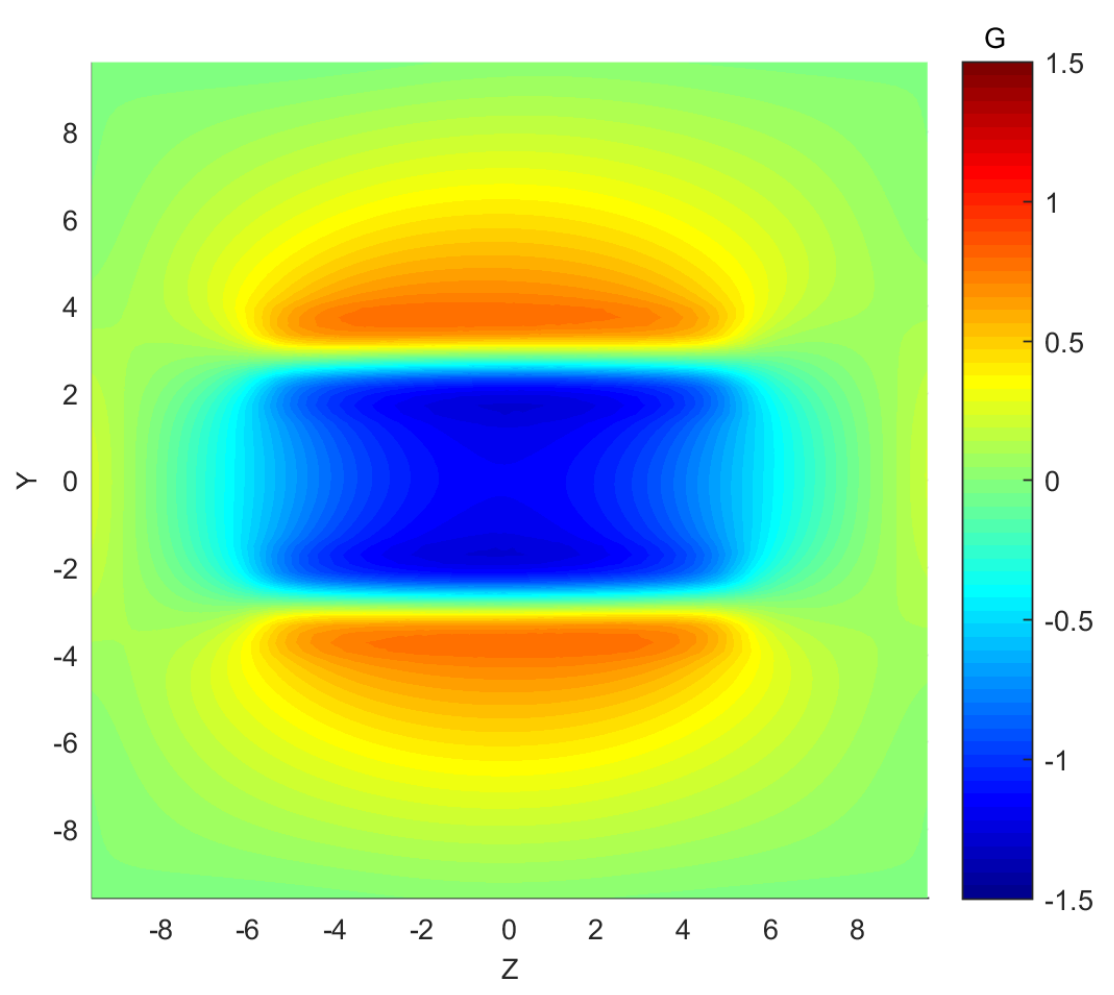}}
\caption{Stress fields $\sigma_{11}$ on $z=0$ plane and $\sigma_{13}$ on $x=0$ plane for the martensite lenticular inclusion in an austenite matrix.}
\label{fig:lenticular_stress_ma}
\end{figure}

Figure \ref{fig:lenticular_stress_ma} shows the stress component $\sigma_{11}$ on $z=0$ plane and $\sigma_{13}$ on $x=0$ plane for the lenticular martensite-austenite transformation. Due to the dislocation density $\bfalpha$ along the interface, the stress is not zero along the interface. 

Another case of theoretical interest is a lenticular martensite variant transformation, where the interface is compatible. In this calculation, we adopt the Ni-Mn-Ga material from \cite{kaufmann2011modulated}, whose orientation angle between two (stress-free) variants is $11.6^{\circ}$. We assume the misdistortion between the inclusion and the matrix along the curved interface to be $tan(11.6^{\circ})\bft \otimes \bfn$, with $\bft$ being a unit vector parallel to the curved interface and $\bfn$ being the interface normal vector. Thus, the eigenwall field $\bfS$ is non-zero within the curved layers (top and bottom boundaries of the inclusion) and can be written as 
\[
\bfS = 
\begin{cases} tan(11.6^{\circ})\bft \otimes \bfn \otimes \bfn  & \text{in the layer}\\
\bf0 & \text{otherwise}.
\end{cases}
\]

\begin{figure}
\centering
\subfigure[Stress $\sigma_{11}$ for the lenticular martensite transformation viewed on $z=0$ plane in the finite deformation setting.]{
\label{fig:lenticular_stress_finite_11}
\includegraphics[width=0.45\linewidth]{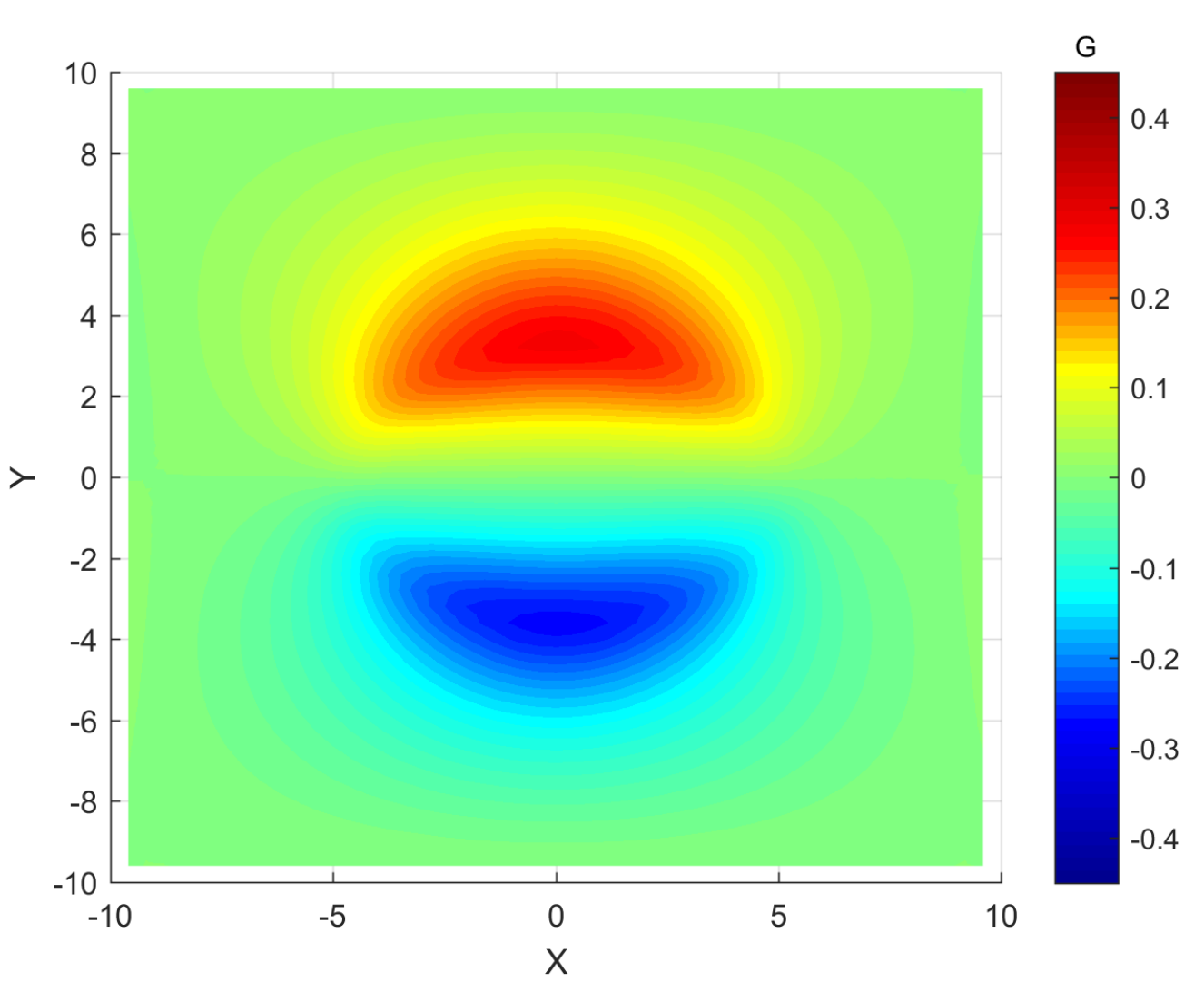}}\qquad
\subfigure[Stress $\sigma_{13}$ for the lenticular martensite transformation viewed on $x=0$ plane in the finite deformation setting.]{
\label{fig:lenticular_stress_finite_13}
\includegraphics[width=0.45\linewidth]{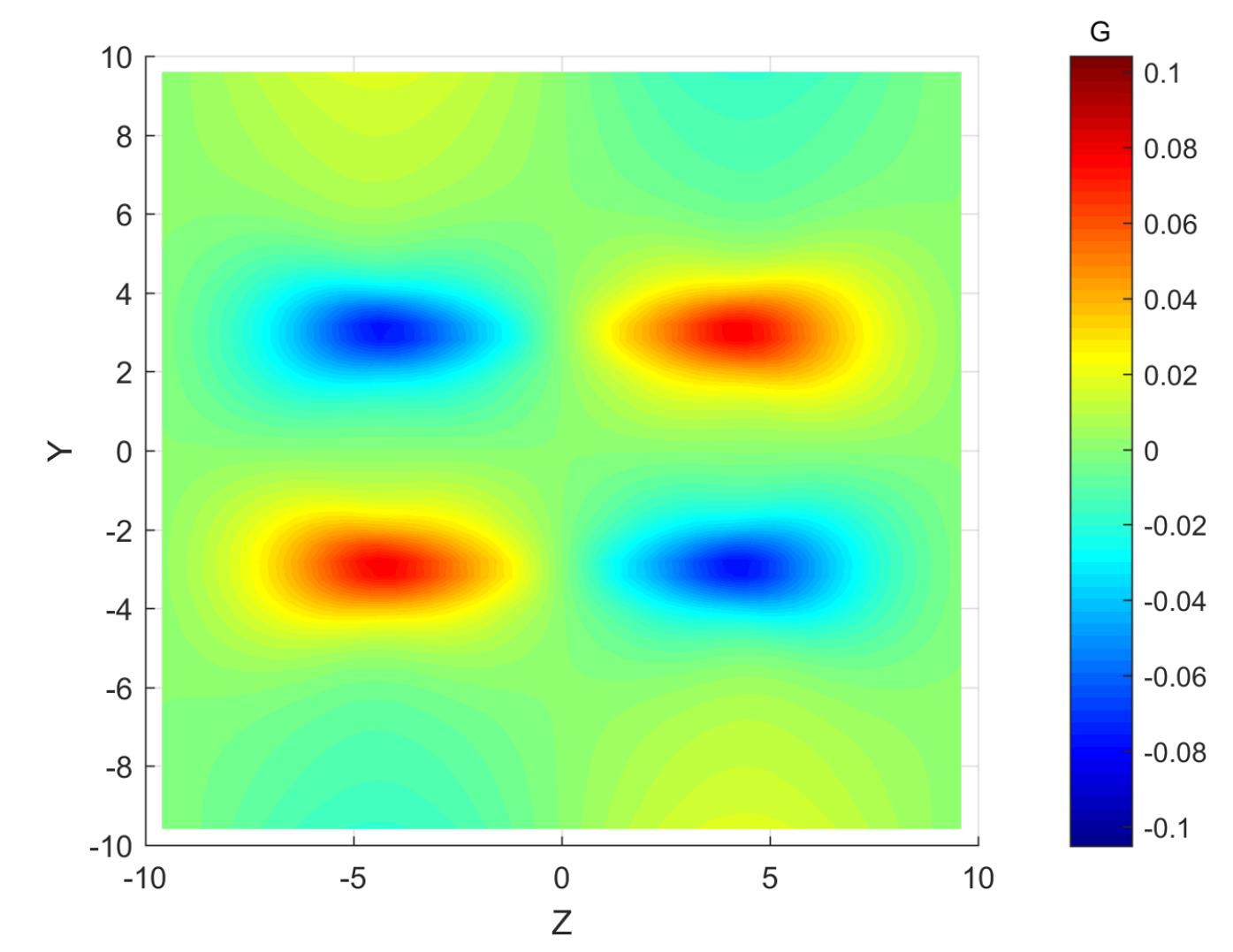}}
\caption{Stress fields $\sigma_{11}$ on $z=0$ plane and $\sigma_{13}$ on $x=0$ plane for the lenticular martensite transformation.}
\label{fig:lenticular_stress}
\end{figure}

Figure \ref{fig:lenticular_stress} shows the stress components $\sigma_{11}$ on $z=0$ plane and $\sigma_{13}$ on $x=0$ plane for the lenticular martensite transformation. Although we do not prescribe the dislocation density $\bfalpha$ due to the compatible interface, the $\bfPi$ is no longer localized at the terminating cores based on the reasoning in (\ref{eqn:append_curvy}) in Appendix \ref{sec:append_pi}. Thus, the stress field along the interfaces is non-zero, as shown in Figure \ref{fig:lenticular_stress_finite_11}. 

We now calculate the fields of a needle shaped inclusions of one martensite variant in another, motivated by the observations in \cite{SEINER20112005}. As opposed to the previous case of a curved interface carrying a rank-one misdistortion at all points, but inducing stresses due to the development of a g.disclination distribution along it, in this example the flat parts of the interface carry no defects, but a stress is developed because the normal to the curved parts of the interface do not agree with the normal direction required by the misdistortion to be compatible (note that this is different from the austenite-martensite transformation described earlier where no flat compatible interface exists). Thus, a dislocation density field needs to be specified along the interface and we specify it in the form 
\[
\bfalpha = \left(\frac{\bfW_2-\bfW_1}{t}\otimes \bfn\right):\bfX,
\]
where $\bfn$ is the interface normal pointing outwards from the inclusion, $t$ is the layer thickness, and $\bfW_1$ and $\bfW_2$ are i-elastic distortions specified in \eqref{eqn:material_para_martensite}.

\begin{figure}
\centering
\subfigure[Illustration of the needle inclusion of one martensite variant in another.]{
\label{fig:needle_shape}
\includegraphics[width=0.45\linewidth]{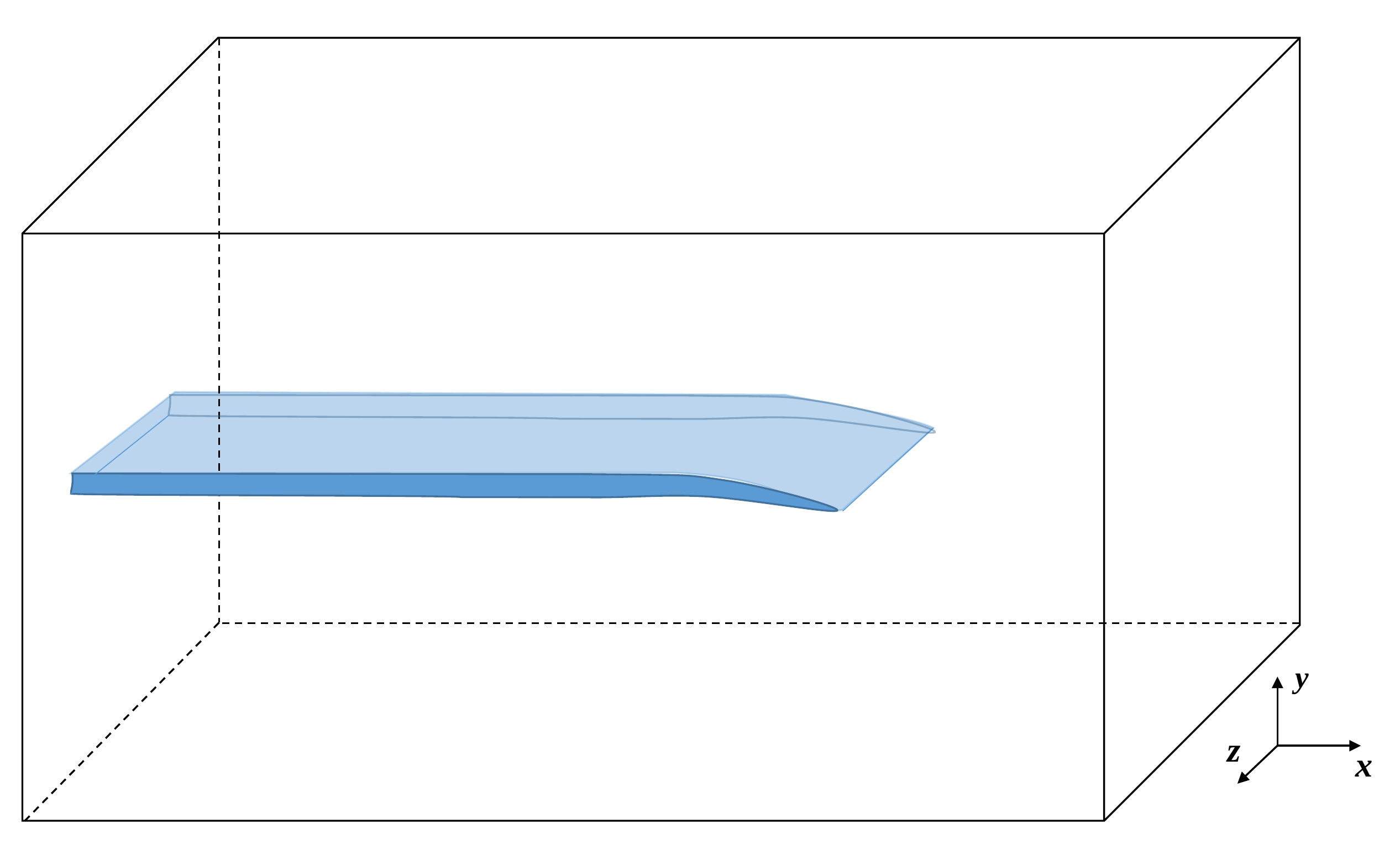}}\qquad
\subfigure[Magnitude of the stress $\bfsigma$ for the needle inclusion viewed on $z=0$ plane in finite deformation setting.]{
\label{fig:needle_stress}
\includegraphics[width=0.45\linewidth]{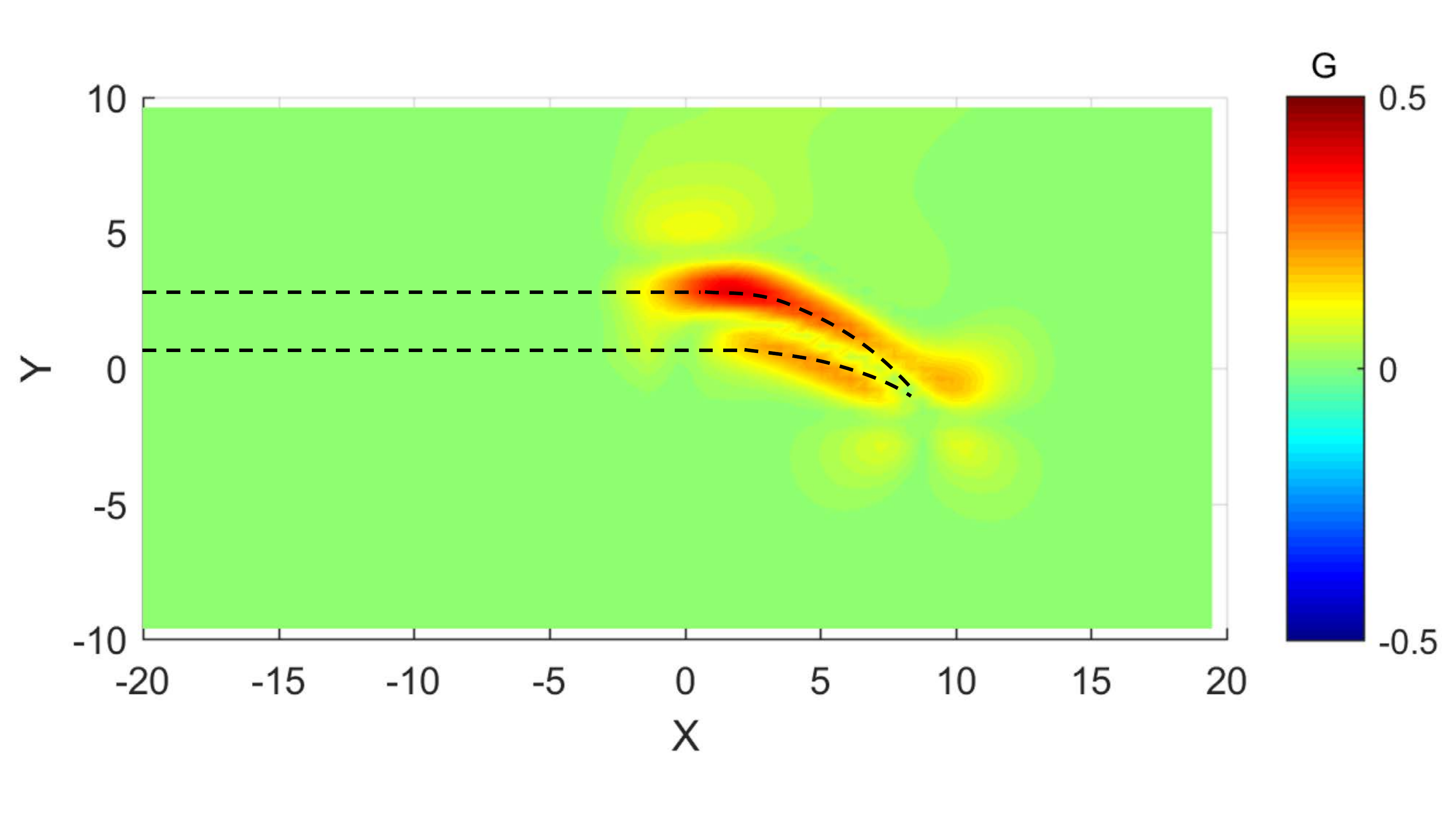}}
\caption{The needle martensite inclusion configuration and magnitude of the stress $\bfsigma$ on $z=0$ plane. The black dash lines represent the top and bottom interfaces of the needle inclusion. The stress field is localized along the curved interface.}
\label{fig:needle}
\end{figure}

Figure \ref{fig:needle_shape} shows the needle inclusion configuration of our calculation and the Figure \ref{fig:needle_stress} shows the $L^2$-norm of $\bfsigma$ for the needle inclusion viewed on $z=0$ plane with finite deformation setting. The stress is localized along the curved interface due to the dislocation density generated from the incompatibility.

\section{Contact with the classical elastic disclination theory}\label{sec:layer_core}

We show here the circumstances in which g.disclination theory reduces exactly to DeWit's \cite{dewit1973theory} defect model, including uniqueness assertions for the stress in both models. Due to the classical theory being established for small deformations, our considerations here are restricted to the small deformation case.

Recall the governing relation $\curl \bfS = \bfPi$. A single isolated g.disclination can be specified by specifying $\bfS$ as an eigenwall field with support in a terminating layer, with appropriate decay properties in a core cylinder at its termination that results in a non-vanishing $\bfPi$ field being defined there. As in \eqref{eqn:s_decompose}, $\bfS = \bfS^{\perp} + \grad \bfZ^s$. We can decompose $\bfS^{\perp}$ into two part, $\bfS^{\perp} = \bfS^{{\perp}skw}+\bfS^{{\perp}sym}$, where $\bfS^{{\perp}skw}$ is a  third-order tensor tensor skew in the first two indices and $\bfS^{{\perp}sym}$ is symmetric in the first two indices:
\begin{equation*}
\bfS^{{\perp}skw}_{ijk} = \frac{1}{2} (\bfS^{\perp}_{ijk}-\bfS^{\perp}_{jik}); \ \ \ \ \bfS^{{\perp}sym}_{ijk} = \frac{1}{2} (\bfS^{\perp}_{ijk}+\bfS^{\perp}_{jik}).
\end{equation*}
Then we have
\[
\bfPi = \curl \bfS^{\perp} = \curl (\bfS^{{\perp}skw} + \bfS^{{\perp}sym}) = \curl \bfS^{{\perp}skw} + \curl \bfS^{{\perp}sym},
\]
and we define
\begin{equation*}
\bfPi^{skw} := \curl \bfS^{{\perp}skw}; \ \ \ \bfPi^{sym} := \curl \bfS^{{\perp}sym}
\end{equation*}
so that
\[
\bfPi = \bfPi^{skw} +\bfPi^{sym}.
\]

It can be checked from the definitions \eqref{eqn:s_decompose} and \eqref{eqn:S_decomposition} that if $\bfS$ is skew in its first two-indices, then $\bfS^\perp_{sym} = \bf0$. The same conclusion holds if $\bfPi$ is skew in its first two indices. 

Recall the dislocation density $\bfalpha$ defined in \eqref{eqn:dislocation}
\begin{equation}\label{eqn:alpha_trans}
\begin{split}
& \bfalpha = \bfS:\bfX + \grad\bfW:\bfX \Rightarrow \bfalpha = \bfS^{\perp}:\bfX + \grad \hat{\bfW} :\bfX \\
& \alpha_{il} = -\epsilon_{lkj}\hat{W}_{ij,k}+\epsilon_{jkl}S^{\perp}_{ijk} \\
& \Rightarrow \epsilon_{rqi}\alpha_{il,q} = -\epsilon_{rqi}\epsilon_{lkj}\hat{W}_{ij,kq}+\epsilon_{rqi}\epsilon_{jkl}S^{\perp}_{ijk,q} 
\Rightarrow \epsilon_{rqi}\alpha^T_{li,q} = - \epsilon_{rqi}\epsilon_{lkj}\hat{W}_{ij,kq} + \epsilon_{ljk}\epsilon_{rqi}S^{\perp}_{ijk,q}.
\end{split}
\end{equation}
Substituting $\bfS^{\perp} = \bfS^{{\perp}skw}+\bfS^{{\perp}sym}$ into the second term of \eqref{eqn:alpha_trans}, we have
\begin{equation}\label{eqn:epeps}
\epsilon_{ljk}\epsilon_{rqi}S^{\perp}_{ijk,q} = \epsilon_{ljk}\epsilon_{rqi}(S^{{\perp}skw}_{ijk,q}+S^{{\perp}sym}_{ijk,q}).
\end{equation}
Since $\bfS^{{\perp}skw}$ is skew in the first two indices, there exists a second order tensor $\bfomega$ such that
\begin{equation}\label{eqn:S_rotation}
S^{{\perp}skw}_{ijk}=\epsilon_{ijs}\omega_{sk}
\end{equation}
so that
\begin{equation}\label{eqn:temp_pi}
\Pi^{skw}_{ijk} = \epsilon_{knm}S^{{\perp}skw}_{ijm,n} \ \Rightarrow \ \Pi^{skw}_{ijk} = \epsilon_{knm}\epsilon_{ijs}\omega_{sm,n} \ \Rightarrow \ \epsilon_{ijq} \Pi^{skw}_{ijk} = \epsilon_{knm}\omega_{qm,n}.
\end{equation}
Equations \eqref{eqn:epeps}, \eqref{eqn:S_rotation}, and \eqref{eqn:temp_pi} yield
\begin{equation}\label{eqn:temp1}
\epsilon_{ljk}\epsilon_{rqi}S^{\perp}_{ijk,q} = \epsilon_{ljk}\epsilon_{rqi}\epsilon_{ijs}\omega_{sk,q} + \epsilon_{ljk}\epsilon_{rqi}S^{{\perp}sym}_{ijk,q}.
\end{equation}
Using \eqref{eqn:temp1} and \eqref{eqn:temp_pi} to note that
\begin{equation}\label{eqntemp2}
\epsilon_{rqi}\epsilon_{ljk}\epsilon_{ijs} \omega_{sk,q} = \epsilon_{rqi}[\delta_{li}\delta_{ks} - \delta_{ls}\delta_{ki}]\omega_{sk,q} = \epsilon_{lrq}\omega_{kk,q}+\epsilon_{ijl}\Pi^{skw}_{ijr},
\end{equation}
we have
\begin{equation}\label{eqn:second_part}
\epsilon_{ljk}\epsilon_{rqi}S^{\perp}_{ijk,q} = \epsilon_{lrq}\omega_{kk,q}+\epsilon_{ijl}\Pi^{skw}_{ijr} +\epsilon_{ljk}\epsilon_{rqi}S^{{\perp}sym}_{ijk,q}.
\end{equation}

For small deformations, $\hat{\bfW} = \bfI - \hat{\bfU}^e$ and we decompose $\hat{\bfU}^e$ into symmetric and skew parts, $\hat{\bfU}^e= \hat{\bfepsilon}^e + \hat{\bfOmega}^e$.
Then we have
\begin{equation}\label{eqn:first_term}
\frac{1}{2} \left( \epsilon_{rqi}\epsilon_{lkj}\hat{W}_{ij,kq} + \epsilon_{lqi}\epsilon_{rkj}\hat{W}_{ij,kq} \right) = - \epsilon_{rqi}\epsilon_{lkj}\hat{\epsilon}^e_{ij,kq}.
\end{equation}
Therefore, substituting \eqref{eqn:second_part} and \eqref{eqn:first_term} into \eqref{eqn:alpha_trans} and taking the symmetric part, we have
\begin{equation}\label{eqn:source}
\left[ \curl \left( \bfalpha^T \right) \right]_{sym} - (\bfPi:\bfX)_{sym} - \left( \curl\left[\left(\bfS^{{\perp}sym}:\bfX\right)^T\right]\right)_{sym} = inc(\hat{\bfepsilon}^{e}),
\end{equation}
where $inc$ is the St. Venant compatibility operator. When $\bfS^{{\perp}sym} = \bf0$, $\bfPi = \bfPi^{skw}$ and \eqref{eqn:source} becomes
\begin{equation}\label{eqn:source_skw}
\left[ \curl (\bfalpha^T) \right]_{sym} - (\bfPi:\bfX)_{sym}  = inc(\hat{\bfepsilon}^{e}),
\end{equation}
which indicates that $inc (\hat{\bfepsilon}^{e})$ is sourced by the defect density fields $\bfalpha$ and $\bfPi$. The linear elastic stress field $\bfT = \bfC \hat{\bfepsilon}^{e}$, with $\bfC$ having the minor symmetries, satisfies equilibrium
\begin{equation}\label{eqn:stress_skw}
\divergence (\bfC:\hat{\bfepsilon}^e) = \bf0.
\end{equation}
When $\bfS^{{\perp}sym}=\bf0$, DeWit's disclination density $\bftheta$ can be defined as $\bfPi:\bfX$ and equations \eqref{eqn:source_skw} and \eqref{eqn:stress_skw} become exactly DeWit's model \cite{dewit1973theory}. 

Thus, we have shown that the stress and $\hat{\bfU}^e_{sym}$ of any solution of small deformation g.disclination theory \eqref{eqn:small_deformation_summary_hat} satisfies the equations of DeWit's theory when $\bfS^{\perp sym} = \bf0$. 

It is shown in Appendix \ref{sec:append_uniqueness} that \eqref{eqn:source_skw} and \eqref{eqn:stress_skw} suffice to uniquely determine the stress field in finite bodies when $\bfC$ is positive-definite (possibly spatially inhomogeneous and with arbitrary anisotropy), when the left-hand-side of \eqref{eqn:source_skw} and statically admissible applied boundary tractions are prescribed data. Hence, for this data, solutions for stress and $\bfU^e_{sym}$ exactly match solutions for the same quantities from DeWit's model.

In Appendix \ref{sec:append_uniqueness} we also prove uniqueness of solutions to linear g.disclination theory and show that for identical prescribed data corresponding to pure disclinations, dislocations and applied tractions, g.disclination theory produces more information than classical disclination theory.

\section{Conclusion} \label{sec:conclusion}
G.disclination theory \cite{acharya2015continuum} is reviewed and computationally implemented in the limited context where the dislocation density field $\bfalpha$ and either the eigenwall field $\bfS$ or g.disclination density field $\bfPi$ are given as input data. The theory deals with discontinuities in elastic distortion involving defects beyond translational dislocations and rotational disclinations.

A numerical scheme based on the Least Squares and Galerkin Finite element methods for solving the g.disclination theory is developed. Both the small deformation (linear) and finite deformation (nonlinear) settings are considered. Various grain and phase boundary problems, including dislocations and disconnections, are solved. By comparing results from our model with the results of classical linear defect theory due to DeWit \cite{dewit1973theory} for both the single disclination and the single dislocation, we have demonstrated that our model is capable of recovering the essential beyond-core features of Volterra defects. Contact has also been made with the Eshelby cut-weld interpretation of a single disclination, at finite deformations. The necessity of accounting for finite deformation theory in many problems related to defects with high misorientations has been demonstrated.

Future work will involve the development of computational tools for the analysis of the full dynamical theory of defect evolution presented in \cite{acharya2015continuum}. Interestingly, the results of this paper seem to suggest that it may very well be within the reach of the dynamical model to deal with non-convex surface energies typical of physically measured grain boundary energies, and to deal with phase transformation problems at large deformations without the use of non-convex elastic stress-strain relationships.

\section*{Acknowledgments}
CZ and AA acknowledge support from grant NSF-DMS-1434734. AA also acknowledges support from grants NSF-CMMI-1435624 and ARO W911NF-15-1-0239.

\appendix

\section*{Appendices}

\section{The fields $\hat{\bfW}$, $\bfW$, and kinematic constraints on $\hat{\bff}$} \label{sec:append_3}

In this Appendix we outline some physical thought-experiments for understanding the fields $\hat{\bfW}$ and $\bfW$, and guidelines for the kinematic constraints on the field $\hat{\bff}$ for the unique solution of  \eqref{eqn:disc_num_finite} when physically expected. The treatment is necessarily non-rigorous  (given the scope of the undertaking), but we nevertheless provide it to lay out our intuition behind the various mathematical constructs used in the paper.

On the current configuration, $\hat{\bfW}$ is to be physically understood at any given point $\bfx$ by the relaxation of a small neighborhood of atoms around $\bfx$ (our interpretation of this procedure is explained in \cite[Sec. 5.4.1]{acharya2015continuum}, with $\bfW(\bfx)$ there to be interpreted as $\hat{\bfW}(\bfx)$ here). We assume that the relaxation always takes small neighborhoods to a state that is the `closest' zero-energy state for the neighborhood from its state in the (generally) stressed current configuration. Let the arbitrarily chosen point where the condition $\bfH^s(\bfx_0)= \bf0$ is imposed be $\bfx_0$. Thus $\hat{\bfW}(\bfx_0) = \bfW(\bfx_0)$. This process of relaxation generates a relaxed shape of the local neighborhood around $\bfx_0$. We will refer to this shape as the reference \emph{tile}. We physically interpret $\bfW(\bfx)$ at any point $\bfx$ as follows:

\begin{itemize}
\item Select a small shape around $\bfx$ in the current configuration.
\item `Measure' the traction acting on the shape through its boundary in the current configuration.
\item Calculate the traction that needs to be applied on the the reference tile to fit into the current shape.
\item Compare two traction fields. If they match, then the shape under consideration is one admissible choice, and the deformation gradient from the current shape to the reference tile is one admissible value of $\bfW(\bfx)$.
\item Given a current configuration and $\hat{\bfW}(\bfx_0)$, in general there can be a set of admissible $\bfW(\bfx)$ for each $\bfx$ in the current configuration. For example, consider a stress-free twin boundary in the current configuration with $\bfx_0$ being in one variant of martensite; then $\bfW(\bfx)$ for $\bfx$ lying in an adjoining variant can be $\bfI$ or correspond to the twinning shear deformation between the two variants. The actual $\bfW(\bfx)$ is decided by further physical considerations, e.g. the  microstructure in the current configuration like the presence of boundaries or defects (of course, the mathematical theory is designed to predict  a definite evolution for the $\bfW$ field).
\end{itemize}

The above procedure allows one to define the fields $\bfS$ and $\bfalpha$, at least in principle. Our theory requires the specification of hard constraints on the field $\hat{\bff}$ for a nominally unique solution to the system \eqref{eqn:disc_num_finite}. Recall that $\hat{\bfW}(\bfx_0) = \hat{\bfchi}(\bfx_0)+\grad \hat{\bff} (\bfx_0) - \bfH^s(\bfx_0)$. Given $\bfS$ and $\bfalpha$, $\hat{\bfchi}(\bfx_0)$ and $\bfH^s(\bfx_0)$ are known, and thus $\grad\hat{\bff}(\bfx_0)$ is known. The kinematic constraints on $\hat{\bff}$ may be generated as follows: choose $\hat{\bff}(\bfx_0)$ arbitrarily; then using $\grad\hat{\bff}(\bfx_0)$, determine $\hat{\bff}(\bfx_0+\delta\bfx)$ around $\bfx_0$ for a small $\delta \bfx$. Then $\hat{\bff}(\bfx_0)$ and $\hat{\bff}(\bfx_0+\bfdelta)$, for appropriately chosen values of $\bfdelta$, can serve as the conditions on $\hat{\bff}$ for eliminating `rigid-body deformation' modes.

As illustrations of some of these ideas, consider the through twin boundary discussed in Section \ref{sec:twin_through}. Figure \ref{fig:layer_skew_through_deformation_2} is the current configuration. For the through twin boundary, $\hat{\bfW}$ is the identity field, and thus the closest-well, stress-free reference is compatible with, and identical to, the current configuration. On the other hand, the elastic reference (Fig. \ref{fig:layer_skew_through_deformation_3}) obtained by mapping the current configuration by $\bfW$ is also compatible, but now represents a compatible shearing across the twin boundary.

\begin{figure}
\centering
\subfigure[The reference configuration mapped by $\bfW$ field for the compatible terminating twin.]{
\includegraphics[width = 0.4\linewidth]{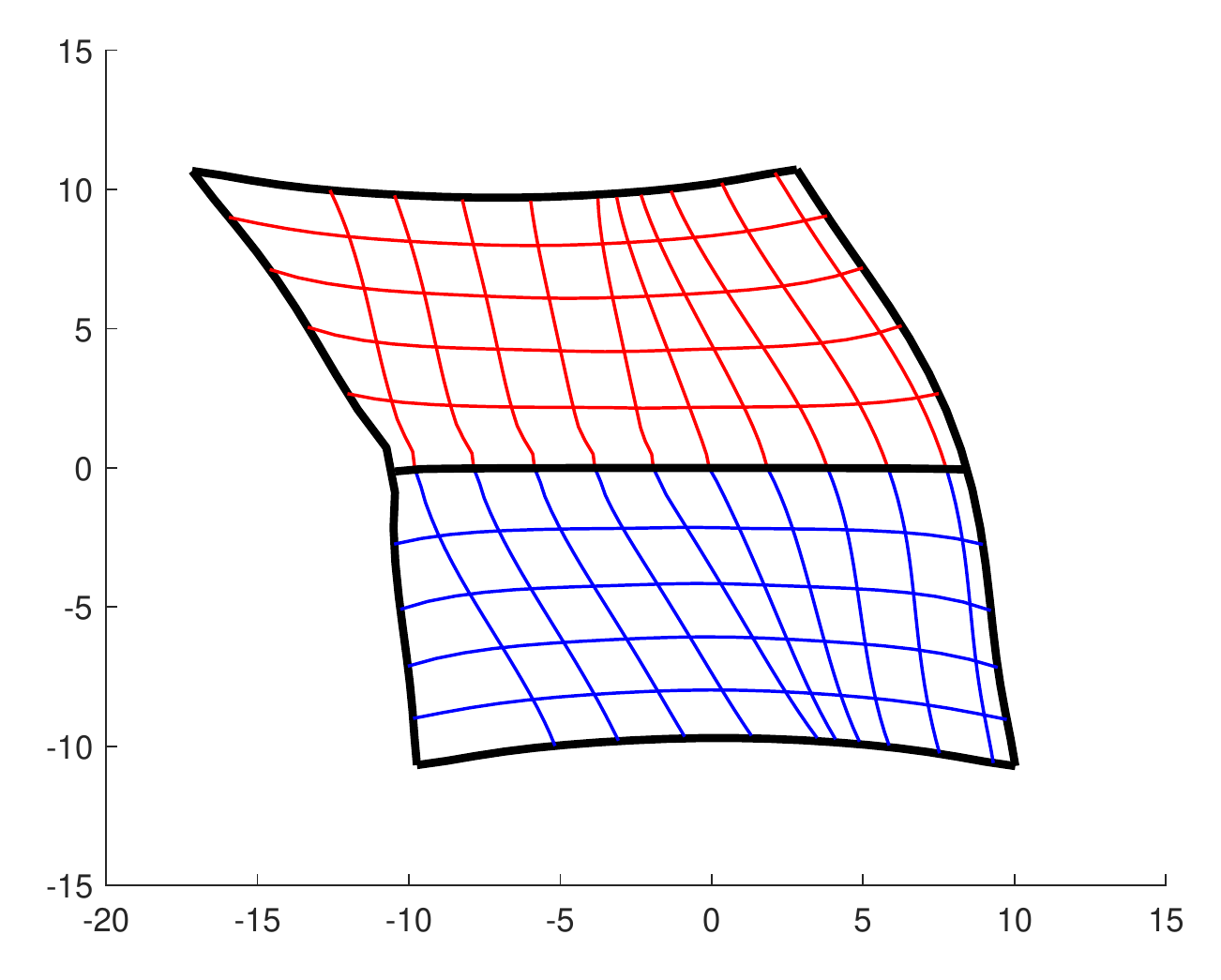}
\label{fig:terminate_twin_deformation_2}
}\qquad
\subfigure[The current configuration for the compatible terminating twin.]{
\includegraphics[width = 0.4\linewidth]{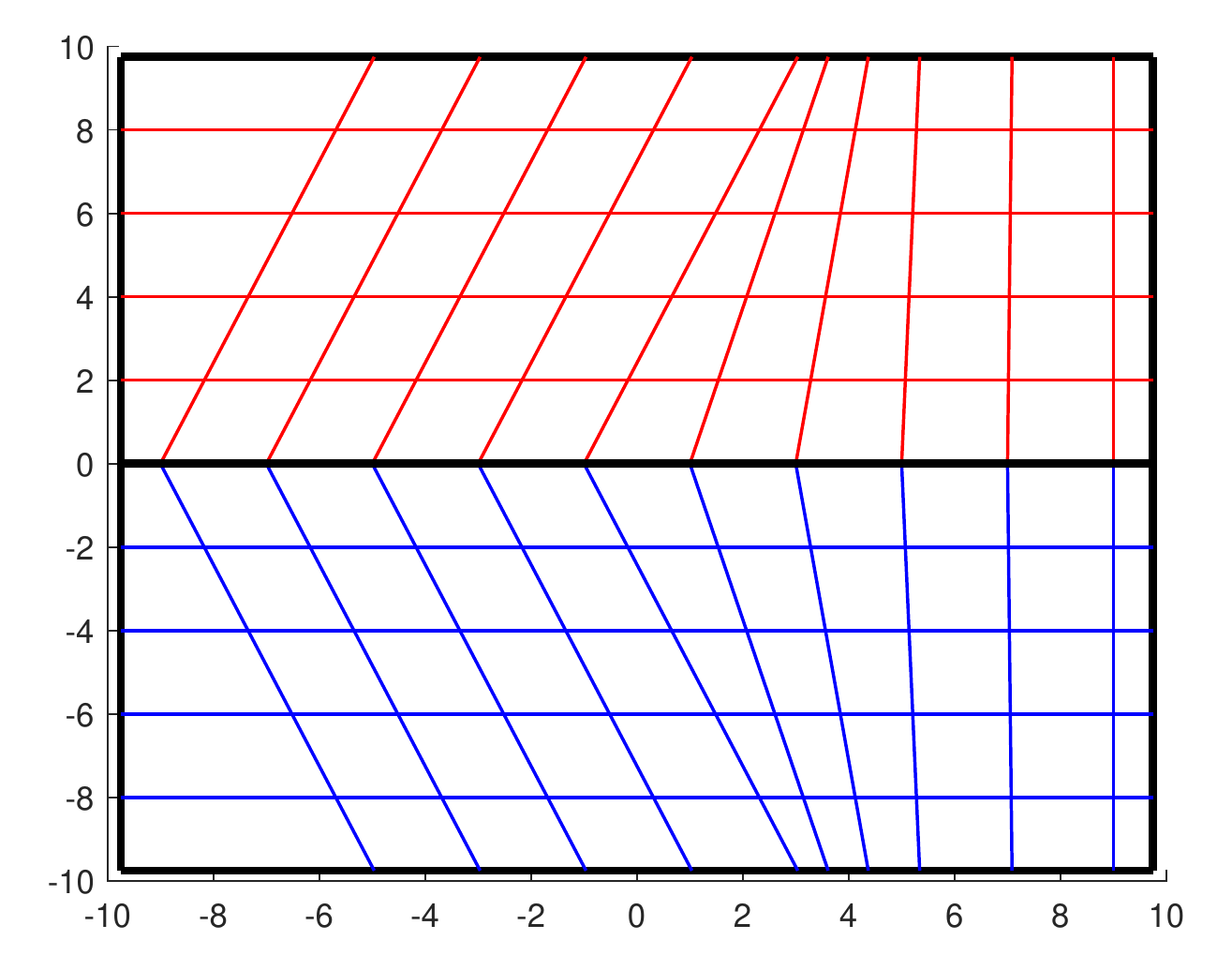}
\label{fig:terminate_twin_deformation_1}
}
\caption{The reference configuration mapped by $\bfW$ and the current configuration of a compatible terminating twin. Since the current configuration has nonzero stress, the reference configuration of the terminating twin is different from the one of the through twin. The displacement at the left bottom and the vertical displacement at the right bottom are fixed to eliminate the rigid motion.}
\label{fig:terminate_twin_deformation}
\end{figure}

\begin{figure}
\centering
\includegraphics[width = 0.6\linewidth]{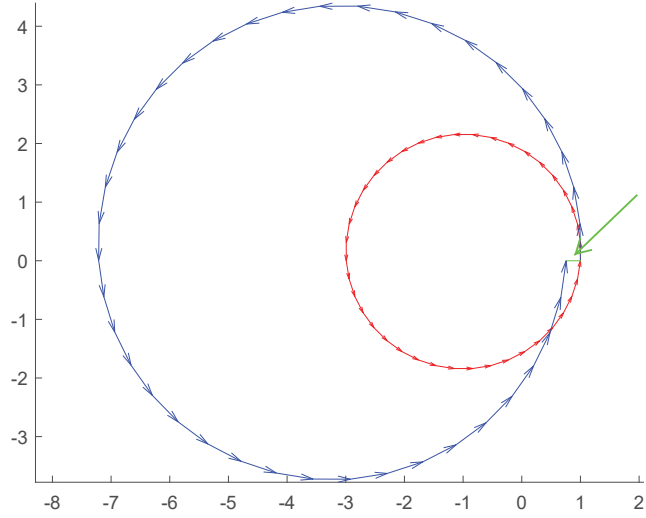}
\caption{Vectors along a circle enclosing the core in the current configuration are mapped by $\hat{\bfW}$. Red arrows are the vectors in the current configuration and blue arrows are the vectors in the reference configuration mapped by $\hat{\bfW}$. The circle in the reference configuration mapped by $\hat{\bfW}$ is compressed, and the green arrow points to the gap representing the incompatibility of $\hat{\bfW}$.}
\label{fig:circle_compressed}
\end{figure}

For the case of the terminating twin discussed in Section \ref{sec:twin_terminate}, the current and elastic reference configurations are shown in Figure \ref{fig:terminate_twin_deformation}. Figure \ref{fig:terminate_twin_deformation_2} is the (compatible) elastic reference configuration obtained by mapping the current configuration in Figure \ref{fig:terminate_twin_deformation_1} by the $\bfW$ field. 
Since the cwi-elastic field $\hat{\bfW}$ is incompatible on the current configuration, Figure \ref{fig:circle_compressed} shows the image of a series of vectors along a circle enclosing the core, mapped by $\hat{\bfW}$. The red arrows correspond to the closed circuit on the current configuration and the blue arrows represent the image of the circuit under $\hat{\bfW}$. Since the body is compressed as discussed in Figure \ref{fig:layer_skew_end_stress}, the blue circle is larger than the red circle. Furthermore, since $\hat{\bfW}$ is incompatible, there is a gap between the start and end of the mapped circuit as shown by the green arrow in Figure \ref{fig:circle_compressed}. However, because of the fact that we are dealing with a (g.)disclination in this case and not a dislocation core, this gap would not be a constant for all loops surrounding the defect core, as can also be mathematically understood by the delocalized nature of the $\bfS^\perp$ field.

\section{Construction of $\bfS$ and $\bfPi$}\label{sec:append_pi}

Recall Figure \ref{fig:disclination_ini} and the definition of the g.disclination density $\bfPi = \curl \bfS$ from (\ref{eqn:Pi}), whose components are given as
\begin{eqnarray*}
(\curl S)_{irm} = e_{mjk}S_{irk,j} = e_{mjk}a_{,j}\left( \frac{W_1-W_2}{t} \right)_{ir}\nu_k + e_{mjk}a\left(\frac{W_1-W_2}{t}\right)_{ir}\nu_{k,j}.
\end{eqnarray*}
Namely,
\begin{equation}\label{eqn:append_pi}
\curl\bfS = \left( \frac{\bfW_1-\bfW_2}{t} \right) \otimes (\grad\, a \times \bfnu) + a \left( \frac{\bfW_1-\bfW_2}{t} \right) \otimes \curl\bfnu.
\end{equation}

To calculate $\curl \,\bfnu$, we first consider $\grad\, \bfnu$ (on the 3-d layer) from (\ref{eqn:append_s}) and (\ref{eqn:normal}) given by
\begin{equation}\label{eqn:append_nablas}
\grad\, \bfnu  = \frac{\partial \bfnu}{\partial \bfs} = \frac{\partial \bfnu}{\partial \xi^1} \otimes \frac{\partial \xi^1}{\partial \bfs} + \frac{\partial \bfnu}{\partial \xi^2} \otimes \frac{\partial \xi^2}{\partial \bfs} + \bf0 \otimes \frac{\partial \xi^3}{\partial \bfs},
\end{equation}
where $\left(\frac{\partial \xi^i}{\partial \bfs}\right)$ is the dual basis corresponding to $\left(\frac{\partial \bfs}{\partial \xi^i}\right)$, and $\frac{\partial \bfs}{\partial \xi^i} \cdot \frac{\partial \xi^j}{\partial \bfs}=\delta_{i}^{j}$. In addition, based on the definition of $\bfs$ in (\ref{eqn:append_s}), we have
\begin{eqnarray*}
\frac{\partial \bfs}{\partial \xi^i} = \frac{\partial \bfx}{\partial \xi^i}+\xi^3\frac{\partial \bfnu}{\partial \xi^i} \qquad i=1,2.
\end{eqnarray*}
Furthermore, since $\bfnu \cdot \bfnu = 1$,
\[
\frac{\partial \bfnu}{\partial \xi^i}\cdot \bfnu = 0 \qquad \text{for $i=1,2$}.
\]
Thus, for $i = 1,2$ and $j = 1,2,3$,
\begin{equation}\label{eqn:app_man}
\begin{split}
&\bfnu \cdot \frac{\partial \bfs}{\partial \xi^i} = \bfnu \cdot \frac{\partial \bfx}{\partial \xi^i} + \xi^3 \frac{\partial \bfnu}{\partial \xi^i} \cdot \bfnu  = 0 \\
&\Rightarrow \frac{\partial}{\partial \xi^j} \left(\bfnu \cdot \frac{\partial \bfs}{\partial \xi^i}\right) = 0 \\
&\Rightarrow \frac{\partial \bfnu}{\partial \xi^j} \cdot \frac{\partial \bfs}{\partial \xi^i} + \frac{\partial \bfs}{\partial \xi^i \partial \xi^j} \cdot \bfnu = 0 \\
& \Rightarrow  \frac{\partial \bfnu}{\partial \xi^j} \cdot \frac{\partial \bfs}{\partial \xi^i} = -\frac{\partial \bfs}{\partial \xi^i \partial \xi^j} \cdot \bfnu
\end{split}
\end{equation}
Since $\frac{\partial \bfs}{\partial \xi^i \partial \xi^j}$ is symmetric with respect to $i$ and $j$, for any $i = 1,2,3$ and any $j= 1,2,3$ we have
\begin{equation}\label{eqn:append_sym1}
\frac{\partial \bfnu}{\partial \xi^j} \cdot \frac{\partial \bfs}{\partial \xi^i} = \frac{\partial \bfnu}{\partial \xi^i}\cdot \frac{\partial \bfs}{\partial \xi^j},
\end{equation}
(which can also be independently checked from (\ref{eqn:app_man})$_4$ for $i = 3$). But this implies that $\grad \, \bfnu$ is a symmetric tensor since $\grad\, \bfnu$ may be expressed as
\[
\grad\, \bfnu  = \left  \{ \frac{\partial \bfs}{\partial \xi^i}\cdot \left( \grad\, \bfnu \frac{\partial \bfs}{\partial \xi^j}\right)\right\} \frac{\partial \xi^i}{\partial \bfs} \otimes \frac{\partial \xi^j}{\partial \bfs} = \left( \frac{\partial \bfnu}{\partial \xi^j} \cdot \frac{\partial \bfs}{\partial \xi^i} \right) \frac{\partial \xi^i}{\partial \bfs} \otimes \frac{\partial \xi^j}{\partial \bfs}.
\]
Since $\curl \bfnu  = -\grad\, \bfnu :\bfX$, we have $\curl \bfnu = \bf0$ (the computation can be done in an orthonormal basis if desired by noting that symmetry of the (covariant or contravariant) components of a tensor is a property that is invariant to choice of basis, whether orthonormal or not). Consequently, (\ref{eqn:append_pi}) implies 
\[
\bfPi = \curl \bfS = \left( \frac{\bfW_1-\bfW_2}{t} \right) \otimes (\grad\, a \times \bfnu).
\]

It is important to note that the $\curl$ senses gradients only in the longitudinal directions of the layer and is insensitive to the (large) gradient in $\bfS$ that exists across the external surfaces of the layer, transverse to the $\xi^3$ direction. To see this, we may assume an extension of the function $\bfnu$ beyond the boundaries of the actual layer along the $\xi^3$ coordinate and assume $a$ to be a smooth function of the form $a(\xi^1, \xi^3) = \hat{a}(\xi^1) b(\xi^3)$ with $b$ such that it goes to $0$ rapidly across the layer boundaries from a constant value of $1$ in the layer. Then $\\curl \bfnu = \bf0$ in the transition layer for $a$ for the same reasons as before, and $\grad \, a \times \bfnu = \frac{\partial a}{\partial \xi^1} \, \frac{\partial \xi^1} {\partial \bfs} \times \bfnu + \frac{\partial a}{\partial \xi^3} \, \frac{\partial \xi^3} {\partial \bfs}\times \bfnu = \frac{\partial a}{\partial \xi^1} \, \frac{\partial \xi^1} {\partial \bfs} \times \bfnu$ since $\frac{\partial \xi^3} {\partial \bfs} = \bfnu$, the large values of $\frac{\partial a}{\partial \xi^3}$ in the transition layer is not sensed by the expression.

Define $D$ as the integral of $\bfPi$ over any area patch threaded by the core, such as the area patch $A$ enclosed by the black dashed line in Figure \ref{fig:disclination_ini}:
\begin{equation}\label{eqn:D}
\bfD := \int_{A} \bfPi d\bfa ,
\end{equation}
where $\hat{\bfn}$ is the unit normal vector of the core surface $A$. After substituting $\bfPi$ in (\ref{eqn:Pi}), we have
\begin{eqnarray*}
\bfD = \int_A \left[ \frac{\bfW_1-\bfW_2}{t}\otimes (\grad\, \bfa \times \bfnu) \right ] d\bfa \\
\bfD = \frac{\bfW_1-\bfW_2}{t} \int_A (\grad\, \bfa \times \bfnu) \cdot  d\bfa
\end{eqnarray*}

With the parametrization in Figure \ref{fig:disclination_ini}, we have 
\begin{eqnarray*}
\grad\, a = \frac{\partial a}{\partial \xi^1} \frac{\partial \xi^1}{\partial \bfs}, \\
d\bfa = \left(\frac{\partial \bfs}{\partial \xi^1} \times \bfnu\right)d\xi^1 d\xi^3,
\end{eqnarray*}
so that
\begin{equation*}
\bfD = \frac{\bfW_1-\bfW_2}{t} \int_{\xi^3=0}^{\xi^3=t} \int_{\xi_1=0}^{\xi^1=c} \frac{\partial a}{\partial \xi^1}\left[ \frac{\partial \xi^1}{\partial \bfs} \times \bfnu \right] \cdot \left[ \frac{\partial \bfs}{\partial \xi^1} \times \bfnu \right] d\xi^1 d\xi^3.
\end{equation*}

Note that 
\begin{eqnarray*}
\left[ \frac{\partial \xi^1}{\partial \bfs} \times \bfnu \right] \cdot \left[ \frac{\partial \bfs}{\partial \xi^1} \times \bfnu \right] = \left[ \left(\frac{\partial \xi^1}{\partial \bfs} \times \bfnu \right) \times \frac{\partial \bfs}{\partial \xi^1}\right ] \cdot \bfnu \\
= \left[ \bfnu \left( \frac{\partial \xi^1}{\partial \bfs} \cdot \frac{\partial \bfs}{\partial \xi^1}\right) - \frac{\partial \xi^1}{\partial \bfs}\left( \bfnu \cdot \frac{\partial \bfs}{\partial \xi^1} \right) \right] \cdot \bfnu.
\end{eqnarray*}
Since $\frac{\partial \xi^1}{\partial \bfs} \cdot \frac{\partial \bfs}{\partial \xi^1}=1$ and $\bfnu \cdot \frac{\partial \bfs}{\partial \xi^1} =0$ as shown in (\ref{eqn:app_man})$_1$, we have
\begin{equation}\label{eqn:temp_d}
\left[ \frac{\partial \xi^1}{\partial \bfs} \times \bfnu \right] \cdot \left[ \frac{\partial \bfs}{\partial \xi^1} \times \bfnu \right] = 1.
\end{equation}
With $a(\xi^1)$ given in (\ref{eqn:a_candidate}), and substituting (\ref{eqn:temp_d}), $\bfD$ can be written as 
\begin{equation}
\bfD = \frac{\bfW_1-\bfW_2}{t} \frac{tc}{c} = \bfW_1 - \bfW_2.
\end{equation}
Thus, we obtain $\bfD = \bfW_1-\bfW_2$ and therefore the equation
\begin{equation*}
\curl \bfY = \bfPi
\end{equation*}
implies that the integral of $\bfY$ along any curve encircling the core is $\bfW_1-\bfW_2$ (by noting that $\divergence \bfPi = \bf0$ and applying the divergence theorem on a `cylinder' with the surface $A$ as one end cap and any arbitrary surface as end-cap with the constraint that its boundary is a curve that encircles the core).

In the case of phase boundaries, the misdistortion $\bfW_1-\bfW_2$ can be written as $\bfc \otimes \bfnu$ representing a shear difference. Then the eigenwall field $\bfS$ takes the form $a (\bfc \otimes \bfnu \otimes \bfnu)$ and therefore 
\begin{equation}\label{eqn:append_curvy}
(curlS)_{irm} = e_{mjk} S_{irk,j} = e_{mjk}a_{,j}\left( \frac{c_i \nu_r}{t} \right)\nu_k + e_{mjk}a\left(\frac{c_i \nu_r}{t}\right)\nu_{k,j} + e_{mjk}a\left(\frac{c_i \nu_k}{t}\right)\nu_{r,j}.
\end{equation}
Based on the same argument to go from (\ref{eqn:append_nablas}) to (\ref{eqn:append_sym1}), the second term is zero and the first term is non-zero only in the core. If the layer where $\bfS$ has support is flat, then the last term vanishes and $\bfPi$ is localized in the core. However, the additional third term is non-vanishing along the layer when the layer is curved, serving as a non-zero source of $\bfPi$ distribution along the whole layer.

\section{Specification of Anisotropic stiffness tensor} \label{sec:append_2}

For the anisotropic stiffness tensor, the elastic constants for the cubic crystal of $Cu$ and $Ag$ are adopted from \cite{simmons1971single}. The stiffness tensor $\bfC$ can be written as 
\[
\bfC = \tilde{C}_{ijkl} \tilde{\bfe}_i \otimes  \tilde{\bfe}_j \otimes  \tilde{\bfe}_k \otimes  \tilde{\bfe}_l,
\]
where $\tilde{\bfe}_i$ is the $i$th principle direction of $\bfC$ and $\tilde{C}_{ijkl}$ is the elastic constant. Denote the transformation from any orthogonal basis $\{\bfe_i\}$ to $\{\tilde{\bfe_i}\}$ as $\bfR$. Namely, the component of $\bfR$, $R_{ij}$, can be define as $\tilde{\bfe}_i\cdot \bfe_j$. Then, we have
\begin{eqnarray*}
\bfe_m \cdot \left(\left(\left(\bfC\bfe_t\right) \bfe_s\right)\bfe_n\right) &=& \tilde{C}_{ijkl}(\tilde{\bfe_i}\cdot\bfe_m)(\tilde{\bfe_j}\cdot\bfe_n)(\tilde{\bfe_k}\cdot\bfe_s)(\tilde{\bfe_l}\cdot\bfe_t) \\
\Rightarrow C_{mnst} &=& \tilde{C}_{ijkl}R_{im}R_{jn}R_{ks}R_{lt}
\end{eqnarray*}

In the incoherent grain boundary disconnection case discussed in Section \ref{sec:anisotropic_grain}, $\bfR$ is a rotation matrix with a rotation angle along $\bfe_3$ axis. The rotation angle of the top part is $22.5^\circ$ and the rotation angle of the bottom part is $-22.5^{\circ}$.

\section{Uniqueness results in linear g.disclination and classical disclination theory}\label{sec:append_uniqueness}

Recall the governing equations \eqref{eqn:small_deformation_summary_hat} of linear g.disclination theory:
\begin{equation}\label{eqn:fb_stress_linear}
\begin{split}
& \curl \, \hat{\bfU}^e = -\bfS^{\perp}:\bfX + \bfalpha\\
& \divergence\, (\bfC:\hat{\bfU}^e)  = \bf0 \\
& (\bfC:\hat{\bfU}^e) \bfn = \bft \qquad \text{on the boundary}.
\end{split}
\end{equation}
We assume that the elasticity tensor $\bfC$ has minor symmetries and is positive definite, possibly anisotropic and spatially inhomogeneous with sufficient smoothness. We also assume the body to be simply-connected.

Now assume there is another solution $\hat{\bfU}^{e'}$ that also satisfies \eqref{eqn:fb_stress_linear}, and define  $\delta \hat{\bfU}^e := \hat{\bfU}^e-\hat{\bfU}^{e'}$. Then, since \eqref{eqn:fb_stress_linear} is linear, we have
\begin{equation}\label{eqn:fb_stress_diff}
\begin{split}
& \curl \, \delta\hat{\bfU}^{e} = \bf0\\
& \divergence\, (\bfC:\delta\hat{\bfU}^{e})  = \bf0 \\
& (\bfC:\delta\hat{\bfU}^{e}) \bfn = \bf0 \qquad \text{on the boundary},
\end{split}
\end{equation}
which implies $\delta \hat{\bfU}^{e}=\bf0$ up to a spatially uniform skew tensor field, this being the standard Neumann proof of linear elasticity, since $\delta\hat{\bfU}^e$ is now a gradient. Thus, $\hat{\bfU}^e = \hat{\bfU}^{e'}$ up to a constant skew tensor and $\hat{\bfU}^e_{sym} = \hat{\bfU}^{e'}_{sym}$. Thus, $\hat{\bfU}^e$ and $\hat{\bfU}^{e'}$ lead to the same stress field, and their skew parts are also essentially uniquely determined up to a constant difference.

On the other hand, the elastic strain of the dislocation and disclination problem in DeWit's model \cite{dewit1973theory} is obtained from 
\begin{equation}\label{eqn:dewit_strain}
\begin{aligned}
\left[ \curl (\bfalpha^T) \right]_{sym} - \bftheta_{sym}  &= inc(\bfepsilon^{e}) \\
\divergence(\bfC:\bfepsilon^e) &= \bf0\\
(\bfC:\bfepsilon^e) \bfn &= \bf0 \qquad \text{on the boundary},
\end{aligned}
\end{equation}
where $\bfepsilon^e$ is the elastic strain, $\bfalpha$ is the dislocation density and $\bftheta$ is the disclination density. $\bfC$ is the stiffness tensor (with the same properties stipulated above). 
Consider another solution $\bfepsilon^{e '}$ that also satisfies \eqref{eqn:dewit_strain}, and denote $\delta \bfepsilon^e=\bfepsilon^{e'}-\bfepsilon^e$. Then we have
\begin{equation*}
\begin{aligned}
inc(\delta \bfepsilon^{e}) & = \bf0\\
\divergence(\bfC:\delta \bfepsilon^e) &= \bf0\\
(\bfC:\delta \bfepsilon^e) \bfn &= \bf0 \qquad \text{on the boundary}.
\end{aligned}
\end{equation*}
The first equation in the above set and the simply-connected body implies that $\delta \bfepsilon^e$ is a symmetrized gradient of a vector field, by St. Venant's compatibility theorem. Then again, this becomes a standard Neumann uniqueness proof in linear elasticity theory and we have $\delta \bfepsilon^e = \bf0$, and thus $\bfepsilon^e_{sym} = \bfepsilon^{e'}_{sym}$. Therefore the stress field from $\bfepsilon^e$ and the stress from $\bfepsilon^{e'}$ are identical, namely the stress field calculated from \eqref{eqn:dewit_strain} is unique.

However, we note one important difference between the two models. For identically specified data, note that g.disclination theory determines the closest-well elastic rotation essentially uniquely, whereas classical disclination theory is completely silent about such determination. This is particularly relevant in the dislocation-only case where there can be no ambiguity in the definition of the elastic rotation in incompatible linear theory.

\newpage
\bibliography{bibtex}
\bibliographystyle{amsalpha}
\end{document}